\RequirePackage[l2tabu]{nag}		
\documentclass[a4paper,11pt,leqno,openbib,oldfontcommands,oneside]{memoir} 

\usepackage{ifpdf}
\ifpdf
\fi
\ifdraftdoc 
	\usepackage{draftwatermark}				
	\SetWatermarkScale{0.3}
	\SetWatermarkText{\bf Draft: \today}
\fi
\newsubfloat{figure}
\newsubfloat{table}
\settrimmedsize{297mm}{210mm}{*}
\settypeblocksize{634pt}{448.13pt}{*} 
\setulmargins{4cm}{*}{*} 
\setlrmargins{*}{*}{1.5} 
\setmarginnotes{17pt}{51pt}{\onelineskip} 
\setheadfoot{\onelineskip}{2\onelineskip} 
\setheaderspaces{*}{2\onelineskip}{*} 
\checkandfixthelayout
\usepackage{fouriernc}
\usepackage[T1]{fontenc}

\OnehalfSpacing 

\setsecnumdepth{subsection} 
\maxsecnumdepth{subsubsection}

\usepackage{calc,soul,fourier}
\makeatletter 
\newlength\dlf@normtxtw 
\newsavebox{\feline@chapter} 
\newcommand\feline@chapter@marker[1][4cm]{%
	\sbox\feline@chapter{%
		\resizebox{!}{#1}{\fboxsep=1pt%
			\colorbox{gray}{\color{white}\thechapter}%
		}}%
		\rotatebox{90}{%
			\resizebox{%
				\heightof{\usebox{\feline@chapter}}+\depthof{\usebox{\feline@chapter}}}%
			{!}{\scshape\so\@chapapp}}\quad%
		\raisebox{\depthof{\usebox{\feline@chapter}}}{\usebox{\feline@chapter}}%
} 
\newcommand\feline@chm[1][4cm]{%
	\sbox\feline@chapter{\feline@chapter@marker[#1]}%
	\makebox[0pt][c]{
		\makebox[1cm][r]{\usebox\feline@chapter}%
	}}
\makechapterstyle{daleifmodif}{

	\renewcommand\printchapternum{\null\hfill\feline@chm[2.5cm]\par}

} 
\makeatother 
\chapterstyle{daleifmodif}

\makepagestyle{myvf} 
\makeoddfoot{myvf}{}{\thepage}{} 
\makeevenfoot{myvf}{}{\thepage}{} 
\makeheadrule{myvf}{\textwidth}{\normalrulethickness} 
\makeevenhead{myvf}{\small\textsc{\leftmark}}{}{} 
\makeoddhead{myvf}{}{}{\small\textsc{\rightmark}}
\newcommand{\clearemptydoublepage}{\newpage{\thispagestyle{empty}\cleardoublepage}}

\makeindex

\usepackage{import}

\usepackage{lipsum}					
\usepackage{amsfonts} 					
\usepackage[centertags]{amsmath}			
\usepackage{stmaryrd}					
\usepackage{amssymb}					
\usepackage{amsthm}					
\usepackage{newlfont}					
\usepackage{layouts}					
\usepackage{graphicx}					
\usepackage{longtable,rotating}			
\usepackage[utf8]{inputenc}			
\usepackage{colortbl}					
\usepackage{wasysym}					
\usepackage{mathrsfs}					
\usepackage{float}						
\usepackage{verbatim}					
\usepackage{upgreek }					
\usepackage{latexsym}					
\usepackage{natbib}		
\setcitestyle{authoryear,open={(},close={)}} 
\usepackage{url}						
\usepackage[spanish,english]{babel}		
\usepackage{color}                    				
\usepackage[dvipsnames]{xcolor}
\definecolor{frenchblue}{rgb}{0.0, 0.45, 0.73}
\definecolor{lapislazuli}{rgb}{0.15, 0.38, 0.61}
\definecolor{mediumpersianblue}{rgb}{0.0, 0.4, 0.65}
\usepackage[colorlinks=true,
		     allcolors=mediumpersianblue]{hyperref}              
\usepackage{memhfixc}					
\usepackage{enumerate}					
\usepackage{footnote}					
\usepackage{microtype}					
\usepackage{rotfloat}					
\usepackage{alltt}						
\usepackage[version=0.96]{pgf}			
\usepackage{tikz}						
\usetikzlibrary{calc}
\usetikzlibrary{patterns}
\usetikzlibrary{arrows,shapes,decorations,
		       automata,backgrounds,
		       petri,topaths}				
\usepackage{epstopdf}
\usepackage{pgfplots}
\pgfplotsset{compat=newest}
\usepackage{longtable}

\newcommand{\pgftextcircled}[1]{                                                                    
    \setbox0=\hbox{#1}%
    \dimen0\wd0%
    \divide\dimen0 by 2%
    \begin{tikzpicture}[baseline=(a.base)]%
        \useasboundingbox (-\the\dimen0,0pt) rectangle (\the\dimen0,1pt);
        \node[circle,draw,outer sep=0pt,inner sep=0.1ex] (a) {#1};
    \end{tikzpicture}
}
\newcommand{\blackged}{\hfill$\blacksquare$}
\newcommand{\whiteged}{\hfill$\square$}
\newcounter{proofcount}

\let\oldsqrt\sqrt
\def\sqrt{\mathpalette\DHLhksqrt}
\def\DHLhksqrt#1#2{%
\setbox0=\hbox{$#1\oldsqrt{#2\,}$}\dimen0=\ht0
\advance\dimen0-0.2\ht0
\setbox2=\hbox{\vrule height\ht0 depth -\dimen0}%
{\box0\lower0.4pt\box2}}
\newcommand{\mycaption}[2][\@empty]{
	\captionnamefont{\scshape} 
	\changecaptionwidth
	\captionwidth{0.9\linewidth}
	\captiondelim{.\:} 
	\indentcaption{0.75cm}
	\captionstyle[\centering]{}
	\setlength{\belowcaptionskip}{10pt}
	\ifx \@empty#1 \caption{#2}\else \caption[#1]{#2}
}
\newcommand{\mysubcaption}[2][\@empty]{
	\subcaptionsize{\small}
	\hangsubcaption
	\subcaptionlabelfont{\rmfamily}
	\sidecapstyle{\raggedright}
	\setlength{\belowcaptionskip}{10pt}
	\ifx \@empty#1 \subcaption{#2}\else \subcaption[#1]{#2}
}
\usepackage{lettrine}
\newcommand{\initial}[1]{%
	\lettrine[lines=3,lhang=0.33,nindent=0em]{
		\color{gray}
     		{\textsc{#1}}}{}}
\theoremstyle{plain}

\theoremstyle{plain}

\theoremstyle{plain}
\theoremstyle{definition}

\theoremstyle{plain}

\theoremstyle{plain}

\theoremstyle{plain}

\newcommand\T{\rule{0pt}{2.6ex}}       
\newcommand\B{\rule[-1.2ex]{0pt}{0pt}}

\def\xmm {\emph{XMM--Newton}}
\def\cxo {\emph{Chandra}}

\def\nustar {\emph{NuSTAR}}
\def\rst {\emph{ROSAT}}
\def\swift {\emph{Swift}}
\def\nicer {\emph{NICER}}

\def\psra{PSR~J0205$+$6449}
\def\psrb{PSR~B2334$+$61}
\def\velajr{CXOU~J085201.4$-$461753}

\newcounter{subsubsubsection}[subsubsection]
\def\subsubsubsectionmark#1{}

\def\subsubsubsection{\@startsection
      {subsubsubsection}{4}{\z@} {-3.25ex plus -1
      ex minus -.2ex}{1.5ex plus .2ex}{\normalsize\bf}}
\def\l@subsubsubsection{\@dottedtocline{4}{4.8em}
      {4.2em}}
\begin{document}
\frontmatter
\pagenumbering{roman}
%
%
%
%
%
%
\begin{titlingpage}
\begin{SingleSpace}
\calccentering{\unitlength} 
\begin{adjustwidth*}{\unitlength}{-\unitlength}
\vspace*{5mm}
\begin{center}
{\large\textsc{Universitat Autònoma de Barcelona}}\\
{\large\textsc{Departament de Física}}\\
\vspace{5mm}
{\includegraphics[width=4cm]{logos/logo_uab.png}}
\vspace{15mm}

{\HUGE Unveiling the Physics of Neutron Stars:}\\[4mm]
{\Large \emph{A 3D expedition into MAgneto-Thermal evolution \\ [2mm] in Isolated Neutron Stars with ``MATINS''}}\\
\vspace{7mm}
{\Large By}\\
\vspace{2mm}
{\Large\textsc{Clara Dehman}}\\

\vspace{1cm}
{ADVISORS:}\\
\vspace{2mm}
{\large\textsc{Prof.~Jos\'e Antonio Pons}}\\
{\large\textsc{Prof.~Nanda Rea}}\\
{\large\textsc{Dr.~Daniele Vigan\`o}}\\
\vspace{7mm}
{TUTOR:}\\
\vspace{2mm}
{\large\textsc{Prof.~Luis Font Guiteras}}

\vspace{2cm}
{\large Astrophysics Department\\
\large Institute of Space Science, ICE-CSIC/IEEC}\\
\vspace{1cm}

{\large{A thesis submitted for the degree of}\\
\emph{Doctor of Philosophy in Physics}}\\
\vspace{1cm}
{{Barcelona\\ November 2023}}
\end{center}
\begin{flushright}
{\small }
\end{flushright}
\end{adjustwidth*}
\end{SingleSpace}
\end{titlingpage}

\clearemptydoublepage
\chapter*{}

\vspace{\fill}

Clara Dehman: \emph{Unveiling the Physics of Neutron Stars: A 3D expedition into MAgneto-Thermal evolution in Isolated Neutron Stars with ``MATINS''}. \\
Copyright $\copyright$ November 2023. All rights reserved. 
\thispagestyle{empty} 
\clearpage
\clearemptydoublepage

\chapter*{Committee Members}
\begin{SingleSpace}

\subsection*{Examining Board Members:}

Prof.~Konstantinos Gourgouliatos, \href{mailto:kngourg@upatras.gr} {\color{lapislazuli}kngourg@upatras.gr}  \\
University of Patras, Greece. \\\\
Prof.~Laura Tolos, \href{mailto:tolos@ice.csic.es}{\color{lapislazuli}tolos@ice.csic.es}  \\
Institute of Space Sciences, Spain. \\\\
Dr.~Michael Gabler, \href{mailto:michael.gabler@uv.es}{\color{lapislazuli}michael.gabler@uv.es}   \\
University of Valencia, Spain.

\subsection*{External Report Authors:}
Prof.~Axel Brandenburg, \href{mailto:brandenb@nordita.org}{\color{lapislazuli}brandenb@nordita.org}  \\
NORDITA -- Nordic Institute for Theoretical Physics, Sweden. \\\\
Prof.~Rosalba Perna, \href{mailto:rosalba.perna@stonybrook.edu}{\color{lapislazuli} rosalba.perna@stonybrook.edu}  \\
Stony Brook University, USA.



\end{SingleSpace}
\clearpage
\clearemptydoublepage

\chapter*{}
\begin{flushleft} 
\hfill \emph{To my parents,\\ \hfill whose unwavering love, sacrifice, and encouragement \\ \hfill have been my guiding light throughout this academic journey.}
\end{flushleft}
\thispagestyle{empty} 
\clearpage
\clearemptydoublepage
\chapter*{Acknowledgements}
\begin{SingleSpace}

\noindent

\initial{T}his thesis work would not have been possible without the assistance of many individuals from various institutions, to them goes my gratefulness. Foremost, I express my appreciation to my advisors, José Pons, Nanda Rea, and Daniele Viganò. Their support and guidance have played a crucial role in helping me develop a scientific methodology. I extend my gratitude to José for his constant presence during times of need. I am also particularly grateful for the opportunity to visit the University of Alicante, which contributed to the advancement of the research detailed in this manuscript. I am thankful to Nanda for lending an ear whenever I had concerns on the progress of my PhD, and for her financial support. I acknowledge Daniele for his attentive supervision and for ensuring timely completion of tasks. I am also grateful for his support when I encountered challenges with paperwork. 

I would like to thank Michael Gabler, Konstantinos Gourgouliatos, and Laura Tolos, for accepting to be members of my PhD thesis committee. I acknowledge Pablo Cerdà, Anthea Fantina, and Cristina Manuel for their willingness to step in as substitutes in case of unforeseen circumstances. I extend my gratitude to Axel Brandenburg, Rosalba Perna, and Grigoris Maravelias for providing favorable assessments on my PhD thesis. I would also like to thank Axel for enabling multiple visits at Nordita during my PhD in the context of the visiting PhD fellow program. These visits significantly contributed to my understanding of physics of turbulence and Magnetohydrodynamic.

I would like to acknowledge the financial support provided by the ERC Consolidator Grant 'MAGNESIA' (No. 817661, PI: N.\,Rea), as well as the partial funding from Nordita. I'm also grateful to the Institute of Space Sciences in Barcelona and the Universitat Autònoma de Barcelona for giving me the opportunity to pursue my PhD studies. My appreciation extends to the research centers that invited me to share my work through seminars, including the Flatiron Institute, the University of Barcelona, the University of Alicante, the Ganil facility, the University of Goethe, Nordita, and the University of Valencia.

To my colleagues and friends, who made this journey more pleasant: Abubakr, Kostas, Michele, Rajath, Vanessa, Albert, Simran, Claudia, Daniel, Charlie, Tuner, Alice, Celsa, Alessio, Stefano, Francesco, Emilie, and Christine, thank you. More friends I have met at the University of Alicante, including Petros, Jorge, and Pantelis, and at Nordita, including Gustav, Ramkishor, Protiti, Salome, Ludovico, Yutong, and Nikhil. Equally, I am thankful to Betsy, Harini, Janny, Anastasia, and Rosa for all the fun evenings. Barcelona has been the perfect place to be in the last four years thanks to its enchanting atmosphere and people. To my childhood friends, Carol, Daed, Mozaya, Reem, and Nourhane, whose support has been a source of strength.

Most importantly, to my family members: Nawal, Youssef, Michelangelo, Marlon, Jennifer, Peter, Chris, Stefania, Mario, Nonno Tuccio, Loris, and Chicho, your unwavering love, support, and belief in me have been a source of strength and motivation. I dedicate this work to you all, the dearest people to me, acknowledging your profound contributions to my academic journey.

\end{SingleSpace}
\clearpage
\clearemptydoublepage
%
%
%
%

\chapter*{Journal Articles Related to this Thesis.}
\begin{SingleSpace}
\initial{T}he present manuscript summarizes and contextualizes with more details the results obtained in the bulk of these papers. Please note that all rights are reserved. A subset of these articles has been included in this thesis, incorporating slight modifications in accordance with the author's retained rights.

\begin{enumerate}

\subsection*{Articles in Theoretical \& Computational Astrophysics:}

\item \underline{C.~Dehman}, D.~Vigan\`o, N.~Rea, J.A.~Pons, R.~Perna \& A.~Garcia-Garcia: $2020$, \small{\textbf{On The Rate of Crustal Failures in Young Magnetars}}, \emph{Astrophys.~J.~L., $902$ L$32$}
(\href{https://arxiv.org/abs/2010.00617}{\underline{arXiv:2010.00617}},\href{https://ui.adsabs.harvard.edu/abs/2020ApJ...902L..32D/abstract}{\underline{ADS}},\href{https://iopscience.iop.org/article/10.3847/2041-8213/abbda9}{\underline{DOI}}).

\item \underline{C.~Dehman}, D.~Vigan\`o, J.A.~Pons \& N.~Rea: 2022, \small{\textbf{3D code for MAgneto$-$Thermal evolution in Isolated Neutron Stars, MATINS: The Magnetic Field Formalism}}, \emph{Mon.~Not.~Roy.~Astron.~Soc., $518$, $1222$} 
(\href{https://arxiv.org/abs/2209.12920}{\underline{arXiv:2209.12920}},\href{https://ui.adsabs.harvard.edu/abs/2023MNRAS.518.1222D/abstract}{\underline{ADS}},\href{https://doi.org/10.1093/mnras/stac2761}{\underline{DOI}}).

\item \underline{C.~Dehman}, J.A.~Pons, D.~Vigan\`o \& N.~Rea: 2023, \small{\textbf{How bright can old magnetars be? Assessing the impact of magnetized envelopes and field topology on neutron star cooling}}, \emph{Mon.~Not.~Roy.~Astron.~Soc.~L., $520$, $42$} (\href{https://arxiv.org/abs/2301.02261}{\underline{arXiv:2301.02261}},\href{https://ui.adsabs.harvard.edu/abs/2023MNRAS.520L..42D/abstract}{\underline{ADS}},\href{https://academic.oup.com/mnrasl/advance-article/doi/10.1093/mnrasl/slad003/6973208?utm_source=authortollfreelink&utm_campaign=mnrasl&utm_medium=email&guestAccessKey=b4196a19-bd30-4b1e-acf5-de9a7c3c9b3c}{\underline{DOI}}).

\item \underline{C.~Dehman}, D.~Vigan\`o, S. Ascenzi, J.A.~Pons \&  N. Rea: 2023, \small{\textbf{3D evolution of neutron star magnetic-fields from a realistic core-collapse turbulent topology}}, 
\emph{Mon.~Not.~Roy.~Astron.~Soc. $523$, $5198$} (\href{https://arxiv.org/abs/2305.06342}{\underline{arXiv:2305.06342}},\href{https://ui.adsabs.harvard.edu/abs/2023arXiv230506342D/abstract}{\underline{ADS}},\href{https://doi.org/10.1093/mnras/stad1773}{\underline{DOI}}).

\item F.~Anzuini, A.~Melatos, \underline{C.~Dehman}, D.~Vigan\`o \& J.A.~Pons: 2022, \small{\textbf{Fast cooling and internal heating in hyperon stars}}, \emph{Mon.~Not.~Roy.~Astron.~Soc., $509$, $2609$}
(\href{https://arxiv.org/abs/2110.14039}{\underline{arXiv:2110.14039}},\href{https://ui.adsabs.harvard.edu/abs/2022MNRAS.509.2609A/abstract}{\underline{ADS}},\href{https://doi.org/10.1093/mnras/stab3126}{\underline{DOI}}).

\item F.~Anzuini, A.~Melatos, \underline{C.~Dehman}, D.~Vigan\`o \& J.A.~Pons: 2022, \small{\textbf{Thermal luminosity degeneracy of magnetized neutron stars with and without hyperon cores}}, \emph{Mon.~Not.~Roy.~Astron.~Soc., $515$, $3014$}
(\href{https://arxiv.org/abs/2205.14793}{\underline{arXiv:2205.14793}},\href{https://ui.adsabs.harvard.edu/abs/2022MNRAS.515.3014A/abstract}{\underline{ADS}},\href{https://doi.org/10.1093/mnras/stac1353}{\underline{DOI}}).

\item J.~Urb\`an, P.~Stefanou, \underline{C.~Dehman} \& J.A.~Pons: 2023, \small{\textbf{Modelling Force-Free Neutron Star Magnetospheres using Physics-Informed Neural Networks}}, \emph{Mon.~Not.~Roy.~Astron.~Soc., $524$, $32$} (\href{https://arxiv.org/abs/2303.11968}{\underline{arXiv:2303.11968}},\href{https://ui.adsabs.harvard.edu/abs/2023arXiv230311968U/abstract}{\underline{ADS}},\href{https://doi.org/10.1093/mnras/stad1810}{\underline{DOI}}).

\item D.~Vigan\`o, A.~Garcia-Garcia, J.A.~Pons, \underline{C.~Dehman} \& V.~Graber: 2021,
\small{\textbf{Magneto-Thermal Evolution of Neutron Stars With Coupled Ohmic, Hall and Ambipolar Effects Via Accurate Finite-Volume Simulations}}, \emph{Comp.~Phys.~Comm., $265$, $108001$}
(\href{https://arxiv.org/abs/2104.08001}{\underline{arXiv:2104.08001}},\href{https://ui.adsabs.harvard.edu/abs/2021CoPhC.26508001V/abstract}{\underline{ADS}},\href{https://www.sciencedirect.com/science/article/pii/S0010465521001132}{\underline{DOI}})

\item S.~Ascenzi, D.~Vigan\`o, \underline{C.~Dehman}, J.A.~Pons, N.~Rea, R.~Perna, \small{\textbf{3D code for MAgneto\,--Thermal evolution in Isolated Neutron Stars, MATINS: thermal evolution and lightcurves}}, under peer review in \emph{MNRAS }(\href{https://arxiv.org/abs/2401.15711}{\underline{arXiv:2401.15711}},\href{https://ui.adsabs.harvard.edu/abs/2024arXiv240115711A/abstract}{\underline{ADS}})

\subsection*{Articles in Observational Astrophysics:}
\item A.~Marino$^*$, \underline{C.~Dehman}$^*$, K.~Kovlakas$^*$, N.~Rea$^*$ et al., \emph{\color{darkgray}``These authors contributed equally to this work''.} \small{\textbf{Constraints on the dense matter equation of state from young and cold isolated neutron stars}}, accepted for publication in \emph{Nature Astronomy }(\href{https://arxiv.org/abs/2404.05371}{\underline{arXiv:2404.05371}},\href{https://ui.adsabs.harvard.edu/abs/2024arXiv240405371M/abstract}{\underline{ADS}})

\item N.~Rea, F.~Coti Zelati, \underline{C.~Dehman} et al.: 2022,
\small{\textbf{Constraining the nature of the 18 min periodic radio transient GLEAM-X J162759.5-523504.3 via multi-wavelength observations and magneto-thermal simulations}}, \emph{Astrophys.~J., $940$, $72$} 
(\href{https://arxiv.org/abs/2210.01903}{\underline{arXiv:2210.01903}},\href{https://ui.adsabs.harvard.edu/abs/2022ApJ...940...72R/abstract}{\underline{ADS}},\href{https://iopscience.iop.org/article/10.3847/1538-4357/ac97ea}{\underline{DOI}}).

\item P.~Esposito, N.~Rea, A.~Borghese et al.: $2020$, \small{\textbf{A Very Young Radio-loud Magnetar}}, \emph{Astrophys.~J.~L., $896$ L$30$}
(\href{https://arxiv.org/abs/2004.04083}{\underline{arXiv:2004.04083}},\href{https://ui.adsabs.harvard.edu/abs/2020ApJ...896L..30E/abstract}{\underline{ADS}},\href{https://iopscience.iop.org/article/10.3847/2041-8213/ab9742}{\underline{DOI}}).

\item F.~Coti Zelati, A.~Borghese, G.L.~Israel et al.: $2021$, \small{\textbf{The New Magnetar SGR J1830-0645 in Outburst}}, \emph{Astrophys.~J.~L., $907$, L$34$}
(\href{https://arxiv.org/abs/2011.08653}{\underline{arXiv:2011.08653}},\href{https://ui.adsabs.harvard.edu/abs/2021ApJ...907L..34C/abstract}{\underline{ADS}},\href{https://iopscience.iop.org/article/10.3847/2041-8213/abda52}{\underline{DOI}}).

\item A.Y.~Ibrahim, A.~Borghese, N.~Rea et al.: 2023, \small{\textbf{Deep X-ray and radio observations of the first outburst of the young magnetar Swift J1818.0-1607}}, \emph{Astrophys.~J., $943$, $20$} (\href{https://arxiv.org/abs/2211.12391}{\underline{arXiv:2211.12391}},\href{https://ui.adsabs.harvard.edu/abs/2023ApJ...943...20I/abstract}{\underline{ADS}},\\\href{https://iopscience.iop.org/article/10.3847/1538-4357/aca528}{\underline{DOI}}).


\subsection*{Articles in Nuclear Physics:}
\item \underline{C.~Dehman}, M.~Centelles, X.~Vi\~nas, \small{\textbf{Impact of Hot Inner Crust on Compact Stars at Finite Temperature}}, under peer review in \emph{A\&A }(\href{https://arxiv.org/abs/2401.16957}{\underline{arXiv:2401.16957}},\href{https://ui.adsabs.harvard.edu/abs/2024arXiv240116957D/abstract}{\underline{ADS}})

\item A.M.~Stefanini, G.~Montagnoli, M.~D'Andrea, M.~Giacomin: $2021$, \underline{C.~Dehman} et al., \small{\textbf{New Insights into Sub-Barrier Fusion of $^{28}$Si $+$ $^{100}$Mo}}, \emph{J.~of Phys.~G, $48$, $055101$}
(\href{https://ui.adsabs.harvard.edu/abs/2021JPhG...48e5101S/abstract}{\underline{ADS}},\href{https://iopscience.iop.org/article/10.1088/1361-6471/abe8e2}{\underline{DOI}}).

\subsection*{Upcoming Publicly Accessible GitHub Repository for The \emph{MATINS} Code:}

\begin{center}
\large\href{https://github.com/csic-ice-magnesia/MATINS}{https://github.com/csic-ice-magnesia/MATINS.}
\end{center}

\end{enumerate}

\end{SingleSpace}
\clearpage
\clearemptydoublepage
\renewcommand{\contentsname}{Table of Contents}
\maxtocdepth{subsection}
\tableofcontents*
\addtocontents{toc}{\par\nobreak \mbox{}\hfill{\bf Page}\par\nobreak}
\clearemptydoublepage
\mainmatter
\chapter{Abstract}
\begin{SingleSpace}
\initial{T}his doctoral thesis undertakes a comprehensive investigation of the long-term evolution of the internal, strong magnetic fields found within isolated neutron stars. These astronomical entities stand as the most potent magnetic objects in the universe. Within the context of neutron stars, their magnetic influence extends beyond their surface, encompassing the magnetized plasma in their vicinity, referred to as the magnetosphere. This overarching magnetic configuration significantly impacts the observable characteristics of the highly magnetized neutron stars, commonly known as magnetars. Conversely, the magnetic fields within their interiors undergo prolonged evolution spanning from thousands to millions of years. This magnetic evolution is intrinsically linked to the concurrent thermal evolution. The diverse range of observable phenomena associated with neutron star underscores the complex and three-dimensional nature of their magnetic topology, thereby requiring sophisticated numerical simulations. These simulations are crucial not only for explaining observed burst mechanisms and the creation of surface hotspots, but also for comprehending the intricate interplay between magnetic fields and temperature under the extreme conditions inherent in these cosmic objects.

A central focus of this thesis involves a thorough exploration of state-of-the-art three-dimensional coupled magneto-thermal evolution models. This marks a pioneering achievement as we conduct, for the first time, the most realistic three-dimensional simulation to date, spanning the first million years of a neutron star's life using the newly developed code \emph{MATINS}, which adeptly accounts for both Ohmic dissipation and Hall drift within the neutron star's crust. Our simulations incorporate highly accurate temperature-dependent microphysical calculations and adopt the star's structure based on a realistic equation of state. To address axial singularities in three-dimensional simulations, we employ cubed-sphere coordinates. We also account for corresponding relativistic factors in the evolution equations and utilize a state-of-the-art envelope model from existing literature, in addition to an initial magnetic field topology derived from proto-neutron star dynamo simulations. Within this framework, we quantitatively simulate the thermal luminosity, timing properties, and magnetic field evolution, pushing the boundaries of numerical modeling capabilities. 

The structural framework of this thesis reflects the unfolding of a comprehensive astrophysical narrative. It commences with Chapter~\ref{chap: neutron stars}, delving into the origin and formation of neutron stars while introducing pivotal attributes such as the equation of state, mechanical structure, magnetic fields, and observational properties. This thematic progression then transitions to Chapter~\ref{chap: magneto-thermal evolution}, where the concept of magneto-thermal evolution is elaborated upon, with a keen focus on the microscopic coupling of magnetic and thermal evolution. Chapter~\ref{chap: envelope}, which builds upon the research conducted by \cite{dehman2023b}, expands the scope to consider magnetized envelope models with varying compositions, interrogating the role of magnetic field topology in shaping magnetar luminosity. Subsequently, Chapter~\ref{chap: Comparison with observations} facilitates a comparative analysis between theoretical simulations and empirical observations. It places particular emphasis on understanding the behavior of cold and young observed pulsars, thus contributing to the imposition of constraints on the nuclear equation of state and allowing the exclusion of approximately 75\% of them. This chapter is based on the work by Dehman et al. (2023, in preparation) and \cite{MDKR2023}. The subsequent thematic chapter, Chapter~\ref{chap: outburst}, which draws from the work by \cite{dehman2020}, undertakes a comprehensive exploration of magnetic-driven activities and the frequency of crustal failures in newly born magnetars. These investigations are conducted within a 2D axisymmetric framework.

The narrative subsequently shifts to Chapter~\ref{chap: MATINS}, where the innovative three-dimensional numerical code \emph{MATINS} is introduced. This code is purposefully designed for MAgneto-Thermal Evolution in Isolated Neutron Stars \citep{dehman2022}. Within this chapter, we conduct an in-depth exploration of the cubed-sphere technique and the formalism associated with magnetic fields within this framework. Moving on to Chapter~\ref{chap: 3DMT}, we present the results of the first three-dimensional coupled magneto-thermal simulations, incorporating a highly realistic initial magnetic field topology inspired by a global model of the magneto-rotational instability in a proto-neutron star. These last two chapters, Chapter~\ref{chap: MATINS} and Chapter~\ref{chap: 3DMT}, are built upon the work by \cite{dehman2022,dehman2023c}. Lastly, Chapter~\ref{chap: conclusions} serves as the conclusion of this thesis, summarizing the main findings and insights garnered throughout the course of this comprehensive study.

\end{SingleSpace}
\clearpage
\clearemptydoublepage
\let\textcircled=\pgftextcircled
\chapter{Neutron Stars}
\label{chap: neutron stars}

\initial{T}he discovery of neutron stars stands as a monumental milestone in our understanding of the cosmos. It was during the early $20^{th}$ century that astronomers began to unravel the secrets hidden within the depths of the universe. One such revelation occurred in 1932 when James Chadwick, a British physicist, made a groundbreaking discovery that would revolutionize our knowledge of stellar remnants. Using experimental techniques, Chadwick successfully identified the presence of an uncharged subatomic particle, which he aptly named the neutron \citep{chadwick1932}. This newfound particle, with its neutral charge, offered a missing link in the understanding of atomic structure.

As the scientific community delved deeper into the mysteries of the universe, they soon realized the potential existence of incredibly dense stellar remnants known as neutron stars. The idea of these celestial objects originated from the work of Austrian astrophysicist Fritz Zwicky and American physicist Walter Baade in the 1930s \citep{baade1934}. Their calculations revealed that the remnants of massive stars, following a supernova explosion, could collapse under gravity's immense force and become incredibly dense. 
It's worth highlighting that well before the discovery of neutrons, physicist Lev Landau had already envisaged the preliminary concept of a giant nucleus star \citep{yakovlev2013}. 
The 1930s witnessed the emergence of further scholarly inquiries exploring the possibility of normal stars with a degenerate core \citep{chandrasekhar1935}. Particularly noteworthy is the work of \cite{oppenheimer1939}, who formulated the equation of state (EoS) for a relativistic neutron gas, laying the essential groundwork for subsequent explorations in this field.

It was not until 1967, however, that the first concrete evidence of a neutron star emerged. Jocelyn Bell Burnell, a British astrophysicist, and her advisor Antony Hewish were conducting observations using a radio telescope when they stumbled upon an unusual signal. The signal, which repeated every $1.337$\,s, puzzled the researchers. After ruling out any terrestrial interference, they considered the possibility of an extraterrestrial source. Burnell and Hewish had unwittingly discovered the first pulsar, a rapidly rotating neutron star emitting beams of electromagnetic radiation as it spins \citep{bell1968}. In 1968, Gold proposed that the pulsations were generated by rapidly spinning neutron stars instead of pulsating white dwarfs \citep{gold1968}. This hypothesis was quickly validated by the detection of a pulsar in the Crab supernova remnant \citep{staelin1968}.

Since Bell's fortuitous discovery, nearly 3389 pulsars have been observed in our Galaxy \citep{manchester2005}\footnote{The ATNF Pulsar Catalogue \url{http://csiro.au/}.}. This remarkable finding set the stage for a new era of astrophysics. Neutron stars, with their extreme densities and intense magnetic fields, became an object of intense study. Scientists employed a wide array of observational tools, from radio telescopes to X-ray and $\gamma$-ray detectors, to unravel the intricacies of these exotic objects. The subsequent discoveries of different phenomenological neutron star categories besides classical pulsars, further solidified our understanding of their properties and provided insights into fundamental physics.

\section{Origin and stellar evolution}
\label{sec: stellar evolution}

Neutron stars defy our understanding of matter and gravity, emerging from the collapse of massive stars. These remnants pack immense mass into a small volume, resulting in mind-boggling densities. With intense magnetic fields, rapid rotation, and powerful radiation beams, neutron stars offer insights into fundamental physics and have implications for astrophysics and gravitational wave astronomy. 

Neutron stars owe their existence to supernova explosions, which are spectacular events marking the explosive end of a massive star's life, typically those with initial masses ranging from $8$ to $20$ times the mass of the Sun. These massive stars form through the gravitational collapse of molecular clouds, where gravity pulls together gas and dust into a dense core. After the formation phase, massive stars enter the main sequence, where they spend the majority of their lives. During this phase, nuclear fusion occurs in the star's core, converting hydrogen into helium through the proton-proton chain or the Carbon-Nitrogen-Oxygen cycle. The energy released from fusion generates an outward pressure that balances the inward gravitational force, maintaining the star's equilibrium. As the massive star consumes its nuclear fuel, it undergoes a series of nuclear reactions to maintain its equilibrium. It burns helium into heavier elements such as carbon, oxygen, and neon through various fusion processes. This evolution continues, with successive shells of the star fusing heavier elements until reaching iron.

When the star's core exhausts its nuclear fuel and reaches iron, a critical point is reached. Iron cannot undergo fusion to release energy. As a result, the core loses its ability to generate an outward pressure that counters gravity's inward pull. The core collapses under its own weight in a fraction of a second. The core collapse triggers a cataclysmic event known as a supernova. The collapsing core releases an enormous amount of gravitational potential energy, leading to a powerful rebound and shock-wave that propagates outward. The shock-wave disrupts the star's outer layers, ejecting them into space in a brilliant explosion. This explosive event disperses heavy elements and enriches the surrounding interstellar medium.

During the core collapse, the density and pressure reach extreme values. Protons and electrons are squeezed together, merging to form neutrons. This process, known as neutronization, occurs when the density surpasses nuclear saturation density, around $2.8 \times 10^{14}$\,g\,cm$^{-3}$. 
The highly dynamical process consists of three main stages, as follows. (i) For about one second after the core bounce, the system consists of a relatively cool central region surrounded by a hot mantle, collapsing and radiating off neutrinos quickly, while still also accreting material. (ii) Over the next $\sim 20$\,s, a slowly developing state of the proto-neutron star can be identified; the system first deleptonizes and heats up the interior parts of the forming star, later it cools down further through abundant production of thermal neutrinos that abandon the neutron star core draining energy from the interior. The proto-neutron star is born extremely hot and liquid ($T \approx 10^{10}-10^{11}$\,K), with a relatively large radius of $\sim 100$\,km. (iii) After several minutes, it becomes transparent to neutrinos and shrinks to its final radius \citep{burrows1986,keil1995,pons1999}, marking the birth of a neutron star. The temperature drops below the melting point of a layer where matter begins to crystallize, forming the crust. The melting temperature depends on the local value of the density, therefore the growth of the crust is gradual and it lasts from hours to months after birth. The outermost layer of a neutron star, i.e., the envelope, remains liquid and sometimes is called the ocean. The neutron star starts its long-term cooling, dominated first by neutrinos and later ($t\gtrsim 10^5$\,yr) by photon emission from the surface.

The density of neutron stars is so staggering that a teaspoon of their material would weigh millions of tons, far surpassing any substance found on Earth. This density is a result of the fundamental physics that governs the behavior of matter under extreme conditions. The newly formed neutron star typically has a mass ranging from about $1.3$ to $2.2$ times the mass of the Sun. Its size is significantly small, with a typical radius of around $10$ to $13$\,km. Neutron stars are in a state known as neutron degeneracy, in which the neutron degeneracy pressure, related to the Pauli principle, counteracts the gravity. This is the last option that dense matter has to prevent a collapse to a black hole, and happens only in neutron stars. As a matter of fact, in other astrophysical bodies like giant planets and white dwarf the electron degeneracy pressure is enough to counteract the gravity. Neutron star's gravitational pull is immensely strong, distorting space-time and bending light around it.

\section{Equation of state}
\label{sec: EOS}

The EoS is crucial for understanding the properties and behavior of neutron stars. It represents the relationship between pressure, density, and composition of matter within these incredibly dense objects, where densities can reach several times that of atomic nuclei. Determining the EoS is essential for understanding the internal structure, mass-radius relationship, and overall behavior of neutron stars. It also influences various astrophysical phenomena such as stability, cooling rates, rotational properties, and gravitational wave signals emitted during mergers with other neutron stars or black holes.

The precise form of the EoS is still a subject of active research and theoretical modeling. However, several theoretical approaches and computational techniques are employed to construct the EoS of neutron stars. These include relativistic mean-field models, quantum chromodynamics calculations, nuclear many-body theories, and astrophysical observations that provide constraints on the properties of neutron stars. By comparing theoretical models with observational data, such as mass and radius measurements of neutron stars, scientists strive to refine and validate the EoS. For a comprehensive review of the EoS of neutron stars, we recommend referring to \cite{burgio2020}.

The EoS depends on various factors, including the interactions between neutrons, protons, electrons, and possibly other exotic particles. At lower densities, the EoS can be described by the physics of dense nuclear matter. The interactions between nucleons (neutrons and protons) are typically described by effective nuclear forces, such as the Skyrme or relativistic mean field models. These models take into account the strong nuclear force that binds nucleons together and the Pauli exclusion principle that limits the number of particles in a given quantum state. The EoS also depends on the presence of electrons. In the outer layers of neutron stars, electrons contribute significantly to the pressure through degeneracy pressure. At even higher densities, electrons may combine with protons to form neutrons and neutrinos through the process of $\beta$-decay, leading to a transition to a neutron-rich matter.

As the density increases, the nature of matter within neutron stars becomes more uncertain. It is believed that at higher densities, exotic particles, such as hyperons (baryons containing strange quarks), meson condensates, or quark matter, may appear. The behavior of these particles and their interactions with nucleons are still areas of active investigation.

Neutron stars are composed of degenerate matter, characterized by temperatures below the Fermi temperature throughout their entire existence. This specific temperature range emphasizes the supremacy of quantum effects over thermal effects, as governed by Fermi statistics. Consequently, the EoS for neutron stars can be approximated as that of zero temperature, effectively disregarding thermal contributions for the majority of their lifespan. 

\section{Tolman-Oppenheimer-Volkoff equations}
\label{sec: TOV}

A neutron star is born hot at a temperature of approximately $10^{10}-10^{11}$\,K and contains a significant number of leptons. However, within a few days of its birth, the temperature decreases to a few $10^9$\,K. As a result, the Fermi energy ($\varepsilon_F$) of all particles in the star is much greater than the kinetic thermal energy in the majority of its volume, except for the thin outermost layers, which are only a few meters thick. The dominant contribution to the pressure in the star is therefore provided by degeneracy pressure, while the thermal and magnetic contributions to the pressure are negligible in most parts of the star. Consequently, it is a reasonable approximation to describe the state of matter as cold nuclear matter in $\beta$-equilibrium, resulting in an effectively barotropic EoS ($P = P(\rho)$) that is used to calculate the underlying mechanical structure. The radial profiles describing the energy-mass density and chemical composition can be calculated once and then held constant as a background model during the simulations of magneto-thermal evolution.

To a very good approximation, we can assume that the mechanical structure of neutron stars is spherical. However, there may be significant deviations from spherical symmetry under certain conditions, such as the presence of ultra-strong magnetic fields (with strengths exceeding $10^{17}$\,G) or extremely rapid rotation (with periods shorter than a few milliseconds). However, our study does not consider such extreme cases. Neutron stars are relativistic objects in which the effects of general relativity play a crucial role. The general relativistic corrections are approximately 20\% at the surface, corresponding to a redshift correction factor $\zeta$, and they decrease inversely with distance. The structure of these stars is governed by the Tolman-Oppenheimer-Volkoff (TOV) equations \citep{oppenheimer1939}, which describe hydrostatic equilibrium under the assumption of a static interior Schwarzschild metric.
\begin{equation}
    ds^2= -c^2 e^{2 \zeta(r)} dt^2 + e^{2 \lambda(r)} dr^2 + r^2 d\Omega^2, 
    \label{eq: standard static metric}
\end{equation}
where $\lambda(r) = - \frac{1}{2} ln\big[1- \frac{2 G}{c^2}\frac{m(r)}{r^2} \big]$ accounts for the space-time curvature, where 
\begin{equation}
    m(r) = 4\pi \int_0^r \rho(\Tilde{r})\Tilde{r}^2 d\Tilde{r}\,,
    \label{eq: gravitational mass}
\end{equation}
is the enclosed gravitational mass within radius $r$, $\rho$ is the mass-energy density, $G$ is the gravitational constant, and $c$ the speed of light. The other relativistic factor that appear in eq.\,\eqref{eq: standard static metric} is the lapse function $e^{2\zeta(r)}$, where $\zeta$ represents the dimensionless gravitational potential. This function accounts for redshift corrections and is determined by the equation
\begin{equation}
\frac{d \zeta (r) }{d r } = \frac{G }{c^2}  \frac{m(r)}{r^2} \Bigg(1+ \frac{4 \pi r^3 P}{C^2 m(r)} \Bigg) \Bigg(1- \frac{2G}{c^2} \frac{m(r)}{r} \Bigg)^{-1},
  \label{eq: lapse function}  
\end{equation}
with the boundary condition $e^{2 \zeta(R)} = 1- 2GM/c^2 R$ at the stellar radius $r=R$ and $M=m(R)$ is the total gravitational mass of the star. The flat space-time is recovered by the limit $\zeta(r) \rightarrow 0$, $\lambda(r) \rightarrow 0$, i.e., enough far away from the compact object.

The mechanical structure of a spherically symmetric star is described by the Tolman-Oppenheimer-Volkoff equation
\begin{equation}
\frac{dP(r)}{dr} = - \Bigg( \rho + \frac{P}{c^2} \Bigg) \frac{d \zeta(r)}{dr},
    \label{eq: pressure TOV}
\end{equation}
where $P(r)$ is the pressure profile. To close the system of equations, it is necessary to supplement eq.\,\eqref{eq: pressure TOV} with the EoS, discussed above.

\section{Structure}
\label{sec: structure}

\begin{figure}
    \centering
\includegraphics[width=.95\textwidth]{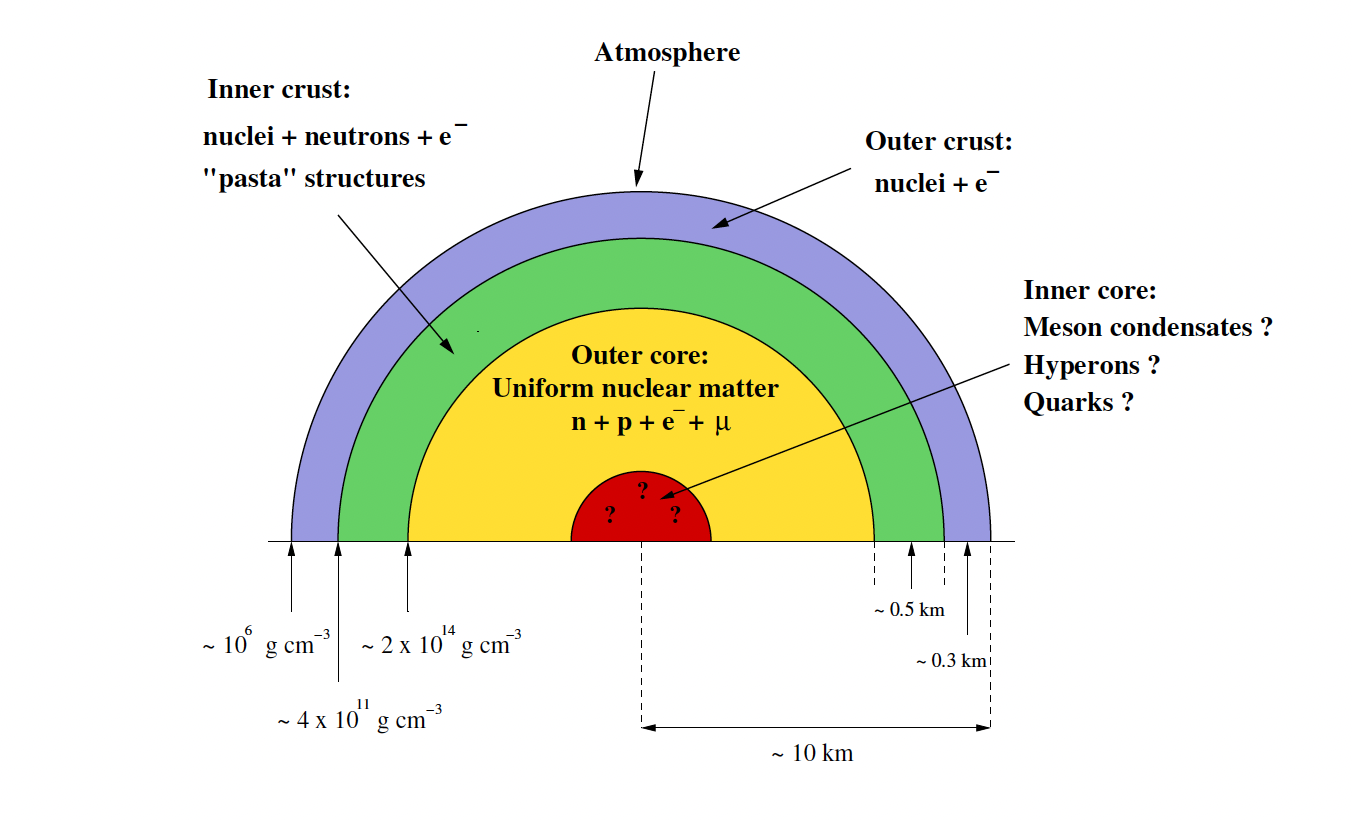}
    \caption[Neutron star structure.]{Figure adopted from \cite{burgio2020}. A schematic cross section of a neutron star illustrating the various regions discussed in the text. The different regions shown are not drawn on scale.}
    \label{fig: NS composition}
\end{figure}

The overall structure of a neutron star is determined by selecting an EoS and solving the TOV equations. These calculations reveal that the density undergoes significant variations across several orders of magnitude from the center to the outermost region of the neutron star, resulting in a highly stratified matter distribution. Beneath a thin stellar atmosphere or a condensed surface, the interior of a neutron star can be divided into three main regions: the core, the crust, and the ocean (envelope). Each region exhibits distinct physical properties (e.g., \cite{haensel2007}). A schematic representation of a neutron star's structure is provided in Fig.~\ref{fig: NS composition}. It is important to note that the precise locations of the transitions between the envelope, crust, and core depend on the specific model being used. Here, we emphasize the physical aspects that characterize these transitions. It should also be noted that the regions between these interfaces do not have uniform radial thickness throughout the star, as the majority of the star's volume is occupied by the core, where densities exceed those of atomic nuclei, as illustrated in Fig.~\ref{fig: NS composition}.

\subsection{Core}

The core of a neutron star constitutes approximately 99\% of the total mass and occupies about 70\% to 90\% of the star's volume, depending on the total mass and the specific EoS. The outer core of a neutron star refers to the internal region where the mass density ranges approximately from $0.5 \rho_0$ to $2\rho_0$, with $\rho_0 = 2.8 \times 10^{14}$\,g\,cm$^{-3}$ being the nuclear saturation density (representing the typical density of a heavy atomic nucleus). Within this region, the matter forms a uniform liquid consisting of neutrons along with a certain fraction of protons and leptons, namely electrons and muons ($npe\mu$ matter). This composition allows the system to maintain $\beta$-equilibrium. 

Deep within the core, in neutron stars of significant mass ($M \gtrapprox 1.5M_\odot$), there exists an inner core that occupies the central part and exhibits densities $\rho \gtrapprox 2\rho_0$. The inner core is characterized by an uncertain composition, which is expected to be more complex than just neutrons, protons, electrons, and muons. Due to the challenges associated with the calculation of its composition and properties, our understanding of the inner core is limited and highly dependent on the specific details of the theoretical model used to describe collective fundamental interactions \citep{shapiro1983,haensel2007}. Various hypothetical options have been proposed within different models, which include: 
\begin{enumerate}
\item Hyperonization of matter: This involves the emergence of different hyperons, primarily $\Lambda$- and $\Sigma^-$-hyperons, resulting in the formation of $npe\mu$ matter.
\item Pion condensation: This entails the formation of a Bose condensate involving collective interactions with the properties of $\pi$-mesons.
\item Kaon condensation: This involves the formation of a similar condensate of $K$-mesons.
\item Deconfinement: This refers to a phase transition to quark matter.
\end{enumerate}
The last three options are often referred to as exotic, and for further details, we direct the reader to Chapter~7 of \cite{haensel2007}. While this book does not delve into the specifics of exotic matter, it does explore the impact of hyperonic matter on the cooling of a neutron star.

\subsection{Crust}

\begin{figure}
    \centering
    \includegraphics[width=.95\textwidth]{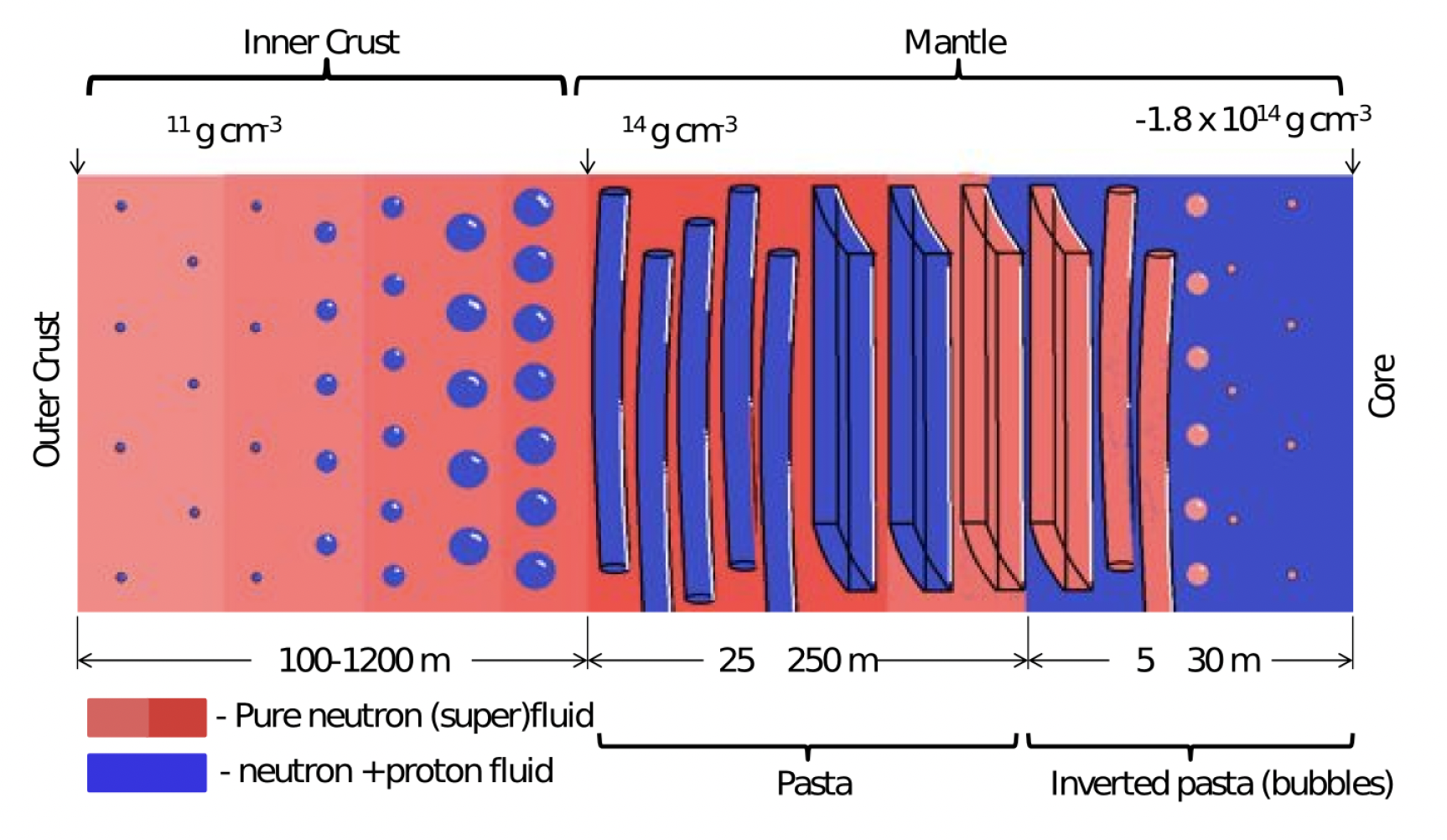}
    \caption[The mantle.]{Figure adopted from \cite{newton2013}. 
    At sufficiently high densities, the nuclei within the neutron star crust have the potential to deform and connect in specific directions, giving rise to the formation of elongated structures known as "gnocchi," "spaghetti," and "lasagne" of nuclear matter. Moreover, these structures can even extend into tube-like formations known as "anti-spaghetti" and "anti-gnocchi," or create a Swiss cheese-like pattern. These unique configurations, collectively referred to as nuclear pasta phases, can exist as a layer at the bottom of the neutron star crust, often referred to as the mantle.}
    \label{fig: mantle}
\end{figure}

As we move from the core to the exterior, both density and pressure decrease. When the density drops below approximately $1.5\times 10^{14}$\,g\,cm$^{-3}$ (half saturation density), matter inhomogeneities set in. Positive charges cluster together, forming individual clusters with a charge denoted as $Z$. These clusters arrange themselves into a solid lattice, minimizing Coulomb repulsion. The lattice, constituting only a small fraction of the star's overall mass and radius, is embedded in a gas of free neutrons and a background of electrons such that the whole system is charge neutral. We refer to this region as the inner crust of the star.

Towards the innermost part of the crust, the matter exhibits tube-like structures and voids resembling Swiss cheese, where there are holes in the homogeneous matter. This analogy can be likened to anti-spaghetti and anti-gnocchi formations, as these configurations aim to minimize their energy \citep{baym1971a,lorenz1993}. This intriguing phenomenon is referred to as the inverted pasta phases. As the density decreases, the nuclei may adopt non-spherical shapes, such as rod-like and plate-like structures, forming what is known as pasta phases (analogous to the shapes of gnocchi, spaghetti, and lasagne) within the nuclear matter \citep{pethick1995}. These distinct phases, illustrated in Fig.~\ref{fig: mantle}, mark the transition towards the uniform nuclear matter in the core. The region where nuclear pasta occurs is typically considered to form a mantle with anisotropic kinetic properties \citep{pethick1998}. The thermodynamic stability of the pasta phase and the existence of the mantle depend on the specific model of nuclear interactions.

At lower densities, neutrons are finally confined within the nuclear clusters, and the matter consists of a lattice composed of neutron-rich nuclei that are permeated by a degenerate electron gas. This particular region is referred to as the outer crust \citep{baym1971b}. It extends from an inner boundary, marked by the so-called neutron drip density of approximately $4 \times 10^{11}$\,g\,cm$^{-3}$, to an outer boundary with a density of around $10^{9}$\,g\,cm$^{-3}$ (depending on the temperature). Within the outer crust, the nuclei are arranged into a crystalline lattice.

The crust of a neutron star experiences extreme conditions in terms of density, temperature, and magnetic field, which cannot be replicated in laboratory settings. Despite being a relatively thin layer, typically around $\sim 1$\,km in thickness, the crust plays a crucial role in various astrophysical phenomena observed in neutron stars. These include pulsar glitches, quasi-periodic oscillations in soft $\gamma$-ray repeaters (SGRs), and thermal relaxation in soft X-ray transients (SXT), which are transient X-ray source with rapid and intense X-ray outbursts \citep{haensel2007,chamel2008,piro2005,strohmayer2006,steiner2009,sotani2013,newton2013,piekarewicz2014}. Understanding these phenomena depends on considering deviations from the assumption of a homogeneous fluid throughout the star and underscores the importance of properly treating the crust.

\subsection{Envelope and atmosphere}
\label{subsec: envelope}

The outermost layer of a neutron star, referred to as the envelope or ocean, typically extends to a depth ranging from a few meters to around hundred meters, depending on the temperature. The envelope consists of a liquid phase with a relatively low density ($\rho \lessapprox 10^9$\,g\,cm$^{-3}$). It constitutes a negligible fraction of the total mass of the neutron star, approximately on the order of $10^{-7}$ to $10^{-8}$\,M$_{\odot}$.

The envelope is characterized by steep gradients in temperature, density, and pressure, making it the region with the most rapid thermal relaxation timescale. As we descend into the envelope, the density gradually decreases until reaching values around $\rho \sim 1$\,g\,cm$^{-3}$, where a gaseous atmosphere of a few centimeters may exist.

In terms of composition, the outer crust and the ocean of a neutron star are relatively simple. They primarily consist of electrons and nuclei, which can be effectively treated as point-like particles. The thermal composition of the envelope is believed to leave a discernible imprint on the observed spectra and can be influenced by the star's history and environment (for a detailed study, refer to Chapter~\ref{chap: envelope}). For instance, neutron stars in binary systems have the potential to accrete light elements from their companion stars.

\section{Magnetic fields}
\label{sec: magnetic field in NS}

Another striking feature of neutron stars is their magnetic fields, which are among the most powerful observed in the universe. These magnetic fields can be billions of times stronger than Earth's magnetic field, generating powerful electromagnetic radiation and particle streams. Consequently, they serve as an additional reservoir of energy. The origin of these intense magnetic fields in neutron stars is predominantly attributed to the conservation of magnetic flux during progenitor and dynamo processes. The magnetic fields' intensity is further amplified during the core collapse dynamo, driven by differential rotation and convection \citep{Aloy_2021, reboul2021, raynaud20}.

The precise magnetic field configuration during a neutron star's formation remains largely unknown. Recent magneto-hydrodynamic (MHD) simulations exploring the magneto-rotational instability (MRI) within core-collapse supernovae \citep{obergaulinger2014,raynaud20,Aloy_2021,reboul2021} have unveiled a complex scenario in which the magnetic energy of the proto-neutron star is distributed across various spatial scales. These simulations indicate that the majority of the magnetic energy resides in small to medium-sized magnetic structures, encompassing both toroidal and poloidal components. Notably, these findings deviate significantly from the oversimplified dipole+twisted torus configurations often deduced from MHD equilibrium studies \citep{ciolfi2013}.

Neutron stars also exhibit remarkable rotational speeds. As a consequence of angular momentum conservation during core collapse, the initial star's rotation experiences a dramatic surge, resulting in neutron stars that can whirl around hundreds of times each second. Since the pioneering work of \cite{gold1968}, pulsars have been widely regarded as rapidly rotating neutron stars with powerful surface magnetic fields. 

The primary manifestation of the magnetic field in neutron stars is the dissipation of rotational energy  driven by the electromagnetic torque. By observing the pulsar's rotational period $P$ (typically spanning from $P \sim 0.001$ to $12$\,s) and its first time derivative in time $\Dot{P}$ (typically ranging from  $\Dot{P} \sim 10^{-16}$ to $10^{-12}$\,ss$^{-1}$), it is possible to estimate the strength of the surface magnetic field using the magnetic dipole model \citep{shapiro1983}. For instance, millisecond pulsars may reveal surface fields within the range of $10^8$ to $10^9$\,G, standard pulsars around $10^{12}$\,G, and magnetars  with even higher values, spanning from $10^{14}$ to $10^{15}$\,G.

The classical vacuum magnetic dipole model, commonly employed to determine the magnetic field strength at the star's surface, is described by the following relation: 
\begin{equation}
B_p \approx \Bigg(\frac{3I P \Dot{P}}{2\pi^2  R^6}\Bigg)^{0.5},
    \label{eq: PPdot}
\end{equation}
where $B_p$ is the strength of the magnetic field at the pole, $R$ is the radius of the neutron star,
and $I$ is the moment of inertia of the star.

Neutron stars undergo a long-term evolution over the course of millions of years. In the solid crust, electrons serve as the primary charge carriers, and the magnetic field undergoes evolution through Ohmic dissipation and Hall drift, as explained in \S\ref{subsec: eMHD}. The magnetic evolution within the core is more uncertain due to its multi-fluid nature, where neutron and protons can be in superfluid and superconductive phases, respectively. Furthermore, ambipolar diffusion can influence the magnetic field's evolution within the star's core \citep{passamonti2016,passamonti2017}. For further discussion on the magnetic field evolution within the interior of a neutron star, we direct the reader to \S\ref{sec: magnetic field evolution}.

\section{Observations}
\label{sec: observations and classification}

\begin{figure}
    \centering
    \includegraphics[width=.9\textwidth]{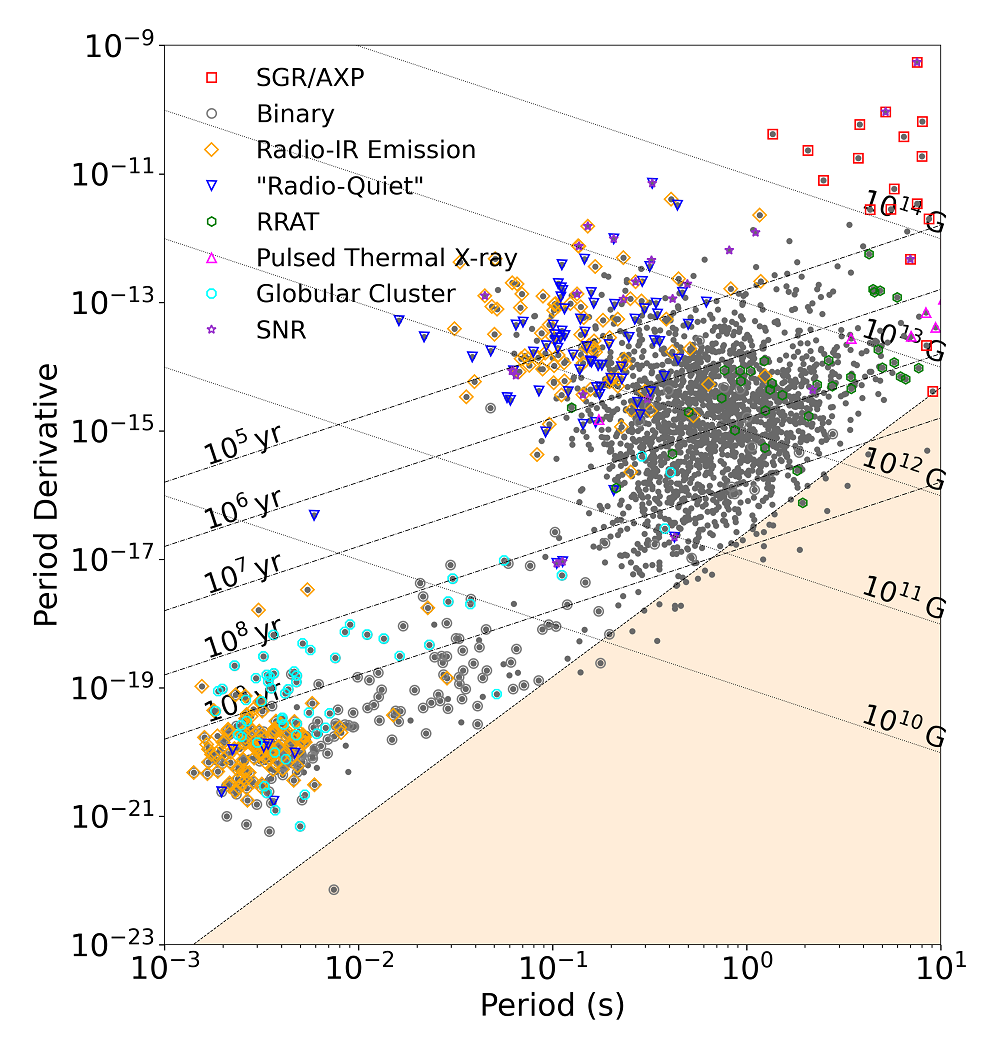}
    \caption[The mantle.]{The P/P-Dot diagram displays the entire pulsar population categorized by their rotation periods and slow-down rates. This plot reveals distinct sub-populations across all classes and frequency bands. The lightly shaded region in the diagram corresponds to the `death valley', where pulses are no longer detected for many of these objects. Image credit: Rami Mandow.}
    \label{fig: PPdot chap1}
\end{figure}

In recent decades, astronomical and technological advancements across all wavelengths have substantially broadened our understanding of neutron stars. Observations in the radio, X-ray and $\gamma$-ray bands have facilitated the identification of thousands of astrophysical sources as neutron stars in diverse environments. These environments include both isolated neutron stars and those accompanied by a variety of companions, some associated with supernova remnants and wind nebulae, while others exist in globular clusters or open fields.
Neutron stars have been detected in additional bands such as optical, infrared, ultraviolet, and $\gamma$-ray. For a comprehensive presentation of the different classes and frequency bands within the neutron star population, refer to Fig.~\ref{fig: PPdot chap1}. These multi-band observations provide valuable insights into the properties and behavior of neutron stars, allowing astronomers to indirectly infer information about their composition, magnetic fields, temperature, and emission mechanisms. Let us now provide a concise overview of the phenomenology of neutron stars across various energy bands.

\subsection{Radio band}
\label{subsec: radio band}

Neutron stars, commonly detected as regular radio pulses, exhibit fascinating characteristics that have been extensively studied in the field of astrophysics. Known as pulsars, these neutron stars emit radio waves, allowing astronomers to precisely measure their rotational period. By monitoring the regularity of the radio pulses and folding hundreds or thousands of them, the spin period of the neutron star can be determined with extraordinary accuracy, rivaling even the most advanced atomic clocks on Earth. This level of precision, reaching up to ten digits, is rarely achieved in astrophysics.
In addition to determining the spin period, prolonged observations enable astronomers to accurately measure the star's spin-down rate, known as the period derivative. By precisely studying the timing and characteristics of these radio pulses, researchers can investigate the long-term evolution of pulsar rotation periods.

Radio observations also provide valuable insights into the magnetic fields of pulsars. The intense magnetic fields of these stars influence their rotation and emission properties. Moreover, radio observations have uncovered intriguing phenomena associated with neutron stars. Pulsar glitches, observed in the majority of pulsars, involve sudden periodic spin-ups believed to be linked to the unpinning of superfluid neutron vortices in the star's crust \citep{anderson1975}. These glitches exhibit diverse behaviors in terms of spin-up amplitude and rotational properties during and after the post-glitch recovery.
Furthermore, certain pulsars display extended emission regions known as pulsar wind nebulae. Radio observations contribute to mapping these nebulae and studying the interaction between the pulsar wind and its surrounding environment. This research sheds light on particle acceleration and the dynamics of pulsar winds.

Other notable irregularities include timing noise, which refers to significant higher derivatives of spin periods. Additionally, there are phenomena such as mode changing, nulling, or intermittency \citep{drake1968,huguenin1970,kramer2006,mcLaughlin2006,wang2007,lyne2010}, all likely associated with sudden changes in magnetospheric radio emission. Changes in average pulse shapes and amplitudes are particularly linked to global magnetospheric reorganizations \citep{lyne2010}.

\subsection{X-ray band}
\label{subsec: X-ray band}

Neutron stars can generate intense X-ray emission from their hot surface or from the surrounding accretion disks if they are in binary systems. X-ray observations of neutron stars provide valuable information about their thermal emission, spectral characteristics, and variations over time. X-ray observatories, such as
\xmm, \cxo, \nustar, \nicer, \swift, and \rst\,, have provided detailed images and spectra, revealing information about the star's temperature, accretion processes, and possible neutron star atmospheres.

X-ray binaries represents a very wide field of research, but in this work we will focus only on isolated neutron star. More than $100$ isolated neutron stars are seen in X-rays, and they are phenomenologically quite heterogeneous, allowing for their classification into various categories. These classifications shed light on the remarkable diversity found among these captivating objects. Let's delve into some of the different classes of neutron stars which are relevant for our evolutionary models here developed. 
\begin{itemize}
   \item Rotation-powered pulsars (RPPs) constitute the predominant population among detected neutron stars, with over 2000 identified sources. These objects are commonly observed in the radio spectrum, and their emission is powered by the conversion of a fraction of their rotational energy into non-thermal radiation. 
   RPPs can be categorized into two main classes: standard pulsars and millisecond pulsars. The latter exhibit extremely short rotational periods, typically in the millisecond range, and are believed to have undergone a recycling process within binary systems involving accretion phases. 
   Only a limited proportion of RPPs have been observed through X-ray detection. These X-ray-emitting sources display a power-law component attributed to rotation; however, they also exhibit thermal emissions.
  Within the subset of X-ray-emitting pulsars, there exist both youthful and aged ($\gg$\,Myr) ones. The X-ray emissions from the nearby elderly pulsars might be attributed to the bombardment of their polar caps by magnetospheric currents as they interact with returning particles. For a comprehensive overview of X-ray RPP observations, refer to the work by \cite{kaspi2016}.

  \item Magnetars represent a unique class of intensely luminous, isolated neutron stars, distinguished by their extraordinarily powerful magnetic fields, which range between $10^{14}$ to $10^{15}$\,G. These fields are substantially more potent than those found in conventional neutron stars. Magnetars are characterized by their prolonged rotational periods, typically spanning several seconds, a feature detailed in contemporary reviews such as \cite{turolla2015,esposito2021}. Beyond being highly magnetized neutron stars, magnetars manifest as Anomalous X-ray Pulsars (AXPs) and Soft Gamma Repeaters (SGRs), often emitting energy that surpasses that produced by their spin-down. The intense magnetic fields of magnetars are responsible for a variety of high-energy events. These include persistent emissions in both thermal and non-thermal high-energy forms, as well as bursts of X-rays and flares of gamma rays.
 
  \item X-ray Dim Isolated Neutron Stars (XDINSs) also known as "The Magnificent Seven" (see for a review \cite{turolla09}). These relatively old ($10^5-10^6$\,yr) and nearby ($<$ 1\,kpc) sources are radio-quiet, exhibit faint X-ray emission, display purely thermal spectra and have distances that are well-constrained, usually determined through parallax measurements. Due to their proximity, the X-ray radiation is not much absorbed by the interstellar medium, so that these sources emerge as the most promising candidates for studying in great detail the thermal spectra of neutron stars.  
  
  \item Central Compact Objects (CCOs) are X-ray sources located at the centers of supernova remnants. They are much dimmer than magnetars but significantly brighter than XDINSs \citep{deluca2017}. 
  Although a dozen have been observed, the timing properties, i.e., period and period derivatives, have been measured for only three of them. The magnetic field strength inferred from these measurements is of the order of $10^{10}$\,G. Such a low intensity of the dipolar component of the magnetic field cannot explain the bright luminosity observed in these objects. A way to explain the low field intensity is the hidden magnetic field scenario \citep{ho2011,vigano12b}, which involves an episode of fall-back accretion after the supernova explosion. That results in an external magnetic field (the surface field) lower than the hidden or buried internal one, which might re-emerge once accretion stops. 
\end{itemize}
Each class of neutron star presents unique challenges and opportunities for scientific exploration. 
Moreover, the X-ray band plays a crucial role in studying neutron stars, particularly in investigating their magneto-thermal evolution (see for a review \cite{pons2019}). This emission is influenced by the complex interplay between the star's strong magnetic field and its thermal properties. As the star undergoes cooling, its surface temperature decreases, resulting in alterations to the observed X-ray emission.

By studying the X-ray emission from neutron stars and monitoring its evolution over time, astronomers can gain insights into the thermal behavior, magnetic field decay, and overall aging of these cosmic objects.

\subsection{Gamma-ray band}
\label{subsec: gamma band}

$\gamma$-ray observations of neutron stars provide a window into the extreme and energetic phenomena associated with these compact stellar remnants (for a review see \cite{grenier2015}).
Some neutron stars emit $\gamma$-rays, which are the highest energy form of electromagnetic radiation \citep{cheng1977,smith2023}. Neutron stars emit $\gamma$-rays through various mechanisms, including the acceleration of charged particles in their strong magnetic fields and the interaction of high-energy particles with their surroundings. $\gamma$-ray telescopes, such as NASA's Fermi Gamma-ray Space Telescope, have detected pulsating $\gamma$-ray emissions from a subclass of neutron stars known as $\gamma$-ray pulsars, which reveal valuable information about their rotation and magnetic field properties.
Additionally, $\gamma$-ray observations have unveiled other fascinating phenomena associated with neutron stars, such as magnetars that can produce intense and short-lived bursts of $\gamma$-rays. The increasing statistics of high-energy pulsars helps to constrain the emission models, providing information about the region of magnetospheric emission and the involved electromagnetic processes, like synchro-curvature radiation and inverse Compton \citep{cheng1996,cotizelati2020,iniguez2022}. Very recently, a few pulsars have been observed in TeV (the upper range of $\gamma$-ray) \citep{albert2021}.

\subsection{Ultraviolet, optical and infrared bands}
\label{subsec: ultraviolet, optical and infrared bands}

Neutron stars exhibit intrinsic faintness in the ultraviolet, optical, and infrared wavelengths. However, recent technological advancements have enabled the identification of counterparts for a subset of pulsars, amounting to a few tens of cases (see \cite{mignani2012} and references within). Similar to other energy bands, the analysis of pulse profiles provides valuable information regarding the location of the emission region. Although spectroscopy and polarization measurements are limited to only a handful of cases, they can offer constraints on the energy distribution and density of magnetospheric particles.

In the optical and ultraviolet bands, astronomers can infer the anisotropic thermal map of the surface, which aids in constraining cooling models as these bands encompass the majority of thermal emission from cold stars with ages on the order of millions of years. Optical observations, benefiting from good angular resolution, facilitate measurements of proper motions and parallaxes, thereby providing reasonable estimates of distances. Furthermore, optical and infrared observations prove useful in examining the presence of debris disks surrounding isolated neutron stars.

\subsection{Gravitational waves}
\label{subsec: gravitational waves}

The detection of gravitational waves from neutron stars has opened an entirely new avenue for studying these objects. When neutron stars merge or undergo asymmetrical events, they generate gravitational waves, which cause ripples in space-time. Advanced detectors like LIGO and Virgo have successfully captured signals from these events, known as binary neutron star systems \citep{ligovirgo2017}, providing valuable insights into their masses, spins, and EoS. 

Moreover, there has been recently a surge of interest in ``Continuous Gravitational Waves'', which pertain to emissions from rapidly spinning pulsars with spin frequencies at the optimal sensitivity range of LIGO/Virgo detectors (tens/hundreds of Hz). For a comprehensive review of this exciting field, we recommend referring to the work by \cite{piccinni2022}.

These detections enable us to explore matter at extreme densities and test theories of gravity under the most challenging conditions. When combined with other electromagnetic signals, such as $\gamma$-rays and radio waves, these gravitational wave observations enhance our understanding of the astrophysical processes associated with neutron stars and their role in the evolution of the universe.

\clearemptydoublepage

\let\textcircled=\pgftextcircled
\chapter{Magneto-Thermal Evolution}
\label{chap: magneto-thermal evolution}

\initial{A}fter a few minutes from birth, neutron stars settle into a stratified, hydrodynamically static configuration, where neither convection nor differential rotation occurs. If they are isolated, they will have no further relevant energy source, leading to the decay of kinetic (rotation), magnetic, and thermal energy over the long term. In fact, the rich multi-wavelength phenomenology observed during the active life of neutron stars, such as beamed non-thermal emission, thermal X-ray emission, transient bursts, and flux enhancements, ultimately relies solely on the rotationally-powered electromagnetic torque and the rearrangement of the proto-neutron star's huge magnetic energetic inheritance, which occurs through long-term dissipation and sporadic abrupt re-configurations.

To quantitatively assess the long-term evolution of a neutron star's magnetic fields in the crust, as well as its internal temperature, we can simplify the MHD equations into a system of only two interconnected evolution equations: the Hall induction equation and the heat diffusion equation. These equations are extensively described in the review by \cite{pons2019}. Both equations are influenced by the star's composition, which is determined by the chosen EoS \citep{aguilera2008,pons2019,anzuini2022,vigano2021,dehman2022,dehman2023b,dehman2023c}. Additionally, the magnetic field induces anisotropic heat transport, being more efficient along rather than across the magnetic field lines. The dissipation of electric currents, responsible for sustaining the magnetic field (Joule heating), further heats the internal layers, slowing down the long-term cooling. Such dissipation rate depend in turn on the temperature, via electric conductivity. These and other minor effects, reviewed below in more detail make the magnetic and thermal evolution strongly coupled.

Such magneto-thermal evolution in isolated neutron stars has undergone extensive exploration through 2D simulations \citep{pons2007,aguilera2008,pons2009,vigano2012,vigano2021}. These models have effectively revealed the general properties of the isolated neutron star population \citep{vigano12b,vigano2013,pons2013,gullon2014,gullon2015}. Furthermore, recent efforts have been dedicated to investigating the magnetic evolution without the constraints of axial symmetry \citep{wood2015,gourgouliatos2016,degrandis2020,degrandis2021,igoshev21a,igoshev21,dehman2022,dehman2023c}.

In this chapter, we lay the foundation of this thesis by presenting the formalism governing the magnetic and thermal evolution. The detailed discussion on the magnetic field evolution can be found in \S\ref{sec: magnetic field evolution}, where we explore the underlying physical processes occurring in different regions of the star. Moving forward, in \S\ref{sec: neutron star cooling}, we provide a comprehensive review of the theory of neutron star cooling. We emphasize the key microphysical components that are essential for solving the induction and heat diffusion equations in \S\ref{sec: microphysics}. Lastly, in \S\ref{sec: MT evolution}, we describe the coupling between magnetic and thermal evolution, and provide a brief introduction to the main structure of the numerical codes used in the upcoming chapters.

\section{Magnetic field evolution in the interior of neutron stars}
\label{sec: magnetic field evolution}

The interior of a neutron star contains a complex multi-fluid system with different coexisting species exhibiting distinct average hydrodynamical velocities. In the crust, nuclei form a solid lattice, possibly having a plastic behavior \citep{lander2016}, while the fluid made by relativistic electrons carries currents, sustaining the magnetic field. In the inner crust, there exists a third component: neutrons, which can be in a superfluid phase and become partially decoupled from the heavy nuclei, forming a neutral fluid. In the core, the situation becomes even more complicated with the coexistence of neutrons and protons that can both be in superfluid and superconducting phases, respectively. A fully consistent set of evolution equations that can be used to simulate the evolution has not been developed yet. Consequently, to represent this system, different levels of approximation are used, gradually incorporating the relevant physics. This section provides an overview of the theory, aiming to represent the most significant processes governing the magnetic field evolution in a relatively simplified mathematical form. In our analysis, we adopt the background model introduced in \S\ref{sec: TOV} and account for the relativistic corrections.

Let's start from the basis. The evolution of any electromagnetic field is governed by the Maxwell equations, which incorporate relativistic corrections and are represented as follows:
\begin{enumerate}
    \item Gauss's Law for electric fields:
     \begin{equation}
    \boldsymbol{\nabla} \cdot \boldsymbol{E} = 4 \pi \rho_q.
\end{equation}

\item Amp\`ere-Maxwell's Law for magnetic fields:
\begin{equation}
   \frac{1}{c} \frac{\partial \boldsymbol{E}}{\partial t} = \boldsymbol{\nabla} \times ( e^{\zeta} \boldsymbol{B}) - \frac{4 \pi}{c} e^{\zeta} \boldsymbol{J}.
   \label{eq: ampere law}
\end{equation}

\item Gauss's Law for magnetic fields, which is the solenoidal constraint representing the absence of magnetic monopoles:
\begin{equation}
\boldsymbol{\nabla} \cdot \boldsymbol{B} = 0.
\end{equation}

\item Induction or Faraday's Law:
\begin{equation}
   \frac{1}{c} \frac{\partial \boldsymbol{B}}{\partial t} = - \boldsymbol{\nabla} \times ( e^{\zeta} \boldsymbol{E}).
   \label{eq:induction_general} 
\end{equation}
\end{enumerate}
Here, $\boldsymbol{\nabla}$ operators account for the metric factors (the $e^\lambda$ term seen in \S\ref{sec: TOV}), $c$ represents the speed of light, the $\zeta$ term was introduced in \S\ref{sec: TOV}, $\boldsymbol{E}$ and $\boldsymbol{B}$ denote the electric and magnetic fields, respectively. $\rho_q$ stands for the electric charge density, and $\boldsymbol{J}$ represents the current density. 

MHD investigates the dynamics of a conducting, magnetized medium. A fundamental assumption in MHD is that the timescale of variation of the electromagnetic field is significantly larger than the typical timescale of collisions within the plasma. As a consequence, in the Amp\`ere-Maxwell equation (eq.\,\eqref{eq: ampere law}), the displacement current, represented by the left-hand side term, becomes negligible in comparison to the right-hand side terms. This implies that the conducting fluid can almost instantaneously respond to any changes in $\boldsymbol{B}$ to establish a current, so that $\partial\boldsymbol{E}/\partial t \approx 0$ always. As a consequence, the current in MHD is:
\begin{equation}
\boldsymbol{J} =  e^{-\zeta} \frac{c}{4\pi}
\boldsymbol{\nabla} \times (e^{\zeta}\boldsymbol{B}).
\label{eq: electric current}
\end{equation}
Therefore, in MHD the Maxwell equations reduce to the solenoidal constraint and to the general induction equation (\ref{eq:induction_general}).
In order to solve the latter, we must express the electric field $\boldsymbol{E}$ in terms of other variables, such as constituent component velocities and the magnetic field itself (\S\ref{subsec: eMHD} and \S\ref{subsec: B-evol neutron star core}). This prescription can be achieved through a generalized Ohm's law, incorporating the electrical conductivity represented by a tensor $\hat{\sigma}$, as derived by \cite{yakovlev1990,shalybkov1995}:
\begin{equation}
    \boldsymbol{J} = \hat{\sigma} \boldsymbol{E}~.
    \end{equation}

The magneto-thermal evolution of neutron stars is primarily governed by three processes (elaborated in \S\ref{subsec: eMHD} and \S\ref{subsec: B-evol neutron star core}): Ohmic dissipation, Hall drift (relevant mostly in the crust), and ambipolar diffusion (relevant mostly in the core) \citep{goldreich1992,shalybkov1995,cumming2004,passamonti2016}. These effects are thought to play a dominant role, however, it is essential to note that other factors can also become significant under specific conditions. For instance, there are theoretical arguments proposing additional slow-motion dynamical terms, such as plastic flow \citep{beloborodov2014,lander2016,lander2019,gourgouliatos2021}, as well as magnetically induced superfluid flows \citep{ofengeim2018} or vortex buoyancy \citep{muslimov1985,konenkov2000,elfritz2016,dommes2017}. Typically, all these effects can be effectively introduced as advective terms of the form $\boldsymbol{E} = - \boldsymbol{v}/c \times \boldsymbol{B}$, with $\boldsymbol{v}$ representing an effective velocity. The latter may depend on $\boldsymbol{B}$ itself, or on the temperature gradient. Additionally, thermoelectric effects, exemplified by the Biermann Battery phenomenon, which describe how thermal flux gives rise to carrier movement and thus generates a current,  have also been proposed to become significant in regions with large temperature gradients \citep{geppert1991,geppert1995,wiebicke1991,wiebicke1992,wiebicke1996,gourgouliatos2022}. 

Despite their potential significance, the majority of these additional terms are not included in most existing literature, and detailed numerical simulations on these aspects are scarce. However, it is possible that some of these effects may play a more crucial role than previously thought and warrant careful revisiting. In this thesis, we will focus on highlighting the principal characteristics of the most standard and better-understood physical processes.


\subsection{Relativistic Hall-MHD}
\label{subsec: eMHD}

In the crust of a neutron star, the electric field $\boldsymbol{E}$ is determined by the electrical current density $\boldsymbol{J}$ and the advection of the magnetic field by the charged component of the fluid, specifically the electrons with velocity $\boldsymbol{v}_e$. The electric field is expressed as follows:
\begin{equation}
    \boldsymbol{E} = \frac{\boldsymbol{J}}{\sigma} - \frac{\boldsymbol{v}_e}{c}\times \boldsymbol{B},
\end{equation}
where the electric conductivity $\sigma$ (further elaborated in \S\ref{subsec: conductivity}) actually represents the longitudinal (parallel to the magnetic field) component of the general conductivity tensor $\hat{\sigma}$, given by $\sigma = e^2 n_e \tau_e/m_e^*$, where $n_e$ denotes the electron number density, $\textit{e}$ is the electric charge, $\tau_e$ is the electron relaxation time, and $m_e^*$ is the effective mass of electrons in the relaxation time approximation. The electron velocity in the crust is simply proportional to the electric current
\begin{equation}
    \boldsymbol{v}_e = - \frac{\boldsymbol{J}}{\textit{e}n_e}.
\end{equation}

Then, the Hall induction equation, considering the temperature- and density-dependent magnetic diffusivity $\eta_b = c^2/(4\pi \sigma)$, the Hall-prefactor $f_h = c/(4\pi \textit{e} n_e)$, and the current density definition (eq.\,\eqref{eq: electric current}) can be written as:
\begin{equation}
\frac{\partial \boldsymbol{B}}{\partial t} = - \boldsymbol{\nabla} \times \bigg[ 
\eta_b \boldsymbol{\nabla} \times ( e^{\zeta} \boldsymbol{B}) + f_h \big[ \boldsymbol{\nabla} \times (e^\zeta \boldsymbol{B})
\big] \times \boldsymbol{B} \bigg].
\label{eq: induction equation no Rm}
\end{equation}
The first term on the right-hand side corresponds to the Ohmic (dissipative) effect, and the magnetic diffusivity $\eta_b$ depends on the temperature, which points to the importance of the coupling of the magnetic and the thermal evolution. The magnetic energy converted into heat per unit volume per unit time is given by: 
\begin{equation}
    Q_J = \frac{\boldsymbol{J} \cdot \boldsymbol{J}}{\sigma}= \frac{4 \pi}{c^2} \eta_b \boldsymbol{J} \cdot \boldsymbol{J}. 
    \label{eq: joule heating}
\end{equation}
This term enters in the heat transfer equation (as shown in \S\ref{sec: neutron star cooling}), contributing to the prolonged heat retention of strongly magnetized neutron stars compared to weakly magnetized ones \citep{pons2019}.

The second term of eq.\,\eqref{eq: induction equation no Rm}, represents the non-linear Hall term. The latter arises due to the Lorentz force acting on the electrons and exhibits no temperature dependence. However, its magnitude varies significantly in the crust due to its inverse dependence on the electron density. Furthermore, the Hall term tends to drive electric currents towards the boundary between the crust and core. At this interface, the presence of high impurity content and pasta phases may lead to rapid dissipation of the magnetic field, resulting in an enhancement of the Joule heating and a substantial reduction in spin-down \cite{pons2013}. Additionally, as we will see in Chapter~\ref{chap: MATINS} and Chapter~\ref{chap: 3DMT}, the Hall term acts to redistribute magnetic energy across all scales, counterbalancing Ohmic dissipation at the smallest scales. Over a considerable period, around ${\cal O}(10^4)$\,yr, the crustal magnetic topology approaches a slowly varying configuration, with the energy distributed over a broad range of multipoles. This process induces a Hall cascade with a slope of approximately $\sim l^{-2}$ for the intermediate and small scales \citep{dehman2022}.

Factoring out the magnetic diffusivity $\eta_b$ from eq.\,\eqref{eq: induction equation no Rm}, we can alternatively express the Hall induction equation as follows:
\begin{equation}
\frac{\partial \boldsymbol{B}}{\partial t} = - \boldsymbol{\nabla} \times \Bigg( 
\eta_b \bigg\{ \boldsymbol{\nabla} \times ( e^{\zeta} \boldsymbol{B}) + \omega_B\tau_e \big(\boldsymbol{\nabla} \times (e^\zeta \boldsymbol{B})\big) \times \boldsymbol{b}
\bigg\} \Bigg).
\label{eq: induction equation}
\end{equation}
Here, $\boldsymbol{b}$ is the unit vector along the direction of the magnetic field.
This form of the induction equation explicitly shows that the magnetization parameter $\omega_B \tau_e$, with $\omega_B=\textit{e}B/m_e^* c$ (representing the gyro-frequency of electrons with charge $-e$ and effective electron mass $m_e^*$ moving in a magnetic field strength $B$), plays the role of the magnetic Reynolds number, denoted as $R_m = \frac{f_h B}{\eta_b}$. This parameter determines the degree of anisotropy in the heat equation. Moreover, it acts as an indicator of the relative importance between the Hall and Ohmic terms. When it greatly exceeds unity, the Hall drift dominates, and the purely parabolic diffusion equation transitions to a hyperbolic one. Generally, as we move from the interior towards the surface, $\omega_B \tau_e$ increases due to the drop in density and conductivity. It is essential to exercise caution when interpreting analytical estimates of the Ohmic or Hall timescales (often done in literature in a simplistic way), as they can vary significantly due to local conditions, sometimes differing by several orders of magnitude.

It is worth noting that the Hall-MHD description of the crust is based on the concept that ions are confined within the crustal lattice, while only electrons remain mobile. However, molecular dynamics simulations \citep{horowitz2009} reveal that the matter behaves elastically up to a certain maximum stress level. Beyond this threshold, the magnetic stresses, quantified by the Maxwell term $\hat{M} = \frac{B_i B_j}{4\pi}$, cannot be compensated by the elastic response. A more rigorous condition for this is the von Mises criterion (applied e.g. in \citealt{lander2015}). Crustal failures studied in Chapter~\ref{chap: outburst} are treated in the most simplified manner as star-quakes. This study examines the frequency and energetic of internal magnetic rearrangements by evaluating the accumulated stress, which is proposed to be the cause of magnetar outbursts \citep{pons2011,perna2011} and a possible scenario to explain fast-radio bursts (FRB) \citep{dehman2020}. This behaviour can be described as crustal failure, where the low material densities allow for the propagation of sudden fractures or failures. 

Rather than experiencing sudden failures, materials subjected to very slow shearing forces may display an alternative behavior, entering a plastic regime characterized by a gradual plastic flow instead of abrupt crustal failures. Despite the different dynamics, the energetic arguments related to the release of energy due to the accumulation of magnetic stresses remain similar. Recent simulations illustrate this plastic flow assumption under Stokes flow, where a viscous term balances magnetic and elastic stresses \citep{lander2019}. The comparison between Ohmic and Hall evolution shows significant plastic-like motions in the star's external layers. Similar arguments have been proposed to explain heat deposition by visco-plastic flow and the propagation of thermo-plastic waves \citep{beloborodov2014}. Depending on which hypotheses we make, the interpretation of velocities in the advective term ($\boldsymbol{v} \times \boldsymbol{B}$) of the induction equation requires a proper physical and mathematical approach. However, \cite{lander2019} show how including a simplified treatment of the plastic behaviour in the Hall induction equation produce only a slight reduction of the Hall dynamics, without changing dramatically the global picture. Therefore, the Hall induction equation employed in this thesis (and in most of the dedicated literature) can be considered a fair enough approximation.

\subsection{Magnetic field evolution in neutron star cores}
\label{subsec: B-evol neutron star core}

The evolution of the magnetic field in the core is more uncertain due to its multifluid nature and the occurrence of proton superconductivity. This complexity arises because there may be regions with protons in either the normal phase or the superconducting phase, the latter being of type-II \citep{baym1969,sedrakian2019} or type-I (leading to magnetic field expulsion due to the Meissner effect). Recent calculations \citep{wood2020} predict phase coexistence in mesoscopic regions (larger than the flux tubes but smaller than the macroscopic length-scales), introducing several length and time-scales into the problem.

When superconductivity is neglected, the typical time-scales for Ohmic dissipation and Hall advection exceed the cooling time-scales, resulting in minimal changes to the magnetic field in the stellar core \citep{elfritz2016,dehman2020,vigano2021}. However, the inclusion of ambipolar diffusion could partially speed up the dynamics under certain conditions \citep{castillo2020}. This is due to its cubic dependence on the magnetic field, making it a dominant process influencing the evolution of magnetars during the first $10^3-10^5$\,yr, although this remains a topic of debate. A more consistent approach, incorporating hydrodynamic effects in the superfluid/superconducting core, could substantially accelerate the evolution, as recently discussed in terms of estimated time-scales by \cite{gusakov2020} (see also references therein).
Moreover, an initial complex magnetic topology can also reduce the typical length- and time-scales, compared to the usually assumed purely dipolar fields, adding further intricacy to the understanding of the magnetic field evolution in the core.


The short way to incorporate ambipolar diffusion is to generalize the form of the electric field by introducing the ambipolar velocity $\boldsymbol{v}_a$:
\begin{equation}
\boldsymbol{E} = \frac{\boldsymbol{J}}{\sigma} + \frac{1}{\textit{e}n_ec} \boldsymbol{J} \times \boldsymbol{B} - \frac{\boldsymbol{v}_a}{c}\times \boldsymbol{B},
    \label{eq: electric field ambipolar}
\end{equation}

In the simplest case, the system achieves $B$-equilibrium more rapidly than it evolves, and the ambipolar velocity is directly proportional to the Lorentz force
\begin{equation}
    \boldsymbol{v}_a = f_a \boldsymbol{J}\times \boldsymbol{B},
\end{equation}
where $f_a$ is a positive-defined drag coefficient. We also note that, alternatively, the ambipolar term can be written as \citep{pons2019}:
\begin{equation}
    -\frac{\boldsymbol{v}_a}{c} \times \boldsymbol{B} = \frac{f_a}{c} \bigg[B^2\boldsymbol{J} - (\boldsymbol{J}\cdot\boldsymbol{B})\boldsymbol{B}
    \bigg] = \frac{f_a}{c} B^2 \boldsymbol{J}_{\perp}.
\end{equation}
The term takes on an explicit form resembling a resistive-like component, characterized by a coefficient dependent on $B^2$. It specifically affects currents perpendicular to the magnetic field ($\boldsymbol{J}_\perp$), aligning the magnetic field with the current and inducing a force-free configuration, defined by $\boldsymbol{J}\times\boldsymbol{B}=0$. Importantly, the impact of this term strongly depends on the magnetic geometry, not just its strength. It does not influence the current flowing along magnetic field lines.

In this thesis, we primarily focus on the magnetic field evolution within the crust of a neutron star. In the core, we adopt two different approaches: either a simplified and convenient treatment of $B=0$, or we consider a large-scale field in the core (Chapter~\ref{chap: outburst}) that decays over timescales much longer than the crustal timescales, effectively rendering the core magnetic field as nearly fixed. Although we anticipate significant dynamics in the core, their timescales are likely to be longer than those in the crust, even when accounting for all the effects. Therefore, simulating the evolution of crustal magnetic fields can provide a reasonable understanding of the overall behavior.


\section{The physics of neutron star cooling}
\label{sec: neutron star cooling}

The X-ray spectra detected from several isolated neutron stars display a notable thermal component, thought to be emanating from their surface. In certain cases, independent estimates of a star's age are also available, enabling us to examine the correlation between temperatures and age. This provides an indirect approach to test the physics of the neutron star's interior through a neutron star cooling model.

The study of neutron star cooling holds significant importance in developing realistic evolution models. When these models are compared with observations of thermal emissions from neutron stars at various ages \citep{page2004,yakovlev2004a,yokovlev2008,page2009,potekhin15b}, they offer valuable insights into several aspects. These include the chemical composition, magnetic field strength, and the topology of the regions responsible for generating these radiations. Moreover, they can even shed light on the properties of matter at higher densities deep within the star, and help constrain the nuclear EoS, as we will demonstrate in Chapter~\ref{chap: Comparison with observations}.


After the proto-neutron star stage (as described in \S\ref{sec: stellar evolution}), the neutron star becomes transparent to neutrinos, and heat is primarily carried away by neutrino emission from the entire stellar body, as well as by thermal photon emission from the surface. The core of a neutron star becomes isothermal after a few decades due to its high thermal conductivity. On the other hand, the crust experiences slower thermal relaxation, typically taking around $t \sim 100-200$\,yrs. As a result, the crust's temperature during these very early stages is higher than that of the core, and the thermal luminosity of the star does not accurately reflect the thermal evolution of the core.

Eventually, the relaxation period of the crust ends when the cooling wave propagating from the core reaches the stellar surface \citep{gnedin2001}. Consequently, the thermal luminosity drops significantly by various orders of magnitude, depending on factors such as thermal conductivity, heat capacity of the crust layers, and whether neutrons are in a superfluid state \citep{lattimer1994,gnedin2001}.

The internal temperature evolution is governed by the heat diffusion equation \citep{shibanov1996} as follows: 
\begin{equation}
     c_{v}  \frac{\partial (e^{\zeta} T)}{\partial t} + \boldsymbol{\nabla}\cdot(e^{2\zeta} \boldsymbol{F}) = e^{2\zeta}(Q_{J} - Q_{\nu}) ,
    \label{eq:thermal_evolution}   
\end{equation}
where $c_{v}$ represents the heat capacity per unit volume of nucleons and leptons, $Q_J$ is the heating power per unit volume and, in the case of Ohmic heating included in this thesis, it is given by eq.\,\eqref{eq: joule heating}. The source term is given by the neutrino emissivity per unit volume $Q_\nu$, which accounts for the energy losses through neutrino emission. Both terms are dependent on the temperature. The internal, local temperature is denoted by $T$.

The heat flux $\boldsymbol{F}$, in the Fourier's law approximation, is given by:
\begin{equation}
   \boldsymbol{F} = - e^{-\zeta} \hat{\kappa} \cdot \boldsymbol{\nabla}(e^{\zeta}T),
   \label{eq:heat flux}
\end{equation}
where $\hat{\kappa}$ represents the thermal conductivity tensor.

\section{Microphysics}
\label{sec: microphysics}

In this section, we revisit the main microphysical ingredients necessary for simulating the long-term magneto-thermal evolution of neutron stars. Our focus here will mainly be on the crustal properties, providing a concise overview. For more comprehensive reviews, readers may refer to works such as \cite{chamel2008} and \cite{page2012}.

For a comprehensive computation of the various microphysical ingredients necessary for the heat diffusion equation and the induction equation, we highly recommend referring to the review by \cite{potekhin2015} and exploring the publicly available codes\footnote{\url{http://www.ioffe.ru/astro/conduct/}} developed by Alexander Potekhin. These public codes have been utilized to compute the microphysical ingredients in our simulations, providing valuable insights and tools for studying the intricate physics governing neutron star interiors.

\subsection{Superfluidity and superconductivity}
\label{subsec: superfluid and superconductivity}

While we have a good understanding of superfluidity properties in laboratory conditions through experiments with cold atoms, the situation becomes much more complex when it comes to pairing in nuclear matter. To study this, intricate theoretical calculations are necessary due to its limited knowledge. The transition to the pairing phase varies depending on the nature of the nucleons involved and the conditions to which they are subjected.

The transition to the superfluid/superconducting phase begins shortly after birth, with a substantial portion of the core becoming superconducting within the first year. Conversely, the epoch of the transition to a neutron superfluid phase in the core is highly uncertain, possibly occurring tens to thousands of years later. The transition to superfluid matter happens at a specific temperature called the critical temperature ($T_c$). Below this critical temperature, fermionic particles undergo the pairing phenomenon. This pairing phenomenon is closely related to the energy required to break a Cooper pair. The strength of the nuclear interaction plays a crucial role in determining the critical temperature ($T_c$) at which this pairing phase transition occurs.

In a neutron star, the Cooper pair is formed by two fermions with a spin of $1/2$, such as neutrons and protons, and they exist in a neutron star in either a $^1S_0$ or $^3P_2$ state.  In the inner crust, there exists a neutron superfluid coupled in the $^1S_0$ state. In the core, there are a neutron superfluid coupled in the $^3P_2$ state and a proton superconductor described with $^1S_0$ pairing. The singlet state $^1S_0$ is obtained through rotation-invariant and isotropic interactions, while the $^3P_2$ state is anisotropic.

The presence of superfluidity and superconductivity in neutron stars has significant effects on their cooling process. Pairing in nuclear matter plays a crucial role in neutron star cooling, with minimal impact on the EoS, but a strong reduction in neutrino emissivities and specific heat \citep{yakovlev1999, yakovlev2001, potekhin2015}. The primary contribution to both quantities comes from particles near the Fermi surface, and their behavior is influenced by the presence of an energy gap in the nucleon dispersion relations. Additionally, superfluidity and superconductivity activate a new neutrino emission channel through "Cooper pair breaking and formation" processes. As the temperature approaches the critical value ($T_c$), this new channel is triggered due to the continuous formation and breaking of Cooper pairs. The efficiency of the Cooper pair breaking and formation process increases as the temperature decreases, reaching a maximum at $T \sim 0.2T_c$, after which it becomes exponentially suppressed \citep{page2004}.

These three effects, namely the suppression of specific heats (\S\ref{subsec: specific heat}), the creation of a new neutrino channel, and the suppression of neutrino emissivity (\S\ref{subsec: neutrino reactions}), collectively influence the cooling process. While the suppression of neutrino emissivity slows down cooling, the other two effects accelerate it. Overall, unless extremely unusual choices are made for the three gap models in the neutron star, cooling is typically accelerated. Since the primary objective of this work is not to explore the vast parameter space encompassing superfluidity models, detailed discussions concerning their impact on cooling models are deferred to dedicated studies (e.g., \cite{page2006}, \cite{aguilera2008}, \cite{page2009}, and \cite{potekhin2015}).

It is claimed that the observed decrease in temperature of the neutron star in Cassiopeia A over a few decades (refer to \cite{elshamouty2013} for data analysis) has been linked to the onset of superfluidity, as discussed in previous works by \cite{shternin2011} and \cite{page2011}. For a recent complete and accurate study, we refer the reader to \cite{posselt2022}, where they observe a decrease in temperature, but it can be ascribed to calibration issues. To analyze this phenomenon, we adopt the parametrization introduced by \cite{ho2015}. For the singlet (isotropic pairing), the parametrization is given by:
\begin{equation}
    k_b T_c \approx 0.5669 \Delta,
\end{equation}
where $k_b$ is the Boltzmann constant and $\Delta$ is the pairing gap. For the triplet (anisotropic pairing), the parametrization is:
\begin{equation}
    k_b T_c \approx 0.1887 \Delta.
\end{equation}

According to most models, proton pairing ceases to exist at high densities, typically around $\rho \gtrsim 10^{15}$\,g\,cm$^{-3}$. The situation becomes considerably more complicated when considering the pairing of neutrons in the core, primarily due to the growing significance of relativistic effects. As of now, no definitive approach to this problem has been established, leaving it open to ongoing investigation. 

While various calculations have been performed, there is general agreement that the nuclear pairing energy, or energy gap, falls within the range of $\Delta = 0.1$ to $1$ MeV (see, for instance, \cite{gezerlis2010}). However, a significant degree of uncertainty remains regarding its dependence on density.

Neutron star cooling calculations require the critical temperature as a function of baryon density. To address this, \cite{kaminker2002} introduced a phenomenological formula that has now become widely used for the Fermi wavenumber ($k_F$) dependence of the energy gap at zero temperature. The formula is as follows:
\begin{equation}
    \Delta(k_{F,N}) = \Delta_0 \frac{(k_{F,N} - k_0)^2}{(k_{F,N} - k_0)^2 + k_1} \frac{(k_{F,N} - k_2)^2}{(k_{F,N} - k_0)^2 + k_3}~. 
\end{equation}
This expression holds true for $k_0 < k < k_2$, with the energy gap $\Delta$ vanishing outside this range. Here, $k_{F,N}$ represents the Fermi wavenumber for each type of nucleon, where $N$ can be either a neutron or a proton. The parameters in the formula play specific roles: $\Delta_0$ regulates the amplitude of $\Delta$, while $k_0$ and $k_2$ determine the positions of the low- and high-density cut-offs, respectively. Additionally, $k_1$ and $k_3$ specify the shape of $\Delta(k_{F,N})$. This formula proves valuable as it allows analytical parameterization for numerical calculations in various models, eliminating the need to rely on tables.

\subsection{Specific heat}
\label{subsec: specific heat}

\begin{figure}
    \centering
    \includegraphics[width=.95\textwidth]{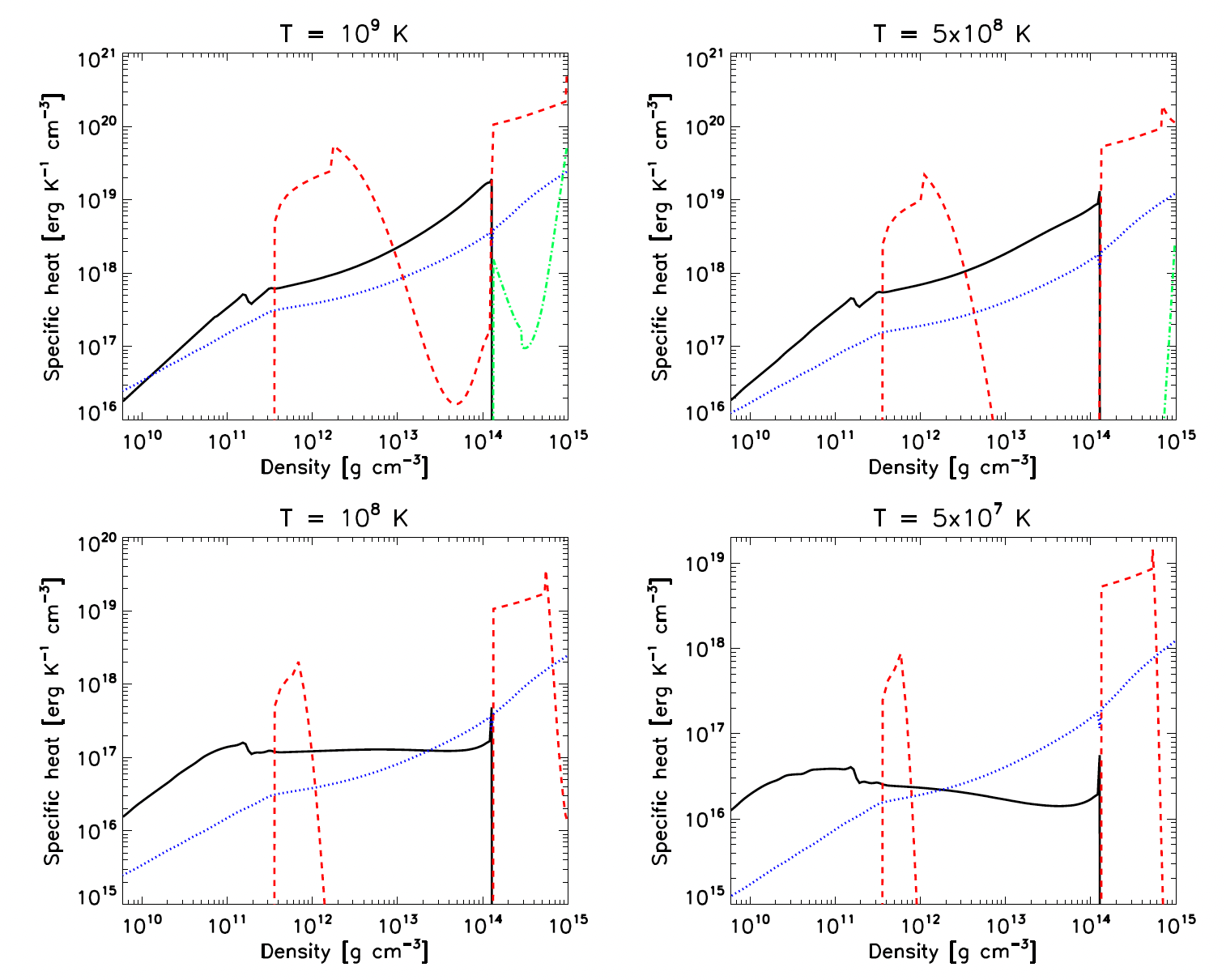}
    \caption[Specific Heat.]{Figure adopted from \cite{pons2019} shows the contributions to the specific heat from neutrons (red dashes), protons (green dot-dashed), electrons (blue dots), and ions (black solid line) as a function of density, spanning from the outer crust to the core. Each panel corresponds to different temperatures, as indicated. The superfluid/superconductive gaps employed in the figure are the same as those in \cite{ho2012}. The neutron star is described with the SLy4 EoS \citep{douchin2001} and has a mass of $1.4$\,M$_{\odot}$.}
    \label{fig: specific heat}
\end{figure}

The specific heat $c_V$ plays an essential role in computing the heat diffusion equation. The majority of the total heat capacity of a neutron star is attributed to the core, which contains most of the mass and is composed entirely of degenerate matter. For degenerate fermions of a species $i$ with number density $n_i$ and Fermi wavenumber $k_{F,i} = (3\pi^2 n_i)^{1/3}$, the heat capacity per unit volume is given by:
\begin{equation}
c_{v,i} = \pi^2 \frac{n_i k_b^2 T}{m_i c^2} \frac{\sqrt{x_{F,i}^2 + 1}}{x_{F,i}^2},  
    \label{eq: specific heat}
\end{equation}
where $x_{F,i} = \hbar k_{F,i} /m_i c$ describes the relativistic effects of the species. Here, $\hbar$ denotes the reduced Planck constant, and it is related to Planck's constant $h$ by the equation $\hbar = h/2\pi$. The contribution of the relativistic electrons is approximated as:
\begin{equation}
c_{v,e} \approx 5.4 \times 10^{19} \bigg(\frac{n_e}{n_0}\bigg)^{2/3} T_9 \,\text{erg cm} ^{-3} \text{K}^{-1},
    \label{eq: electron specific heat}
\end{equation}
and for the non-relativistic nucleons ($N=n,p$), it is given by:
\begin{equation}
c_{v,N} \approx 1.6 \times 10^{20} \frac{m_N^*}{m_N} \bigg(\frac{n_N}{n_0}\bigg)^{1/3} T_9 R^{cv} \,\text{erg cm} ^{-3} \text{K}^{-1},
    \label{eq: nucleon specific heat}
\end{equation}
where $n_0 \approx 0.16$\,fm$^{-3}$ is the nuclear saturation density, $T_9$ is the temperature in units of $10^9$\,K, and $R^{cv}$ is a factor that accounts for the exponential reduction due to the presence of superfluidity \citep{levenfish1994}.

The regions with superfluid nucleons are distinguishable by significant drops in the specific heat. The contribution from protons is consistently negligible. In the outer core, where neutrons are not superfluid, their contribution becomes dominant. The crustal specific heat is determined by the presence of dripped neutrons, the degenerate electron gas, and the nuclear lattice \citep{vanriper1991}. Generally speaking, the specific heat of the lattice makes the most substantial contribution, except in parts of the inner crust where neutrons are not superfluid, or for temperatures $T \lesssim 10^8$\,K, when the electron contribution becomes dominant. Nonetheless, due to the relatively small volume of the crust, its heat capacity is considerably smaller compared to the core contribution.

Fig.~\ref{fig: specific heat} illustrates the diverse contributions to the heat capacity per unit volume for the components of neutron star matter as a function of density, considering various fixed temperatures, namely, $T=10^9, 5\times 10^8, 10^8, 5\times 10^7$\,K. These calculations incorporate the superfluid/superconductive gaps from \cite{ho2012}, particularly utilizing their "deep neutron triplet" model. The computations are performed using the SLy4 EoS and considering a neutron star mass of $M=1.4$\,M$_{\odot}$. In the outer crust, the dominant contribution stems from phonons, which represent vibrating ions. In the inner crust and core, the neutron term dominates. While in the core, the proton term becomes significant, but it never reaches a comparable magnitude to that of neutrons. Consequently, when integrated over the entire star, the dominant component remains the neutron heat capacity.

\subsection{Electric and thermal conductivities}
\label{subsec: conductivity}

Heat transport is governed by thermal conductivity, whereas the transport of momentum by charged particles is regulated by electrical conductivity. These conductivities are determined by the mean free path of the relevant carriers. Scattering processes involving charged particles (such as electrons, protons, and ions) influence both electrical and thermal conductivity, while neutron and phonon processes affect only thermal conductivity. On the other hand, radiative transport involving photons is primarily significant near the surface and will not be discussed here. In the context of neutron star physics, the thermal and electric conductivities are predominantly governed by the electrons due to their high mobility and depend on factors such as electron density, crustal temperature, strength and topology of the magnetic field, and impurity concentration within the crust \citep{fantina2020,carreau2021}.

In situations where the typical energy transferred during a collision is much smaller than $k_b T$, the electron relaxation time approximation is employed. In this approximation, the thermal conductivity tensor can be expressed as:
\begin{equation}
\hat{\kappa}_e = \frac{\pi^2 k_b^2 n_e T}{3 \epsilon_F} \hat{\tau}_e^{th}, 
\end{equation}
and the electrical conductivity tensor as
\begin{equation}
 \hat{\sigma}_e = \frac{n_e c^2 \textit{e}^2}{\epsilon_F}\hat{\tau}_e^{el}.
\end{equation}
Here, $\hat{\tau}_e^{th}$ and $\hat{\tau}_e^{el}$ represent the electron's thermal and electrical relaxation times, respectively, and $\epsilon_F = m_e c^2 \sqrt{x_F^2+1}= m_e^*c^2$ denotes the Fermi energy.

We have introduced the concept of \emph{electron relaxation time} tensors, with their components being $\big(\sum_k \nu_k\big)^{-1}$ -- in other words, the inverse of all the collision frequencies $\nu_k$ associated with the relevant processes involved in the transport of heat or electrical current. In the strongly quantized regime, these tensor components are non-trivial and require precise numerical calculations, taking into account the limited number of occupied Landau levels.

In the presence of a strong magnetic field with direction $\boldsymbol{b} = \boldsymbol{B}/B$, one can express the electron contribution to the electrical currents and anisotropic flux of heat as \citep{urpin1980}:
\begin{equation}
    \boldsymbol{J} = \hat{\sigma} \boldsymbol{E} = \sigma_{||} \boldsymbol{E}_{||} + \sigma_{\perp} \boldsymbol{E}_{\perp} + \sigma_{H} \boldsymbol{b}\times \boldsymbol{E},
    \end{equation}
    \begin{equation}
  \boldsymbol{F} =  - e^{-\zeta}  k_{\perp} \bigg[  \boldsymbol{\nabla} (e^\zeta T)  +  (\omega_B \tau_e)^2 (\boldsymbol{b}\cdot \boldsymbol{\nabla} (e^\zeta T)) \boldsymbol{b} +  \omega_B \tau_e (\boldsymbol{b} \times  \boldsymbol{\nabla} (e^\zeta T)) \boldsymbol{b} \bigg]. 
    \label{eq: anisotropic flux heat}
\end{equation}
Expressing the tensor components in a basis aligned with the magnetic field orientation, one can identify longitudinal, perpendicular, and Hall components. The symmetric part of the $\hat{\sigma}$ tensor is related to the components in the $\boldsymbol{E}$-$\boldsymbol{B}$ plane, specifically $\boldsymbol{E}_{||} = \boldsymbol{b} (\boldsymbol{E}\cdot \boldsymbol{b})$ and $\boldsymbol{E}_{\perp} = \boldsymbol{b} \times (\boldsymbol{E}\times \boldsymbol{b})$. The antisymmetric part is perpendicular to both $\boldsymbol{B}$ and $\boldsymbol{E}$, resulting in a complex structure when the equation is inverted to express $\boldsymbol{E}$ as a function of $\boldsymbol{J}$ and $\boldsymbol{B}$. The same principle applies to $\hat{\kappa}$ and $\boldsymbol{\nabla} (e^\zeta T)$, where the heat flux is also explicitly decomposed into three parts: heat flowing in the direction of the redshift temperature gradient $\boldsymbol{\nabla} (e^\zeta T)$, heat flowing along the magnetic field lines (the direction of $\boldsymbol{b}$), and heat flowing perpendicular to both.

If the quantizing effects are not relevant due to very large number of occupied Landau states, then the following classic relations hold \citep{urpin1980}:
\begin{align}
 &  \tau_{||} = \tau_e,    \nonumber\\
  & \tau_{\perp} = \frac{\tau_e}{1+ (\omega_B \tau_e)^2},    \nonumber\\
  & \tau_H = \frac{\omega_B \tau_e^2}{1 +  (\omega_B \tau_e)^2}.
\end{align}
Here, $\omega_B$ is the gyro-frequency, and $\tau_e$ is the non-magnetic relaxation time.
The conductivities across magnetic field lines are suppressed by a factor of $1+(\omega_B \tau_e)^2$. This leads to anisotropic heat transport. It's worth noting that in this limit, the electric field can be expressed as:
\begin{equation}
    \boldsymbol{E} =  \hat{\sigma}^{-1} \boldsymbol{J},
\end{equation}
where $\hat{\sigma}^{-1}$ is the electrical resistivity tensor. Explicitly, it can be shown that
\begin{equation}
     \boldsymbol{E} = \frac{1}{\sigma} \boldsymbol{J} + \frac{B}{c n_e e} \boldsymbol{J} \times \boldsymbol{b}.
\end{equation}
The first term represents the Ohmic resistivity in the induction equation \eqref{eq: induction equation}, with $\sigma = \sigma_{||}$, is dominated by electrons, and must take into account all the
(usually temperature-dependent) collision processes of the charge carriers. The second term, which is independent of $\tau_e$, gives rise to the Hall effect.

Under weak magnetic field conditions, the electron conductivities are isotropic, and the thermal and electric conductivity tensor, $\hat{\kappa}$ and $\hat{\sigma}$, reduces to a scalar quantity multiplied by the identity matrix. As the background exhibits spherical symmetry, temperature gradients are primarily radial throughout most of the star.

It's important to highlight that the thermal conductivity of the core is significantly higher, by several orders of magnitude, compared to the crust. Consequently, the core remains isothermal, rendering precise values of thermal conductivity in the core unimportant. Thermal gradients can only develop and be maintained in the crust (typically in the first centuries or in the presence of long-term heat sources) and the envelope, where most of the temperature gradient occurs.

In the crust, various dissipative processes contribute to thermal conductivity, including mutual interactions between electrons, lattice phonons (collective motion of ions in the solid phase), impurities (defects in the lattice), superfluid phonons (collective motion of superfluid neutrons), and normal neutrons. The mean free path of free neutrons, limited by interactions with the lattice, is expected to be much shorter than that of electrons. Quantizing effects due to the presence of a strong magnetic field only become significant in the envelope or the outer crust for very large magnetic fields ($B \geq 10^{15}$\,G) and are therefore neglected hereafter.

The electron conductivity accounts for all collision processes of the charge carriers, and it is typically temperature-dependent. 
In the crust of a neutron star, the electrical conductivity typically ranges from $\sigma \sim 10^{22}-10^{25}$\,s$^{-1}$, which is several orders of magnitude larger than in the core, where the electrical conductivity is approximately $\sigma \sim 10^{26}-10^{29}$\,s$^{-1}$ \citep{pons2019}. As a result, the longer Ohmic timescales in the core potentially influence the magnetic field evolution only at much later stages ($t \geq 10^8$\,yr), when isolated neutron stars become too cold to be observed \citep{pons2009,pons2019}.

\subsection{Neutrino reactions} 
\label{subsec: neutrino reactions}

Another crucial ingredient in solving the heat diffusion equation is the neutrino emissivity. Neutrino-emitting reactions take place in a wide range of matter compositions, which differ significantly between the solid crust and the liquid core of a neutron star. The theoretically calculated emissivities are complex functions of temperature and matter composition. In general, neutrino emissivities exhibit a steep dependence on temperature, often expressed as high powers of this factor, whereas their dependence on density remains comparatively weak. Therefore, the higher the temperature, the larger is the energy drained by neutrinos.

Since the first minute after birth, the mean free path of neutrinos has been much longer than the star's radius \citep{burrows1986, pons1999}. As a result, when the star is still hot, the abundant production of neutrinos carries energy away from it. During the first $t \approx 10^5$\,yr of a neutron star's life, the typical cooling curves, represented by the redshifted photon luminosity $L_{\gamma}$ versus stellar age, are strongly influenced by the dominant neutrino emission mechanisms \citep{lattimer1991,prakash1992,yakovlev1999,yakovlev2001,page2004,potekhin2015}. These mechanisms drain energy from both the crust and the core of the neutron star. Table~\ref{table: neutrino emissivity} provides a brief overview of the main neutrino emission channels that are active in the stellar interior.

\begin{table}
    \centering
    \begin{tabular}{|lcccccc|c}
    \hline
      \textbf{Process}   &  & $Q_\nu$ [erg cm$^{-3}$ s$^{-1}$]& & \textbf{Onset} &&   \textbf{Ref} \\
    \hline
    \hline
     \textbf{Core} &&&&&&\\
    \hline 
   \hline
    Modified Urca (n-branch)    &&&& & & \\
      $nn \rightarrow pne\Bar{\nu}_e ,~ pne \rightarrow nn\nu_e$  && $8\times 10^{21} \mathcal{R}^{MU}_n n_p^{1/3} T^8_9  $ &&  && \small$(1)$\\
    Modified Urca (p-branch)    &&&& & &
    \\
       $np \rightarrow pp e\Bar{\nu}_e,~ ppe \rightarrow np\nu_e$           && $8\times 10^{21} \mathcal{R}^{MU}_p n_p^{1/3} T^8_9  $ && \small$Y^c_p = 0.01$ && \small$(1)$ \\
        \hline
  N-N bremstralung &&&&  & &  \\
  $nn \rightarrow nn \nu  \Bar{\nu}$ && \small $7\times 10^{19} \mathcal{R}^{nn}n_n^{1/3} T^8_9 $  && && \small$(1)$ \\
   $np \rightarrow np \nu  \Bar{\nu}$ &&  \small $1\times 10^{20} \mathcal{R}^{np}n_p^{1/3} T^8_9 $  && && \small$(1)$ \\
    $pp \rightarrow pp \nu  \Bar{\nu}$ &&  \small $7\times 10^{19} \mathcal{R}^{pp} n_p^{1/3} T^8_9 $ && && \small$(1)$ \\
  \hline
    e-p Bremsstrahlung &&  &&  && \\
   $ep \rightarrow ep\nu\Bar{\nu}$ && \small$2\times 10^{17} n_B^{-2/3} T^8_9$ &&  &&\small$(2)$ \\
    \hline
   Direct Urca  &&  && && \\
      $n \rightarrow p  e    \Bar{\nu}_e,~ p  e \rightarrow n  \nu_e $ &&  \small$4\times 10^{27} \mathcal{R}^{DU} n_e^{1/3} T^6_9$  && \small$Y^c_p = 0.11$ && \small$(3)$   \\
       $n \rightarrow p  \mu^-    \Bar{\nu}_\mu,~ p  \mu^- \rightarrow n  \nu_\mu $ &&  \small $4\times 10^{27} \mathcal{R}^{DU} n_\mu^{1/3} T^6_9$ && \small$Y^c_p = 0.14$&& \small$(3)$ \\
      \hline
      \hline
       \textbf{Crust} & &&& &&\\ 
      \hline
      \hline
    Pair annihilation  &&&&   & & \\
     $e e^+   \rightarrow    \nu  \Bar{\nu}$ &&  \small$9\times 10^{20} F_\text{pair} (n_e,n_{e^+}) $  && && \small$(4)$ \\ 
       \hline
       Plasmon decay  &&   && && \\ 
       $\Gamma \rightarrow \nu + \Bar{\nu}$ && \small$1\times 10^{20} I_{pl} (T,y_e) $ && && \small$(5)$ \\
       \hline
      $e$-nucleus Bremsstrahlung   &&   && && \\
      $e  (A,Z) \rightarrow   e (A,Z)   \nu  \Bar{\nu}$ && \small$3\times 10^{12} L_{eN} Z \rho_0 n_e T_9^6 $   && && \small$(6)$ \\
      \hline 
         $N-N$ Bremsstrahlung  &&   && && \\
      $ n  n \rightarrow  n  n   \nu  \Bar{\nu}$ && \small$7 \times 10^{19} R^{nn} f_\nu n_n^{1/3} T^8_9 $  && && \small$(1)$ \\
        \hline
    \hline
   \textbf{Core and crust} &&&& &&\\
    \hline 
  \hline
     Cooper Pair Breaking and Formation (CPBF) &&  &&  &&  \\
       $\tilde{B}  \tilde{B} \rightarrow \nu \Bar{\nu} $ && \small$1\times10^{21} n_N^{1/3} F_{A,B} T^7_9$   &&  && \small$ (7)$\\
        \hline
       Electron Synchrotron  &   & &&&& \\
    $e \xrightarrow{B}e    \nu  \Bar{\nu} $ && \small$9\times 10^{14} S_{AB,BC}B_{13}^2 T_9^5 $    && && \small$(8)$\\
      \hline  
    \end{tabular}
    \caption[Neutrino emissivities]{Neutrino processes and their emissivities $Q_{\nu}$ in the core and in the crust of a neutron star, taken from \cite{aguilera2008}. The third column shows the onset for some processes to operate ($Y^c_p$ denotes the critical proton fraction). Detailed functions and precise factors can be found in the references (last column): (1) \cite{yakovlev1995}; (2)\cite{maxwell1979}; (3) \cite{lattimer1991}; (4) \cite{yakovlev1994}; (5) \cite{yakovlev2001}; (6) \cite{haensel1996}; (7) \cite{yakovlev1999}; (8) \cite{bezchastnov1997}. The proton fraction threshold for the modified Urca and DUrca onset is described for the SLy4 EoS.}
    \label{table: neutrino emissivity}
\end{table}

In the case of a sufficiently massive star, the proton concentrations in the core may surpass the minimum threshold required (depending on the EoS) for activating nucleonic direct Urca (DUrca) reactions, leading to \emph{enhanced cooling scenario} \citep{lattimer1991,prakash1992,haensel1994,yakovlev2001}, in which a temperature drop is predicted after tens of years. This process involves the beta decay of neutrons and the electronic capture by protons. However, DUrca reactions are not always allowed. Energy and momentum conservation, combined with the condition of electric charge neutrality in matter, impose that DUrca is permitted only if the Fermi momenta $pF_{n}$, $pF_p$, and $pF_{e,\mu}$ satisfy the triangle condition \citep{lattimer1991,yakovlev1999}. According to this condition, the value of each Fermi momentum should be smaller than the sum of the other two. In neutron star matter, $pF_{n}$ is larger than the other two, and the triangle condition reads \citep{yakovlev2001}:
\begin{equation}
    pF_n <  pF_p + pF_{e,\mu}.
\end{equation}
The given condition for Fermi momenta can also be expressed similarly in terms of baryonic fractions $Y_i = n_i/n_b$. The baryonic fraction $Y_p$ must exceed the value of $Y^c_p = 0.11$ for matter composed of a classical neutrons, protons, and electrons gas. More accurate descriptions report values of $Y^c_p = 0.14$. Typically, DUrca mechanism operates in the innermost regions of the star \citep{lattimer1991,yakovlev1999,yakovlev2001}. 

Neutron stars in which DUrca processes are inactive undergo \emph{standard cooling scenario}, where the total emissivity is dominated by slow processes $\propto T^8$. The standard cooling is mainly governed by the modified Urca, the different Bremsstrahlung processes, and Cooper pair neutrino emissions, resulting in a slower thermal evolution compared to the enhanced regime. The modified Urca, is similar to DUrca but less efficient, involving a mediating particle. In neutron, proton, and electron matter, the modified Urca process is often referred to as the neutron branch and the proton branch \citep{yakovlev1995,yakovlev1999,yakovlev2001}. These reactions have analogs entailing muons instead of electrons \citep{yakovlev2001,raduta2018,raduta2019}. The condition for the modified Urca reactions is less restrictive and is almost always satisfied \citep{yakovlev2001}:
\begin{equation}
    pF_n <  3 pF_p + pF_{e,\mu}.
\end{equation}
Other essential processes include the Bremsstrahlung channels, named for a vague analogy with the electrodynamical process. These channels proceed via weak neutral currents and produce neutrinos of any flavor.

Most of the effective neutrino processes involving superfluid components (such as modified URCA and nucleon Bremsstrahlung) are exponentially suppressed by factors fitted using the mathematical functions $\mathcal{R}$ as described in \cite{yakovlev2001}. The onset of superfluidity opens new channels in neutrino production: the Cooper Pair Breaking and Formation processes (CPBF). According to \cite{leinson2006}, they are suppressed in the crust (in the neutron $^1S_0$ pairing), but they are active in the core (in the proton $^1S_0$ and neutron $^3P_2$ channels). Crustal superfluidity has a minor impact on long-term evolution, affecting only the early relaxation of the crust, a stage lasting about $100$\,yrs, before thermal coupling between the crust and core is achieved \citep{gnedin2001,page2009}. The bulk of energy loss is regulated by the core's physics, where most of the mass is concentrated. Consequently, the total neutrino emissivity is highly sensitive to the superfluid gaps. Notably, the reduction of specific heat $c_V$ due to superfluidity and the Cooper pair channel overcome the suppression of nucleonic neutrino channels, resulting in a faster cooling curve.

A further complication arises from the presence of a strong magnetic field. In that case, the relativistic electrons have a different dispersion relation and they can directly emit neutrino pairs, analogously to the synchrotron emission of photons, with an emissivity proportional to $B$. This process is known as the neutrino synchrotron channel. A few processes, such as pair annihilation and photo-neutrino emissivity, are relevant only in the envelope. 

\section{Coupled magneto-thermal simulations}
\label{sec: MT evolution}

The interconnections between the magnetic and thermal evolution equations manifest at the microphysical level. On one hand, as temperatures decrease due to neutrino emission, thermal and electric conductivities increase, rendering matter more resistant. On the other hand, the evolving magnetic field impacts thermal conductivity both along and across its lines, thus influencing the local temperature. This, in turn, causes significant variations in the surface temperature distribution, which can be observed and constrained through measurements. Simultaneously, the Hall term drives electric currents towards the crust-core boundary, where the presence of high impurities and pasta phases facilitates a more efficient dissipation of the magnetic field \citep{pons2013}. Consequently, the magnetic energy is converted into Joule heating $Q_J$. While the Hall term itself does not directly dissipate magnetic energy, it gives rise to small-scale magnetic structures where Ohmic dissipation is enhanced. To a lesser extent, the magnetic field $\boldsymbol{B}$ also influences $c_V$ and $Q_\nu$.

To investigate the long-term magneto-thermal evolution within the interior of a neutron star, specialized numerical simulations using advanced codes are essential. These simulations serve as powerful tools for understanding the complex interplay between magnetic fields and thermal processes that govern the star's behavior over extended periods. In this thesis, studies were carried out using two magneto-thermal evolution codes: a 2D code available within our collaboration (latest version by \cite{vigano2021}) and a newly developed 3D code named \emph{MATINS}, introduced in Chapter~\ref{chap: MATINS} (\cite{dehman2022}, Ascenzi et al. in preparation). 

\subsection{Background model}
\label{subsec: background model}

Since the induction equation and the heat diffusion equation are coupled at a microphysical level, they must be supplemented by an EoS. This EoS enables the construction of the neutron star's background model, permitting interpolation of the provided tables through various schemes to compute at each point of the star the microphysics, e.g., $\eta_b$, $c_V$, $\hat\kappa$, and $Q_\nu$, essential for our magneto-thermal simulations. Moreover, the superfluid and superconductive models for neutrons and protons, respectively, are taken into account as they significantly impact the cooling timescales through their influence on $c_V$ and $Q_\nu$. Various additional microphysical ingredients play a crucial role in the interior of a neutron star. Notably, the EoS and superfluid models have a significant effect on the cooling process, but they play a comparatively smaller role in the evolution of the magnetic field, especially when compared to the initial topology chosen for the system.

Utilizing the SLy4 EoS sourced from the online public database CompOSE\footnote{\url{https://compose.obspm.fr/}} (CompStar Online Supernovae Equations of State), we have constructed a neutron star model with a radius of $R_\star = 11.7$\,km and a mass of 1.4\,M$_\odot$, achieved by imposing a central pressure of $1.36 \times 10^{35}$\,bar. For this particular choice of EoS, our computational domain covers the range from $R_c = 10.9$\,km to $R = 11.6$\,km, spanning from the crust-core interface to a density of approximately $\rho \sim 10^{10}$\,g$\,$cm$^{-3}$, which we designate as the crust-envelope interface.

\begin{figure}
    \centering
    \includegraphics[width=0.8\textwidth]{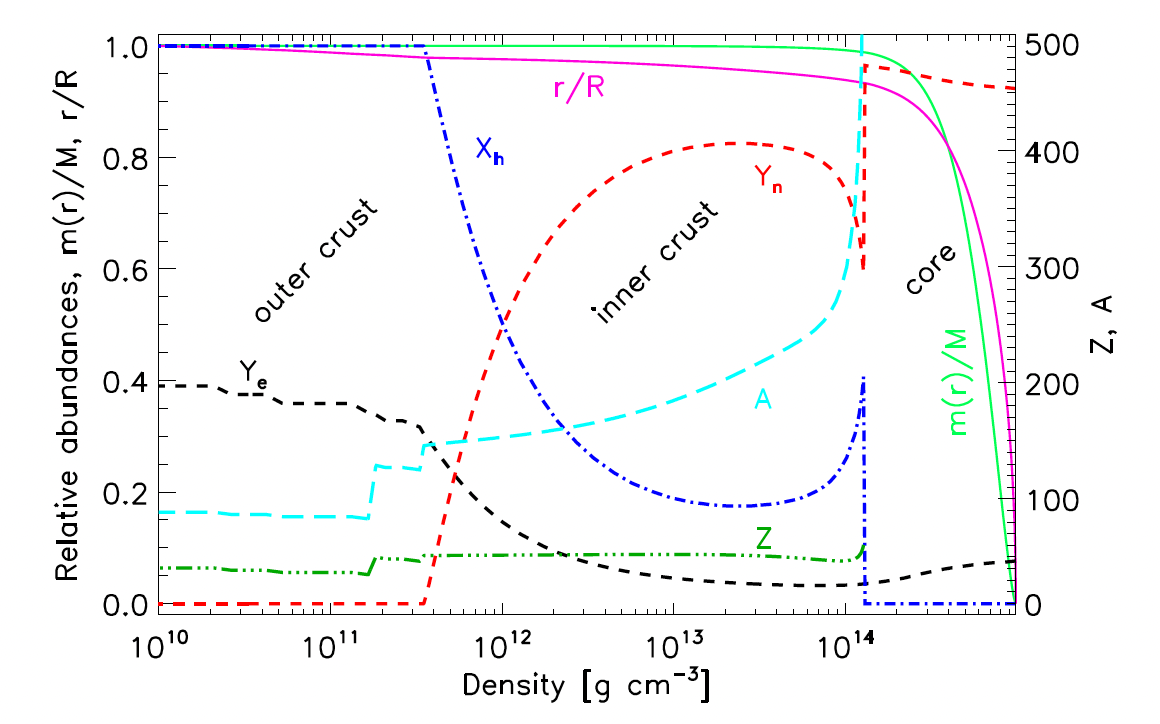}
    \caption{Figure adopted from \cite{pons2019}. It displays the structure and composition of a $1.4$ M$_\odot$ neutron star, with SLy4 EoS. The plot shows, as a function of density from the outer crust to the core, the following quantities: mass fraction in the form of nuclei $X_h$ (blue
dot-dashed line), the fraction of electrons per baryon $Y_e$ (black dashes), the fraction of free neutrons per baryon $Y_n$ (red dashes), the atomic number $Z$ (dark green triple dot-dashed), the mass number $A$ (cyan long dashes), radius normalized to $R$ (pink solid), and the corresponding enclosed mass normalized to the star mass (green solid).}  
\label{fig: composition}
\end{figure}

In Fig.~\ref{fig: composition}, we present a typical profile of a neutron star obtained using the SLy4 EoS \citep{douchin2001}. The SLy4 EoS is among the realistic EoSs that support a maximum mass consistent with observations, approximately $M_\text{max} \sim 2 - 2.2$\,M$_\odot$ \citep{demorest2010, antoniadis2013, margalit2017, ruiz2018, radice2018, cromartie2020}. Fig.~\ref{fig: composition} illustrates the enclosed radius and mass, as well as the fractions of different components, all as functions of density, spanning from the outer crust to the core of the neutron star. At densities $\rho \gtrsim 4\times 10^{11}$\,g\,cm$^{-3}$, neutrons separate from the nuclei, marking the beginning of the inner crust, and, under sufficiently low temperatures, they become superfluid. It's worth mentioning that this representation does not include the envelope and atmosphere of the neutron star. For a more comprehensive discussion, additional details can be found in \cite{haensel2007} and \cite{potekhin15b}.

Throughout most of this thesis, our primary focus remains on a single model: the SLy4 EoS and a mass of 1.4\,M$_\odot$. Nonetheless, it's important to note that we have rigorously tested the two codes with various EoSs and masses. For instance, in Chapter~\ref{chap: Comparison with observations}, we delve into different choices of masses and EoSs to explain the faint luminosity of young neutron stars observed in our galaxy. Below is a compilation of the various EoSs currently available in both codes: SLy2 \citep{danielewicz2009}, SLy4 \citep{douchin2001}, BSK21 \citep{potekhin2013}, BSK22-24-25-26 \citep{pearson2018}, SkMp \citep{bennour1989}, Ska and SKb \citep{kohler1976}, CMF2, and CMF6 \citep{dexheimer2008}. For those interested in exploring these EoSs in more detail, valuable information and data tables are available through the online CompOSE database.

\subsection{Numerical methods}
\label{subsec: numerical methods}

The 2D and the 3D code \emph{MATINS} are implemented in Fortran90/95 and share a common code base. They contain modules devoted to physics (thermal evolution, magnetic evolution and microphysics), data structures and support (grid and constants), utility modules (output), and an external module for utilizing third-party libraries. Both codes feature a fully modular structure, making them easier to maintain, develop, and extend.  The 2D code incorporates an integrated CMake build system. This is not the case for the 3D code, but the latter uses OpenMP to optimize the main loops. The techniques employed in these codes are based on a conservative formulation, where Stokes' and Gauss' theorems are applied to each numerical cell, as elaborated in Chapter~\ref{chap: MATINS}. 

\begin{figure}
\centering
\includegraphics
[width=0.6\textwidth]{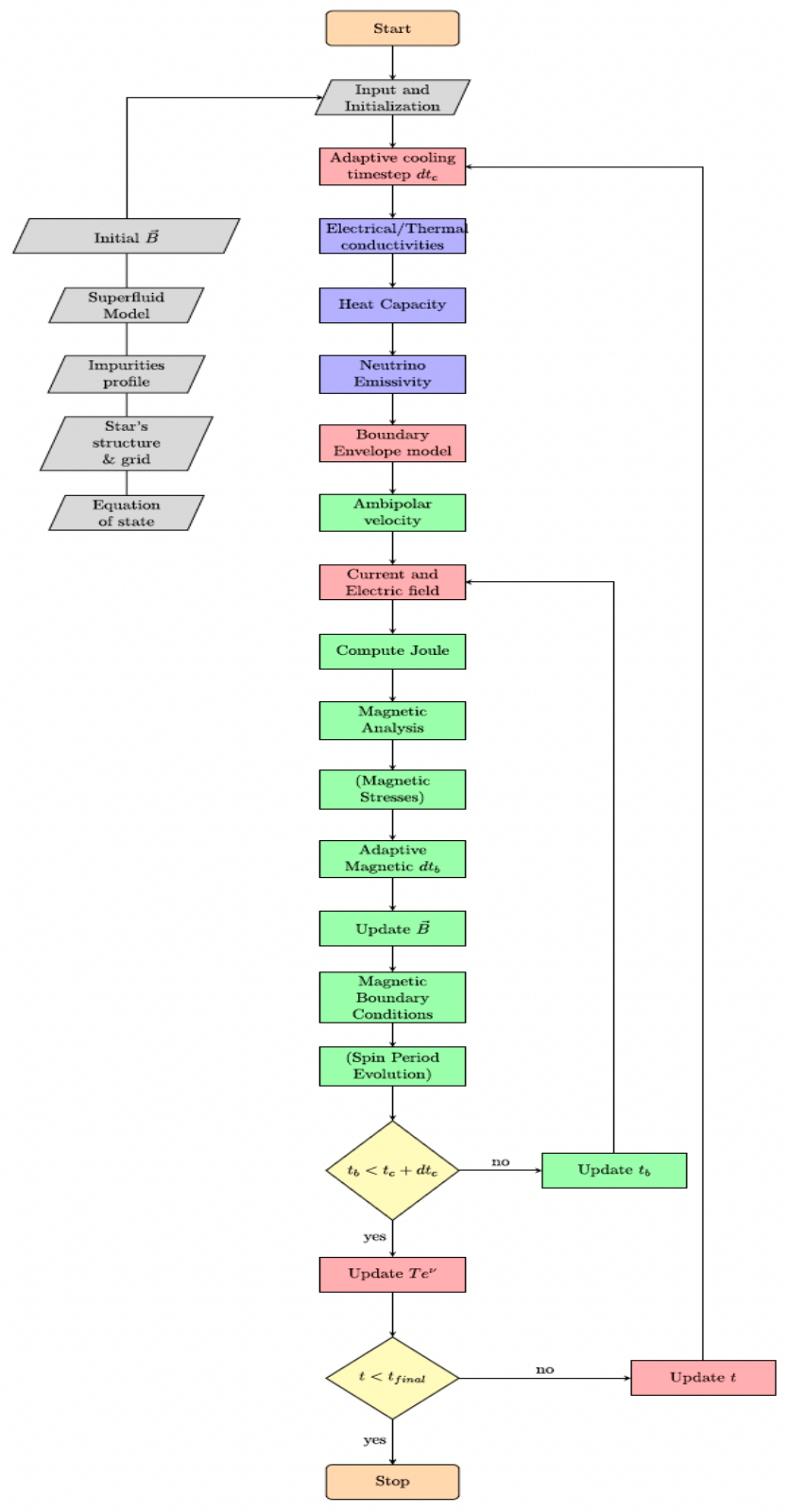}
\caption[Flowchart of the magneto-thermal 2D evolution code]{Figure adopted from \cite{vigano2021}. It represents the flowchart of the magneto-thermal 2D evolution code. Thermal evolution blocks are highlighted in red, microphysics parts in blue, and magnetic evolution steps in green. In parenthesis there are the optional by-products of the calculations.}
\label{fig: structure of the 2D code}
\end{figure} 

Fig.~\ref{sec: structure} provides a flowchart of the main program of the code, adapted from \cite{vigano2021}, and representative of both the 2D and 3D versions. The initialization stage calculates the star's structure for a given EoS and central pressure, while input parameters such as initial temperature, magnetic field strength and topology, impurity parameter, superfluid model, envelope model, and numerical methods are provided. Various fixed quantities and mathematical functions, such as the numerical grid, relativistic factors, geometrical elements, Legendre polynomials (in the 2D code) and spherical harmonics (in the 3D code), are also computed. The code then enters the main loop, which comprises three main parts: microphysical calculations, magnetic field evolution, and thermal evolution. Additionally, rotational evolution is a by-product derived from the evolution of the dipolar component at the surface ($B_\text{pol}^\text{dip}$). Therefore, it can also be performed as a post-process, using a given spin-down formula, which may or may not include inclination angle evolution. Magnetic stresses and the frequency of crustal failure events can also be estimated as by-products.

The evolution of the magnetic field requires the use of an explicit, centered, or upwind scheme, with careful consideration of the non-linear term. Various time advancement methods, such as Euler, Runge-Kutta, and Adams-Bashforth, are employed to advance the magnetic field in time. A comprehensive description of the 3D magnetic evolution equation is provided in Chapter~\ref{chap: MATINS}.

We confine the magnetic field evolution to the crust of a neutron star. Consequently, we apply perfect conductor boundary conditions at the crust-core interface, resulting in the vanishing of the radial magnetic field component (see \S\ref{subsec: inner BC}). At the star's surface, potential boundary conditions (imposing no currents circulating outside the star) are commonly adopted in most studies (see \S\ref{subsec: outer B.C}). However, our recent research \citep{urban2023} takes a different approach by coupling the interior crustal evolution with the magnetosphere using physics-informed neural networks (PINN), with a detailed description provided in Appendix~\ref{appendix: neural networks}. Although this coupling was initially performed in 2D, we are currently working on implementing it in 3D.

The heat diffusion equation can be solved using standard methods for parabolic equations with stiff terms, given the highly nonlinear nature of neutrino emissivities with respect to temperature (see \S\ref{subsec: neutrino reactions}). The Joule term $\propto \sigma_e (T)^{-1}$ can also be considered stiff, although its dependence on $T$ is less pronounced. Such stiffness is effectively managed by implicit methods, which involve linearizing the source term and inverting the tridiagonal block matrix $\mathscr{M}$. This matrix relates the updated set of values $\Tilde{T}_a^{n+1}$ to the previous set $\Tilde{T}_a^{n}$, where $\Tilde{T}\equiv T e^\zeta$ is the redshifted temperature, and $a$ labels each cell. The relationship is expressed as $\mathscr{M}^{ab} \Tilde{T}_b^{n+1} = v_a (\Tilde{T}_a^{n})$, where $v_a$ is the vector that also contains the old temperatures, the sources, and their dependencies on the local temperatures $T_a^n$. The matrix elements arise from the discretization of the problem on a spherical coordinate grid in the 2D code and the cubed-sphere coordinate grid (\S\ref{sec: cubed sphere formalism} and Appendix~\ref{appendix: cubed-sphere formalism}) in the 3D code, using standard centered differences to evaluate the gradients in the heat flux $\boldsymbol{F}$. Temperature values are defined at the center of each cell, where the heating rate and neutrino losses are also evaluated. To facilitate this task, our code utilizes the LAPACK public library \citep{lapack}. For a comprehensive understanding of solving the heat diffusion equation, please refer to \cite{aguilera2008} and Ascenzi et al. 2023 (in preparation).

The core becomes isothermal, with a constant $\Tilde{T}$ after approximately $100$\,yrs. Therefore, in our code, we treat the core evolution as a 1D problem by considering the core as a single radial layer. This approach significantly reduces computational time. We achieve this by considering the correct weighted average of the specific heat and the neutrino emissivity in the core, but evolve only one temperature, considering the thermal conductivity only at the crust-core interface.
On the other hand, to determine the temperature at the star's surface, we must take the star's envelope into consideration, as described in \S\ref{subsec: blanketing envelope}.

It is worth noting that temperatures are not allowed to be smaller than $10^6$\,K because the microphysics implemented are not suitable for such regimes. Therefore, $10^6$\,K is taken as a floor value, which means that the cooling model can follow the star for a maximum of approximately $10^6$\,yrs.

\subsubsection{Adaptive timesteps}
\label{subsec: timestep}

The cooling and magnetic timescales undergo significant variations throughout the star’s life, with the thermal timescales considerably larger than its magnetic counterpart. During the neutron star's cooling process, two effects emerge: first, neutrino emissivities decrease by several orders of magnitude; second, the matter becomes more thermally and electrically conductive. Consequently, employing a fixed timestep spanning Myr-long duration would result in an unnecessarily substantial computational burden. It is therefore advisable to adopt distinct dynamic definitions for the numerical timestep, with one for each equation.

For cooling simulations, one can use a timestep, denoted as $dt_c$, that increases with time, as temperature variations are more significant in the early stages. In our case, we typically employ a phenomenologically increasing timestep, starting with $dt_c = 10^{-2}$\,yr during the initial years when the temperature drops rapidly. We then increase it to $dt_c \sim t/100$\,yrs until it reaches a relatively a large value which is kept uniform, approximately $dt_c \sim 100 - 1000$\,yrs. While this choice may not be finely tuned for optimization, it is a practical implementation that ensures stability in the implicit scheme described above.

The timestep used in magnetic evolution, denoted as $dt_b$, is more complex to determine because the intensity and topology of the magnetic field define the Ohmic and Hall timescales, in addition to being influenced by conductivity and electron density (\S\ref{subsec: discretization}). Any precise assessment of the Courant-limited maximum value for $dt_b$ is challenging due to the non-linearity of the problem. As a matter of fact, the characteristic velocities of the eMHD equations can only be obtained in their linearized version, as discussed in \cite{vigano2019}, which involves perturbations on top of a background field -- unlike our realistic scenario.

It is worth noting that the cooling timesteps are more memory-intensive because they require solving a system of equations with an $N^2$ sparse matrix ($N$ is the total number of points). During each cooling timestep, the microphysics must be calculated at each point. In contrast, the magnetic timesteps are much smaller, typically a fraction of a year, and more evenly spaced in time (\S\ref{subsec: discretization}). Therefore, each cooling timestep encompasses many magnetic timesteps. Currently, we use a timestep that is proportional to time itself for the thermal part. Roughly speaking, it takes about $10^3 - 10^4$ iterations for the thermal part and about $10^7$ iterations for the magnetic part to reach $1$\,Myr of evolution.

\subsubsection{Thermal boundary condition: the blanketing envelope}
\label{subsec: blanketing envelope}

In the low-density region, where the envelope extends approximately $100$\,m further (see \S\ref{subsec: envelope}), the largest temperature gradient is observed. Due to its relatively low density, radiative equilibrium is established much more quickly compared to the interior, and the envelope reaches a stationary state on shorter timescales. The significantly different thermal relaxation timescales between the envelope and the interior (crust and core) make it computationally infeasible to attempt long-term ($\approx$ Myr) cooling simulations on a numerical grid that includes all layers up to the star's surface. As a result, the outer layer is effectively treated as a boundary condition for the thermal evolution. Consequently, the typical approach is to separately compute stationary envelope models and then employ a phenomenological fit to predict the value of the local surface temperature ($T_s$). This prediction depends on several factors, including the magnetic field's geometry and intensity, the chemical composition of the envelope (which is uncertain), the crust/envelope boundary density, as well as the temperature at the base of the envelope ($T_b$).

A widely used classical $T_b-T_s$ expression, derived from fitting a set of heavy-element envelopes for various $T_b$ values and surface gravity $g_{14}$, is the non-magnetized envelope described by \cite{gudmundsson1983}:
\begin{equation}
T_{b8} = 1.288 \Bigg[\frac{T_{s6}^4}{g_{14}} \Bigg]^{0.455}, 
    \label{eq: Tb-Ts Gud}
\end{equation}
where $g_{14}$ is the surface gravity in units of $10^{14}$\,cm\,s$^{-1}$, $T_{b8}$ is $T_b$ in $10^8$\,K, and $T_{s6}$ is $T_s$ in $10^6$\,K. More elaborate relations, incorporating different values for the crust/envelope boundary density, magnetic field intensity and geometry, and chemical composition, are discussed in the literature. These diverse envelope models are included in both codes. For a detailed examination of the impact of these various envelopes on the cooling process, we refer the reader to Chapter~\ref{chap: envelope}.

The \emph{bolometric luminosity}, as seen by an observer at infinity, is derived by assuming blackbody emission from each patch of the neutron star's surface at the temperature $T_s$:
\begin{equation}
L_{\gamma} = 2 \pi \int_0^{\pi} \int_0^{2\pi} \sigma_{sb} R^2_{\infty} T^4_{\infty} \sin\theta d\theta d\phi.
\label{eq: luminosity}
\end{equation}
Here, $T_\infty = e^\zeta T_s$ and $R_\infty = e^{-\zeta} R_{\star}$ are the redshifted surface temperature and the radius seen at infinity, respectively. $\sigma_{sb} = 5.67 \times 10^{-5}$\,erg cm$^{-2}$\,s$^{-1}$\,K$^{-4}$ is the Stefan-Boltzmann constant. 

Instead of pure blackbody emission, alternative emission models have been proposed, such as magnetic atmospheres \citep{vanadelsberg2006,ho2007,ho2008,Suleimanov2009,Suleimanov2012} or condensed surface models \citep{vanadelsberg2005,perezazorin2005,medin2007,potekhin2012}. In our simulations, surface radiation is typically approximated as blackbody radiation. Although this might be a simplification, it has been demonstrated that it does not significantly impact our results \citep{potekhin15b}. In general, different emission models influence the spectrum's characteristics in various ways, thus playing a crucial role in interpreting observations as we deduce physical properties through spectral fitting to specific emission models.
\clearemptydoublepage
\let\textcircled=\pgftextcircled
\chapter{Magnetized Envelope Models}
\label{chap: envelope}

\initial{R}esearchers have long hoped that comparing observations of direct thermal emission from the surface of neutron stars to theoretical cooling models (i.e., the long-term evolution of thermal luminosity and temperature) could yield valuable information about the star's interior, such as the nuclear EoS and chemical composition \citep{tsuruta1971,yakovlev2004,page2006,potekhin2015,pons2019}.

This chapter builds upon the research conducted by \cite{dehman2023b}. Our primary objective is to demonstrate that the envelope models, which are incorporated into multidimensional simulations "solely" as boundary conditions (\S\ref{subsec: blanketing envelope}), play a crucial role in establishing a connection between the internal properties and the observable quantities (e.g., effective temperature and luminosity), particularly for highly magnetized objects like magnetars. To achieve this, we thoroughly compare various envelope models from the existing literature, aiming to assess their impact on the evolution of theoretical cooling models under varying magnetic field intensities and geometries. Through this comprehensive analysis, we seek to determine the optimal conditions that enable us to employ observational X-ray data to effectively constrain cooling models and derive essential parameters of the neutron star.

\section{Envelope models}
\label{sec: Envelope models}

Among the many early studies of the thermal structure of neutron stars, we must mention the seminal works of \cite{tsuruta1971}, \cite{gudmundsson1983}, \cite{hernquist1985} and \cite{schaaf1990}, who pointed out that regions with tangential magnetic field are much colder than the regions where the field is nearly radial (see \cite{yakovlev1994} and references therein for a review of the early works). The $T_b-T_s$ relation for the envelope model proposed by \cite{gudmundsson1983}, assuming a non-magnetized envelope composed of iron and iron-like nuclei, is expressed in eq.\,\eqref{eq: Tb-Ts Gud}.
Later, \cite{potekhin1997} constructed a more general fit valid for different compositions. These envelope models incorporated enhanced calculations of the EoS and opacities within the outer layers of neutron stars. In particular, the so-called accreted envelopes contain layers of different chemical elements (H, He, C, O shells) created from accreted matter from the supernova fallback material. For the partially accreted envelopes, the $T_b-T_s$ relation proposed by \cite{potekhin1997} is expressed as follow:
\begin{equation}
    T^4_{s,6} = \frac{a T_{s6,\text{Fe}}^4 + T_{s6,a}^4}{a+1},
\end{equation}
where $ a = \big[ 1.2 + (5.3\times 10^{-6}/\varrho)^{0.38}  \big] T_{b9}^{5/3}$. $T_{b9}$ is $T_b$ in unit of $10^9$\,K and 
\begin{equation}
     \varrho = g_{14}^2 \Delta M/M_\odot.
     \label{eq: accretion mass}
\end{equation}
Here, $M_{\odot}$ is the mass of a neutron star and $\Delta M$ is the accreted mass. The surface temperature for a fully accreted envelope is expressed by the following equation:
\begin{equation}
    T^4_{s6,a} = g_{14}\, (18.1 \, T_{b9}\,)^{2.42}.
\end{equation}
The surface temperature for a purely iron (non-accreted) envelope is expressed by the following equation:
\begin{equation}
    T^4_{s6,\text{Fe}} = g_{14} \Bigg[ \bigg(7 \Upsilon\bigg)^{2.25} + \bigg(\frac{\Upsilon}{3}\bigg)^{1.25} \Bigg],
\end{equation}
with $ \Upsilon = T_{b9} - (\,T_*/10^3\,) $ and $T_*= (\, 7\, T_{b9} \, \sqrt{g_{14}}) ^{1/2}$.

The pioneering work of \cite{page1994} and \cite{page1996} introduced realistic surface temperature distributions for neutron stars with dipolar and dipolar $+$ quadrupolar magnetic fields. The latter study specifically provided $T_b - T_s$ relationships associated with these magnetic field configurations. The thermal structure of neutron stars with magnetized envelopes was initially studied by \cite{potekhin2001} and subsequently refined by \cite{potekhin2003}, while considering various compositions instead of an iron-only envelope.
It includes the effect of magnetic fields on the $T_b-T_s$ relation, providing analytical fits valid for a magnetic field strength up to $10^{16}$\,G and arbitrary inclination angles of the field lines with respect to the normal to the surface. Similar studies exploring other field topologies were done by \cite{geppert2004,geppert2006} and \cite{perezazorin2006}.
Subsequent calculations in \cite{potekhin2007} included the effect of the neutrino emissivity in the outer crust. Following that, \cite{pons2009} reexamined the problem of the magnetized envelope with two objectives: (i) improving the microphysical inputs, specifically upgrading the thermal conductivity, as the contribution of ions or phonons to the envelope's thermal conductivity can alleviate the anisotropy of heat conduction \citep{chugunov2007}; (ii) assessing the accuracy of the plane-parallel approximation employed in \cite{pons2009} envelope. When the latter approach is applied to a spherical star, meridional heat fluxes in the envelope are not allowed and, therefore, this approximation may be inaccurate when these fluxes compete with the purely radial ones. The following form has been found to fit the $T_b-T_s$ relation proposed by \cite{potekhin2001,potekhin2003, pons2009}
\begin{equation}
    T_s (B, \vartheta, g, T_b) \approx T^0_s(g,T_b) \, \mathcal{X}(B, \vartheta,T_b),
    \label{eq: Ts Tb pons potekhin}
\end{equation}
where 
\begin{equation}
    \mathcal{X}(B, \vartheta,T_b) =  \bigg[ \mathcal{X}_{||}^\iota(B,T_b) \,  cos^2 \vartheta +\mathcal{X}_{\perp}^\iota(B,T_b)\, sin^2 \vartheta \bigg]^{1/\iota}.
\end{equation}
The functional forms of $\mathcal{X}_{||}$ and $\mathcal{X}_\perp$, as well as the values of $T^0_s$ and $\iota$, are determined by constructing a set of stationary envelope models with varying values of $g$, $T_b$, and $B$, while maintaining a consistent magnetic field geometry (i.e., different values of $\vartheta$ at different latitudes). In this context, $\vartheta$ represents the angle between the magnetic field and the normal to the surface. 

Subsequently, the surface temperature of a partially accreted envelope, as described by \cite{potekhin2003}, can be approximated using interpolation:
\begin{equation}
    T_s = \bigg[\gamma T_{s6,a}^4 + \bigg(1-\gamma\bigg)\, T_{s6,\text{Fe}}^4  \bigg]^{1/4},
\end{equation}
\begin{equation}
    \gamma = \bigg[ 1+ 3.8 \, \big(0.1 \Upsilon\big)^9 \bigg]^{-1} \bigg[ 1+ 0.171 \, \Upsilon^{7/2}\, T_{b9} \bigg]^{-1}, ~~~~ \Upsilon = - \text{log}_{10} \big(10^{6} \varrho \big).
\end{equation}
Here, $\varrho$ is defined in eq.\,\eqref{eq: accretion mass}, and $T_{s6,a}$ and $T_{s6,\text{Fe}}$ follow the same form as defined in eq.\,\eqref{eq: Ts Tb pons potekhin}.

The state-of-the-art models can be found in the thorough review by \cite{potekhin2015}. They present new fits for non-accreted magnetised envelopes, including both the effects of neutrino emission and the effects of non-radial heat transport. The $T_b-Ts$ relation proposed by \cite{potekhin2015}, is described as follows: 
\begin{equation}
    T_s = T_\text{eq} + \big(T_\text{p} - T_\text{eq} \big) \, \frac{(1+ a_1 + a_2)\, cos^2\vartheta}{ 1 + a_1 \, cos\vartheta + a_2 \, cos^2 \vartheta}, 
    ~~~~ a_1 =\frac{a_2 T_{b9}^{1/2}}{3}, ~~~~ a_2 = \frac{10\, B_{12}}{T_{b9}^{1/2} + 0.1\, B_{12} \, T_{b9}^{-1/4}}. 
\end{equation}
Here, $\vartheta$ corresponds to the magnetic colatitude angle, $B_{12}= B_p/10^{12}$\,G. The corrected surface temperature at the pole, is reproduced by the expression
\begin{equation}
    T_\text{p} = T_\text{p}^{(0)} \Bigg[ 1 + \bigg( T_\text{p}^{(0)}\,/\, T_\text{p}^{\text(max)}  \bigg)^4 \Bigg]^{-1/4},
\end{equation}
where $T_\text{p}^{\text(max)}$ is the limiting temperature, at which $T_\text{p}(T_b)$ levels off due to the neutrino emission from the crust and it is approximately given by
\begin{equation}
  T_\text{p}^{\text(max)} = \bigg( 5.2 \, g_{14}^{0.65}  + 0.093 \sqrt{g_{14} B_{12}}\bigg) \times 10^6 \text{\,K}.  
\end{equation}
At the magnetic pole, the effective surface temperature, neglecting neutrino emission from the crust, is approximately given by the expression
\begin{equation}
 T_\text{p}^{(0)} = \bigg[ g_{14} \bigg(T_1^4 + \bigg(1+ 0.15\, \sqrt{B_{12}} \bigg)T_0^4 \bigg)  \bigg]^{1/4} \times 10^6 \text{\,K},  
\end{equation}
with
\begin{equation}
    T_0 = \big(15.7\, T_{b9}^{3/2} + 1.36\, T_{b9} \big)^{0.3796}, ~~~~~ T_1 = 1.13 \, B_{12}^{0.119} T_{b9}^a, ~~~~~ a= 0.337\,/\,(1+ 0.02\, \sqrt{B_{12}}\,).
\end{equation}
The equatorial surface temperature can roughly be evaluated as 
\begin{equation}
   T_{\text{eq}} =  \Bigg[1+ \frac{(1230 \, T_{b9})^{3.35} B_{12} \sqrt{1+ 2\, B_{12}^2} }{ (B_{12}+ 450 \, T_{b9} + 119\,B_{12}\,T_{b9}  )^4 }  + \frac{0.0066 \, B_{12}^{5/2}}{T_{b9}^{1/2} + 0.00258\, B_{12}^{5/2} }      \Bigg]^{-1} ~ T_\text{p} .
\end{equation}
The latter envelope model proposed by \cite{potekhin2015} is the one mostly used throughout this thesis.

\begin{figure}
\includegraphics[width=.49\textwidth]
{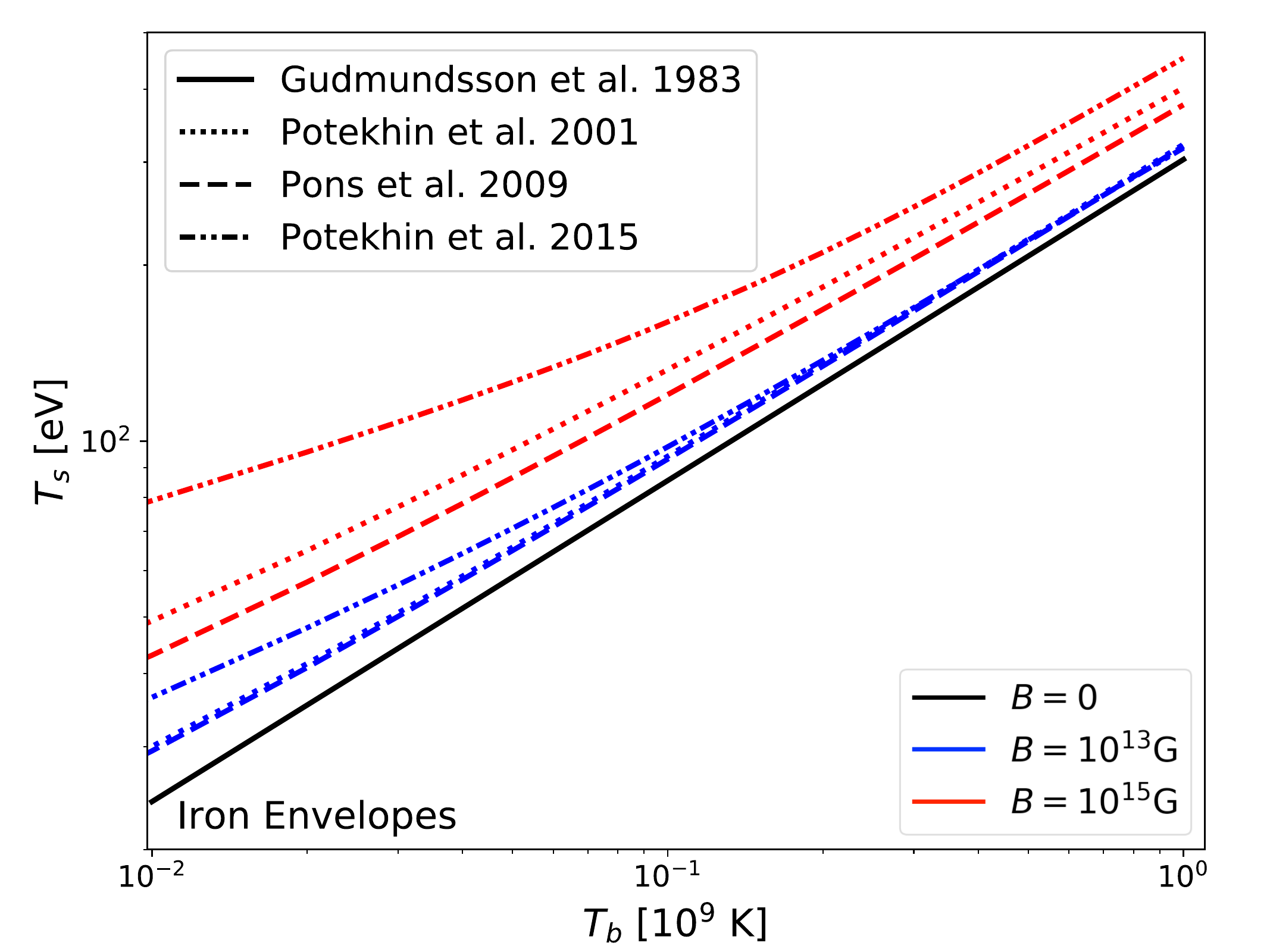}
\includegraphics[width=0.49\textwidth]
{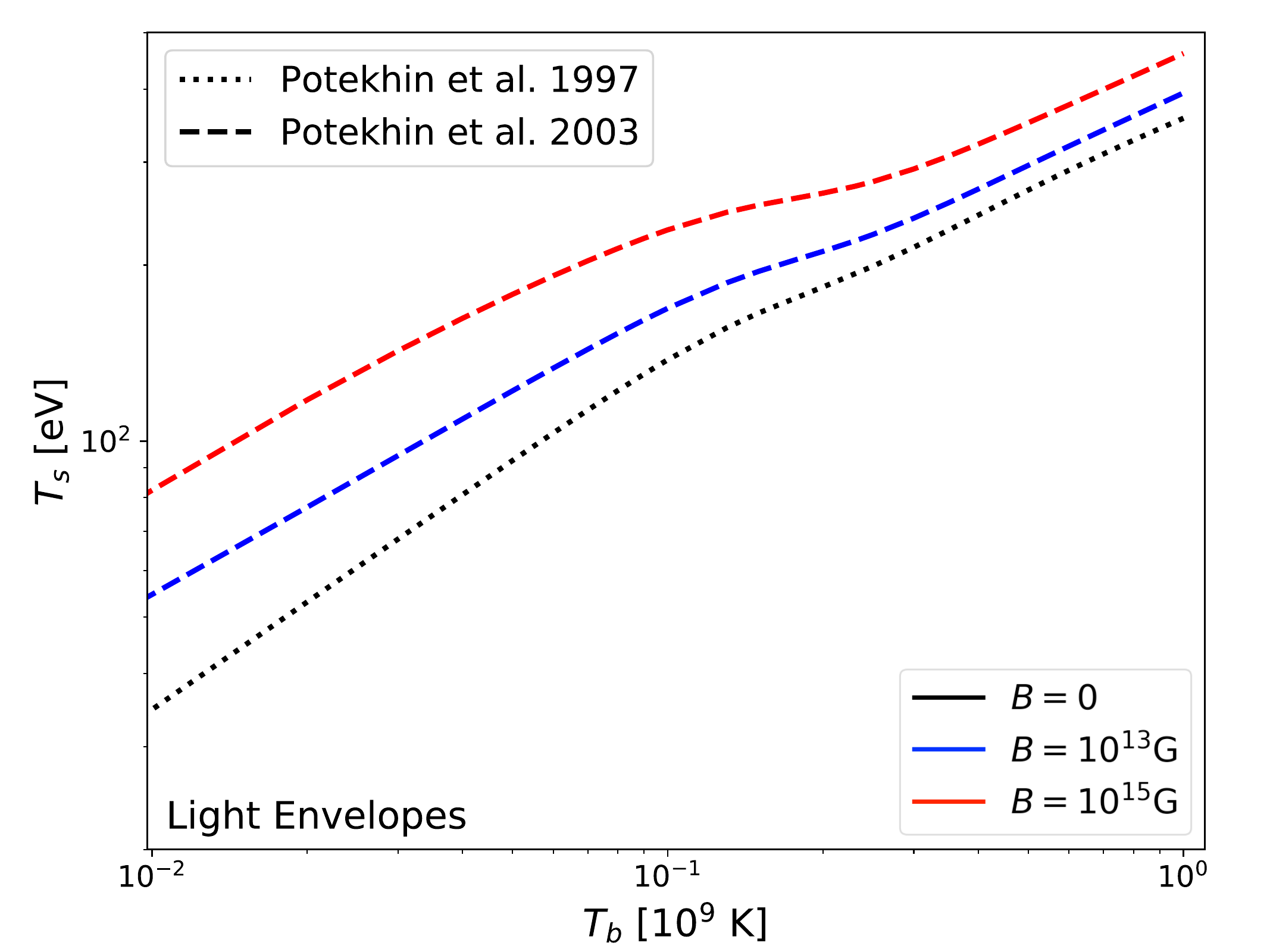}
\caption[The $T_b-T_s$ relations of different envelope models.]{The $T_b-T_s$ relations of different envelope models are presented. The non-accreted models are depicted in the left panel, while the fully accreted ones are shown in the right panel. The studied envelopes include: (i) non-magnetized models in black \citep{gudmundsson1983,potekhin1997}, and (ii) magnetized models in color \citep{potekhin2001,potekhin2003,pons2009,potekhin2015}. In the case of the magnetized models, two distinct values of a purely radial magnetic field strength are considered (suitable for polar $T_s$ if the topology is a simple dipole): $B=10^{13}$\,G (in blue) and $10^{15}$\,G (in red).}
\label{fig: envelope models Tb-Ts}
\end{figure} 

Since neutron stars are observed both as isolated sources or as part of binary systems, it is common to consider two different compositions for the envelope: either iron, arguably expected in the case of catalyzed matter in isolated systems, or light elements, mainly thought as products of accretion from a companion star or in a newly born systems that witness fall-back accretion after a supernova. In this study, we explore four models composed of iron (non-accreted matter) and two models of fully accreted (light) envelopes. Such models can be found in \cite{gudmundsson1983,potekhin1997,potekhin2001,potekhin2003,aguilera2008,pons2009, potekhin2015}.

In essence, an envelope model represents a stationary solution for the heat transfer equation, which is then fitted to establish an empirical relation between the surface temperature ($T_s$) that determines radiation flux and the interior temperature ($T_b$) at the crust/envelope boundary, for a given magnetic field, assumed to be the ones given at the boundary. The envelope model accounts for the sharp temperature gradient that occurs at low density. The location of $T_b$ is typically selected to correspond to a density of about $\rho = 10^{10}$ g cm$^{-3}$. At such low densities, neutrino emission is usually negligible as long as $T_b < 10^9$ K, which occurs within a few decades after the neutron star's birth. Therefore, we can ignore corrections due to neutrino emissivity.

Additionally, we neglect Ohmic dissipation in the envelope since there are no significant currents flowing through it. This is justified by the low conductivity of the material. This choice is consistent with the selection of potential boundary conditions (see \S\ref{subsec: outer B.C}), although magnetospheric currents could flow through the envelope and create hotspots (refer to \S\ref{subsec: FF BC} and \cite{akgun2017, akgun2018} for more details).
Similarly, we disregard any heating caused by accretion or bombardment, which might be relevant in cases like millisecond pulsars or cold, old pulsars, where tiny X-ray hotspots are observed, likely due to such processes. However, for our purposes, which mainly focus on young/middle-aged isolated neutron stars, these neglects are justified.

The $T_b-T_s$ relations of the envelope models under analysis (\S\ref{subsec: blanketing envelope}) are visually represented in Fig.~\ref{fig: envelope models Tb-Ts}. In the left panel, we present the iron envelopes, while the light envelopes are displayed on the right. The studied envelopes include: (i) two models without magnetic field dependence (shown in black), such as \cite{gudmundsson1983} in the left panel (represented by solid lines) and \cite{potekhin1997} in the right panel (represented by dots);
(ii) other magnetised envelopes, for which we show the $T_s$ for two values of a purely radial surface magnetic field strength (i.e., suitable for a pole in a dipolar topology): $B=10^{13}$\,G (in blue) and $B=10^{15}$\,G (in red).
\cite{potekhin2001} is illustrated with dots (left panel), 
\cite{potekhin2003} with dashed lines (right panel), \cite{pons2009} with dashed lines (left panel), and finally \cite{potekhin2015} with dashdotdotted lines (left panel). 

\begin{figure}
\centering
\includegraphics[width=1.\textwidth]
{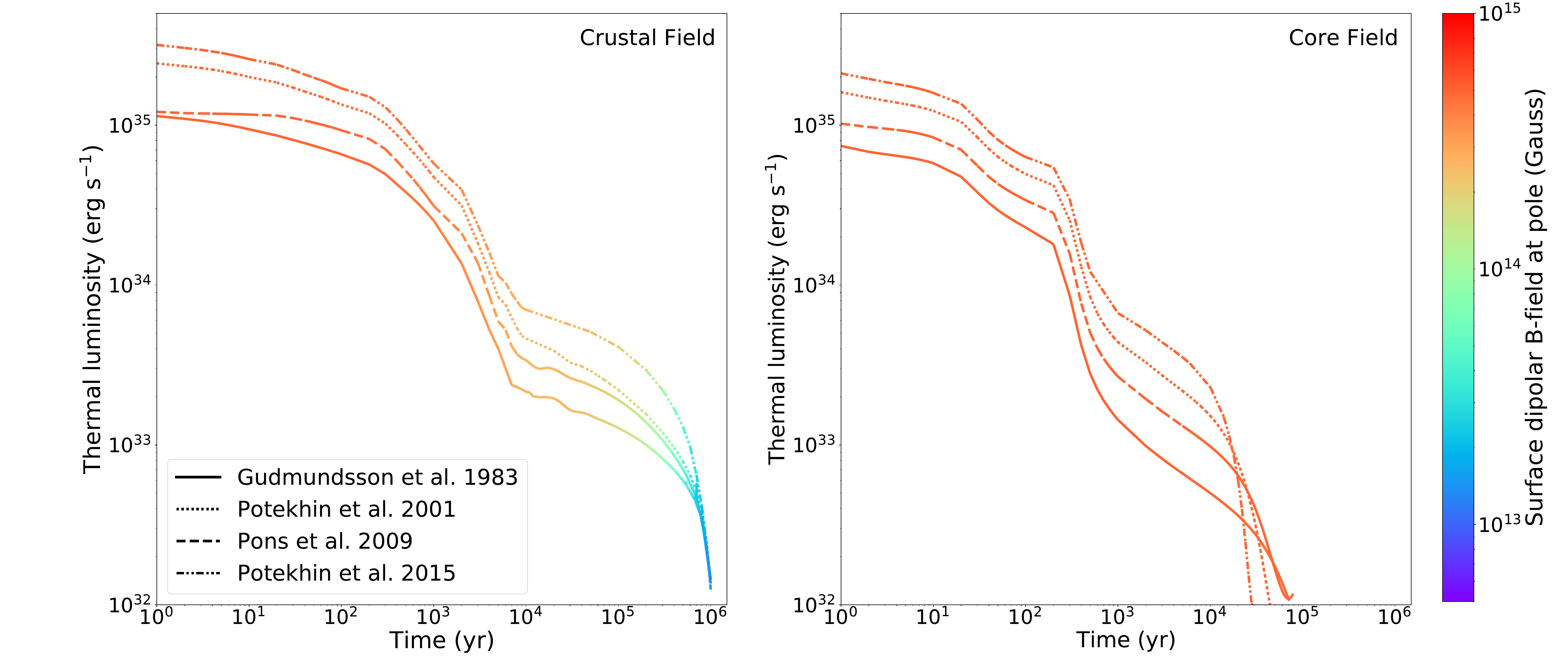}
\caption[Luminosity curves of iron envelopes with an initial magnetic field intensity at the polar surface of $B=5 \times 10^{14}$\,G .]{Luminosity curves of the four studied iron envelopes \citep{gudmundsson1983,potekhin2001,pons2009,potekhin2015} with an initial magnetic field intensity at the polar surface of $B=5 \times 10^{14}$\,G (hereafter, the colorbar denotes its evolution). The left panel corresponds to crust-confined topology, whereas the right one to core-dominant field topology. }
\label{fig: cooling curves}
\end{figure} 


The lowest effective temperature $T_s$ among all models is displayed by \cite{gudmundsson1983}. Every new effect incorporated in later works (composition, magnetic field) results in a higher predicted surface temperatures $T_s$ for a given $T_b$. Let us briefly review the main conclusions from a quick comparison of models.

In general, it is well known that assuming light-elements envelopes appreciably affect the neutron star luminosity \citep{potekhin1997}. 
Compared to iron models, we have a higher $T_s$ for the same $T_b$.
Concerning magnetic fields, as long as the average intensity is $B \lesssim 10^{13}$\,G, we expect a surface temperature similar to the non-magnetised case (for the same given composition).
On the other hand, under magnetar conditions, the general trend is that higher magnetic fields result in higher surface temperatures, all other factors being equal. Interestingly, in the left panel of Fig.~\ref{fig: envelope models Tb-Ts}, it is evident that the more recent calculations, which include more accurate physics (such as state-of-the-art microphysics ingredients, the effects of neutrino emission, and non-radial heat transport), have revised the predicted value of $T_s$ to levels higher than those found in any of the previous studies.

Thus, it is expected that the state-of-the-art envelope models predict different cooling curves from the models used two decades ago. This motivates us to revisit the results for cooling curves and consider different magnetic field topologies and strengths, as an important step in understanding the observational data. 

\section{Neutron star cooling models}
\label{sec: cooling models}

The cooling history of a magnetar is a delicate balance between neutrino and photon emissivity on one side and Joule heating in the star's crust on the other. If the currents are dissipated in the outer crust, the heat deposited is more effectively transported to the surface and has an impact on the star luminosity. On the contrary, heat dissipated in the inner crust or the core is very inefficient in modifying the surface temperature, because it is essentially lost via neutrino emission, as first discussed in \cite{kaminker2006} to explain the high thermal luminosities of magnetars.

To compare the different envelopes existing in the literature,
we have used the 2D magneto-thermal code (the latest version is described in \cite{vigano2021}) to run a set of cooling models using different initial configurations.
The neutron star background model is a $1.4 M_\odot$ neutron star built with the SLy4\footnote{\url{https://compose.obspm.fr/}} EoS \citep{douchin2001}, and we assume the superfluid models of \cite{ho2015}, which is the reason for the abrupt change in the slope of the cooling curves at ages $\sim 300$\,yrs in, e.g., the right panel of Fig.~\ref{fig: cooling curves}.
The rapid cooling during the photon cooling era is also caused by the low core heat capacity, which in turn depends on the assumed pairing details. 
A comprehensive revision of the microphysics embedded in magneto-thermal models can be found in \cite{potekhin2015}.

We considered two families of magnetic field topologies to study in detail the two extreme configurations: (i) crust-confined field consisting of a poloidal dipole and a toroidal quadrupole with steep radial gradients: the radial component of the magnetic field vanishes at the crust-core interface, while the latitudinal ($B_{\theta}$) and toroidal ($B_{\phi}$) components are different from zero; (ii) core-dominated twisted-torus magnetic fields as in \cite{akgun2017}, i.e., a dipolar topology, with the currents circulating almost only in the core, and Gyr-long decay timescales. We stress that, for our purposes, we choose these two topologies mean to cover a wide range of values for the crustal Ohmic dissipation.
For each topology, we consider two different field strengths ($10^{13}$\,G and $5\times 10^{14}$\,G) for the initial value of the dipolar field at the polar surface. The maximum initial toroidal field is fixed to $10^{13}$\,G in all cases. The magnetic field at the surface is always matched continuously with a current-free magnetic field (i.e. the electric currents do not leak into the magnetosphere $\boldsymbol{\nabla} \times \boldsymbol{B}= 0$, with vanishing field at infinity). 

\begin{figure}
\centering
\includegraphics[width=0.7\textwidth]
{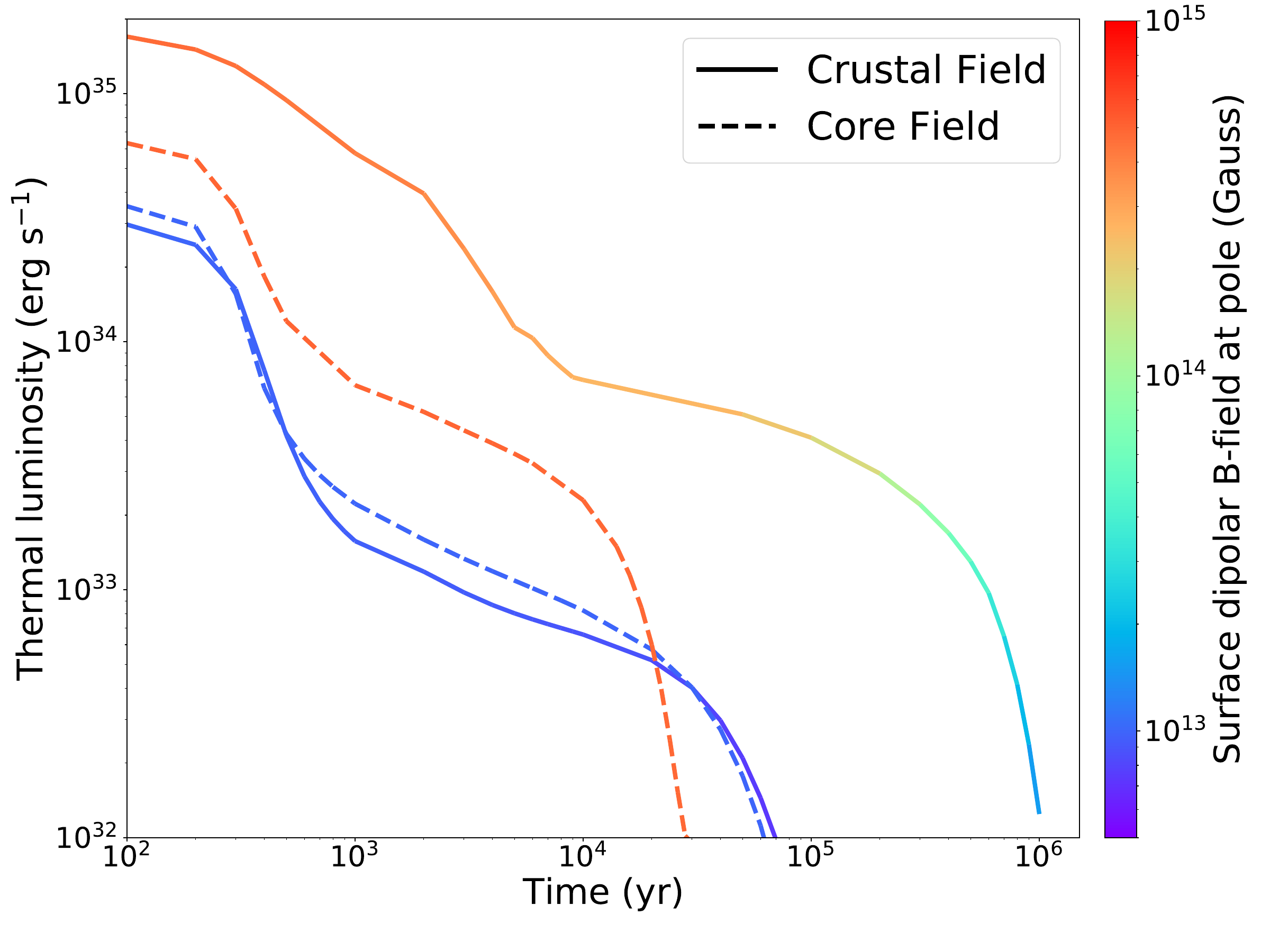}
\caption[Luminosity curves of the latest magnetised iron envelope \citep{potekhin2015}.]{Luminosity curves of the latest magnetised iron envelope \citep{potekhin2015} with different initial magnetic field intensities. Solid lines correspond to crustal field models and dashed-curves to core field ones.}
\label{fig: potekhin et al. 2015 cooling curves}
\end{figure} 

We have used the different envelope models presented in \S\ref{sec: Envelope models} coupled with the neutron star cooling models, to study the dependence of the neutron star cooling curves on the assumed envelope, in two given magnetic field topologies. Magnetar cooling curves obtained using different iron envelopes and a field strength of $B=5 \times 10^{14}$\,G are shown in Fig.~\ref{fig: cooling curves}. In the left panel, we consider a crustal-confined topology, and in the right panel we have a core-dominant field.
For a high field intensity, e.g., magnetar-like scenario, there are significant qualitative differences between crustal-confined and core-dominant field. Let us summarize the main findings:
\begin{itemize}
    \item At early times, during the neutrino cooling era (say $t < 10^4$\,yr), both models are similar. The interior temperature evolves independently of the envelope model (photon radiation is negligible), and the different $T_b-T_s$ relation translates directly in the surface temperature. Interestingly, the most recent models show the highest luminosities. This is a direct consequence of the results of Fig.~\ref{fig: envelope models Tb-Ts} (left panel).
    \item Later, once we enter the photon cooling era, the situation is inverted. The envelope models that provide a higher surface temperature actually radiate photons (which now govern the evolution) more efficiently, and the star cools down faster.
    \item In this epoch, the difference between crustal-confined and core-threading magnetic field becomes more evident. In the first case, heat dissipation occurs relatively close to the surface, which keeps the stellar crust warmer and delays the drop of the luminosity. In the second case, Joule heating is completely inefficient (currents are mostly in the core), and the effect mentioned above, with a very fast drop of luminosity for high field models becomes evident.
\end{itemize}

\begin{figure}
\centering
 \includegraphics[width=1.\textwidth]{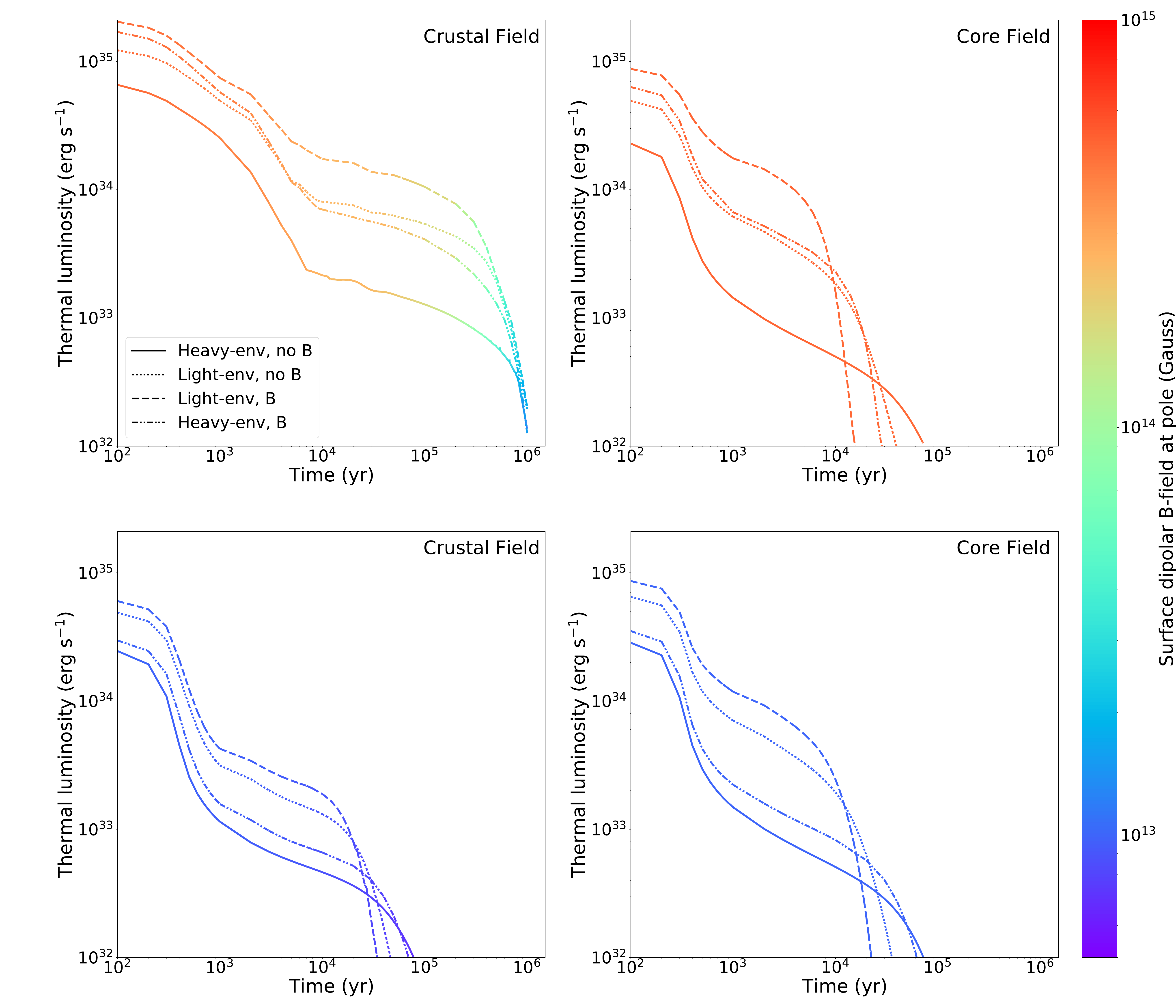}
\caption[Luminosity curves of four studied envelope models: \cite{gudmundsson1983} (Heavy-env, no B), \cite{potekhin1997} (Light-env, no B), \cite{potekhin2003} (Light-env, B) and \cite{potekhin2015} (Heavy-env, B)]{Luminosity curves of four studied envelope models: \cite{gudmundsson1983} (Heavy-env, no B), \cite{potekhin1997} (Light-env, no B), \cite{potekhin2003} (Light-env, B), and \cite{potekhin2015} (Heavy-env, B). On the left-hand side, we show the results of models with crustal-confined magnetic field, and on the right, those with core-dominant field topology. 
The results are represented at two initial magnetic field intensities at the polar surface), e.g., at $B=5 \times 10^{14}$\,G (upper panels) and at $B=10^{13}$\,G (bottom panels). }
\label{fig: comparison with light envelopes}
\end{figure} 


To illustrate more clearly these differences, in Fig.~\ref{fig: potekhin et al. 2015 cooling curves} we compare
cooling curves adopting the \cite{potekhin2015} envelope but now varying both, the field topology and strength. 
We show the results with the two different magnetic field intensities. For a relatively low magnetic field, e.g., $10^{13}$\,G, the crustal-confined and core-dominated simulations have a very similar behavior and the magnetic field does not dissipate much (the curves keep the blue color throughout the evolution). For the strong field case, $5 \times 10^{14}$\,G, the crustal-confined models show a significant dissipation of the magnetic field, e.g., the magnetic field has dissipated from $5\times 10^{14}$\,G (red) to a few $10^{13}$\,G (turquoise) after 1 million year of evolution (colorbar of Fig.~\ref{fig: potekhin et al. 2015 cooling curves}). As a consequence, the significance of Joule heating is crucial in the crust-confined model, whereas it is nearly insignificant for the core-dominated model being considered here. This is due to the significantly lower intensity of crustal currents and the much longer Ohmic timescales of the core currents. Additionally, the minimal energy dissipation in the core converts into neutrinos. The most relevant difference is that core-threaded field simulations with magnetized envelopes show faster cooling after $10^4$\,yr than low field models.
Therefore, the observational appearance of a magnetar at late times essentially depends on where currents are located and how much magnetic flux penetrates the core.

We also note that, for a strong enough magnetic field in the core of a neutron star, e.g., $B= 10^{15}$\,G, an additional cooling channel via neutrino synchrotron \citep{kaminker1997} is activated. It provides further cooling of the neutron star. However, we found that this effect is subdominant.  

We now extend our analysis to accreted (light element) envelopes. The results of the comparison between light and heavy elements are shown in Fig.~\ref{fig: comparison with light envelopes}. 
In the left panels, we display the results for models with crustal-confined magnetic fields and in the right panels those for core-dominant fields, for the two dipolar intensities
$B=5 \times 10^{14}$\,G (upper panels) and at $B=10^{13}$\,G (bottom panels). 

The same qualitative features discussed for iron envelopes are valid, but with luminosities shifted to slightly higher values (up to an order of magnitude) for accreted envelopes during the neutrino cooling epoch. Instead, it drops faster as soon as we enter in the photon cooling era. That is due to the even higher $T_s$ resulting from light elements in the envelope. A strong magnetic field enhances this effect. In the top-right panel of Fig.~\ref{fig: comparison with light envelopes}, we clearly see how high luminosities ($> 10^{34}$\,erg/s) are kept for about $10^4$\,yrs, but then quickly drop below $10^{32}$\,erg/s (and therefore objects become undetectable) after a few tens of kyr. Simulations using "old" non-magnetized models \citep{gudmundsson1983} do not allow to capture this behaviour.

\section{Discussion}
\label{sec: discussion envelope}

Understanding how young and middle-aged magnetars cool is of great importance for a correct interpretation of the observational data. In this work, we have revisited the cooling curves of neutron stars, focusing on the effect of the assumed envelope models (typically used as a boundary condition), and considering two extreme field topologies, crustal and core field. We noticed that the $T_b-T_s$ relation is very sensitive to the magnetic field strength. Although for relatively low magnetic fields, different magnetized iron envelopes predict similar effective surface temperatures. For relatively strong magnetic fields in the magnetar range there are substantial differences. For a given temperature at the base of the envelope ($T_b$), the most recent models (that incorporate better microphysics) predict surface temperatures ($T_s$) about a factor 2-3 higher than their predecessors. Since the flux scales as $T_s^4$, this correction significantly enlarges the photon luminosity, which in turn leads to a very fast transition from a high luminosity epoch (during the neutrino cooling era) to a very low luminosity (soon after we enter the photon cooling era).
This trend is very pronounced in models where the magnetic field threads the star's core and most electric currents circulate there. Conversely, in crustal-confined models, the additional energy released by Joule heating close to the star surface is very effective and governs the energy balance equation, which counterbalances the effect. 
Thus, depending on where the bulk of the electrical currents circulates, one can expect middle-aged magnetars which are relatively bright or sources with very low luminosities ($<10^{32}$ erg/s) which persistent emission is essentially undetectable as X-ray sources.
We stress again that a $10^4$ yr, high field neutron star with a core field (light and heavy element envelopes) can actually be much cooler than a similar neutron star with a pulsar-like field, only because of the effect of magnetic field in the envelope (see Fig.~\ref{fig: comparison with light envelopes} upper-right panel). This has potentially strong implications for population synthesis studies of the pulsar and magnetar populations because observational biases introduced by the lack of detectability of some class of sources affect the predictions of birth rates and field distributions. We plan to incorporate these effects in future works.

To briefly compare our results with observational data, one should only concentrate on objects with "Real ages" and that are at the extremes of our cooling curves: A) 1E 2259+586 (middle-aged magnetar) can only be explained with a crustal-field and magnetized light elements. B) All XDINS cannot be explained with core-fields. They necessarily need that the crustal-field has a strong component but the envelope can be light or heavy, magnetic or non-magnetic. C) CCOs are in an age and luminosity range that do not allow distinguishing between envelope models or magnetic topology. D) Middle-age faint pulsar such as PSR B0656+14 might be explained only with the fast decay of the light element envelope curves, since for low magnetic field neutron stars, light envelopes might produce cooler neutron stars than heavy elements for older ages ($>10^4$\,yr), regardless of the field configuration (see Fig.~\ref{fig: comparison with light envelopes} bottom panels). Ultimately, the existence of strongly magnetized neutron stars with detectable thermal emission at later times would be a strong argument in favor of a crustal magnetic field.

Our study highlights the importance of treating carefully all ingredients in the complex theory of neutron star cooling. Boundary conditions neglecting the role of the envelope, or using non-magnetized envelopes, can lead to discrepancies as large as one order of magnitude relative to observational data. On the other hand, an accurate estimation of surface luminosity is important to constrain any source property (e.g. surface B-field or age).

\subsubsection*{Corresponding scientific publications}
\underline{C.~Dehman}, J.A.~Pons, D.~Vigan\`o \& N.~Rea: 2023, \textbf{How bright can old magnetars be? Assessing the impact of magnetized envelopes and field topology on neutron star cooling}, \emph{Mon.~Not.~Roy.~Astron.~Soc.~L., $520$, $42$} (\href{https://arxiv.org/abs/2301.02261}{\underline{arXiv:2301.02261}},\href{https://ui.adsabs.harvard.edu/abs/2023MNRAS.520L..42D/abstract}{\underline{ADS}},\href{https://academic.oup.com/mnrasl/advance-article/doi/10.1093/mnrasl/slad003/6973208?utm_source=authortollfreelink&utm_campaign=mnrasl&utm_medium=email&guestAccessKey=b4196a19-bd30-4b1e-acf5-de9a7c3c9b3c}{\underline{DOI}}).

\clearemptydoublepage
\let\textcircled=\pgftextcircled
\chapter{Confronting Observations with Theoretical Models to Constrain the EoS}
\label{chap: Comparison with observations}

\initial{T}iming measurements of neutron stars reveal a diverse range of inferred magnetic field strengths, spanning from $10^8$\,G in millisecond pulsars to $10^{15}$\,G in magnetars. An effective approach for assessing the surface magnetic field involves observing thermal radiation in the soft X-ray band. X-ray observatories such as \xmm, \cxo, \nustar, \nicer, \swift, and \rst\, have successfully identified compact objects emitting thermal radiation. In this regard, neutron stars observed in X-rays, appearing as isolated entities, manifest significant phenomenological heterogeneity.

As extensively detailed in \S\ref{subsec: X-ray band}, neutron stars exhibit a diverse range of characteristics and are categorized according to the primary energy source driving their emissions. RPPs are powered by the conversion of a fraction of their rotational energy into non-thermal radiation. Nevertheless, in some instances, they also display thermal emissions. Magnetars are the most powerful magnets in the present universe \citep{turolla2015,esposito2021}. Their bright X-ray luminosity, long periods, and the occurrence of observed bursts and outbursts are attributed to the restless dynamics and dissipation of a strong magnetic field ($\sim 10^{14}$\,G) near the star's surface \citep{thompson1995}. Radio-quiet XDINSs represent a category of relatively aged, nearby cooling neutron stars, distinguished by their distinctly clear thermal emissions. CCOs constitute a small collection of enigmatic, radio-quiet sources. In certain cases, they combine a very weak external magnetic field with a relatively high luminosity, along with indications of an anisotropic surface temperature distribution \citep{gotthelf2013}.

In the context of these diverse neutron star characteristics, a significant discovery by \cite{keane2008} demonstrated that the galactic rate of core-collapse supernovae is smaller than the sum of the birth rates of the different neutron stars populations. This raises questions about attributing all neutron star formation to core-collapse supernovae, as they alone cannot account for the entire neutron star population in our Galaxy. \cite{keane2008} propose various solutions to this issue, one of which suggests that different neutron star populations are related through an evolutionary process. If this holds true, we only need to consider a single birthrate for these populations, which helps resolve (or at least alleviate) the issue with core-collapse supernovae rates. Indeed, magneto-thermal cooling models are capable of establishing this evolutionary link. In this context, one of the main theoretical tasks is to explain the diverse phenomenology of their X-ray emission, population, and evolutionary paths \citep{vigano2013}. Other pieces of evidence support the unified picture of neutron stars, such as RPPs exhibiting magnetar behavior \citep{gogus2016} and a continuous spectrum of behavior between RPPs and the so-called rotating radio transient, characterized by varying degrees of intermittency \citep{weltevrede2011}.

\section{Observational data}
\label{sec: observational data}

We are presently engaged in the task of comparing the outcomes generated by magneto-thermal evolution codes with established observational data and precise timing parameters derived from a carefully chosen subset of sources. For meaningful comparisons, it becomes imperative to have access to isolated neutron stars that possess both well-defined ages and precise thermal luminosities. While our Galaxy hosts thousands of isolated neutron stars, only a scant few dozen of these objects meet both criteria.

In the present study, founded upon the work of Dehman et al. (2023, in preparation), we conducted an archival search for known isolated neutron stars that were observed using \xmm\ and/or \cxo. We selected those exhibiting a statistically significant quasi-thermal component in their X-ray spectrum, thereby expanding the original sample established by \cite{vigano2013}. A comprehensive documentation of about 70 sources is presented in Table~\ref{tab:timing-1} and Table~\ref{tab:spectral}. Our classification framework categorizes the sample into four distinct classes, which include:
\begin{itemize}
\item 27 magnetars, for which we exclusively utilize data from their prolonged quiescence phases characterized by predominantly thermal emission. This selection encompasses all the sources employed by \cite{cotizelati2018} (for a comprehensive overview of their observational properties, we direct readers to their work). Furthermore, we expand this collection by including PSR J1846-0248, which exhibited magnetar-like activity during its 2020 outburst \citep{blumer2021}, along with 3XMM J185246.6+003317 \citep{rea14}, PSR J1622-4950 \citep{levin2010}, and more recent additions to the magnetar catalog, such as Swift J1818.0-1607 \citep[e.g.][]{ibrahim2022}, SGR J1830-0645 \citep{cotizelati2021}, and Swift J1555.2-5402 \citep{enoto21}.
\item 13 CCOs. The emissions from these objects are commonly characterized by one or more dominant thermal components (e.g., \cite{Doroshenko2018}). We have selected the confirmed CCOs with sufficiently detailed spectral information and a known age associated with the supernova remnant (SNR). The sample of sources identified with such a criterion includes the sample used in \cite{vigano2013}, with the addition of 9 sources lacking available timing parameters, namely, $P$ and $\dot{P}$. 
\item 20 RPPs. The X-ray emission from these objects can be dominated by thermal emission or be mostly entirely non-thermal, especially for very young and energetic neutron stars. The selected sources represent a subset of the larger list of known RPPs (for an observational overview of X-ray pulsars, see \citealt{becker09}), filtered with the requirement that the blackbody-like component in the X-ray spectrum was dominant or at least statistically significant. With respect to the previous catalogue by \cite{vigano2013}, we are adding: 1RXS J1412+7922 \citep{zane11}, PSR J0205+6449 \citep{Slane2004}, PSR J1357-6429 \citep{Chang2012}, PSR J1741-2054 \citep{Marelli2014}, PSR B1822-08 \citep{Hermsen2017}, PSR J1957+5033 \citep{Zyuzin2021}, PSR J0554+3107 \citep{Tanashkin2022}. 
\item 7 XDINS, commonly referred to as "The Magnificent Seven" (for a comprehensive review, see \cite{turolla09}). These nearby and X-ray faint sources exhibit purely thermal spectra and possess accurately determined distances, often obtained through parallax measurements. Consequently, they serve as ideal subjects for our study.
\end{itemize}

\subsection{Timing properties and age estimates}

If both the spin period $P$ and the period derivative $\dot{P}$ of the source are known, the characteristic age $\tau_c = P/2\dot{P}$ can serve as an approximate measure of the true age. These two ages align only if the initial period was shorter than its current value and if the surface dipolar magnetic field $B_p$, which contributes to the magnetic torque (as indicated in eq.\,\eqref{eq: PPdot}), has remained constant throughout the neutron star's lifetime, as demonstrated in eq.\,\eqref{eq: characteristic age}. 
\begin{equation}
\tau_\text{real} = \tau_c - \frac{P_0^2}{2 P \dot{P}} - \frac{\int_0^{\tau_\text{real}} \left[ (K B_p^2)(t') - (K B_p^2)\text{now} \right] dt'}{(K B_p^2)\text{now}}~.
\label{eq: characteristic age}
\end{equation}
However, this scenario is less common. Typically, for middle-aged and older objects, the characteristic age $\tau_c$ tends to exceed the real age due to the decay of the magnetic field over time \citep{vigano2013}. It is worth noting that for the majority of CCOs, both the spin period $P$ and the period derivative $\dot{P}$ remain unknown. Consequently, measured magnetic field values and age estimates are lacking for these objects.

On the other hand, when the object is situated within a SNR, a kinematic age can also be deduced by analyzing the expansion of the nebula (for a comprehensive review, see \cite{allen2004}). For a few other nearby sources, such as certain XDINSs and a handful of magnetars, the proper motion, with an association to a birthplace, can offer an alternative estimation of the actual age \citep{Tetzlaff2011, Tendulkar2013}. 

Regarding the timing properties and kinematic age, we have gathered the most up-to-date and reliable information available in the literature, sourcing from the ATNF catalogue\footnote{The ATNF Pulsar Catalogue \url{http://csiro.au/}.} \citep{manchester2005} and the McGill online magnetar catalogue\footnote{The McGill online magnetar catalogue \url{http://www.physics.mcgill.ca/~pulsar/magnetar/main.html}.}. In Table~\ref{tab:timing-1}, we present the sources in our study, along with their established timing properties, characteristic age, and alternative age estimate, when available.
Furthermore, we provide the calculated value of the surface dipolar magnetic field at the pole $B_p$, based on the standard dipole-braking formula (eq.\,\eqref{eq: PPdot}). 

Notably, for RPPs, timing properties exhibit stability over several decades, enabling precise measurements of $P$ and $\dot{P}$. However, in the case of magnetars, the presence of timing noise is considerably more pronounced. In some instances, different values of $\dot{P}$ have been reported \citep{tong2013b}, differing even by orders of magnitude (as seen in Table 2 of the online McGill catalogue). Consequently, it is prudent to exercise caution when considering these values, particularly for objects with the highest $\dot{P}$ values.

\subsection{Luminosity and temperature inferred from spectral analysis}

Luminosities and temperatures can be derived through spectral analysis, but achieving precise determinations is often challenging. Luminosity calculations are inevitably affected by uncertainties in distance measurements. On the other hand, inferred effective temperatures depend on the choice of an emission model (such as black-body versus atmosphere models, composition, presence of a condensed surface, etc.), which, especially in scenarios involving strong magnetic fields, can introduce significant theoretical uncertainties. Often, multiple models can fit the data equally well, lacking a clear, physically motivated preference for one over the others. This situation can lead to inferred effective temperatures differing by up to a factor of two. Moreover, photoelectric absorption from the interstellar medium introduces an additional source of error in temperature determinations. This is due to the correlation between the hydrogen column density $n_H$ and the obtained temperature value from spectral fits. Different choices of absorption models and metal abundances can also yield divergent temperature outcomes. In the common scenario of inhomogeneous surface temperature distributions, an approximation with two or three regions at different temperatures is typically employed. Lastly, when dealing with data containing a limited number of photons and/or pronounced absorption features, the temperature becomes poorly constrained by the fit, adding a large statistical error to the systematic one. Due to all these factors, the temperatures derived from spectral fits present challenges in direct comparison with the physical surface temperatures obtained from cooling models.

Considering the aforementioned factors, the luminosity emerges as a preferable metric for comparing data to theoretical models. Since it is an integrated quantity, it effectively mitigates the impacts of anisotropy and spectral model selection. The principal source of uncertainty associated with luminosity often stems from imprecise knowledge of the source distance. In the most challenging instances, distance measurements are known with errors on the order of a few units, leading to potential luminosity uncertainties spanning up to one order of magnitude. Additionally, interstellar absorption predominantly affects the energy range in which the majority of middle-aged neutron stars emit ($E \lesssim 1$\,keV). This characteristic makes it easier to detect hotter sources (such as magnetars) or those in closer proximity (like XDINSs). Analogous to the temperature scenario, opting for distinct absorption models and chemical abundances can introduce further systematic errors in luminosity determination. However, in the most unfavorable circumstances, the relative error remains around 30\%, rendering it a secondary error source in comparison to the distance uncertainty. For a more comprehensive understanding of the data reduction and analysis procedures undertaken in this study, please refer to Appendix~\ref{appendix: observational data}.

\section{Unifying the diversity of isolated neutron stars}
\label{sec: unification picture}
 
\subsection{Pulsar spin-down properties}
Timing measurements of pulsars show a positive time derivative of their spin period, $\dot{P}$ (i.e., a negative time derivative of the angular velocity, $\dot{\Omega}$). The loss of rotational energy is
˙\begin{equation}
    \dot{E} = I \Omega \dot{\Omega} = - 3.95 \times 10^{46} I_{45}\frac{P}{\dot{P}}  \text{\,erg\,s}^{-1}, 
    \label{eq: rotational energy loss}
\end{equation}
where $I = \int_V \rho(r) r^2 dV$ is the moment of inertia of the star and $I_{45} = I/(10^{45} \text{g\,cm}^{2})$. The energy balance equation between radiation and rotational energy losses (eq.\,\eqref{eq: rotational energy loss}), reads
\begin{equation}
    I \Omega \dot{\Omega} = \frac{B_p^2 R_\star^6 \Omega^4}{6c^3} f_x.
\end{equation}
Differences in the radiation mechanism are included in the factor $f_{x}$: magnetic dipole radiation losses scale as sin$^2 \mathcal{X}$ for vacuum or $1.5(1+\text{sin}^2\mathcal X)$ for force-free magnetosphere \citep{spitkovsky2006}, with $\mathcal{X}$ is the angle with the rotation axis. In general, the previous balance can be written in terms of the spin period $P = \Omega\,/\,2\pi$ and its derivative (same as eq.\,\eqref{eq: PPdot}): 
\begin{equation}
    P \dot{P} = K B_p^2, 
    \label{eq: PPdot chap4}
\end{equation}
with 
\begin{equation}
    K = f_x \frac{2 \pi^2}{3} \frac{R_\star^6}{I c^3} = 2.44 \times 10^{-40} f_{x} \frac{R_6^6}{I_{45}} \text{\,s\,G}^{-2},  
    \label{eq: K}
\end{equation}
where $R_6^6 = R_{\star}/(10^6$\,cm). In literature, the fiducial values $I_{45} = I$, $R_6 =1$ and $f_x=1$ (vacuum orthogonal rotator) are commonly used. 

Throughout this work, we will obtain the long-term evolution of $B_p(t)$ from simulations for a given neutron star model (fixed $R_\star, I$ and $f_x$). Integrating in time eq.\,\eqref{eq: PPdot chap4}, we will obtain the corresponding evolution of timing properties. We will always consider $K$ constant, ignoring the possible time variation of two quantities in eq.\,\eqref{eq: K}: the angle, $f_x = f_{x}(t)$, and the effective moment of inertia, $I = I(t)$. The latter could vary during the early stages, with the growth of the superfluid region in the core, rotationally decoupled from the exterior \citep{glampedakis2011,ho2012}. We will also neglect other possible mechanisms to the spin-down, like strong particle winds \citep{tong2013}. We also neglect the spin-down by gravitational radiation, because it can be efficient only during the first minutes or hours of a neutron star life, when rotation is sufficiently fast and the mass quadrupole moment large enough (see e.g., \cite{haskel2006} and references therein).

Model dependencies enter in the spin-down factor $K$, eq.\,\eqref{eq: K}, as $f_x R_6^6/I_{45}$, where $R_6$ and $I_{45}$ depend on EoS and star mass. To quantify such variations, \cite{lattimer2001} considered the moments of inertia resulting from many EoS. \cite{bejger2002} revised it, finding a correlation between $I, M$ and $R_\star$, expressed by the following fit:
\begin{equation}
    I = a(x)\, M R_\star^2,
\end{equation}
\begin{equation}
    a(x) = \begin{cases}
x\,/\,(0.1 + 2x) ~~~ x\leq 0.1 \\
2(1+5x)\,/\,9 ~~~~ x>0.1 
    \end{cases} ,
\end{equation}
where $x$ is the dimensionless compactness parameter:
\begin{equation}
    x= \frac{M}{M_{\odot}} \frac{\text{km}}{R_\star}~. 
\end{equation}
The value $a(x) = 0.4$ corresponds to a constant density sphere, but realistic stars are expected to have a lower value, since the mass is concentrated towards the center. Most EoSs predict radii in the range $8-15$\,km, and masses $1-2$ M$_\odot$. Considering the possible range of $x \sim 0.1-0.2$, the corresponding values are $a(x) \sim 0.1-0.2$ (see Fig.~1 in \cite{bejger2002}).

For higher order multipoles, the magnetic field decreases faster with distance, therefore they have significantly lower torques compared with the dipole: their contribution can be safely neglected.

\subsection{P-Pdot diagram and evolutionary tracks}

\begin{figure}[ht]
    \centering
        \includegraphics[width=\textwidth]{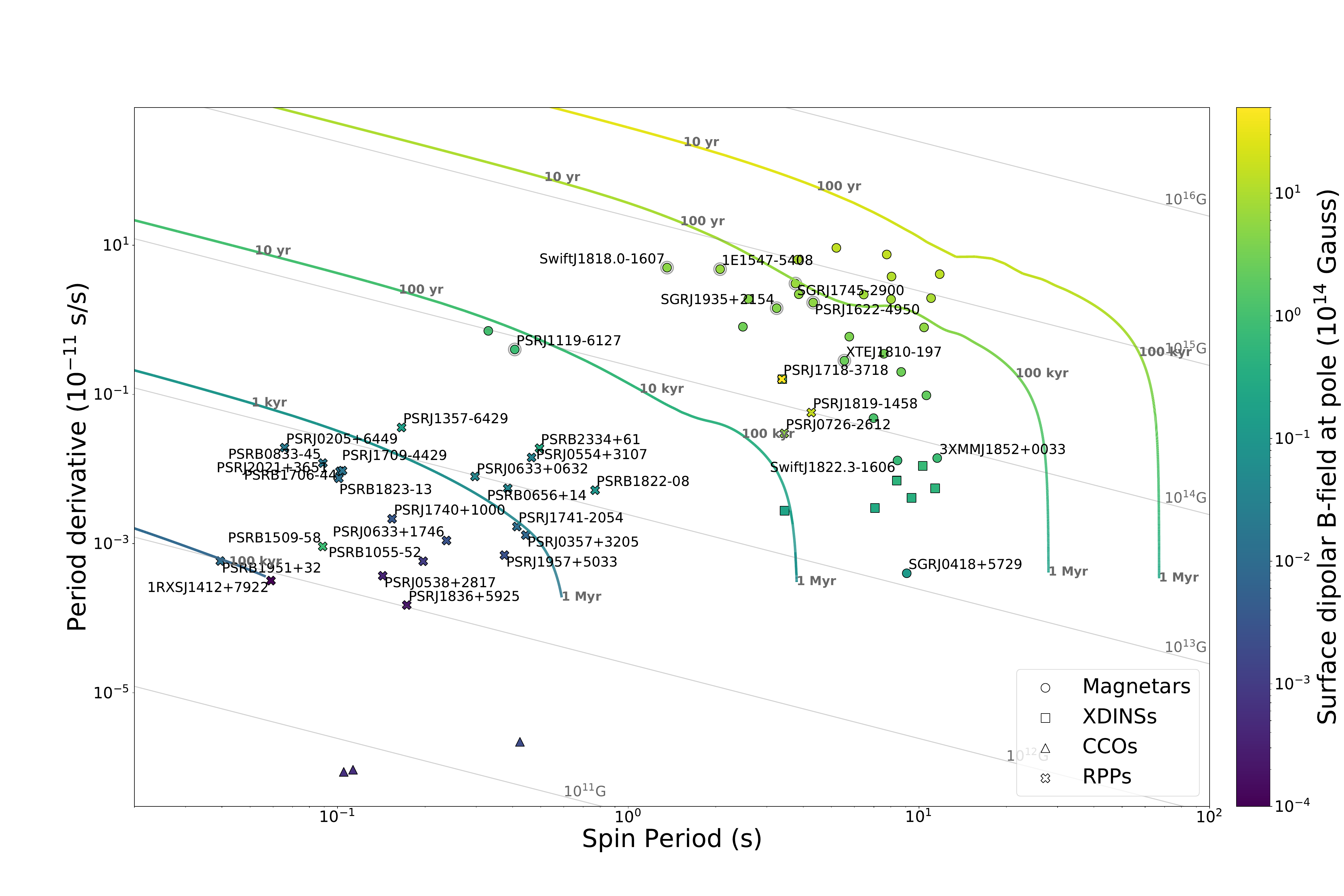}
    \caption{Evolutionary tracks in the $P-\dot{P}$ diagram are plotted using the SLy4 EoS with a mass of M=$1.4$\,M$_{\odot}$ and a heavy magnetized envelope model \citep{potekhin2015}.}
    \label{fig: P Pdot}
\end{figure}

To understand the observational data and precise timing parameters derived from a carefully chosen subset of sources, we conducted a series of 2D magneto-thermal simulations. These simulations involve exploring various initial configurations using a neutron star background model with a mass of $1.4$\,M$_\odot$ constructed with the SLy4 EoS, and assuming the superfluid models as described in \cite{ho2015}. For the envelope model, we utilized both the heavy model described in \cite{potekhin2015} and the light model described in \cite{potekhin2003}. We performed these simulations with five different initial magnetic field values for the surface dipolar field at the pole: $1\times10^{12}$\,G, $1\times10^{13}$\,G, $1\times10^{14}$\,G, $1\times10^{15}$\,G, and $3\times10^{15}$\,G. Additionally, we considered a quadrupolar toroidal field of one order of magnitude higher intensity.
 
In Fig.~\ref{fig: P Pdot}, we present the evolutionary tracks obtained from magneto-thermal simulations in the P-Pdot diagram, showcasing various initial values of magnetic field strength. These tracks are then compared to the observed timing properties of X-ray pulsars. We represent Magnetars with circles, RPPs with stars, XDINSs with squares, and the CCOs with triangles. It's worth mentioning that magnetars with grey contours correspond to radio-powered sources, and as discussed in \S\ref{sec: observational data}, we only have timing data available for three out of the 13 observed CCOs. The black solid lines on the graph represent the evolutionary paths a star would follow if there were no magnetic field decay. We have marked different ages ($t=10$\,yr, $100$\,yr, $10$\,kyr, $100$\,kyr, and $1$\,Myr) above the tracks obtained from magneto-thermal simulations. The colorbar in Fig.~\ref{fig: P Pdot} indicates the intensity of the dipolar poloidal field, $B_p$. It's important to note that $B_p$ remains relatively constant during an initial epoch, typically up to $\sim 100$\,kyr, depending on the initial magnetic field strength. This stability is reflected in the unchanged color of the curves during the initial phase. Stronger initial magnetic fields decay more quickly than weaker ones. Any change in the color of the curve over time corresponds to a change in the intensity of the dipolar field at the star's surface.

Comparing these tracks with observational data, one can observe common evolutionary links among groups of isolated neutron stars. This is evident for neutron stars with magnetar-like emission, represented with circles, and XDINSs, represented with squares. However, for the upper evolutionary track corresponding to the high magnetic field of $10^{15}$\,G, we notice that some magnetars are located at the beginning of this track, but there are no sources found at its end. This observation can be explained by two arguments: (i) The complex magnetospheric configuration differs from the available theoretical models. Considering the bundle of currents in the magnetosphere could potentially yield a higher torque with lower magnetic field intensity, which may offer a more accurate explanation for the evolution of neutron stars. (ii) Furthermore, it's worth noting that this group of neutron stars does not appear to have older descendants, which contrasts with other magnetars. This observation may be attributed to the magnetic field topology; in the case of very high magnetic fields and the concentration of currents in the star's core, where Joule dissipation is weak, the X-ray luminosity is expected to decline quite early, typically around $100$\,kyr. Consequently, older magnetars may exhibit faint emissions (see Chapter~\ref{chap: envelope}). Additionally, population synthesis studies \citep{gullon2014,gullon2015} suggest that very few objects are expected in this specific region of interest. Therefore, it's possible that due to observational limitations, we are not detecting any of these sources. 

Finally, it's worth noting that when considering all known pulsars detected across various energy bands (see ATNF catalogue), a seemingly sharp limit of around $\sim 12$\,s emerges in the distribution of spin periods. Our sample, as shown in Table~\ref{tab:timing-1}, highlights that the spin periods of most magnetars cluster around $\sim 10$\,s. The prevailing explanation for this phenomenon is the gradual decay of the magnetic field as the neutron star ages \citep{colpi2000}. As a result, the spindown rate becomes too slow to lead to longer rotation periods while the star remains bright enough to be observable as an X-ray pulsar. According to this scenario, low-field magnetars and XDINSs can be considered as older magnetars whose external dipolar magnetic fields have decayed to typical values around $10^{13}$\,G \citep{turolla2011,rea2010}. Our simulations have confirmed this behavior.

\subsubsection{Thermal X-ray luminosity}

\begin{figure}[ht]
    \centering
    \includegraphics[width=\textwidth]{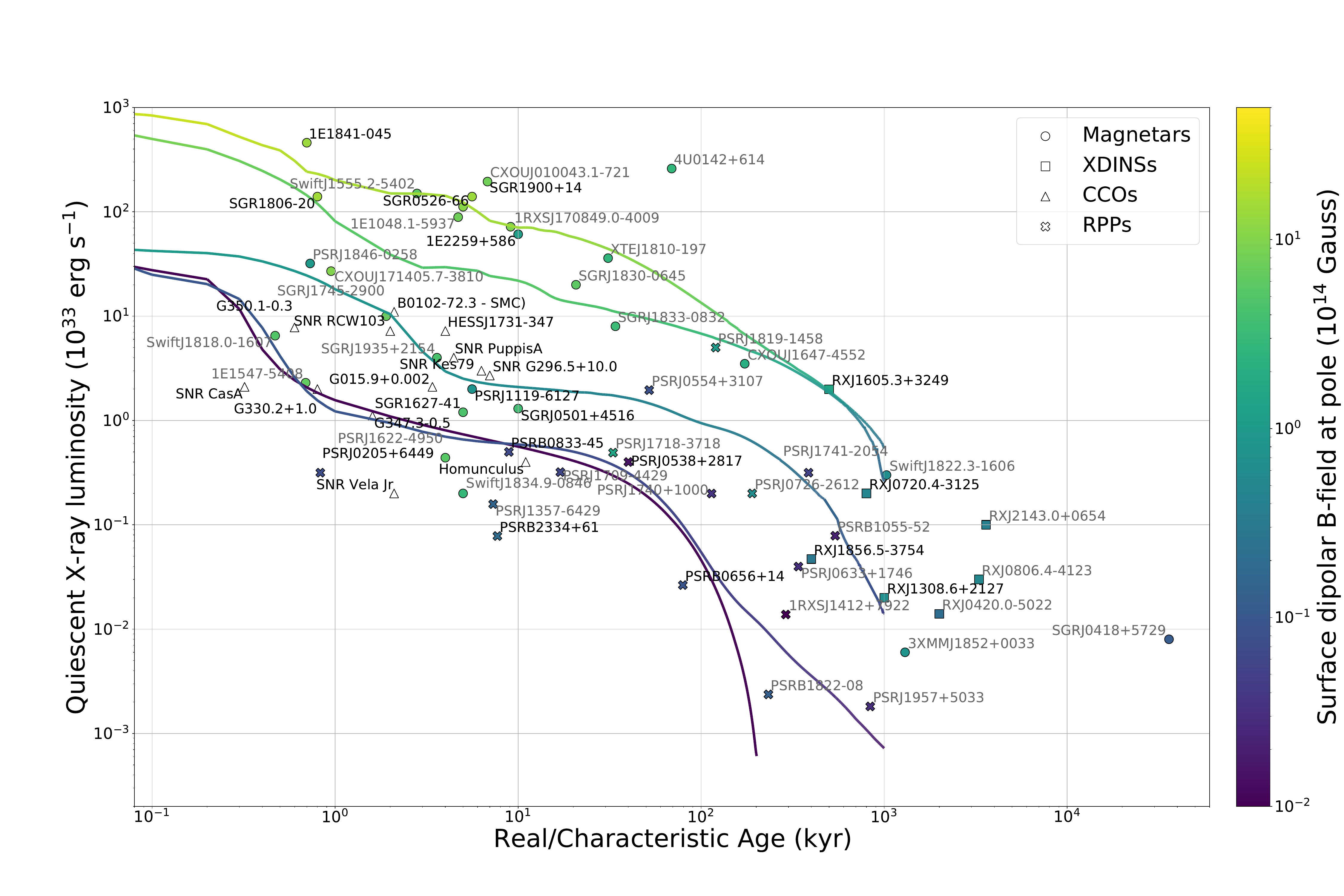}
    \caption{Comparison of observational data with theoretical cooling curves. These curves represent the results obtained using the SLy4 EoS for a neutron star mass of M=$1.4$\,M$_{\odot}$ and a heavy envelope model \citep{potekhin2015}.}
    \label{fig: Luminosity curves}
\end{figure}

In Fig.~\ref{fig: Luminosity curves}, we present the cooling curves along with the X-ray luminosity extracted from observational data (Table~\ref{tab:spectral}). Objects shown in black represent those for which we have the real age determined through kinematic measurements, while objects in grey represent those for which we have the characteristic age derived from timing properties.

Our theoretical simulations encompass the majority of the studied sources. What's particularly intriguing is that objects with higher inferred magnetic fields are systematically falling on the cooling curve with initial magnetic field strengths on the order of $10^{14}-10^{15}$\,G. This presents compelling evidence in favor of the scenario where magnetic field decay drives their bright luminosity. Magnetic fields exceeding $B \geq 10^{14}$\,G possess sufficient strength to significantly heat the crust and fuel the observed X-ray radiation. Notably, the cooling timescale for strongly magnetized objects is approximately one order of magnitude longer compared to weakly magnetized neutron stars.

\begin{figure}[ht]
    \centering
    \includegraphics[width=\textwidth]{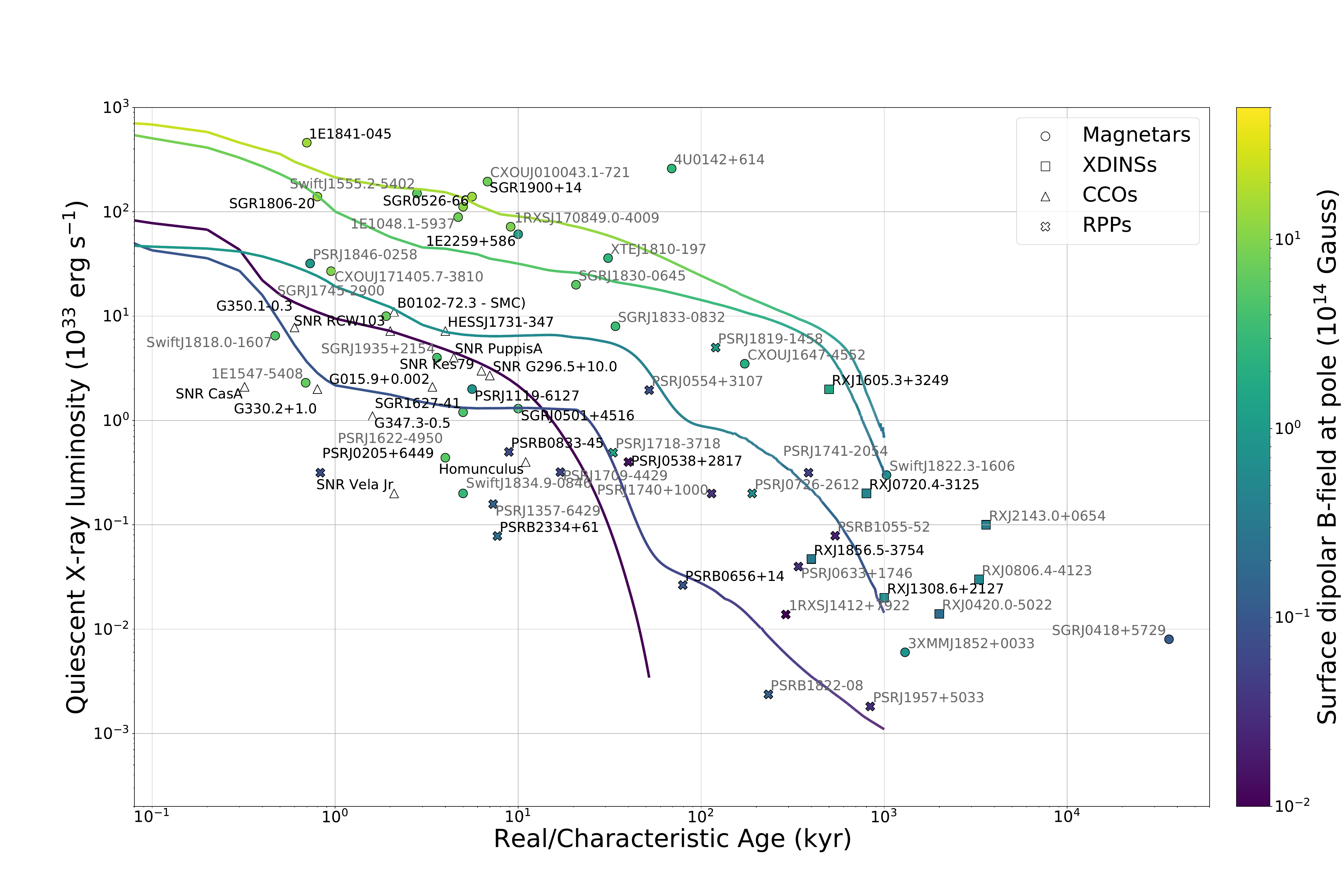}
    \caption{The same as shown in Fig.~\ref{fig: Luminosity curves}, but in this case, considering a light envelope model \citep{potekhin2003}.}
    \label{fig: Luminosity curves light env}
\end{figure}

The cooling curves using iron envelopes \citep{potekhin2015} with $B_p = 10^{15}$\,G barely reach the very high X-ray luminosity of the observed magnetars. This observation raises the possibility that these relatively young sources might have light element envelopes, or it's conceivable that these objects are born with even higher magnetic fields, reaching several $10^{15}$\,G. Therefore, we utilize simulations conducted with the light envelope model \citep{potekhin2003}, and the results are presented in Fig.~\ref{fig: Luminosity curves light env}. It's worth highlighting that the luminosity predicted by the light envelope model can be significantly brighter, up to an order of magnitude, in terms of X-ray luminosity. The detailed discussion can be found in Chapter~\ref{chap: envelope}. However, when dealing with these highly magnetized objects, caution is paramount. Firstly, it's essential to note that the luminosity curves presented in this figure specifically reflect internal heat. For a more comprehensive analysis, one must consider heating from the magnetosphere, where magnetospheric currents create hotspots on the stellar surface, each approximately $1$\,km in radius. These hotspots appear to play a crucial role in explaining the extreme brightness of highly magnetized magnetars, as neither the higher cooling curves with $B_p=3\times 10^{15}$\,G nor the light envelope model can account for it. Furthermore, Joule dissipation in the envelope layer should be taken into account. This dissipation leads to additional magnetic field decay and, as a result, heats the neutron star's surface.

For neutron stars with magnetic fields $B_p \leq 10^{14}$\,G, the presence of the magnetic field is not expected to have a significant impact on luminosity. Standard cooling curves for $B_p<10^{14}$\,G can adequately account for all these weakly magnetized sources by adjusting the star's mass and/or EoS. Considering the uncertainties in inferred ages and luminosities, the cooling curves for weakly magnetized neutron stars align with all observational data, with a few noteworthy exceptions. To begin with, CCOs exhibit very low $\dot{P}$ values (below the shown range), indicating the presence of a weak external magnetic field. Given our knowledge that they are young, being surrounded by SNRs, their minimal spin-down rate implies that their periods have experienced little change since their birth. The weak magnetic field inferred from timing properties appears to contrast with the relatively high luminosities they exhibit, which are significantly higher than those of standard RPPs. Light-element envelope models (see Fig.~\ref{fig: Luminosity curves light env}) can reconcile their age and large luminosity. This situation aligns with the hidden magnetic field scenario \citep{ho2012,vigano12b}. 

On the other hand, three sources with known real ages stand out as significantly colder by about an order of magnitude compared to other objects of similar young ages. These sources include two RPPs, namely \psra\, ($P=70$\,ms, $B_p = 7\times10^{12}$\,G, and age=841\,yr \citep{Kothes2013}) and \psrb\, ($P=490$\,ms, $B_p = 2\times 10^{13}$\,G, and age=7700\,yr \citep{YarUyaniker2004}), along with a CCO, \velajr\, (2500-5000\,yr \citep{Allen2015}). Regarding the pulsar PSR\,B0656$+$14, initially classified among the extremely cold sources, our simulations in Fig.~\ref{fig: Luminosity curves light env} demonstrate that at older ages ($>10^4$\,yr), PSR\,B0656$+$14 can be explained by using light envelopes (as discussed in Chapter~\ref{chap: envelope}). A detailed study of these faint objects is addressed in \S\ref{sec: cold and young NS}.

\section{Cold and young neutron stars: constraints on the EoS}
\label{sec: cold and young NS}

The secular cooling of neutron stars is influenced by the EoS, mass, magnetic field, and composition of the envelope, which can vary from star to star. Neutron star cooling results from a combination of neutrino emission from the dense core and thermal photons emitted from the outer layers. By measuring the surface temperatures of numerous objects across a wide age range, neutron star cooling models (and consequently, the EoS) can be effectively constrained \citep{page2004}.

Historically, cooling curves have been categorized into two theoretical groups: the standard or minimal cooling curves and the enhanced cooling curves (refer to \S\ref{subsec: neutrino reactions}). So far, all measured surface temperatures of neutron stars have consistently, albeit marginally, aligned with the standard cooling scenario \citep{page2004, potekhin2015}. Although indications of enhanced cooling have been presented, the lack of well-constrained spectral energy distributions, exact ages, or precise distances has impeded firm EoS constraints in that regard. 

In this study, building upon the work submitted by \cite{MDKR2023} to the Nature Astronomy journal, we undertake a comprehensive examination of the three exceptionally cold, young, and nearby neutron stars identified as outliers in Fig.~\ref{fig: Luminosity curves}, namely, \velajr, \psra\, and \psrb. To reconcile theoretical models with observed data, the existence of enhanced cooling processes is necessary, enabling us to impose constraints on the neutron star EoS.

\subsection{Simulations}
\label{subsec: cold simulations}

As hyperons are known to contribute to fast cooling \citep{anzuini2021}, we were particularly interested in including hyperons into the star's core. To account for the presence of hyperons, we employed the 2D magneto-thermal code \citep{anzuini2021, anzuini2022}, that is an extension of the original 2D code developed by \cite{pons2007, aguilera2008, vigano2012, vigano2021}, suitably modified to calculate the magneto-thermal evolution in stars containing both nuclear and hyperon matter. The code smoothly combines the GM1'A EoS \citep{gusakov2014} with the SLy4 EoS \citep{douchin2001} in the crust, while considering the influence of hyperon matter on the star's microphysics. The simulations performed using this EoS are referred to as GM1A. One crucial modification in the microphysics, as introduced in \cite{anzuini2021, anzuini2022} in comparison to \cite{vigano2021}, involves the superfluid gap model, enabling accurate descriptions of stars containing hyperons. Moreover, the most significant effect of this study arises from the inclusion of the hyperon DUrca channel and the Cooper pair of hyperons channel. These two neutrino cooling channels, especially the hyperon DUrca one, lead to enhanced cooling in young neutron stars. For a more comprehensive understanding of these modifications and their implications, we direct the reader to \cite{anzuini2021, anzuini2022}.

We conducted a comprehensive set of 81 simulations, exploring three distinct EoSs that encompass various cooling channels (see \S\ref{subsec: neutrino reactions}). The EoSs used were SLy4 \citep{douchin2001}, which does not activate enhanced cooling processes, BSK24 \citep{pearson2018}, and GM1A \citep{gusakov2014,anzuini2021,anzuini2022}, which, for certain masses, involve enhanced cooling processes such as nucleons DUrca and hyperons DUrca. The latter channel is activated only for the GM1A EoS. Moreover, we considered three different masses (1.4, 1.6, and 1.8\,M$_\odot$) and nine initial magnetic field values for the surface dipolar field, ranging from $1\times10^{12}$\,G to $7\times10^{13}$\,G at the pole. To avoid Joule heating and focus on the cold neutron stars scenario, we did not include a toroidal field. That is due to the fact that the dissipation of the magnetic field in the highly resistive crust could hide the effect of the enhanced cooling mechanisms for stars with magnetic fields above $10^{13}$\,G \citep{aguilera2008}. 
Furthermore, we solely employed an iron envelope model, as alternative compositions containing light elements would predict a thermal luminosity approximately one order of magnitude brighter than that projected by the iron-envelope model at these ages (see Chapter~\ref{chap: envelope}). For the simulations performed using the SLy4 and BSK24 EoSs, we utilize the code from \cite{vigano2021}, whereas for the GM1A EoS, we utilize the code from \cite{anzuini2021, anzuini2022}.

\subsection{Statistical analysis}
\label{subsec: statistical}

\begin{figure}[htp!]
    \includegraphics[width=\columnwidth]{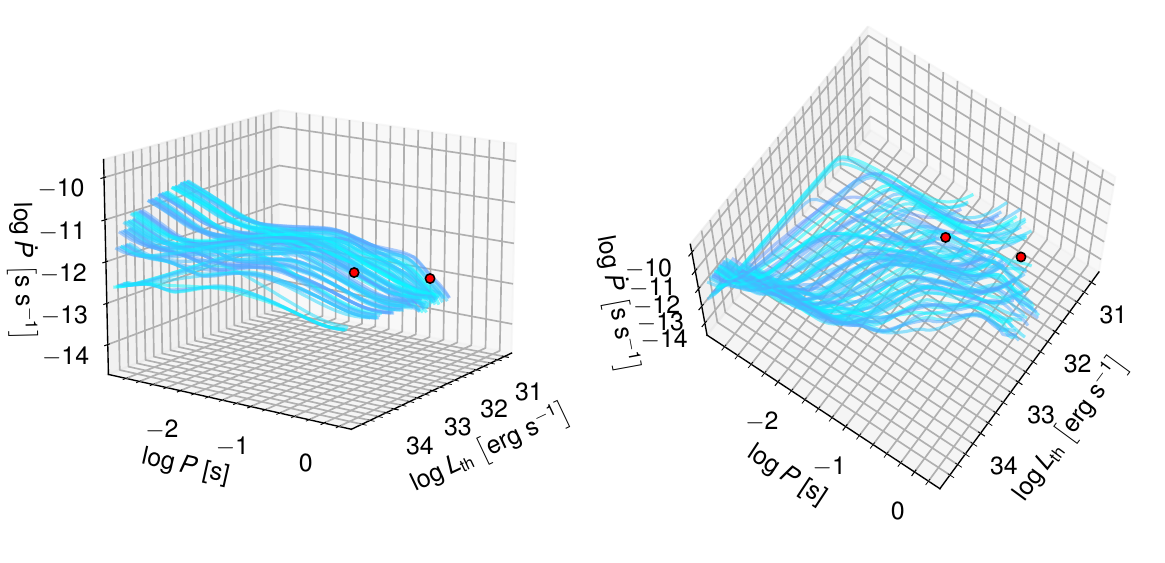}
    \caption{The tracks of our simulations in the $P{-}\dot{P}{-}L_\text{th}$ space in a time span of $100$\,kyr (cyan lines) from two different view angles (left and right panels). The positions of \psra~ and \psrb~are denoted with red discs.}
    \label{fig:3dplots}
\end{figure}

To constrain the nature of each source, we start comparing their observed parameters, $L_\text{th}$, $P$ and $\dot{P}$, against the tracks of the 81 simulations in the same 3D parameter space, each containing 128 points corresponding to various times from the formation of the neutron star up to ${\sim}100$\,kyr age ($0, 1.3, 1.5, 1.8, \ldots, 81200, 97400$\,yr). Ideally, the most probable parameters (EoS, mass, and $B_p$) of a given source are those of the simulated neutron star sharing the same observed features. However, due to the finiteness of the simulations, and the uncertainty on the luminosity (we ignore the uncertainties on $P$ and $\dot{P}$ since the relative uncertainty on $L_\text{th}$ is significantly higher), there is no simulation passing exactly through the positions of the sources in the feature space. Additionally, multiple simulations may be found in the vicinity of the sources (see Fig.~\ref{fig:3dplots}). Consequently, we opt for a machine learning model that given the observational and simulation data, can identify the most probable simulations, and as a consequence, the posterior distribution of the parameters, namely, the EoS, mass and $B_p$. We tried with two approaches, deep-learning and classification, explained in detail in the next paragraphs.

\subsubsection{The deep learning approach}
\label{txt:deeplearning}

The simulation parameters are either categorical (EoS) or continuous (mass and $B_p$). If the categorical and continuous parameters could be constrained by mutually exclusive sets of features (e.g., if the EoS could be constrained only by $L_\text{th}$, while mass and $B_p$ only by $P$ and $\dot{P}$), then their estimation could rely on independent classification and regression models on the separate sets of inputs and output. However, this is not the case in the simulations: we need to employ a machine learning approach that performs both classification and regression simultaneously. Using \textsc{TensorFlow} \citep{tensorflow}, we developed a multilayer perceptron (MLP) neural network that predicts the parameters, with a loss function being the sum of the loss for the EoS, and the loss for the values of mass and $B_p$. The architecture is summarised as an input layer of size 3 (for the three features), fully-connected hidden layers (with rectified linear unit activation functions), and two output layers connected to the last hidden layer: (i) the classification output layer of size 3 and softmax activation function (for the classification probabilities for BSK24, SLy4, GM1A), and the (ii) regression output layer of size 2 and linear activation function (for the mass and $B_p$). We tried different numbers and sizes of hidden layers, loss functions (e.g., mean squared error of the regression output, and cross entropy for the classification output). We found that the models were converging slowly (up to 10000 epochs) with poor results: the best accuracy in predicting the EoS class was ${\sim}65\%$ which is not significant with respect to a \emph{dummy} classifier (33\% accuracy if randomly assigning one of the three EoSs) and the approach in \S\ref{txt:machine} (90\% accuracy). We attribute this failure to the small number of simulations, which being computationally expensive are hampering any efforts to train a successful neural network for our task.

\subsubsection{The classification approach}
\label{txt:machine}

Interestingly, classification can be seen as \lq{}discretised\rq{} regression: instead of estimating the mass of a neutron star, we can classify it into distinct mass classes, $1.4, 1.6$ and $1.8$\,M$_\odot$. However, training three classifiers for EoS, mass and $B_p$ is not optimal since this ignores the interplay between the parameters in the evolution of neutron stars imprinted in the feature space. In our case, each simulation corresponds to a combination of EoS, mass and $B_p$ classes, and therefore there is only one class: the \emph{simulation class} which can be modelled as the ID of the simulation. This is a simple classification task which can be easily carried out with standard, well-understood machine learning classifiers that are easily trained on small data sets. Using a probabilistic classification algorithm to predict the simulation class, we can predict the classification probability of each simulation for each observed source. Then, the posterior of each parameter to have a specific value is simply the sum of all the classification probabilities of the simulations sharing the same value. For example, the posterior of the EoS of a source being BSK24 is the sum of the classification probabilities of all the simulations (classes) for which EoS is BSK24, predicted on the features of the source:
\begin{equation}
P(\text{BSK24}) = \sum_{k=1}^{n} P(\text{BSK24}=\text{EoS}_k) \pi_k,
\end{equation}
where $P(\text{BSK24}=EoS_k)$ is either 1 or 0, denoting whether the BSK24 EoS was used in the $k$-th simulation, $\pi_k$ is the prior on the class which in our case is the classification probability of the $k$-th simulation when predicting the class of the observation, and $n$ is the total number of simulations. The same approach is used for the continuous parameters as well (mass and $B_p$), with the output being still the marginal probabilities at the distinct values used in the simulations.

We stress that one should be careful in the interpretation of the various classification metrics (e.g., accuracy). The number of simulation classes depends on the choices for the range and resolution in the initial conditions, which are generally restricted due to technical and modelling difficulties (e.g., computational cost, available EoS models, etc.). Ideally, a large number of simulations could be run, leading to a paradox: due to the continuous nature of the mass and $B_p$, the classification probability would approach 0 even if the \lq{}correct\rq{} model is present in the training. Consequently, the absolute scale of the accuracy of the trained classifier is not an estimate of the performance of the methodology. Conversely, the relative accuracy between different algorithms (or hyperparameters) measures their relative ability to learn the feature space given the simulation choices and the observational uncertainties.

\subsubsection{Selection of classifier and hyperparameters}

To select the classification algorithm, we design a cross-validation test bed. We consider 8 different classifiers offered by the \textsc{scikit-learn} package \citep{sklearn}: k-nearest neighbour, random forest, decision tree, logistic, support vector (SVC), nu-SVC, MLP, and Gaussian process classifier, and multiple hyperparameter choices for each (3-10 different values for a key hyperparameter such as $k$ for the $k$-nearest neighbour, or the kernel for the Gaussian process classifier).

We set aside $^1/_6$ of the data as a test data set that will be used to estimate the accuracy of the classifier with the best hyperparameters. The remaining data set is separated in 5 folds of equal fractions ($^1/_6$ of the original data). A 5-fold cross-validation approach is adopted to measure the accuracy of the different classifiers and hyperparameter choices. However, the test samples fall very close to the training samples since they follow distinct curves in the feature space. Consequently, most classifiers will exhibit very high performance which does not reflect the accuracy when applied in real data which are subject to measurement uncertainties. For this reason, we \lq{}disturb\rq{} the cross-validation samples by adding Gaussian noise in the decimal logarithm of the luminosity of the simulations (we ignore the uncertainties on $P$ and $\dot{P}$ since they are negligible), to simulate the presence of uncertainty. We train, cross-validate, and test the classifiers at six different scales for the uncertainty: 0 (no disturbance), 0.05, 0.10, 0.15, 0.2, 0.25, and 0.30\,dex, a range that includes the uncertainty on $L_\text{th}$ in our sources (close to $0.2$\,dex). 

For each classifier and $L_\text{th}$ uncertainty level, we use the cross-validation technique to optimise for the hyperparameters. Then, using the test data set we measure the accuracy score, i.e., the fraction of test data points that the algorithm was able to match to their original track. 

Trying all classification algorithms initially, we found that the $k$-nearest neighbours and random forest classifiers presented the highest accuracy scores. Additionally, they are computationally efficient during both training and prediction. For this reason, we focus on these two classification methods from now on.

\begin{figure}[htp!]
    \includegraphics[width=\columnwidth]{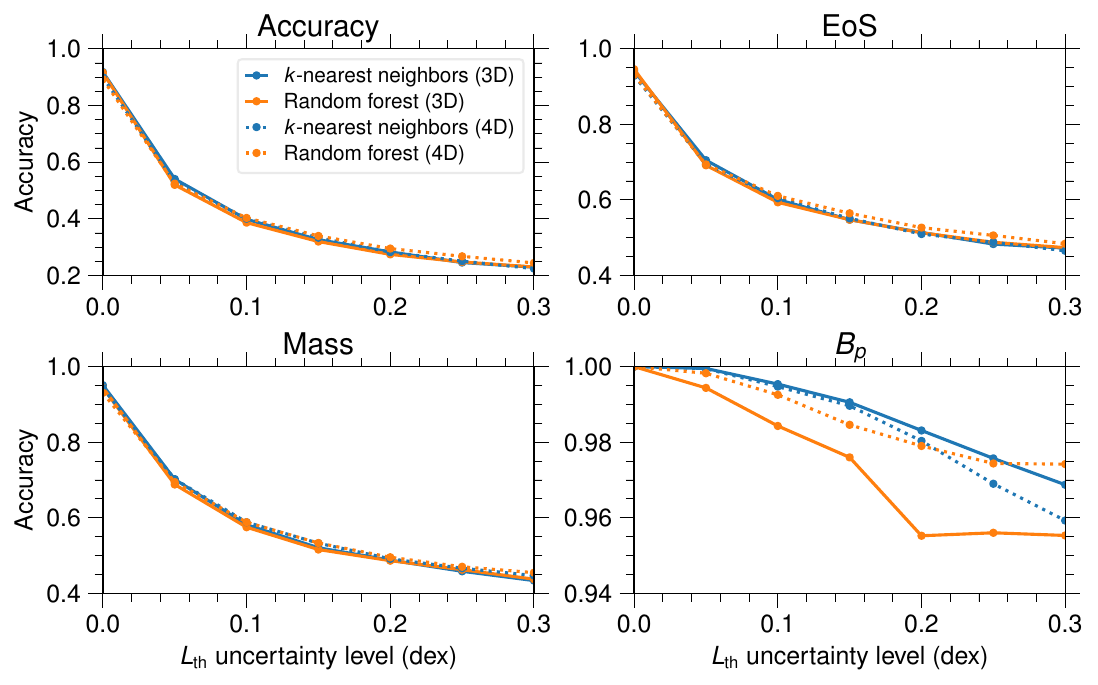}
    \caption{The accuracy score (top left panel; \emph{Accuracy}) of the best-performing 3D (solid lines) and 4D (dotted lines) classifiers, $k$-nearest neighbours (KNN; blue) and random forest (RF; orange), as a function of the uncertainty scale in the $L_\text{th}$. In the rest of the panels we show the marginalised accuracy for the \emph{EoS}, \emph{Mass} and $B_p$ (see titles), i.e., the accuracy in matching the test data points to a model sharing the same value for each parameter.}
    \label{fig:accuracy}
\end{figure}

In the top left panel of Fig.~\ref{fig:accuracy}, with solid lines, we show the accuracy of the $k$-nearest neighbour and the random forest classifier, as a function of the $L_\text{th}$ uncertainty level. Both algorithms perform equally well. For the prediction of the properties of the observed source, we select the random forest classifier because it performs better at high uncertainties (${\gtrsim}0.2$\,dex). Additionally, the random forest classifier has two attractive properties: it's not sensitive to the scale of the features, and it is intrinsically a probabilistic algorithm.

Finally, since we are interested in the ability to predict the physical properties of the pulsars, we measure the accuracy with respect to the ability to predict them independently. For example, if a test data point corresponds to a model with M$=1.4$\,M$_\odot$, does the predicted model have the same mass (no matter what the EoS or $B_p$)? In the top right, bottom left and bottom right panels of Fig.~\ref{fig:accuracy}, we show the marginalised accuracy for the EoS, mass and $B_p$, respectively (with solid lines). We find that the magnetic field is easier to be learned (${>}95\%$ accuracy), while the EoS and mass are sometimes mismatched ($43-95\%$ accuracy), especially at high uncertainty levels. The fact that the accuracy score is not 100\% even with no added noise in the luminosity, reflects the degeneracy between the models: as it can be seen in Fig.~\ref{fig:3dplots}, the tracks of the different simulations often occupy the same regions of the feature space. Here, we remind that the accuracy score should be used only for comparisons of algorithms, and not as a measure of performance of the methodology.

\subsubsection{Predictions accounting for the uncertainty on the luminosity}
\label{txt:asymmetric}

We use the random forest classifier that has been optimised for the level of $L_\text{th}$ uncertainty in our observations, retrained using all the available simulation data (without separation to training, validation, and test data sets \citep{Tsamardinos22}). However, the uncertainty of the luminosity is not used directly during prediction. To take into account the observational uncertainty of a given source's luminosity, we use a Monte Carlo approach: we sample the error distribution of the luminosity 100,000 times, predict the properties of the sources, and sum the results. To sample the error distribution of the luminosity, we use Monte Carlo uncertainty propagation: we model the error distributions of the flux and the distance, draw samples from them, and calculate the luminosity samples. This is to avoid standard uncertainty propagation\footnote{We note that applying the standard uncertainty propagation formula (by averaging the low and high error bars) resulted to ${\sim}20\%$ differences in the resulting classification probabilities with respect to the Monte Carlo uncertainty propagation using asymmetric error distributions as outlined below.} for three reasons \citep{Singh21}: (i) we have high relative uncertainties in the quantities involved; (ii) the flux and distance confidence intervals are not symmetric (especially in the case of the distance of \psrb); (iii) the classifier operates in log-space where even symmetric error bars are transformed into asymmetric ones.

\begin{figure}[htp!]
    \includegraphics[width=\columnwidth]{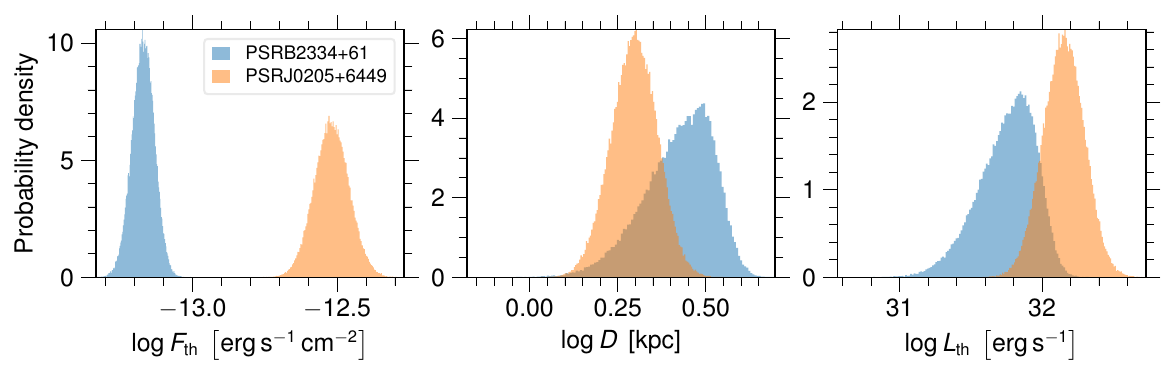}
    \caption{Histograms of samples from the distributions of the flux ($F_\text{th}$; left), distance (middle) and luminosity ($L_\text{th}$; right) in logarithmic space, for both sources (PSRB2334+61 with blue, and PSRJ0205+6449 with orange). The shapes of the error distributions illustrate the asymmetry in the independent variables (flux and distance) and the Monte Carlo-propagated luminosity.}
    \label{fig:lx_samples}
\end{figure}

First, we represent the flux and distance uncertainties using the binormal distribution which is flexible enough to represent asymmetric distributions. A binormal distribution's probability density function (PDF) is the result of stitching together the opposite halves of two distinct normal distributions with the same mean value (which acts as the mode of the new distribution) but different standard deviations \citep{Wallis14}. Consequently, we represent each flux and distance measurement with a binormal distribution by adopting the measured value as the mode, and fitting for the two standard deviations such that the reported confidence intervals are matching those of the binormal distribution. In Fig.~\ref{fig:lx_samples}, we show the PDFs of the constructed binormal distributions (in the form of histograms of their samples) for the flux and distance (left and middle panel, respectively) of the two sources for which we apply the classification, as well as the derived luminosity error distribution.

We predict the probability of the 81 simulations, 100,000 times for each sample from the $L_\text{th}$ distribution of each source. By summing up the 100,000 results, we find the classification probability for each model. In the top panel of Table~\ref{tab:classprob}, we show the models the highest classification probabilities. In addition, in the table, we report the marginalised probability for the EoS, mass and $B_p$ of each source, which are also shown in Fig.~\ref{fig:posteriors}.

\begin{table*}
    \centering
    \begingroup
    \footnotesize
    \setlength{\tabcolsep}{8pt} 
    \renewcommand{\arraystretch}{0.7} 
    \begin{tabular}{|l|c||l|c|}
    \hline
\multicolumn{2}{|c||}{PSRB2334+61}  &  \multicolumn{2}{c|}{PSRJ0205+6449} \\
 \hline \multicolumn{4}{|c|}{\textsc{I. Five best models and their classification probabilities}} \\ \hline 
GM1A, $1.6\,M_\odot$, $2{\times}10^{13}$G  &  0.4534  &  BSK24, $1.6\,M_\odot$, $7{\times}10^{12}$G  &  0.3995 \\
BSK24, $1.8\,M_\odot$, $1{\times}10^{13}$G  &  0.1070  &  GM1A, $1.6\,M_\odot$, $7{\times}10^{12}$G  &  0.1064 \\
GM1A, $1.8\,M_\odot$, $2{\times}10^{13}$G  &  0.0990  &  BSK24, $1.6\,M_\odot$, $1{\times}10^{13}$G  &  0.0881 \\
BSK24, $1.6\,M_\odot$, $2{\times}10^{13}$G  &  0.0571  &  GM1A, $1.4\,M_\odot$, $7{\times}10^{12}$G  &  0.0688 \\
GM1A, $1.6\,M_\odot$, $3{\times}10^{13}$G  &  0.0558  &  GM1A, $1.6\,M_\odot$, $1{\times}10^{13}$G  &  0.0546 \\
 \hline \multicolumn{4}{|c|}{\textsc{II. Equation of state}} \\ \hline 
BSK24  &  0.2855  &  BSK24  &  \textbf{0.6426} \\
GM1A  &  \textbf{0.6986}  &  GM1A  &  0.3519 \\
SLy4  &  0.0159  &  SLy4  &  0.0055 \\
 \hline \multicolumn{4}{|c|}{\textsc{III. Mass (M$_\odot$)}} \\ \hline 
1.4  &  0.0093  &  1.4  &  0.1259 \\
1.6  &  \textbf{0.6307}  &  1.6  &  \textbf{0.7577} \\
1.8  &  0.3600  &  1.8  &  0.1165 \\
 \hline \multicolumn{4}{|c|}{\textsc{IV. $B_p$ (G)}} \\ \hline 
$1{\times}10^{12}$  &  0.0002  &  $1{\times}10^{12}$  &  0.0118 \\
$3{\times}10^{12}$  &  0.0171  &  $3{\times}10^{12}$  &  0.0363 \\
$5{\times}10^{12}$  &  0.0390  &  $5{\times}10^{12}$  &  0.0710 \\
$7{\times}10^{12}$  &  0.0400  &  $7{\times}10^{12}$  &  \textbf{0.6111} \\
$1{\times}10^{13}$  &  0.1923  &  $1{\times}10^{13}$  &  0.2069 \\
$2{\times}10^{13}$  &  \textbf{0.6213}  &  $2{\times}10^{13}$  &  0.0484 \\
$3{\times}10^{13}$  &  0.0900  &  $3{\times}10^{13}$  &  0.0102 \\
$5{\times}10^{13}$  &  0.0000  &  $5{\times}10^{13}$  &  0.0041 \\
$7{\times}10^{13}$  &  0.0000  &  $7{\times}10^{13}$  &  0.0001 \\
    \hline
    \end{tabular}
    \caption{For the two sources, \psrb~and \psra, the five most probable models sorted according to their classification probability (panel \emph{I}), as well as the marginalised probabilities of the considered EoSs, and values for the mass and magnetic field (panels \emph{II}--\emph{IV}). The highest marginal probabilities are denoted with bold typeface.}
    \label{tab:classprob}
    \endgroup
\end{table*}

\subsubsection{Accounting for the age information}

Since the simulations track the evolution of the properties of the pulsars in time, if the real age of the source is known, it provides an additional constraint. We add the time as another input variable, making the feature space four-dimensional, and then repeat the above analysis: (i) optimise the two classification algorithms for different luminosity uncertainty scales (see dotted lines in Fig.~\ref{fig:accuracy}); (ii) select the random forest classifier optimised for the $0.2$\,dex uncertainty scale (it outperforms the $k$-nearest neighbour classifier at uncertainty scales ${>}0.1$\,dex); (iii) predict the parameters of the two sources using the age estimates in Table~\ref{tab:timing-1}.

\begin{figure*}
    \centering
    \includegraphics[width=\columnwidth]{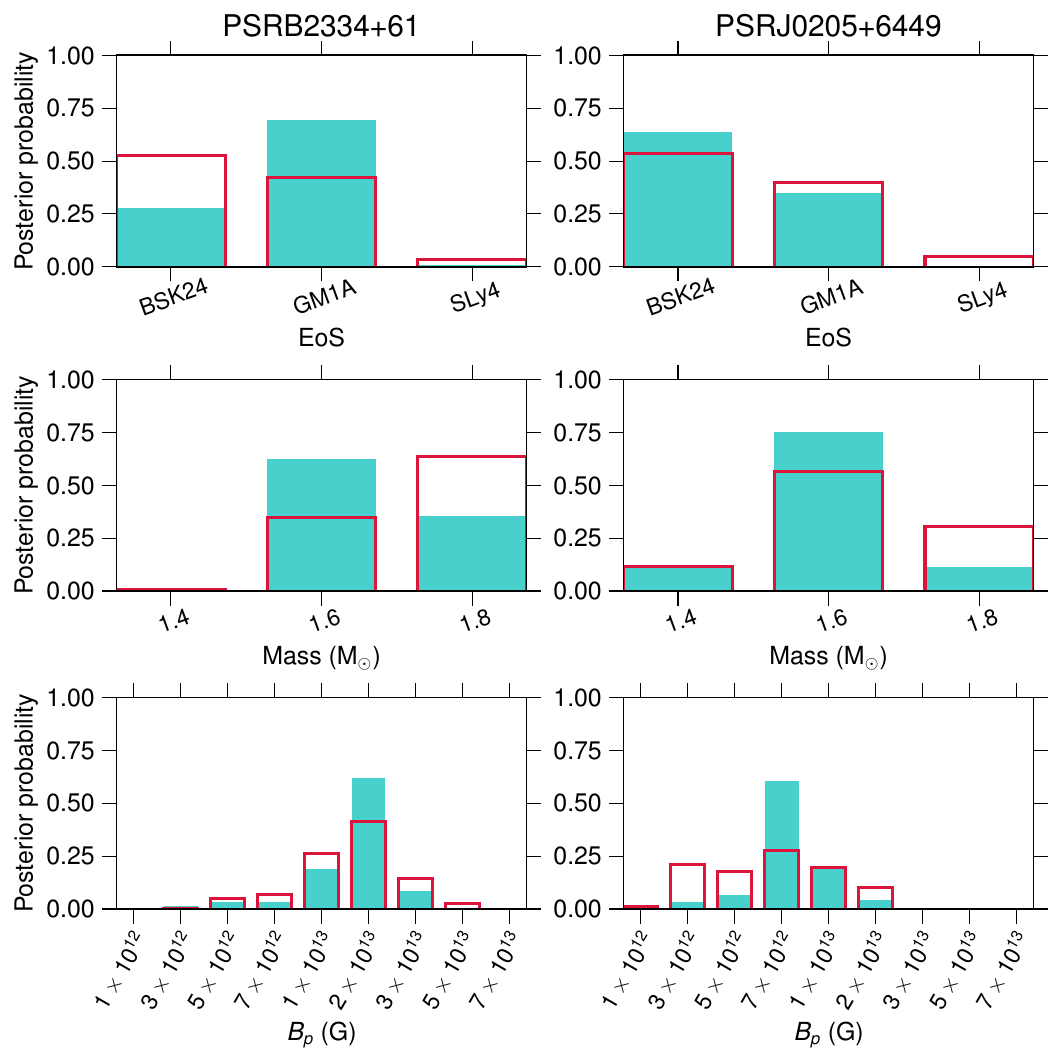}
    \caption{The marginalised probabilities for the EoS, mass and $B_p$ of the two sources, \psrb~and \psra, using the 3D (cyan filled bars) and 4D classifiers (red rectangles).}
    \label{fig:posteriors}
\end{figure*}

Having a 4D feature space, it is impossible to visually inspect its coverage by the simulation tracks. Instead, we ensured that the observed sources fall within the range of the simulation evolutionary tracks by confirming that their positions are inside a simplex in Delaunay hypertetrahedralisation on the simulation data. The predictions of the 4D classifier are shown in Table~\ref{tab:classprob4D} and Fig.~\ref{fig:posteriors} (with red rectangles).

%
These results are for the most part consistent with the 3D classifier. For both sources, the 3D and 4D classifiers indicate the same most probable value for the $B_{p}$. For \psra, the most probable values for the EoS and mass are the same, while for \psra they differ, but not significantly (BSK24 and GM1A EoS, and 1.6 and 1.8\,M$_\odot$ masses have high marginal posteriors ${>}25\%$).

We notice that the posteriors in the 4D case are less ``peaky''.
This is in contrast to our expectation that adding another feature (age) would create larger contrast between the marginal probabilities (effectively lowering the entropy of information). 
Given the small number of EoSs in our simulations, at this stage, we do not consider our method well suited to constrain the EoS itself. Nevertheless, for both sources, our statistical analysis do shows that the SLy4 EoS and the $1.4$\,M$_\odot$ mass scenario are found to be highly improbable with or without considering the age information.

\begin{table*}
    \centering
    \begingroup
    \footnotesize
    \setlength{\tabcolsep}{8pt} 
    \renewcommand{\arraystretch}{0.7} 
    \begin{tabular}{|l|c||l|c|}
    \hline

\multicolumn{2}{|c||}{PSRB2334+61}  &  \multicolumn{2}{c|}{PSRJ0205+6449} \\
 \hline \multicolumn{4}{|c|}{\textsc{I. Five best models and their classification probabilities}} \\ \hline 
GM1A, $1.6\,M_\odot$, $2{\times}10^{13}$G  &  0.1980  &  BSK24, $1.6\,M_\odot$, $7{\times}10^{12}$G  &  0.1203 \\
BSK24, $1.8\,M_\odot$, $1{\times}10^{13}$G  &  0.1641  &  BSK24, $1.8\,M_\odot$, $2{\times}10^{13}$G  &  0.0857 \\
BSK24, $1.8\,M_\odot$, $2{\times}10^{13}$G  &  0.1087  &  BSK24, $1.6\,M_\odot$, $5{\times}10^{12}$G  &  0.0755 \\
BSK24, $1.8\,M_\odot$, $3{\times}10^{13}$G  &  0.0651  &  BSK24, $1.6\,M_\odot$, $1{\times}10^{13}$G  &  0.0723 \\
BSK24, $1.8\,M_\odot$, $7{\times}10^{12}$G  &  0.0629  &  BSK24, $1.6\,M_\odot$, $3{\times}10^{12}$G  &  0.0696 \\
 \hline \multicolumn{4}{|c|}{\textsc{II. Equation of state}} \\ \hline 
BSK24  &  \textbf{0.5323}  &  BSK24  &  \textbf{0.5409} \\
GM1A  &  0.4265  &  GM1A  &  0.4054 \\
SLy4  &  0.0413  &  SLy4  &  0.0537 \\
 \hline \multicolumn{4}{|c|}{\textsc{III. Mass (M$_\odot$)}} \\ \hline 
1.4  &  0.0101  &  1.4  &  0.1226 \\
1.6  &  0.3512  &  1.6  &  \textbf{0.5682} \\
1.8  &  \textbf{0.6387}  &  1.8  &  0.3092 \\
 \hline \multicolumn{4}{|c|}{\textsc{IV. $B_p$ (G)}} \\ \hline 
$1{\times}10^{12}$  &  0.0000  &  $1{\times}10^{12}$  &  0.0157 \\
$3{\times}10^{12}$  &  0.0082  &  $3{\times}10^{12}$  &  0.2136 \\
$5{\times}10^{12}$  &  0.0544  &  $5{\times}10^{12}$  &  0.1813 \\
$7{\times}10^{12}$  &  0.0725  &  $7{\times}10^{12}$  &  \textbf{0.2806} \\
$1{\times}10^{13}$  &  0.2674  &  $1{\times}10^{13}$  &  0.2015 \\
$2{\times}10^{13}$  &  \textbf{0.4195}  &  $2{\times}10^{13}$  &  0.1072 \\
$3{\times}10^{13}$  &  0.1476  &  $3{\times}10^{13}$  &  0.0001 \\
$5{\times}10^{13}$  &  0.0304  &  $5{\times}10^{13}$  &  0.0000 \\
$7{\times}10^{13}$  &  0.0000  &  $7{\times}10^{13}$  &  0.0001 \\

\hline
    \end{tabular}
    \caption{Same as Table~\ref{tab:classprob} but using the 4D classifier which accounts for the age information.}
    \label{tab:classprob4D}
     \endgroup
\end{table*}

\subsection{Discussion}
\label{subsec: constraining nuclear EOS}

\begin{figure*}
\centering
\includegraphics[width=0.9\textwidth]{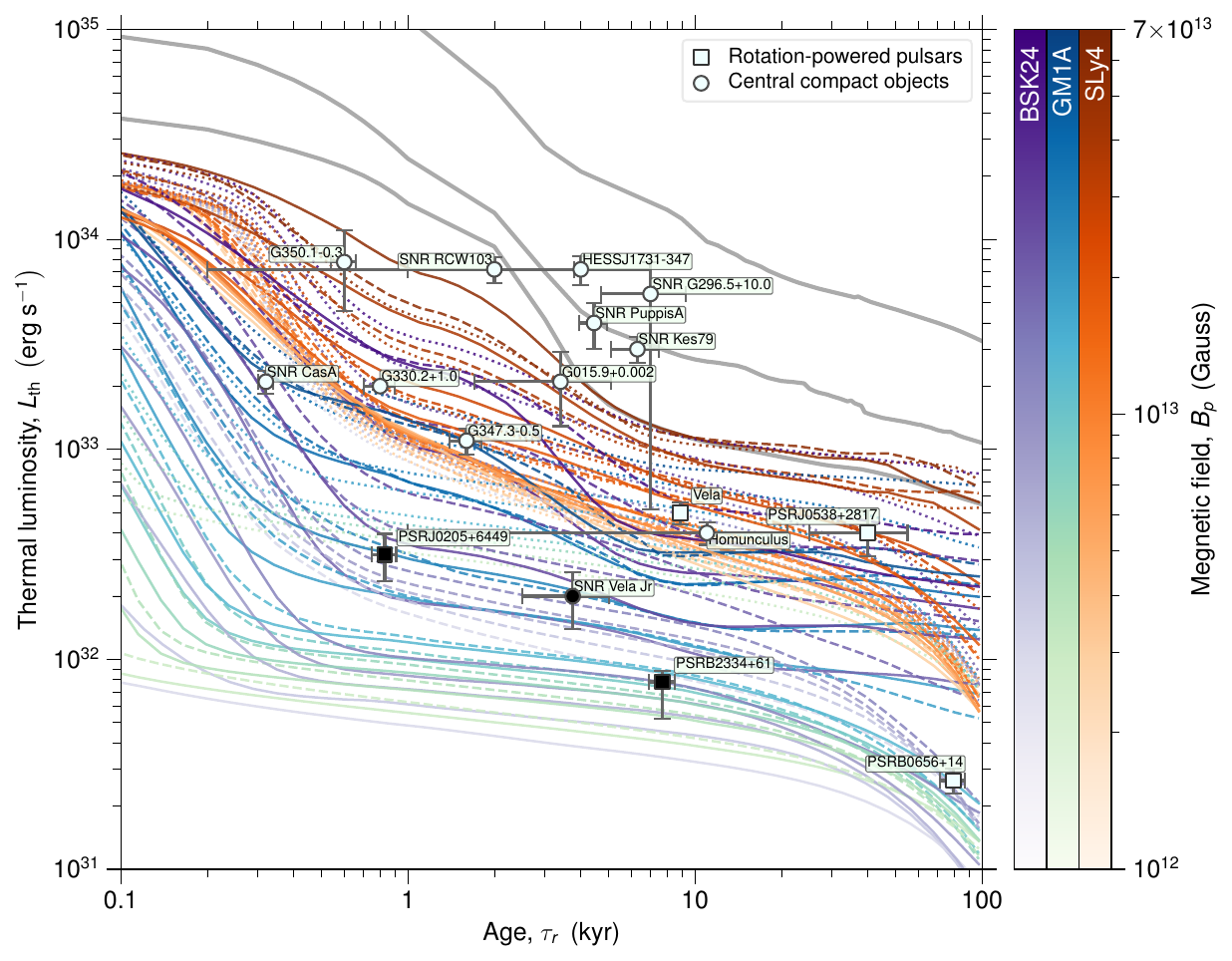}
\caption{Comparison between observational data and theoretical cooling curves. RPPs are represented with squares, while CCOs are denoted with circles. The three sources we study in this work are coloured in black. The 81 theoretical cooling curves used in our analysis are shown for three different EoSs: SLy4 (in orange), BSK24 (in violet), and GM1A (in blue). We considered three distinct masses: 1.4\,M$\odot$ (dots), 1.6\,M$\odot$ (dashed lines), and 1.8\,M$_\odot$ (solid lines). Additionally, we investigated nine initial magnetic field values for the surface dipolar field at the pole ranging from $1 - 70\times10^{12}$\,G. Only for comparison, we also show three grey curves corresponding to stronger magnetic field intensities of $10^{14}$\,G, $3\times 10^{14}$\,G, and $10^{15}$\,G, not used in our statistical analysis for the BSk24 EoS and considering 1.6\,M$_{\odot}$. }
\label{fig: cooling}
\end{figure*}

In Fig.~\ref{fig: cooling}, we present the thermal luminosity of a subset of objects discussed in \S\ref{sec: observational data}, specifically focusing on all the RPPs and CCOs with precise age and distance estimates. This approach avoids reliance on the characteristic age of the pulsar, which is inferred from timing analysis and is known to have large uncertainties. These sources are notably young, with ages ranging between  $800 - 8000$\,yrs, yet their extreme coldness suggests the presence of rapid/enhanced cooling mechanisms. We compare the observational measurements with the magneto-thermal evolutionary tracks of the simulations described in \S\ref{subsec: cold simulations}. Upon initial examination, it becomes evident that some explored scenarios do not match the faint thermal luminosities of these extremely cold sources. Specifically, when assuming a SLy4 EoS (as indicated by the orange curves), the significant drop in surface temperature observed in the three outliers cannot be achieved for any combination of mass and magnetic fields. However, in the exotic scenario involving hyperons in the core (GM1A, indicated by the blue/green curves), the cooling might proceed rapidly enough to be consistent with the observational data. Additionally, for the BSK24 EoS, when the neutron star is massive, e.g., $M \geq 1.6$\,M$_{\odot}$, nucleon DUrca is activated, and the tracks exhibit enhanced cooling compatible with the data.

To pursue a more quantitative approach, we employed machine learning methods (see \S\ref{subsec: statistical}) to find curves that best describe each source. To achieve this, we utilized a 3D space encompassing the thermal luminosity $L_\text{th}$, spin period $P$, and period derivative $\dot{P}$ as parameters for both the observational data and simulations. Furthermore, we extended these simulations into a 4D space, incorporating the age of the sources. However, due to the unknown values of $P$ and $\dot{P}$ for the CCO \velajr, the analysis was restricted to the two RPPs (\psra\, and \psrb). Extending to a 4D space allowed us to verify whether curves explaining the observed $L_\text{th}$ would also predict $P$ and/or $\dot{P}$ compatible with the timing parameters of the sources at the same age.

\begin{figure}
\centering
\includegraphics[width=0.49\textwidth]{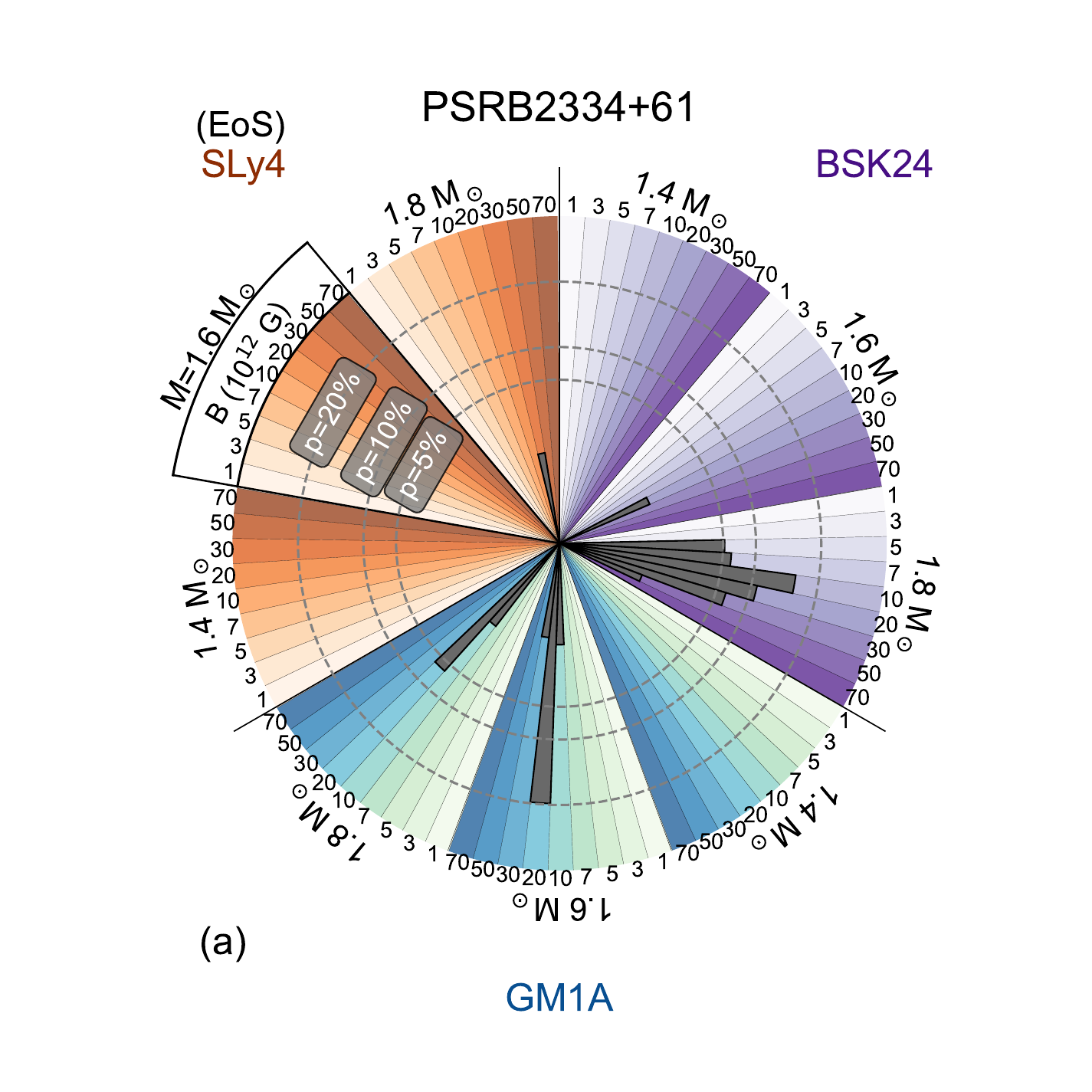}
\includegraphics[width=0.49\textwidth]{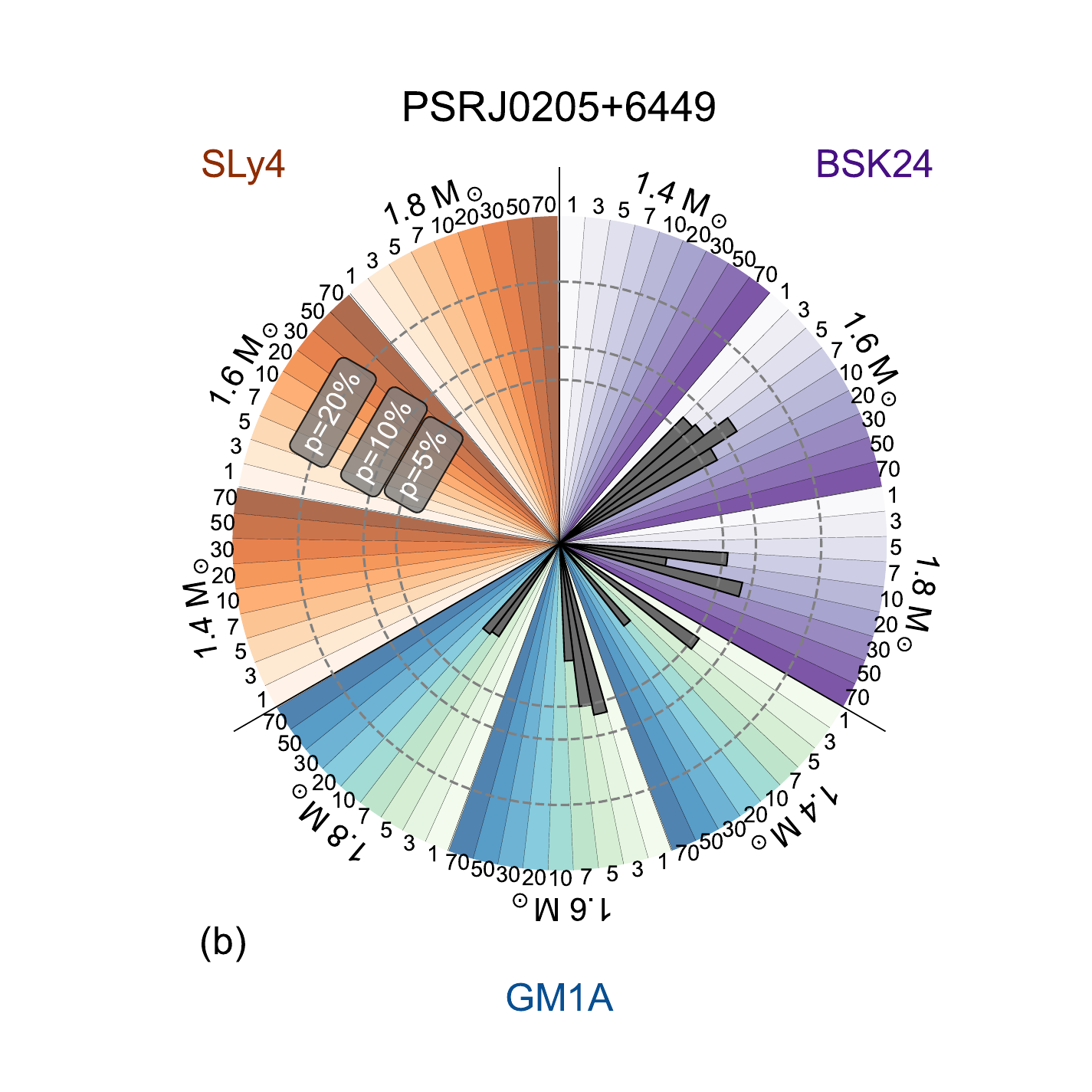}
\caption{Chart pies breaking down the results of our 4D classification method for \psrb\, (left) and \psra\, (right) and to test how well our simulations match our data. Each chart is divided in three sectors for each of the EoS used, marked with different colours (e.g. purple for BSK24, orange for SLy4, green for GM1A). In each sector, three subsections are individuated for the different masses used. Finally, the subsections are broken down further in cloves representing different values of the magnetic field used in the simulations. Superimposed to these cloves, grey histograms are drawn. Their radial length expresses the marginal probability for the simulation-clove to be the real cooling curve followed by the source.}
\label{fig:statistical}
\end{figure}

In Fig.~\ref{fig:statistical}, we present a pie chart summarizing the results of our 4D machine learning simulations (the 3D results are very similar \S\ref{subsec: statistical}). Based on these methods, we can quantitatively rule out the EoS without a physical mechanism to activate enhanced cooling at a young age, specifically the SLy4 EoS.

For \psrb, all cooling curves indicate that the source likely has a relatively high mass ($\sim 1.8 M_{\odot}$ when the age is considered) and a dipolar magnetic field at birth of around $\sim3-5\times10^{13}$\,G, which is compatible with its current value. Although there seems to be a slight preference for the hyperon EoS (GM1A) for this pulsar, it's important to note that conducting more simulations with other enhanced cooling EoSs may yield similar probabilities. Thus, at this stage, we cannot precisely constrain the exact EoS using this technique.

Regarding \psra, the results from applying these methods indicate that both enhanced cooling EoSs, BSK24 and GM1A, with masses around $\sim 1.6$\,M$_{\odot}$, are compatible with its observed parameters. The simulations also suggest an initial magnetic field within $\sim3-10\times10^{12}$\,G, which agrees with the estimated value for this pulsar. For both \psrb\ and \psra, simulations with the SLy4 EoS, considering only standard cooling for the three studied masses, do not provide an acceptable match with the data (with probabilities below 16\%, or below 5.4\% when age is taken into account; see \S\ref{subsec: statistical}) for any mass or magnetic field. Furthermore, this conclusion holds true for \velajr\, when considering only the source's thermal luminosity and age.

\begin{figure}
\centering
\includegraphics[width=0.9\textwidth]{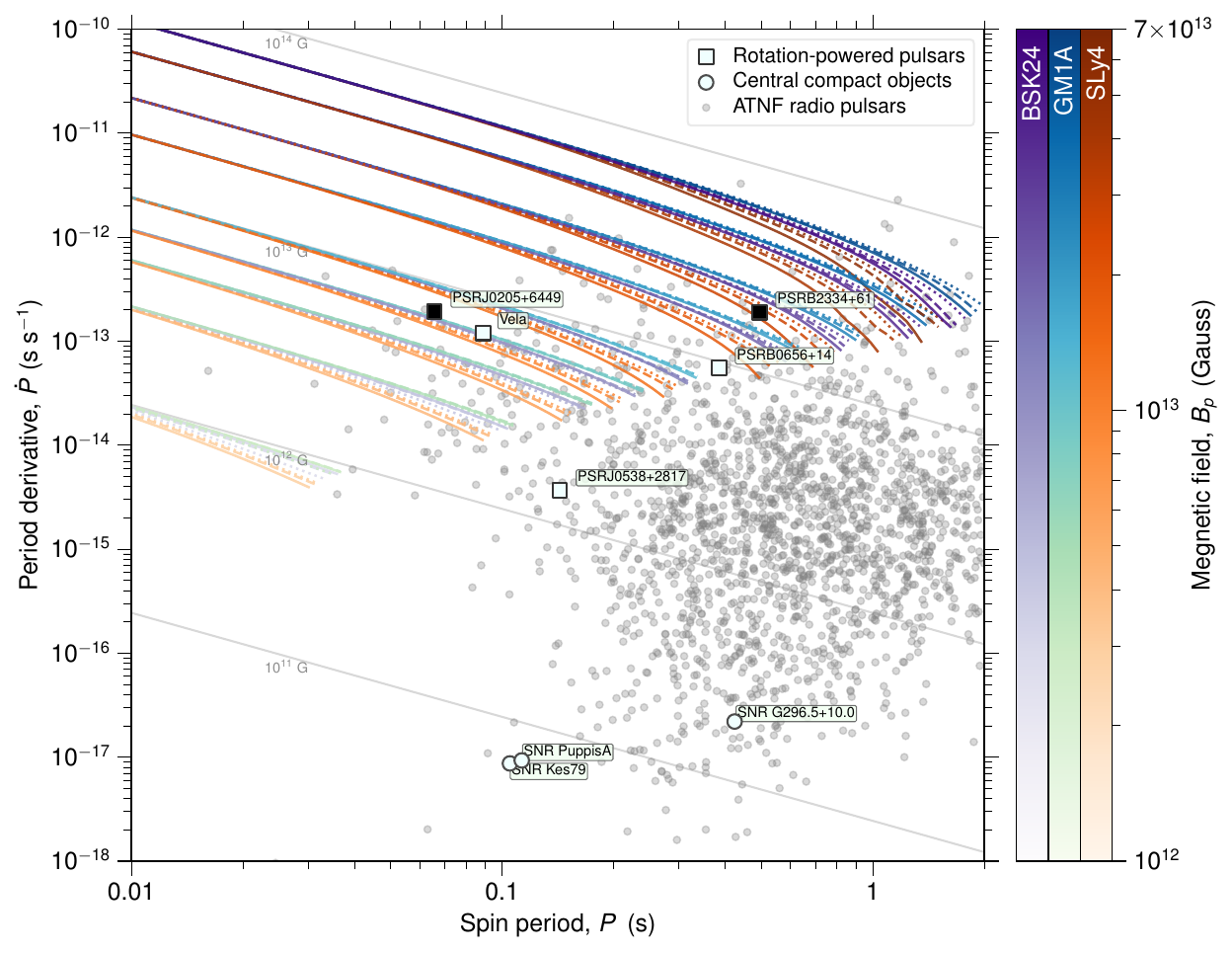}
\caption{Same as Fig.~\ref{fig: cooling} but showing the period $P$ and the period derivative $\dot{P}$.}
\label{fig: p-pdot}
\end{figure}

To provide a comprehensive view, in Fig.~\ref{fig: p-pdot}, we present the same simulations as in Fig.~\ref{fig: cooling}, but this time depicting the evolution of period and period derivative.

Given the uniqueness of the EoS, these results provide compelling evidence that neutron stars cannot be governed by an EoS that is incompatible with the low luminosities observed in \psra, \psrb, and \velajr. This marks the first direct measurement of enhanced cooling. Our findings indicate that only EoSs (and compositions) permitting fast cooling processes in the first few thousand years can be successfully reconciled with the thermal emission from all sources in our sample.

While these results do not delve into a detailed and comprehensive analysis of various possibilities for fast cooling (e.g., hyperons, quarks, pure nucleonic matter with very large symmetry energy), nor does it conduct a systematic comparison with all available observational data, the issue itself triggers thoughtful consideration. We now further support the initial suggestion by \cite{aguilera2008} that DUrca cooling could be concealed by strong magnetic fields, potentially leading to misidentifications in the context of fast cooling. Consequently, we have solid evidence for fast neutrino cooling processes taking place in the interior of many neutron stars, a consideration that any realistic EoS should incorporate. This study contributes significantly to constraining nuclear EoS, as a suitable EoS should elucidate both exceptionally bright objects like magnetars and extremely faint objects at young ages. Taking into account a simplified meta-modeling approach \citep{margueron2017}, the excluded range is estimated to encompass approximately 75\% of the proposed nuclear EoSs.

\subsubsection*{Corresponding scientific publications}
\underline{C.~Dehman}, A.~Marino, K.~Kovlakas, N.~Rea et al., \textbf{An updated unification of the isolated neutron stars diversity via magneto-thermal evolution models}, in preparation. \\\\
A.~Marino$^*$, \underline{C.~Dehman}$^*$, K.~Kovlakas$^*$, N.~Rea$^*$ et al., \emph{\color{darkgray}``These authors contributed equally to this work''.} \textbf{Young and cold neutron stars constrain dense matter equation of state}, submitted to \emph{Nature Astron.} \\ \\
F.~Anzuini, A.~Melatos, \underline{C.~Dehman}, D.~Vigan\`o \& J.A.~Pons: 2022, \textbf{Fast cooling and internal heating in hyperon stars}, \emph{Mon.~Not.~Roy.~Astron.~Soc., $509$, $2609$}
(\href{https://arxiv.org/abs/2110.14039}{\underline{arXiv:2110.14039}},\href{https://ui.adsabs.harvard.edu/abs/2022MNRAS.509.2609A/abstract}{\underline{ADS}},\href{https://doi.org/10.1093/mnras/stab3126}{\underline{DOI}}). \\ \\
F.~Anzuini, A.~Melatos, \underline{C.~Dehman}, D.~Vigan\`o \& J.A.~Pons: 2022, \textbf{Thermal luminosity degeneracy of magnetized neutron stars with and without hyperon cores}, \emph{Mon.~Not.~Roy.~Astron.~Soc., $515$, $3014$}
(\href{https://arxiv.org/abs/2205.14793}{\underline{arXiv:2205.14793}},\href{https://ui.adsabs.harvard.edu/abs/2022MNRAS.515.3014A/abstract}{\underline{ADS}},\href{https://doi.org/10.1093/mnras/stac1353}{\underline{DOI}}).

\clearemptydoublepage
\let\textcircled=\pgftextcircled
\chapter{Crustal Failures in Young Magnetars}
\label{chap: outburst}

\initial{M}agnetars, the most magnetic objects within the neutron star population, show magnetically-driven enhancements of their thermal and non-thermal emission, commonly referred to as outbursts \citep{beloborodov2016,cotizelati2018}. Their irregular and spectacular activity also includes short bursts and flares, which are believed to involve the magnetosphere and may possibly be triggered by interior dynamics.

Potentially related to magnetars are fast radio bursts (FRBs) \citep{margalit2018}, very bright ($\sim$ 50\,mJy - 100\,Jy) millisecond-long bursts in the radio band, known for over a decade. The extragalactic nature of FRBs has been established through the derivation of their distances from radio dispersion measures (DMs). This was further confirmed by identifying their host galaxies, which affirmed their extragalactic origin (for recent reviews, see \cite{petroff2019,cordes2019}). Their short durations, coherent radio emission, and the discovery of several repeating FRBs \citep{spitler2014,fonseca2020}, strengthened their connection with neutron stars, and in particular with strongly magnetized ones \citep{wadiasingh2020,cheng2020}.

Among the tens of theoretical models (see \cite{platts2019} for a list), \cite{lyutikov2016} propose that FRBs are giant pulses coming from young neutron stars spinning at very fast rotation frequencies. On the other hand, many works assume that the star's magnetic energy is the ultimate engine behind the emission. For instance, \cite{lyubarsky2014} proposed maser synchrotron emission arising from far regions, where the magnetar wind interacts with the nebula inflated by the wind within the surrounding medium, and generates a forward and a reverse shock, both able to provide intense, coherent emission frequencies of tens to hundreds of MHz. A variant is the baryonic shell model, originally proposed by \cite{beloborodov2017} and further developed by \cite{metzger2019} and \cite{margalit2020}, in which an ultra-relativistic flare collides with matter ejected from an earlier flare. \cite{kumar2020} propose a model where a relatively small Alfv\'en disturbance at the surface, with amplitude $\delta B\sim 10^{11}$\,G, is able to trigger the FRB emission via an efficient conversion to coherent curvature radiation from plasma bunches, accelerated by local electric fields.

The recent detection of a FRB-like radio bursts simultaneously with a bright hard X-ray burst from the Galatic magnetar SGR\,1935+2154 \citep{chime2020,bochenek2020,mereghetti2020} showed for the first time that magnetar bursts can indeed be associated with FRB-like radio emission.
It has been speculated that the underlying mechanism is the same as for the extragalactic events, for which we see only the very bright events from a much larger volume \citep{margalit2020}.

Several works, e.g., \cite{yuan2020,lu2020}, have proposed a magnetic disturbance from the surface driving the FRB. Regardless of the location and the exact mechanism, in the majority of magnetically-powered FRB models, the ultimate trigger to the event is expected to come from the interior of the neutron star. Internal magnetic fields evolve in the long term due to the Ohmic dissipation and Hall drift in the crust \citep{vigano2013,pons2019,dehman2022,dehman2023c}. In the core, recent significant progress has been made in modeling and understanding the ambipolar diffusion and the evolving MHD equilibrium in superfluid (e.g. \cite{graber2015,bransgrove2017,gusakov2019}) and non-superfluid cores (e.g., \cite{castillo2017,ofengeim2018,dommes2020,castillo2020}), which for extreme values of the magnetic field may proceed on relatively short timescales, thus changing the magnetic field on the crust-core boundary \citep{beloborodov2016}.

In this chapter, we evaluate how often the crust of a magnetar will fail during the first centuries of its life, using an updated version of our magneto-thermal code \citep{vigano2012}. This study improves and extends to the early stage previous results obtained for middle-aged magnetars \citep{perna2011,pons2011}, by including the Hall effect and up-to-date microphysics, but not the mechanisms in the core mentioned above. Since the early-stage dynamics are extremely sensitive to the initial conditions and the field topology, we perform the analysis under different assumptions on the initial magnetic field strength and geometry. We discuss our results in light of the current knowledge of FRB rates, recurrent times, and predictions in the FRB-magnetar models.

\section{Magnetic stresses from magneto-thermal models}

As explained in \S\ref{sec: stellar evolution}, during the initial decades of a neutron star's life, the star rapidly cools from its initial temperature of $10^{10}-10^{11}$\,K to a few times $10^8$\,K. As the temperature drops below the density-dependent melting value, the crust grows as a solid lattice of heavy ions. The crust freezing starts from the inner region a few minutes after birth but takes several years to extend to the outer crust (e.g., Fig.~8 of \cite{aguilera2008b}). The crust is thought to provide an elastic response to stresses \citep{chugunov2010} up to a maximum value:

\begin{equation}
    \sigma^{{max}}= \bigg(0.0195 - \frac{1.27}{\Gamma-71} \bigg)n_i\frac{Z^2 e^2}{a},
    \label{eq: sigma max}
\end{equation}
where $\Gamma=Z^2 e^2 / aT $ is the Coulomb coupling parameter, $a=\big[3/(4\pi n_i) \big]^{(1/3)}$ the ion sphere radius, $n_i$ the ion number density, $Z$ the charge number, $e$ the elementary charge, and $T$ the temperature. A recent calculation \citep{kozhberov2020} confirms the validity of the approximation (eq.\,\eqref{eq: sigma max}).

As the magnetic field evolves, the magnetic tensor $\hat{M}_{ij}(\boldsymbol{x},t)=B_i(\boldsymbol{x},t) B_j(\boldsymbol{x},t)/(4\pi)$ changes. Moreover, the stress quantified as a difference between $M_{ij}$ and an equilibrium $M_{ij}^{eq}$, piles up in the crust. When it reaches the maximum stress locally allowed, $(M_{ij}- M_{ij}^{eq})\simeq \sigma^{max}(\boldsymbol{x})$, the crust fails. After the event, a new equilibrium state is established, the affected region freezes again and the crust starts to respond elastically, storing further stress until the cycle is repeated. 

An evaluation of the frequency of such events, which can potentially trigger magnetar activity, can be done by properly following the magneto-thermal evolution and the accumulated local stresses. Such simulations have been performed by \cite{perna2011} and \cite{pons2011}, considering the age range of the observed sources, $\sim 400-10^5$\,yr. They used the magneto-thermal code by \cite{pons2009}, not considering Hall effects. 


\begin{table}
\centering
	\begin{tabular}{ccccccccccc}					
  \hline
Model & Topology & $B_{p}$  &  $E_{mag}^{cr}$ & $E_{tor}^{cr}/E_{mag}^{cr}$ &$N^{100}$ & $N^{400}$ & $N^{1000}$ & First \\ 
&  &  &  &  & &  &  & event \\ 
 & & [$10^{14}$G] & [$10^{44}$erg] & & & & & [yr] \\ \hline 
 & \textbf{CC}& &  &  &  & & &  \\
 \texttt{CrDip} & PD+TQ & $1$ & $320$ & $\sim 50\%$ & 107 & 192 & 245 & 1.3 \\
\texttt{CrDipL} & PD+TQ & $0.1$ & $3.2$ & $\sim 50\%$ & 0 & 0 & 0 & - \\
\texttt{CrDipH} & PD+TQ & $3$ & $2900$ & $\sim 50\%$ & 2380 & 3315 & 3724 & 0.4 \\
\texttt{CrDipE} & PD+TQ & $10$ & $32000$ & $\sim 50\%$ & 23360 & 30842 & 34890 & 0.1 \\
\texttt{CrMultiL}$^*$ &  PM+TM & $0.1$ &  $9.3$ & $\sim 50\%$ &28 & 41 &  49 & 3.0 \\
\texttt{CrMultiH}$^*$ &  PM+TM & $1$ & $2900$ & $\sim 50\%$ & 15538 & 21397 & 25513 & 0.1 \\
\texttt{CrMultiT}$^*$ & PM+TM & $0.1$ & $480$  & $\sim 99 \%$ &525 &1335 & 1908  &0.7  \\
\texttt{CrDipD} & PD+TD & $1$ & $340$ & $\sim 50\%$ & 88 & 176 & 254 & 1.6 \\
\texttt{CrDipP} &  PD+TQ & $1$ & $160$ & $\sim 1\%$ & 45 &  58 & 67 & 2.6 \\ \hline
& \textbf{CT}& &  &  &  & & &  \\
\texttt{CoDipL} & PD+TD & $1$ & $4.7$ & $\sim 50\%$ & 10 & 10 & 10 & 13.7 \\
\texttt{CoDipH}$^*$ &  PD+TD & $3$ & $88$ & $\sim 50\%$ & 160 & 168 & 174 &1.6  \\
\texttt{CoMulti} &  PM+TD & $1$ & $ 750$ & $ \sim 50 \%$ &1948  &  2265 &  2370 & 0.8 \\
\texttt{CoMultiP} & PM+TD & $1$ & $  260$ & $\sim 1\%$ & 528 & 577 & 597 & 4.0 \\
\texttt{CoMultiPH}$^*$ & PM+TD & $3$ & $2400$ & $ \sim 1 \%$ & 9497 & 9997  & 10460  & 0.8 \\
\texttt{CoMultiPE}$^*$ & PM+TD & $6$ & $ 9400$ & $ \sim 1 \%$ &21090  & 23840  & 29065  & 0.4 \\
        \hline
	\end{tabular}
	\caption[Initial models considered and events simulated.]{Initial models considered and events simulated: topology (where CT is a core-threaded configuration, CC is a crust-confined, P and T are the poloidal and toroidal field, D and Q mean purely dipolar and quadrupolar, and M a mixed multipolar configuration); surface value of the dipolar poloidal component $B_{p}$; magnetic energy $E_{mag}^{cr}$ stored in the crust, the fraction $E_{tor}^{cr}/E_{mag}^{cr}$ stored in the toroidal field; $N^x$ is the number of events shown during the first $x=$100, 400 and 1000\,yrs; the first event is the age at which the crust fails for the first time. Models marked with $^*$ show small local numerical instabilities, that can provide artificial events, quantifiable by up to a maximum of $30\%$ of the shown values $N$, thus not affecting the order of magnitude.}
	  \label{tab: models}
\end{table}

As in \cite{pons2011} and \cite{perna2011}, we assume that after a crustal failure event, the surrounding regions of the crust, being close to the maximum stress, $M_{ij}\geq \epsilon \sigma^{max}$, are affected by the instability and resettle to equilibrium. In our simulations, we set the effective parameter to $\epsilon=0.9$. Note that increasing it to values closer to 1 will lead to more frequent events but smaller involved regions, and vice versa.

Our simulations present a few relevant simplifications: (i) We do not include self-consistent magneto-elastic evolution of the crust \citep{cumming2004,li2016,thompson2017,bransgrove2017,lander2019,gourgouliatos2021}\footnote{The study conducted by \cite{lander2019} took into account plasticity, but did not reset the magnetic field during the evolution. However, this issue was properly addressed in the subsequent study by \cite{gourgouliatos2021}.}, and the local magnetic field reconfiguration likely produced after a crustal failure; (ii) In the core we only consider the Ohmic term, which acts on timescales much longer than in the crust: in practice, the magnetic field does not evolve in the kyr here considered and a current sheet develops at the core surface; (iii) At the neutron star surface we impose a potential magnetic field solution (as in almost all studies, reviewed by \cite{pons2019}).

\begin{figure}
\centering
\includegraphics[width=0.5\textwidth]{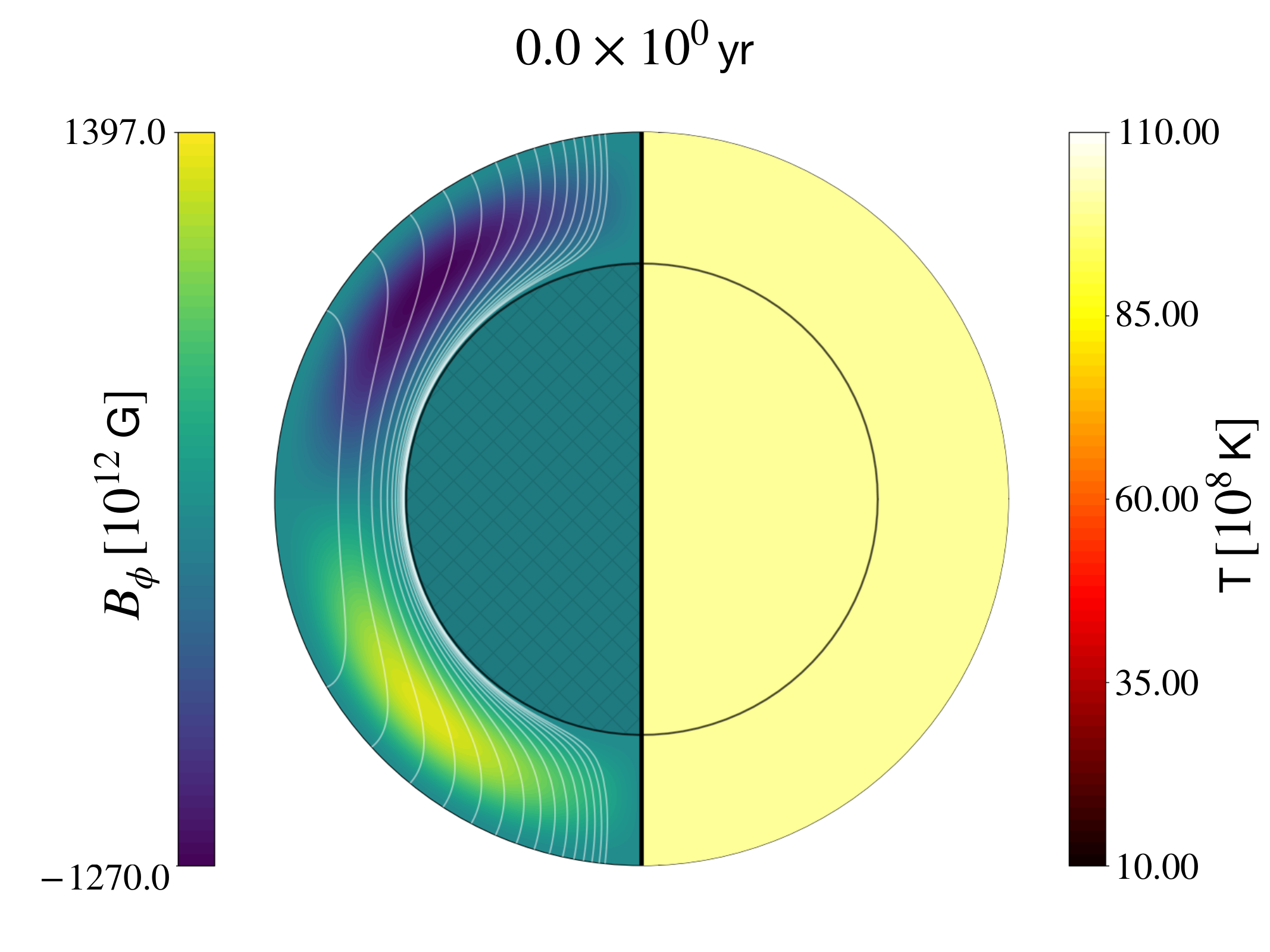}\hfill
\includegraphics[width=0.5\textwidth]{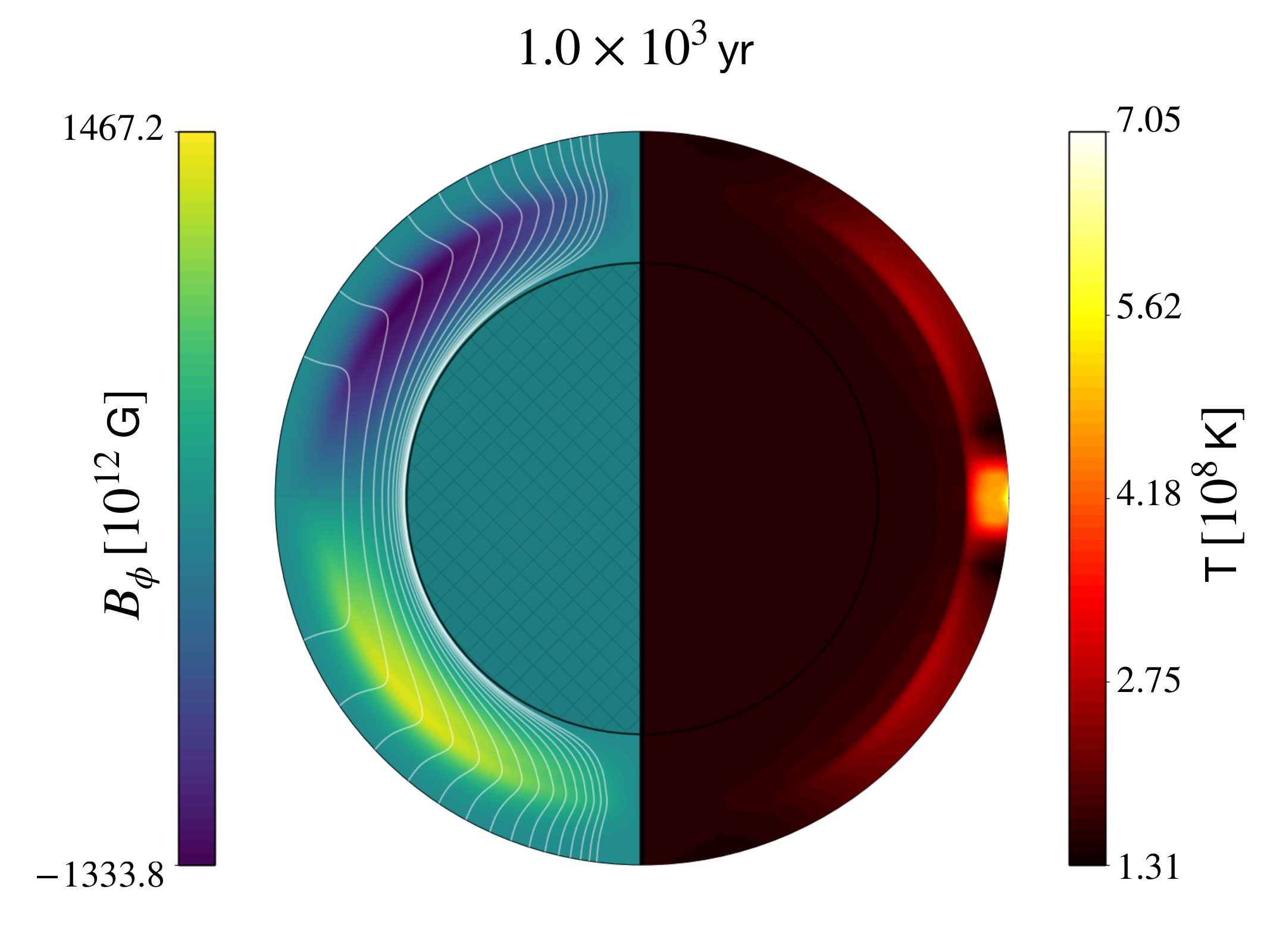}\vspace{0.1cm}
\includegraphics[width=0.5\textwidth]{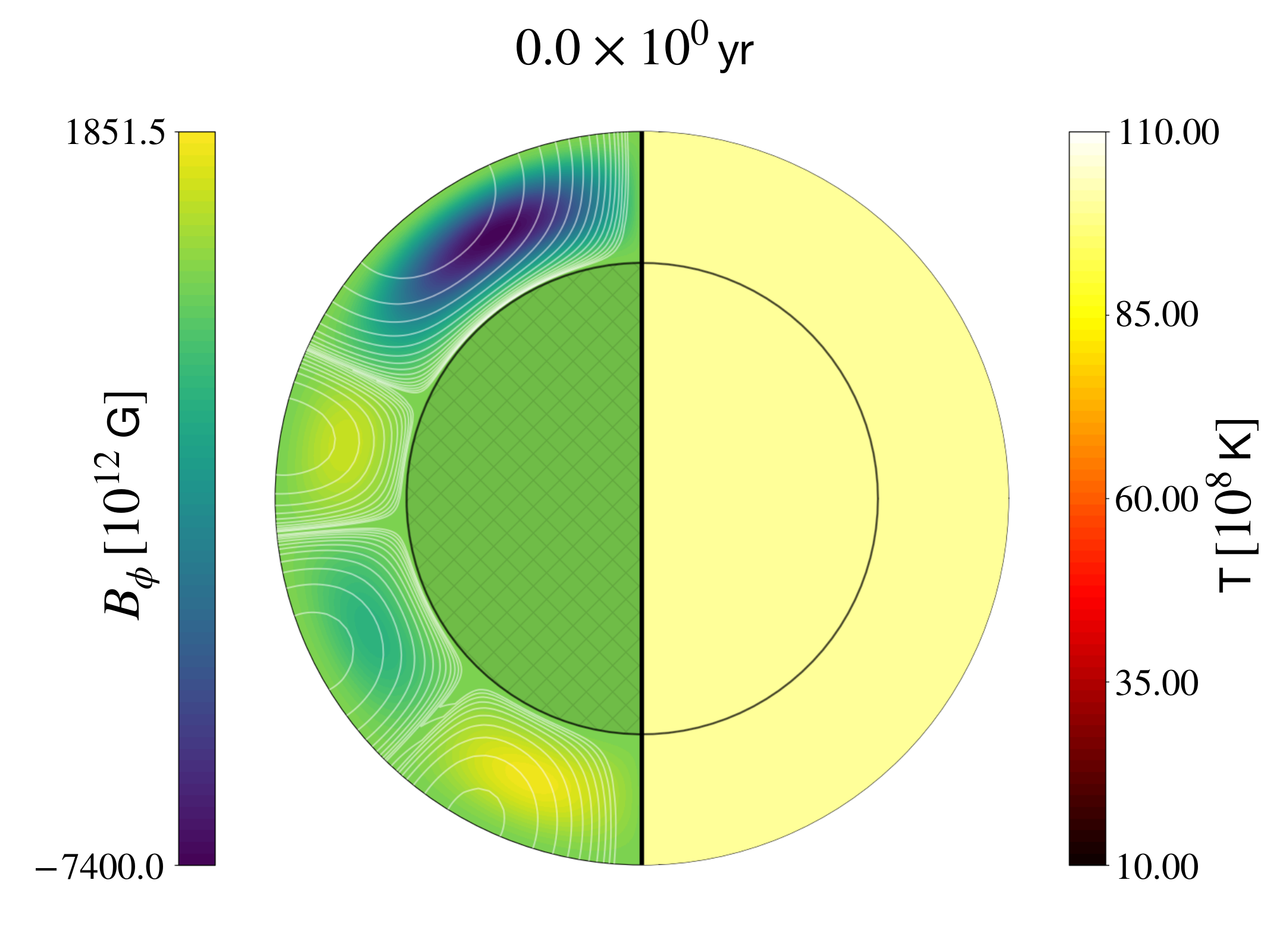}\hfill
\includegraphics[width=0.5\textwidth]{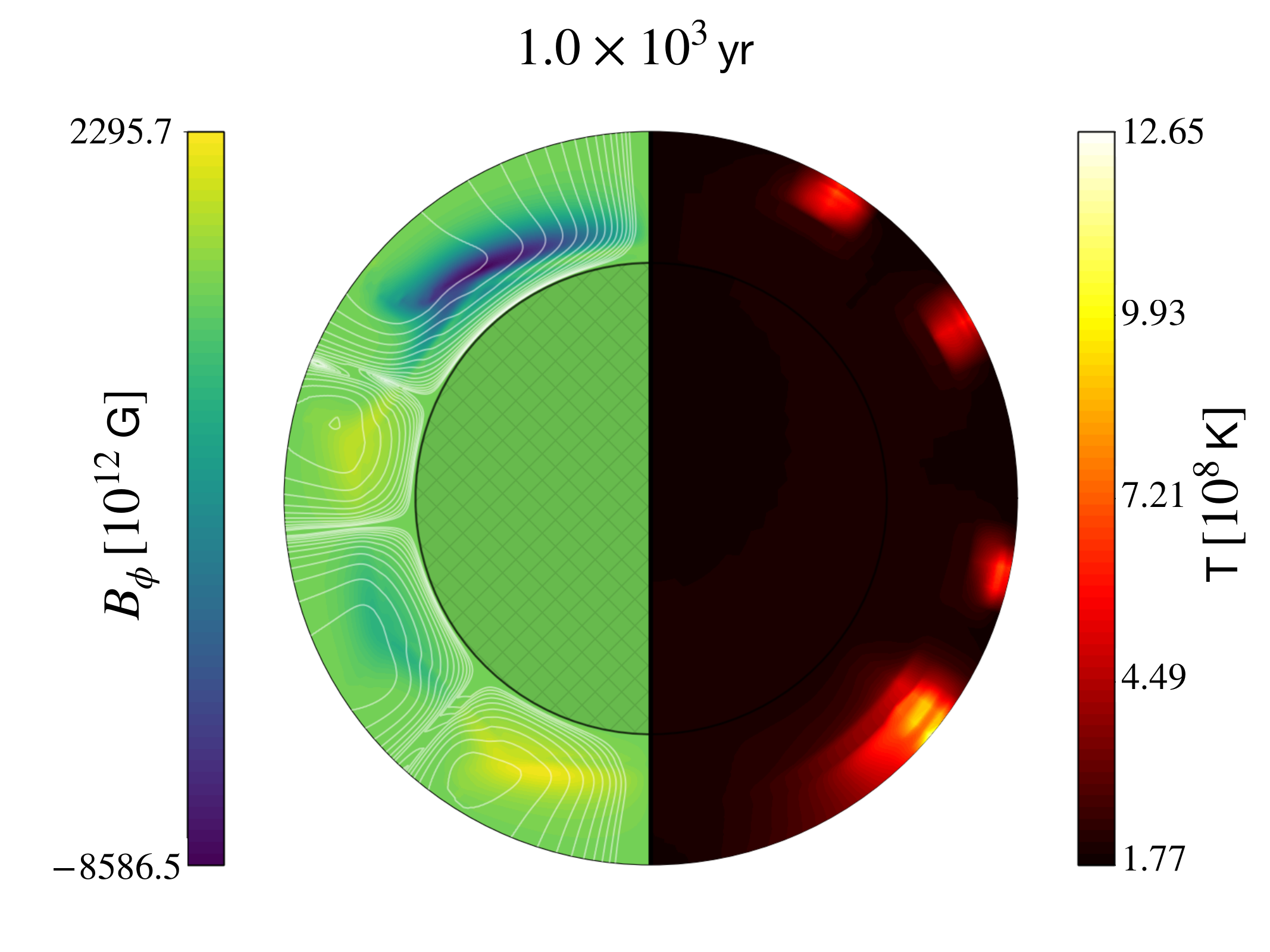}\vspace{0.1cm}
\includegraphics[width=0.5\textwidth]{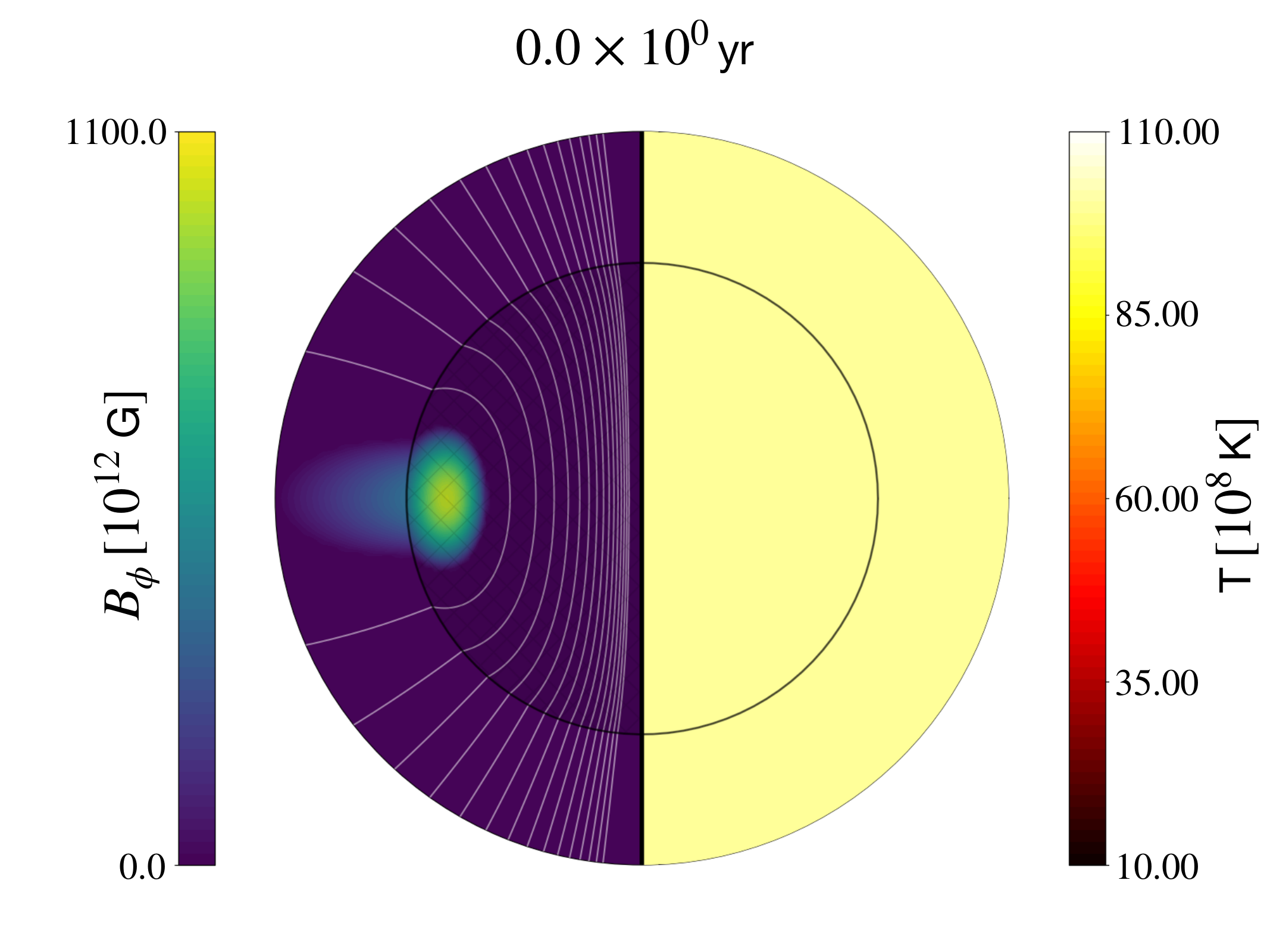}\hfill
\includegraphics[width=0.5\textwidth]{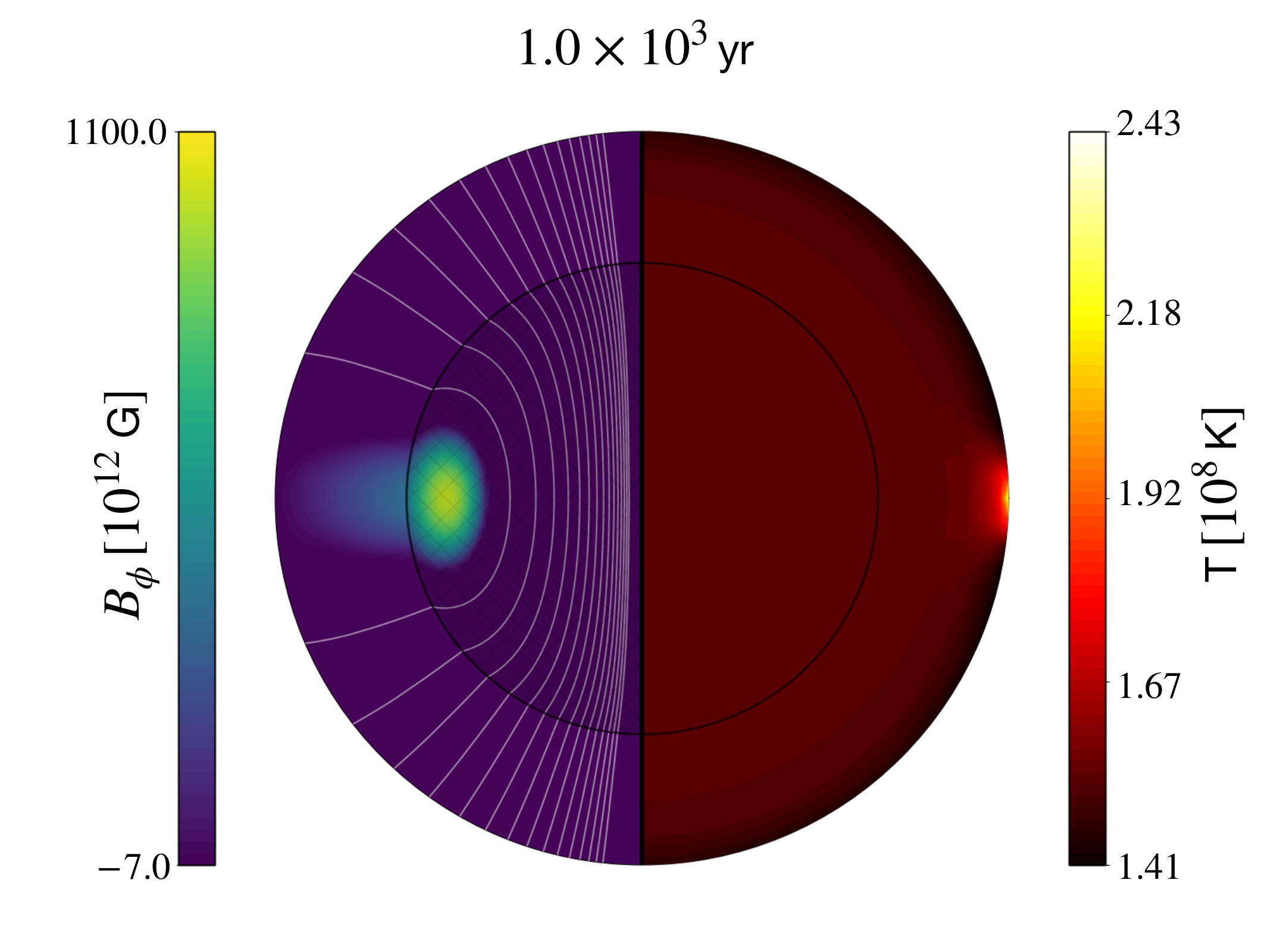}
\caption[Evolution of magnetic field and temperature for the different studied models.]{Evolution of magnetic field and temperature for model {\tt CrDip} (top panels), the high field multipolar model {\tt CrMultiH} (central panels), and the low field core-threaded model {\tt CoDipL} (bottom panels), showing the meridional projection of the magnetic field lines (white lines) and the toroidal field (colors) on the left, and the internal temperature distribution (on the right), at the beginning (left panels) and after 1\,kyr (right panels). The crust has been enlarged by a factor 8 for visualization purposes (thus only apparently bending the lines in the crust-core interface in the bottom panel). Movies of the crustal failure simulations for the different models in table~\ref{tab: models} are illustrated in \small \url{https://www.ice.csic.es/erc-magnesia/crustal-fractures-simulations/}.}
\label{fig:2Dplots}
\end{figure}

One of the primary challenges in considering early ages lies in the extreme sensitivity of the results to the initial magnetic field configuration, both for physical and numerical reasons. Before freezing, the fluid of the magnetized proto-neutron star settles into an MHD equilibrium or stationary state. Various approaches have been proposed in previous works to determine initial configurations, with most of them relying on very large-scale and smooth magnetic fields, often exhibiting a purely dipolar poloidal component. However, it is important to note that nature typically presents more turbulent, off-axis, and non-symmetric magnetic field configurations, similar to those seen in planetary and stellar magnetic fields. For a recent simulation involving 3D coupled magneto-thermal evolution in neutron stars and an initial topology inspired by core-collapse supernovae, please refer to \S\ref{sec: 3DMT}. Nevertheless, in this chapter, based on the work of \cite{dehman2020}, we utilize the 2D magneto-thermal code (\S\ref{sec: MT evolution}) that imposes axisymmetry to restrict the magnetic field topology.

\section{Simulations}
\label{sec:runs}

\subsection{Initial configurations}

Aiming to reduce numerical instabilities which could provoke artificial crustal failures, we have used the latest version of the finite-volume axially-symmetric magneto-thermal code \citep{vigano2021}, improved in stability and efficiency compared to \cite{vigano2012}. It includes numerical methods which are suitable to the finite-volume integral version of the Hall induction equation, and are optimized in terms of stability, accuracy and efficiency. We also employ the most updated microphysical inputs (for a review of cooling and transport see \cite{potekhin2015}).  
We make sure that results are close to the numerical convergence (we have variations of the event rate up to a maximum of $30\%$ for different resolutions, much less than the intra-model variability), using a spatial resolution of $100\times 200$ points in the angular and radial directions, and a fixed timestep of $\sim 10^{-4}-10^{-3}$\,yr for the magnetic advance, and of $10^{-2}$\,yr for the temperature evolution. 

To assess the sensitivity of results upon uncertain initial conditions, we have considered very different topologies. Although not necessarily realistic, the variety considered allows us to explore the range of crustal failure events. They are summarized in Table~\ref{tab: models}, together with the evolution of the events' frequency during the first 1\,kyr of the neutron star's life.

\begin{figure}
\centering
\includegraphics[width=.5\textwidth]{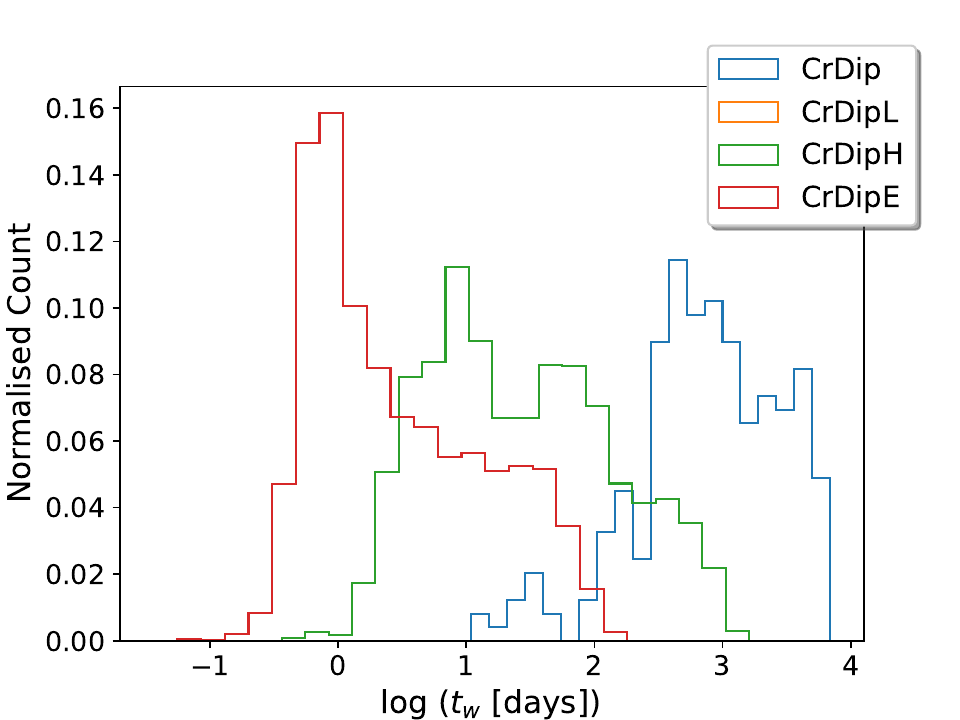}\hfill
\includegraphics[width=.5\textwidth]{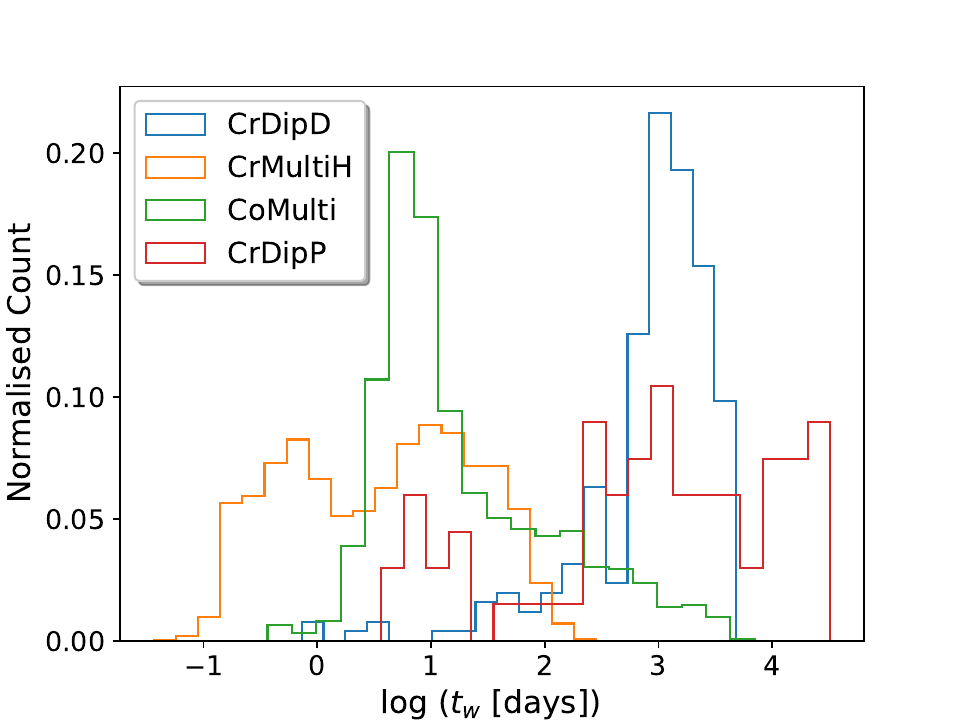}
\caption[Waiting time distributions of the different studied models during the first $1$\,kyr of the neutron star's life.]{Waiting time distributions during the first $1$\,kyr of the neutron star's life. In the first panel, we investigate the effects of the magnetic strength on crustal failures. In the second one, the study is focused on the field topology and on the impact of lowering the toroidal energy.}
\label{fig: WTdistributions crustal-confined}
\end{figure} 

On one hand, we take into account crust-confined models with different weights in the initial multipoles. While it is unlikely that the field is completely expelled from the core, such a kind of topology can be useful to provide an upper limit on the frequency of outburst/burst activity triggered by the dynamics since the field lines are not tied to the highly conductive core and the currents supporting the magnetic field flow entirely in the crust, where spatial- and time-scales are reduced compared to the core. Within these models, we vary the relative fraction of magnetic energy stored in different multipoles of both the poloidal and toroidal components, as follows:
\begin{itemize}
    \item In the fiducial model {\tt CrDip} we consider the equipartition of magnetic energy between a dipolar poloidal magnetic field with polar surface value $B_{ dip}=10^{14}$\,G and a toroidal quadrupolar field having the same magnetic energy as the poloidal one (and a maximum value $B_{\varphi}^{max}=1.27\times 10^{15}$\,G). 
    \item In the low, high, and extreme field models {\tt CrDipL}, {\tt CrDipH} and {\tt CrDipE}, we assume the same topology but with the magnetic field intensity rescaled by a factor 0.1, 3 and 10 respectively. 
    \item In the low and high field models {\tt CrMultiL} and {\tt CrMultiH}, we consider a topology with a similar contribution of the first four multipoles in both the poloidal and toroidal components, with an equipartition of magnetic energy between them. In {\tt CrMultiT}, the quadrupolar toroidal field accounts for $ 99\%$ of the total magnetic energy.
    \item In model {\tt CrDipD} a dipolar (instead of a quadrupolar) toroidal field is considered.
    \item In model {\tt CrDipP} we adopted a poloidal-dominated configuration, with the quadrupolar toroidal field accounting for less than $ 1\%$ of the total magnetic energy.
\end{itemize}

Additionally, we consider different core-threaded configurations:
\begin{itemize}
    \item Model {\tt CoDipL} (low field) and {\tt CoDipH} (high field), consisting of the often used large-scale twisted dipole plus torus configurations, as in \cite{akgun2017}. Again, we set an equal fraction of poloidal and toroidal magnetic energy in the crust.
    \item In model {\tt CoMulti}, we consider a topology with similar contributions from the dipole and the quadrupole in the poloidal field, and a dipolar toroidal field, with an equipartition of magnetic energy between them. On the other hand, for {\tt CoMultiP}, {\tt CoMultiPH}, and {\tt CoMultiPE}, the toroidal magnetic energy accounts for less than $1\%$ of the total magnetic energy, where {\tt H} and {\tt E} refer to high and extreme field models, respectively.
\end{itemize}

Note how models {\tt CrDip}, {\tt CrMultiH}, {\tt CrDipD}, {\tt CrDipP}, {\tt CoDipL}, {\tt CoMulti}, and {\tt CoMultiP} have the same dipolar polar magnetic field $B_{ dip}=10^{14}$\,G, which rules the spin-down and, indirectly, the rotationally powered emission. However the magnetic energy stored in the crust, $E_{mag}^{cr}$, can vary by a few orders of magnitude, depending on the relative weight of the multipolar and toroidal components and on whether currents circulate only in the crust or not.

In Fig.~\ref{fig:2Dplots} we show the initial and evolved (at 1\,kyr) magneto-thermal models {\tt CrDip} (top panels), \textbf{\tt CrMultiH} (central panels), and {\tt CoDipL} (bottom panels). The crustal-confined models show relevant magnetic activity (left hemispheres of each panel), with a local non-trivial rearrangement of the toroidal field distribution (colors) and a deformation of the poloidal field lines (white lines). The temperature distribution (right hemisphere) also evolves to be inhomogeneous, due to the transport anisotropy induced by the local intensity and direction of the magnetic field.
On the other hand, the core-threaded model barely shows any evolution, due to the fact that the currents are mostly concentrated in the core, where the conductivity is very high and no Hall effect is present: the magnetic field lines are rooted and almost frozen in the core. The core-threaded cases give much slower dynamics than the crust-confined ones, due to the large characteristic spatial scales, and the fact that most of the supporting currents circulate in the highly conductive, non-solid core. A realistic configuration could be core-threaded, but consists of small scales, both in the core and the crust, so that the corresponding results should be effectively covered by our exploration of parameters. Note that, in all cases, the value of the dipolar component at the surface, $B_{p}$, decreases at most by a few percent.

\subsection{Event rate}

Focusing on the predicted event rate, the gross number for model {\tt CrDip} is $\sim 250$ events during the first 1\,kyr of the neutron star's life, mostly concentrated in the first century. Furthermore, we emphasize that by changing the overall magnetic strength with respect to model {\tt CrDip}, the frequency decreases for the low-field dipolar model ({\tt CrDipL}), and increases for the high-field ({\tt CrDipE}) and the extreme field ({\tt CrDipH}) model, by $\sim 1$ and $\sim 2$ orders of magnitude, scaling the number of events as $N^{1000}\sim E_{mag}^{cr}/10^{44}$. The same scaling is noticed for the core-threaded models. There are no significant differences between a different topology for the toroidal field (quadrupolar in {\tt CrDip} and dipolar in {\tt CrDipD}), while important differences are noted if the magnetic energy is changed, maintaining the same $B_{p}$: poloidal-dominated {\tt CrDipP}, or considering multipolar configurations ({\tt CrMultiL}, {\tt CrMultiH}, and {\tt CrMultiT}).

For all models, the older the magnetar, the fewer the expected number of events. As shown in Table~\ref{tab: models}, $\sim 40-85\%$ of the events during the first 1000\,yrs are concentrated in the first century, with a gradual decrease. This percentage decreases to $\sim 10-40 \%$ in the next 300\,yrs, whereas in the last 600\,yrs, the percentage oscillates between $\sim 5-20 \%$. This is consistent with the findings for middle-age magnetars by \cite{perna2011,pons2011}: magnetars become less and less active, but they can still show activity even at late stages.

The waiting time distributions are illustrated in Fig.~\ref{fig: WTdistributions crustal-confined}. In the first panel, we compare models {\tt CrDip}, {\tt CrDipL}, {\tt CrDipH}, and {\tt CrDipE}. We find that increasing the strength of the magnetic field results in a shorter waiting time. This is also the case if we consider a multipolar topology (models {\tt CrMultiH} and {\tt CoMulti}), whereas lowering the fraction of the toroidal energy (model {\tt CrDipD}) has a tiny impact on the distribution. However, the waiting time is shifted from $\sim [10:10^4]$ to $ \sim [0.1:300]$~days, by increasing the field strength and considering the multipolar topology.

\begin{figure}
    \centering
        \includegraphics[width=.8\textwidth]{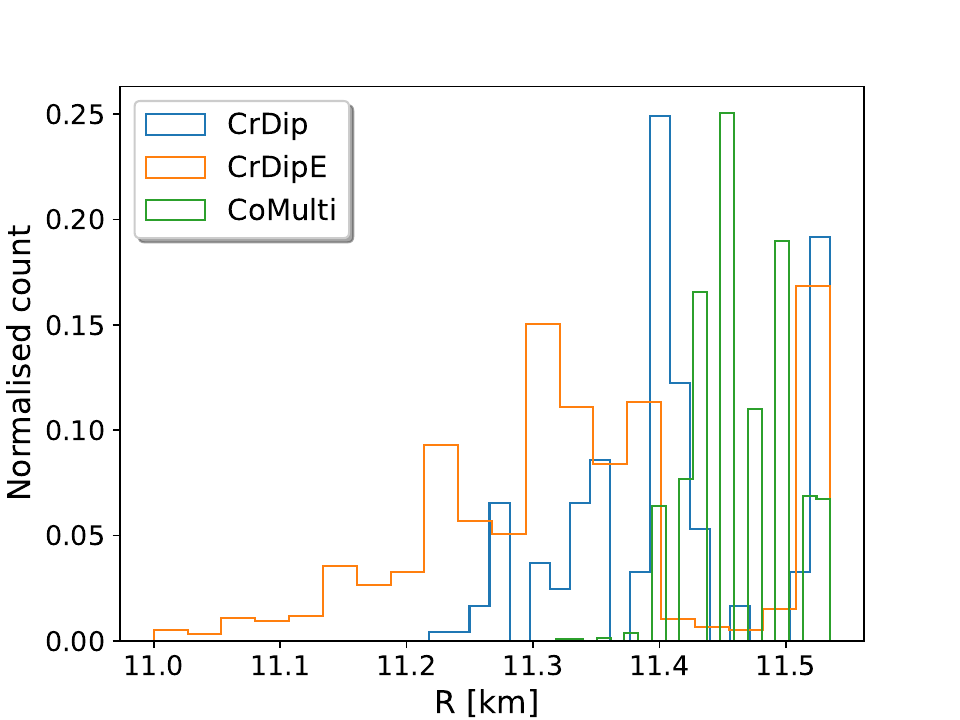}
    \caption[The location distributions of crustal failures as a function of the neutron star radius for the different studied models.]{The location distribution of crustal failures as a function of the neutron star radius during the first 1\,kyr of the neutron star's life. }
    \label{fig: radius distribution 025}
\end{figure}

In Fig.~\ref{fig: radius distribution 025}, we present the crustal failure distributions as a function of the neutron star radius for a set of models, i.e., model {\tt CrDip}, {\tt CrDipE}, and {\tt CoMulti}. Even though such distributions are partially model-dependent, most of the events happen in the outer crust, i.e., $R \sim [11.3: 11.55]$ km, because the maximum stress is smaller in the outermost, lighter layers. A non-negligible fraction of the events takes place in the outermost $50$\,m of the neutron star, i.e., $\sim 25\%$.

\section{Discussion}
\label{sec:discussion}

Ideally, modeling the link between magnetospheric magnetar activity and crustal failure requires: (i) following the internal magnetic dynamics and how the magnetic stresses grow with time, (ii) understanding and quantifying the mechanisms leading to the propagation of the disturbance into the magnetosphere, (iii) describing how the magnetospheric instability/reconnection powers the emission.

This study focuses on the first issue, assessing the crustal failure frequency of newly born magnetars, under different initial topologies, through 2D magneto-thermal simulations, including the Hall term. Moreover, the presence of small-scale structures, very likely if the birth amplification involves turbulent processes, can increase the expected event rate, through initial enhanced Hall dynamics. More factors can further extend the range of events found, among which we must mention the magnetic field boundary conditions (here considered as potential, possibly inadequate for the dense environment around a newborn magnetar), the fixed effective parameter $\epsilon$, controlling how large is the region undergoing failure at the same time.

\begin{figure}
    \centering
    \includegraphics[width=.8\textwidth]{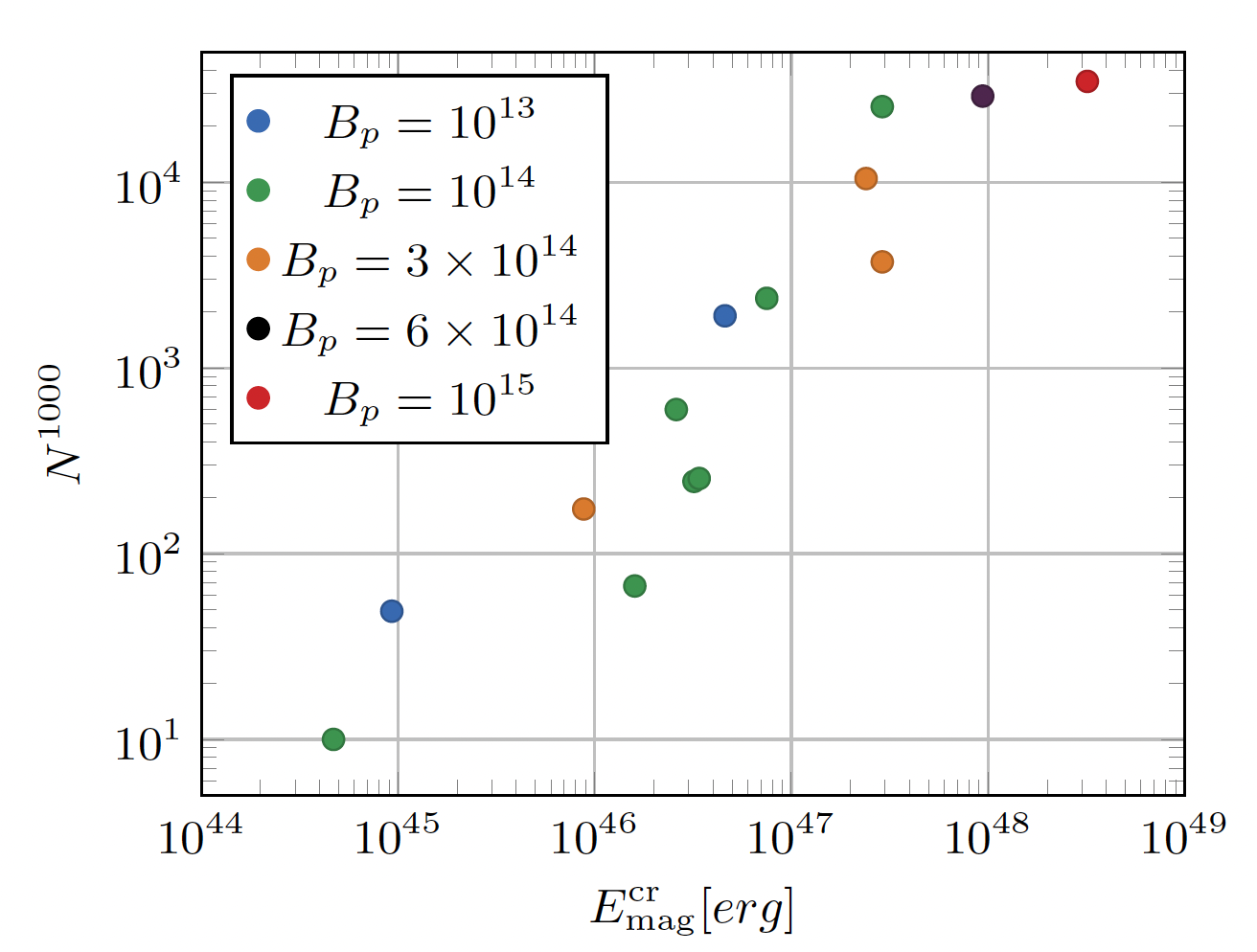}
    \caption[Number of events in the first 1\,kyr as a function of the crustal magnetic energy.]{Number of events in the first 1\,kyr as a function of the crustal magnetic energy. Each point represents a specific model, with colors indicating the value of $B_{p}$ as in the legend.}
    \label{fig:events}
\end{figure}

The main finding is that the event frequency scales linearly with the magnetic energy stored in the crust, with some intrinsic dispersion given by the topology, $E_{mag}^{cr}$ (see Fig.~\ref{fig:events}), which strongly depends on the assumed and highly uncertain field topology.

Given the simplifications in our approach, additional words of caution are needed in the interpretation of the results:\\
(i) We do not include the recent advances in the modeling of the fluid motion in neutron star core (with or without superfluidity), limiting to either confine the field to the crust or to consider only the very slow Ohmic dissipation. In particular, the ambipolar diffusion could drive (slightly) faster changes and higher rates \citep{beloborodov2016}.\\
(ii) Magneto-elastic simulations, and the inclusion of the local field rearrangement and dissipation, could reduce the rates, especially the most extreme ones (thus potentially weakening the linear trend in the upper range). Moreover, the critically stressed crust may behave plastically \citep{kobyakov2014}, leading to Hall drift and thermoplastic waves \citep{beloborodov2014,li2016} and partially hampering the Hall dynamics \citep{lander2019}.\\
(iii) One potential future application of the study on crustal failure using the 3D code described in Chapter~\ref{chap: MATINS} is to enhance our understanding of how three-dimensional dynamics can significantly influence the results. By employing this approach, we can gain valuable insights into the quantitative impact of 3D effects on the outcomes.

We expect that overcoming such caveats above will systematically change the quantitative results, but not the important conclusion that the dipolar field strength is a barely relevant parameter when considering a young magnetar's activity. As a matter of fact, for a given $B_{p}$, our calculations show a very broad range of crustal event rates. Even a relatively modest $B_{p}=10^{14}$ G could potentially provide almost no events (large-scale field penetrating in the core) or $N \sim 1$ event per day of the neutron star life, if most of the magnetic energy is hidden in the form of crustal multipolar and toroidal fields. The expected crustal failure rate is, thus, strongly dependent on the initial topology of the magnetic field. These results shed light on a potential issue of most FRB-magnetars models, which are usually focused on the value of $B_{p}$ only.

\subsection{Comparison with Fast Radio Bursts and Galactic magnetars}

Our calculations have implications both for FRBs (if indeed associated with magnetars), as well as for Galactic magnetars. 
In both scenarios, the observed event rate can be summarized as
\begin{equation}
N_{obs}=N_{failure} \times N_{sources} \times x_{vis}\,,
\end{equation}
where $N_{sources}$ is the number of potential sources (young magnetars within a certain observable volume), $N_{failure}$ is the crustal event rate per star calculated in our study, and $x_{vis}$ is the effective fraction of events that will actually be detected. The latter number is unconstrained and depends on the specific mechanisms ultimately leading to detection, observational biases, and both physical (beaming, distance, absorption, etc.) and instrumental (sensitivity, field of view, band-width, etc.) limits.

In the case of magnetar bursting activity happening in the magnetosphere, $x_{vis}$ depends on the crustal response to failure (one can expect that only the outermost events can further affect the magnetosphere), the propagation of the disturbance outside, the magnetospheric dynamics and the emission mechanism. Besides these fundamental physical uncertainties, emission beaming, distance and absorption will affect $x_{vis}$, resulting in an expected value much smaller than one.

For FRBs, according to \cite{rane2016}, the observed rate is estimated to be $N_{obs} \sim 10^3-10^4$ per day per sky above 4\,mJy per ms.
Considering as a reference the number of galaxies within the distance of the prototypical FRB repeater FRB121112, $\sim 10^7$, and assuming $\sim 10$ magnetar per galaxy
with $\lesssim $ 1000\,yrs, we would expect a potential number of FRB-emitting magnetars of about $N_{sources} \sim 10^8$. Equating these numbers, we obtain a constraint $x_{vis}^{frb}\times N_{failure} \sim 10^{-5}-10^{-4}$ per day. This means that models with high crustal magnetic energy, $E_{mag}^{cr}\gtrsim 10^{46}$\,erg, which have $N_{failure}\sim 1$ per day, are compatible with $x_{vis} \ll 1$.

Similarly, very young ($\lesssim 1$ kyr) Galactic magnetars are expected to be $N_{sources} \sim 10$. The observed rate of outbursts for them (excluding older magnetars) is $N_{obs} \sim 10^{-3}-10^{-4}$ per day \citep{cotizelati2018}. Therefore, $x_{vis}^{galactic}\times N_{failure} \sim 10^{-5}-10^{-4}$, similarly to the FRBs.

This preliminary estimate, although plagued by uncertainties on the total number of sources and the still incomplete statistics, shows that the range of events detected per star, $x_{vis}\times N_{failure}$, in the FRB and Galactic magnetar outburst scenarios, are at least compatible. This points to a possible common origin of crustal triggers, even though they manifest in different ways.

As this chapter comes to a close, it is essential to highlight that the investigation of 3D effects in the activity induced by the strong magnetic fields of magnetars has remained relatively unexplored until now. However, a promising breakthrough has been achieved with the development of the advanced code \emph{MATINS}, which is extensively described in Chapter~\ref{chap: MATINS}. This innovative tool opens up exciting avenues for further research into 3D studies. In the upcoming chapters, we will delve into the realm of three-dimensional investigations, embarking on an exploration of the capabilities and insights offered by our newly introduced code \emph{MATINS}.




\subsubsection*{Corresponding scientific publications}
\underline{C.~Dehman}, D.~Vigan\`o, N.~Rea, J.A.~Pons, R.~Perna \& A.~Garcia-Garcia: $2020$, \textbf{On The Rate of Crustal Failures in Young Magnetars}, \emph{Astrophys.~J.~L., $902$ L$32$}
(\href{https://arxiv.org/abs/2010.00617}{\underline{arXiv:2010.00617}},\href{https://ui.adsabs.harvard.edu/abs/2020ApJ...902L..32D/abstract}{\underline{ADS}},\href{https://iopscience.iop.org/article/10.3847/2041-8213/abbda9}{\underline{DOI}}).
\clearemptydoublepage
\let\textcircled=\pgftextcircled
\chapter{The 3D Code MATINS}
\label{chap: MATINS}

\initial{U}pon witnessing the physical results using the 2D code, we now unveil the expansion to 3D with the introduction of \emph{MATINS}, a new {\it three-dimensional code for MAgneto-Thermal evolution in Isolated Neutron Stars}, based on a finite-volume scheme. This chapter primarily relies on the research conducted by \cite{dehman2022}, where we concentrate on introducing the cubed-sphere coordinates and the magnetic evolution formalism, with a particular consideration of crustal-confined magnetic fields. The \emph{MATINS} code will soon be available to the public, providing a valuable resource for the high-energy astrophysics community focused on exploring various astrophysical aspects related to highly magnetized neutron stars. The code can be accessed at the following URL:
\begin{center}
\href{https://github.com/csic-ice-magnesia/MATINS}{https://github.com/csic-ice-magnesia/MATINS}.
\end{center}

The internal magnetic field evolution of isolated neutron stars has been extensively explored through 2D simulations \citep{pons2007}, which were later coupled to the temperature evolution \citep{aguilera2008,pons2009,vigano2012,vigano2021}. These models successfully explained the general properties observed in the population of isolated neutron star \citep{vigano12b,vigano2013,pons2013,gullon2014,gullon2015}. Recent efforts were devoted to investigate the magnetic evolution without the restrictions of axial symmetry. \cite{wood2015} and \cite{gourgouliatos2016} presented the first 3D simulations of crustal-confined fields employing a pseudo-spectral code adapted from the geo-dynamo code \emph{PARODY} \citep{dormy1998} to the neutron star scenario.
These simulations revealed new dynamics and the creation of long-living magnetic structures at various spatial scales. Even using initially axisymmetric conditions, the growth of initially tiny perturbations breaks the symmetry, and non-axisymmetric modes quickly emerge and grow \citep{gourgouliatos2020}. These non-axisymmetric modes typically have length scales comparable to the crust thickness.

Generally speaking, for sufficiently high enough magnetic fields ($B\gtrsim 10^{14}$\,G), the Hall cascade keeps transferring energy to small scales \citep{gourgouliatos2016}, which in turn enhances Ohmic dissipation and eventually keeps the star hot and X-ray visible for longer timescales, as seen in 2D simulations \citep{vigano2013}. Another interesting result is the formation of magnetic spots on the neutron star's surface \citep{gourgouliatos2018}, using extreme initial configurations previously explored in 2D \citep{geppert2014}. Very recently, \cite{degrandis2020} presented the first 3D magneto-thermal evolution code with increasing physical self-consistency, applied to different sub-classes of neutron stars \citep{igoshev21a,igoshev21,degrandis2021}. However, these simulations employ simplified microphysics and background structure. For a comprehensive review of magneto-thermal evolution models, refer to \cite{pons2019}.

Different systems of coordinates and numerical methods have been employed to address 3D MHD in a shell, in other astrophysical scenarios. In some cases, finite volume/finite difference schemes are employed with Cartesian coordinates using the "star-in-a-box" approach \citep{kapyla2021}. However, this choice is not optimal for several reasons. Firstly, magnetic fields and physical quantities often exhibit rapid variations in the radial direction, making it more convenient to treat the radial coordinate separately from the other two coordinates. Secondly, since the star's surface is approximately spherical (with possible deviations being much smaller than other relevant scales), describing it in Cartesian coordinates becomes computationally expensive compared to coordinate systems that incorporate a radial direction. The need to refine all directions to solve strong radial gradients and the discretization of spherical boundaries onto the Cartesian grid introduce spurious noise, leading to artificial modes that can only be partially mitigated by increasing the resolution (see Appendix A of \cite{vigano2021} for further details).
A more natural choice would be to use spherical coordinates, as done in two dimensions. However, the coordinate system exhibits irregular behavior on the axis, resulting in several numerical limitations, some of which are quite compelling. One option is to exclude the axis from the grid to avoid complications related to coordinate singularities.
Finally, instead of employing finite volume/difference schemes, an alternative approach is to utilize Pseudo-spectral methods that make use of spherical coordinates. However, these methods come with the disadvantage of a complex formalism required to convert the equation into the spherical harmonics space, as well as limitations in capturing the discontinuities generated by the non-linear Hall term (eq.\,\eqref{eq: induction equation}). Therefore, we prefer the use of a finite volume scheme.

In this study, we employ the cubed-sphere coordinates, originally introduced by \cite{ronchi1996}. Codes based on such a grid have been used to simulate various physical scenarios, including for instance: general circulation models for Earth or planets \citep{breitkreuz18,ding19}, general relativity \citep{lehner2005,hebert18,carrasco18,carrasco19}, MHD accretion \citep{koldoba02,fragile09,hossein18}, solar wind \citep{wang19}, seismic waves \citep{vandriel21}, and dynamo in a shell \citep{yin22}.
In this work, we use this unique coordinate system, adapted to the Schwarschild metric, to develop a new code designed to handle the Hall term in the induction equation for low physical resistivity.

\emph{MATINS} stands out due to its distinctive features in comparison to \emph{PARODY}-based published works \citep{degrandis2020, degrandis2021, igoshev21a, igoshev21}. These notable features include: (i) coupled magneto-thermal simulation with the use of the most recent temperature-dependent microphysical calculations, (ii) the use of a star structure coming from a realistic EoS and the inclusion of the corresponding relativistic factors in the evolution equations, (iii) the use of finite-volume numerical schemes discretized over a cubed-sphere grid, allowing for better capture of non-linear dynamics, (iv) flexibility in incorporating new physics, (v) comprehensive documentation and modularity for public use, (vi) optimization and use of OpenMP.

In this chapter, our objective is to investigate the global evolution of the magnetic field in isolated neutron stars, which are relativistic stars where general relativity corrections play a significant role. The structure of these neutron stars is provided by the TOV equations \citep{oppenheimer1939}, which solve the hydrostatic equilibrium assuming a static interior Schwarzschild metric, as described in \S\ref{sec: TOV}. We can either prescribe a simple shell, or build the background model of the neutron star using different models of nuclear EoS at zero temperature. For a detailed description on the numerical structure of \emph{MATINS}, we refer the reader to \S\ref{sec: MT evolution}.

This chapter is structured as follows. In \S\ref{sec: cubed sphere formalism}, we prescribe the cubed-sphere formalism applied to a Schwarschild metric, and the numerical scheme used in the three dimensions magnetic evolution code. In \S\ref{sec: EMHD limit and Hall Induction Equation}, we display the inner and outer magnetic boundary conditions used in this study. Finally, the numerical tests and the comparison with 2D axisymmetric models are presented in \S\ref{sec: numerical test}.

\section{The cubed sphere formalism with the Schwarzschild interior metric}
\label{sec: cubed sphere formalism}

\begin{figure*}
	\centering
	\includegraphics[width=0.9\textwidth]{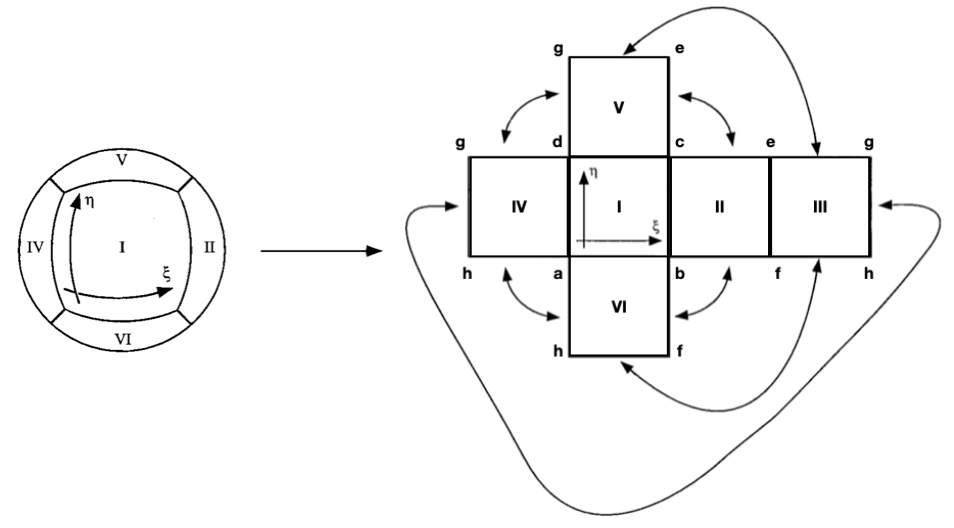}
	\includegraphics[width=0.9\textwidth]{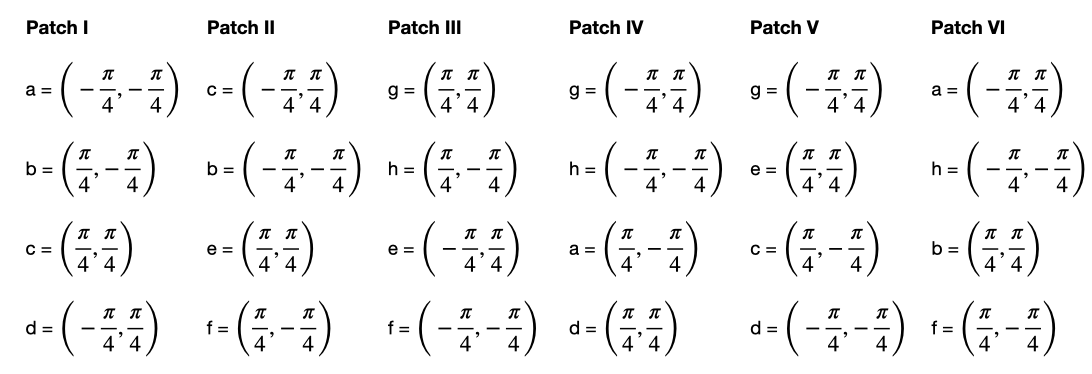}
	\caption[Cubed-Sphere grid.]{Exploded, cubed view of the patches \citep{ronchi1996}. Each patch is identical and is described by the coordinates $\xi$ and $\eta$, both spanning the range $[-\pi/4;\pi/4]$. In the exploded view $\xi$ and $\eta$ grow to the right and upward, respectively, for all patches (only patch I is explicitly drawn here). Arrows identify the 12 edges between patches. The coordinate values $(\xi,\eta)$ of the corners for each of these patches are written in the bottom part as well.}
	\label{fig:full_grid}
\end{figure*}

\subsection{Patches and coordinates}
\label{subsec: definition of patches}

In the cubed sphere formalism, initially introduced by \cite{ronchi1996}, one of the three coordinates corresponds to the radial direction, like in spherical coordinates. This results in the volume being composed of multiple radial layers. As depicted in Fig.~\ref{fig:full_grid}, each layer is covered by six non-overlapping patches that are topologically identical. These patches can be visualized as the result of inflating the six faces of a cube, until it forms a spherical shape. Consequently, each patch is bordered by four patches and is naturally described by two angular-like coordinates, akin to the role of the standard spherical coordinates $\theta$ and $\phi$. We adopt the same notation as the original paper, denoting the patch coordinates as $\xi$ and $\eta$, both in the range $[-\pi/4:\pi/4]$. These two coordinates are orthogonal to the radial direction but are non-orthogonal to each other, except at the patch centers. They uniformly cover the two directions, meaning that the patch shape remains invariant under any rotation of $n\pi/2$ (where $n$ is an integer) around the center of the patch. The transformation relations between the cubed-sphere, spherical, and Cartesian coordinate systems are reported in Appendix~\ref{app:coordinates}.

\subsection{Metric}
\label{subsec: Metric}

We follow the same approach as in \cite{ronchi1996}, but using Schwarzschild interior metric solution of the TOV equation. We introduce the auxiliary variables that will be used in our formalism
\begin{align}
   X &\equiv \tan(\xi), \nonumber\\    
   Y &\equiv \tan(\eta), 
    \nonumber\\  
   \delta &\equiv 1+ X^2 + Y^2,
     \nonumber\\
     C &\equiv (1+ X^2)^{1/2} \equiv  \frac{1}{\cos (\xi)},
 \nonumber\\
 D &\equiv (1+ Y^2)^{1/2} \equiv  \frac{1}{\cos (\eta)}.
   \label{eq: cubed sphere variables}
\end{align}

The metric tensor exhibits the same functional dependence on the auxiliary variables across all patches. In the unit vector basis, it can be expressed as follows:
\begin{equation} 
g_{ij}=
\begin{pmatrix}
1 &0  & 0\\
0 & 1 &  - \frac{XY}{CD} \\
0 &  - \frac{XY}{CD} & 1
\end{pmatrix}.
\label{eq: metric tensor}
\end{equation}
The inverse of the tensor metric reads
\begin{equation}
g_{ij}^{-1}=
\begin{pmatrix}
1 &0  & 0\\
0 & \frac{C^2 D^2}{\delta} &   \frac{CDXY}{\delta} \\
0 &  \frac{CDXY}{\delta} & \frac{C^2 D^2}{\delta}
\end{pmatrix}.
\label{eq: inverse metric tensor}
\end{equation}
In all patches, the radial versor $\hat{e}_r$ is orthogonal to the plane formed by $\hat{e}_\xi$ and $\hat{e}_\eta$ unit vectors, which are not in general orthogonal to each other.

Below, we will employ vectors using either their covariant components, denoted by lower indices, or their contravariant components, denoted by upper indices.
Let us focus first on the geometrical elements (see \S\ref{app: geometrical elements} for a comprehensive derivation). The contravariant components of the infinitesimal length element\footnote{Note that the factor two difference with respect to \cite{ronchi1996} arises because the geometrical elements used in the circulation extend twice the size of the cell (once per each side around a central point, see as an example the red solid lines in Fig.~\ref{fig: overlap region}).} at a given position $\{r,\xi,\eta\}$ are 
\begin{align}
     dl^r(r) &= e^{\lambda(r)}dr, \nonumber\\
      dl^\xi(r,\xi,\eta) &=  \frac{2rC^2D}{\delta} d\xi ,\nonumber\\
       dl^\eta(r,\xi,\eta) &=  \frac{2rCD^2}{\delta} d\eta .
     \label{eq: length elements contravariant}
\end{align}
We define the covariant components of the surface elements in terms of the contravariant length element: 
\begin{align}
   dS_r(r,\xi,\eta)& = \frac{4r^2}{\delta^{3/2}} C^2D^2 d\eta d\xi,  \nonumber\\
    dS_\xi(r,\xi,\eta) & =   \frac{2re^{\lambda(r)} D }{\delta^{1/2}}dr d\eta, \nonumber\\
    dS_\eta(r,\xi,\eta) &=  \frac{2re^{\lambda(r)}C }{\delta^{1/2}} dr d\xi.
   \label{eq: surface elements covariant}
\end{align}
For further details on the derivation of eqs.\,\eqref{eq: surface elements covariant} we refer to the Appendix, in particular eqs.\,\eqref{eq: covariant cross product} - \eqref{eq: covariant cross product metric}.

In our work, we utilize the covariant components of the cross product to compute the covariant surface components (see \S\ref{app: covariant cross product}). These components are then employed in the curl operator within the context of finite volume schemes (see \S\ref{subsec: finite Volume schemes}).

Last, the infinitesimal volume element is obtained by doing the mixed product between the three geometrical lengths:
 \begin{equation}
     dV(r,\xi,\eta) = e^{\lambda(r)} \frac{4r^2 C^2D^2}{\delta^{3/2}} dr d\xi d\eta 
     \label{eq: mixed product}
 \end{equation}

\subsection{Induction equation in neutron star crust}
\label{subsec: finite Volume schemes}
 
We study the non-linear evolution of magnetic fields in neutron star crusts with special attention to the influence of the Hall term, evolving the induction equation: 
\begin{equation}
    \frac{\partial \boldsymbol{B}}{\partial t} = -\boldsymbol{\nabla}\times \Big[ \eta_b \boldsymbol{\nabla}\times (e^{\zeta}\boldsymbol{B}) + \frac{c}{4 \pi e n_e}[\boldsymbol{\nabla} \times (e^{\zeta}\boldsymbol{B})] \times \boldsymbol{B} \Big]. 
   \label{eq: induction equation Rm} 
\end{equation}

The curl operator in the cubed-sphere coordinates, needed to compute $\boldsymbol{J}$ and to advance $\boldsymbol{B}$, can be written in the following concise form in our non-orthogonal metric (applied to a given vector $\boldsymbol{A}$): 
    \begin{equation} 
      (\boldsymbol{\nabla} \times \boldsymbol{A}) = \frac{1}{\sqrt{g} dl^r dl^\xi dl^\eta } \begin{vmatrix}
      dl^r \boldsymbol{e_r} &  dl^\xi \boldsymbol{e_\xi} &   dl^\eta \boldsymbol{e_\eta} \\ 
     dr \frac{\partial}{\partial r} & d\xi \frac{\partial}{\partial \xi} & d\eta \frac{\partial}{\partial \eta} \\ 
      dl^r \boldsymbol{A} \cdot \boldsymbol{e_r} &  dl^\xi \boldsymbol{A} \cdot \boldsymbol{e_\xi}  &  dl^\eta \boldsymbol{A} \cdot \boldsymbol{e_\eta} 
      \end{vmatrix},
      \label{eq: curl operator cubed-sphere}
  \end{equation}
  where $\sqrt{g} =
  \sqrt{\delta}/CD$.
  Explicitly, the components read:
      \begin{align}
      \big(\boldsymbol{\nabla} \times \boldsymbol{A}\big)^r &= 
      \frac{d\xi}{\sqrt{g} dl^\xi dl^\eta}  \Bigg[ \frac{\partial}{\partial \xi} \big(dl^\eta \boldsymbol{A} \cdot \boldsymbol{ e_\eta}  \big) -  \frac{d\eta}{d\xi} \frac{\partial}{\partial \eta} \big(dl^\xi \boldsymbol{A} \cdot \boldsymbol{e_\xi}  \big) \Bigg], 
      \nonumber\\
      &= 
      \frac{d\xi}{dS_r} \Bigg[ \frac{\partial }{\partial \xi}\bigg(dl^\eta A^\eta\bigg)    - \frac{Y}{D} \frac{\partial }{\partial \xi} \bigg( \frac{X}{C} dl^\eta A^\xi  \bigg) 
    \nonumber\\ &   -  \frac{d\eta}{d\xi} \frac{\partial}{\partial \eta}\bigg( dl^\xi  A^\xi\bigg)  +\frac{X d\eta}{C d\xi} \frac{\partial}{\partial \eta} \bigg( \frac{Y}{D}dl^\xi  A^\eta  \bigg)\Bigg],
    \label{eq: stokes theorem radial component}
  \end{align}

 \begin{align}
           \big(\boldsymbol{\nabla} \times \boldsymbol{A}\big)^\xi &= \frac{dr}{\sqrt{g} dl^r dl^\eta}  \Bigg[\frac{d\eta}{dr} \frac{\partial}{\partial \eta} \big(dl^r \boldsymbol{A} \cdot \boldsymbol{ e_r}  \big) -  \frac{\partial}{\partial r} \big(dl^\eta \boldsymbol{A}  \cdot \boldsymbol{ e_\eta}  \big) \Bigg], \nonumber\\
      &= \frac{dr}{dS_\xi} \Bigg[ dl^r \frac{d\eta}{dr} \frac{\partial A^r}{\partial \eta}
       -  \frac{\partial}{\partial r}\big( dl^\eta A^\eta \big) + \frac{XY }{CD} \frac{\partial}{ \partial r} \big(dl^\eta A^\xi \big)  \Bigg],
       \label{eq: stokes theorem xi component}
  \end{align}
  
    \begin{align}
    \big(\boldsymbol{\nabla} \times \boldsymbol{A}\big)^\eta &= \frac{dr}{\sqrt{g} dl^r dl^\xi}  \Bigg[ \frac{\partial}{\partial r} \big(dl^\xi \boldsymbol{A} \cdot \boldsymbol{ e_\xi}  \big) -  \frac{d\xi}{dr} \frac{\partial}{\partial \xi} \big(dl^r \boldsymbol{A} \cdot \boldsymbol{e_r}  \big) \Bigg], \nonumber\\
      &= \frac{dr}{dS_\eta} \Bigg[ \frac{\partial}{\partial r} \big( dl^\xi A^{\xi}\big)  - \frac{XY}{CD} \frac{\partial}{\partial r}\big( dl^\xi A^{\eta} \big)  - dl^r \frac{d\xi}{dr} \frac{\partial A^r}{\partial \xi}  \Bigg],
      \label{eq: stokes theorem eta component}
  \end{align}
where in the second equivalences we apply the Stokes theorem on an infinitesimal surface.

For any field, for output and plotting purposes we calculate the $\theta$ and $\phi$ components, using the transformations detailed in Appendix~\ref{app:coordinates}.

\subsection{Numerical schemes and computational features}
\label{subsec: discretization}

We employ an equally spaced grid in the two angular coordinates of each patch with steps of $d\xi=d\eta$, and a uniform step in the radial coordinate, $dr$, fine enough to adequately sample the large density and field gradients in the crust.

To evolve the magnetic field, we discretize the induction equation (eq.\,\eqref{eq: induction equation}) in the cubed-sphere coordinates, in our shell domain. Using the geometrical elements of \S\ref{subsec: Metric}, we calculate the eqs.\,\eqref{eq: stokes theorem radial component}-\eqref{eq: stokes theorem eta component} in our discretized scheme. We compute the circulation as a second-order accurate line integral along the edges of a cell face and divide it by the corresponding area, like in our previous 2D codes \citep{vigano2012,vigano2021}. The surface around which the circulation is performed includes the area of the four grid cells surrounding each point (therefore, all geometrical elements related to a given point extend one cell size at both sides along the considered direction). A detailed sketch of the circulation is illustrated in red on the left hand side of Fig.~\ref{fig: overlap region}. As noted in previous works (see Appendix A of \citealt{vigano2019}), rising the accuracy of the line integral (for instance, considering the values at the corners of the face) tends to create more numerical instabilities. Therefore, we stick to this second-order recipe.

\begin{figure}
     \centering
     \includegraphics[width=0.9\textwidth]{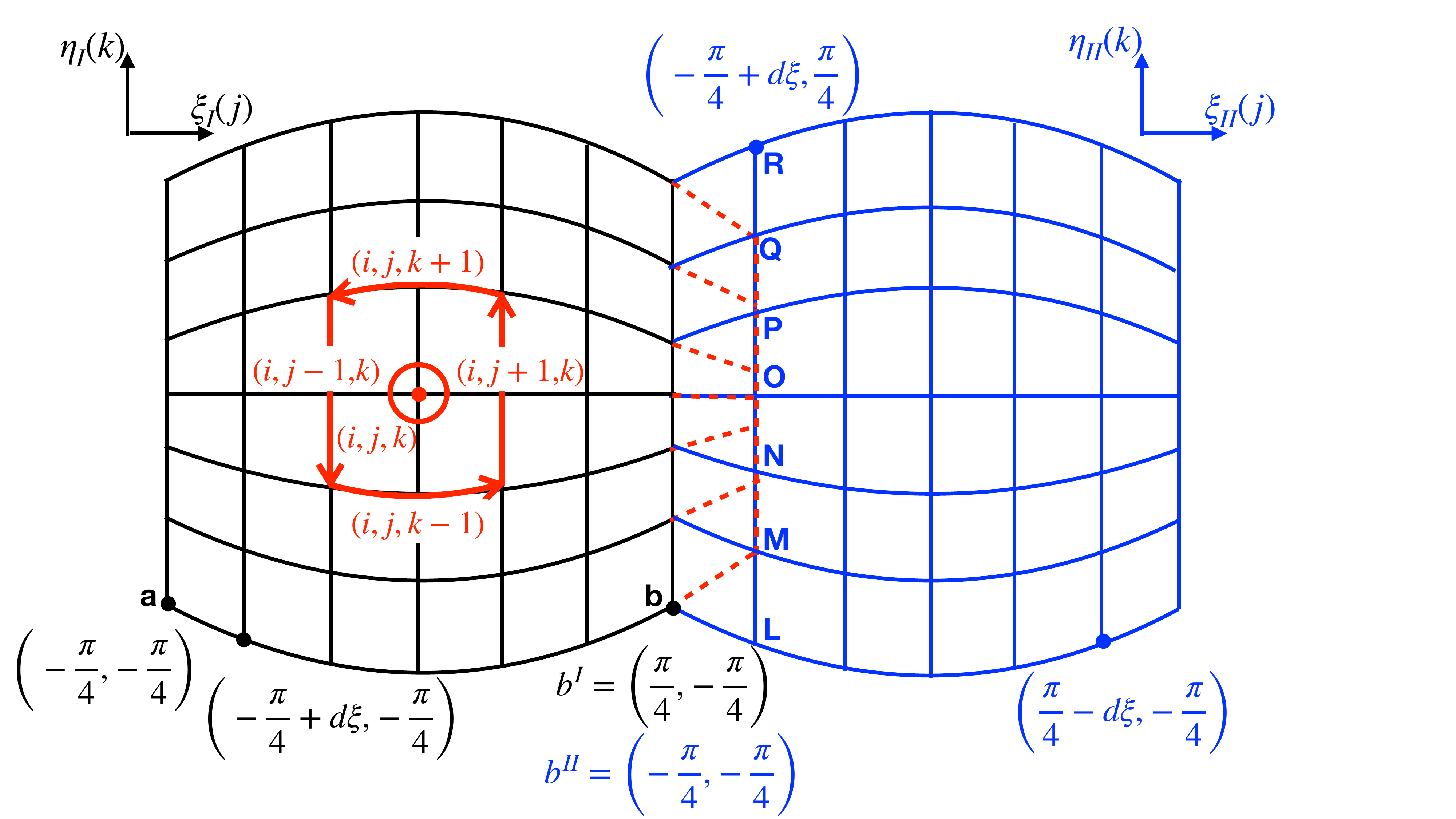}
     \caption{Schematic view of two contiguous equatorial blocks, e.g., patch I (black) and patch II (blue), and the ghosts cells of patch I (endpoints of the red dashes). The view is centered on the common vertical boundary line. The pseudo-horizontal coordinates $\xi$ of the ghost points of one grid, i.e., patch I, coincide with the second points along the $\xi$ coordinates of the last one interior grid points of the contiguous block, i.e., patch II. The ghost points are traced by the red line, and the values of the fields along the pseudo-vertical coordinate, $\eta$, are obtained by interpolations among the adjacent patch points (blue letters). Note that for other pairs of patches, the correspondence of coordinates may be less trivial (see Table \ref{tab:edges} and \S\ref{sec: connection between patches}). A sketch of a centered discretized circulation which extends twice the size of the cell (once per each side around a central point $(i,j,k)$) is displayed on the left hand side of this plot, in red. The circulation shown here is applied to calculate the radial component of the curl operator for a given vector A, i.e., $(\boldsymbol{\nabla} \times \boldsymbol{A})^r$. }
     \label{fig: overlap region}
 \end{figure}

To advance in time, we utilize an explicit fourth-order Runge-Kutta scheme. Although other Runge-Kutta schemes are also implemented in the \emph{MATINS} code, we conducted a thorough exploration of different orders of the Runge-Kutta methods (first-, second-, third-, fourth-, and sixth-order Runge-Kutta methods). Through this investigation, we found that the fourth-order Runge-Kutta method offers the most efficient and most stable performance compared to the other tested methods. More details are provided in Appendix~\ref{appendix: numerical performance}. In explicit algorithms, the stability of the method is constrained by the Courant condition, which limits the timestep to ensure that the fastest wave cannot travel more than one cell length in each time step. An estimate of the maximum allowed timestep for this non-linear system can be written as:
\begin{equation}
    dt^h= k_c {min}\bigg[ \frac{(\Delta l)^2 }{f_h B + \eta}\bigg]_{{points}}
    \label{eq: magnetic timestep}
\end{equation}
where $k_c$ is the Courant number and
$(\Delta l)^2 = [(dl^r)^{-2}+(dl^\xi)^{-2}+(dl^{\eta})^{-2}]^{-1}$ represents the square of the shortest resolved length scale, and the minimum is calculated over all the numerical points of the domain. Note that we are adopting for analogy a definition of the Courant constraint, which is well defined for a system in which the velocity is an independent variable: in that case, the maximum allowed timestep is linearly proportional to the grid size, and the proportionality constant is determined by the method (0.4 for RK4). In our case, the electron velocity is proportional to $\boldsymbol{\nabla}\times \boldsymbol{B}$, hence the (non-rigorous) estimation for the maximum timestep is quadratic with the the grid size, and we don't have an exact number for the value of $k_c$ for a given time-advance method. We instead fine-tune its optimal value by a series of numerical experiments, summarized in Appendix~\ref{appendix: numerical performance}.

The numerical stability of the magnetic evolution in the two codes ({\it MATINS} and the 2D), for a given initial setup, seems comparable: numerical instabilities start to appear at late times, when the star cools down and consequently the dynamics become largely Hall-dominated. This similarity with the 2D is surprising: here we don't employ the upwind-like scheme, the Burgers-like treatment for the toroidal field and the hyper-resistivity, which were all helping the numerical stability in 2D. As discussed in \cite{vigano2012,vigano2021}, in 2D all of them can be formulated and implemented in a compact way, without violating the field divergence and exploiting the axial symmetry, which allows a separation by components of the toroidal and poloidal field. In 3D, applying the same schemes is not possible by construction, and analogous more sophisticated ways to stabilize the code have not been developed so far.

\emph{MATINS} is coded in \texttt{Fortran90} with a modular design approach. In this framework, the microphysics and star's structure modules offer a diverse range of options for EoSs. For in-depth information on the numerical structure of \emph{MATINS} and a comprehensive list of available EoS choices, please refer to \S\ref{sec: MT evolution}.

\emph{MATINS} uses OpenMP to optimize the main loops. The computation bottlenecks are represented by the spherical harmonic decomposition needed in the boundary conditions and by the calculation of the circulation (done twice per each time sub-step). Among the two, the former takes more weight as the resolution increases. The code is faster when compiled with Intel compilers, compared to GNU. To give an idea, for the magnetic evolution simulations starting with $\sim 10^{14}-10^{15}$\,G, here presented, and the typical resolution used, e.g., $N_r=40$ and $N_\xi=N_\eta=43$ per patch, the total computational time for a run of $100$\,kyr is less than 1 day using six i9-10900 processors (2.80\,GHz). For such a simulation, about $\sim 150$ thousand iterations are needed to reach $100$\,kyr of evolution and it takes about $0.24$\,s per iteration. The computational time goes up to 2 days if one utilizes one processor instead of six (i.e., scalability efficiency $2/6\sim 0.3$). Due to the relatively low number of points ($<10^6$ in total for the resolutions used here), the scalability with openMP is decent only up to 6 processors. Therefore, we usually use 6 processors, a number that also takes advantage of the division by 6 patches. The computational cost of the simulations is set by the large number of iterations needed (${\cal O}(10^5)$ for 100\,kyr at the resolution here employed), which is in turn limited by the maximum timestep allowed, eq.\,\eqref{eq: magnetic timestep}. The latter scales with the square of the resolution $\Delta l$: our computational cost rises then with $\sim (\Delta l)^5$. For further details on the scalability of \emph{MATINS}, please refer to Appendix~\ref{appendix: numerical performance}.

Further optimization of the code is still possible and would potentially improve the performance, but will not affect the physical results shown here.

\subsection{Treatment of the edges between patches}
\label{subsec: edges of the patches}

When computing the curl operator introduced in eqs.\,\eqref{eq: stokes theorem radial component}-\eqref{eq: stokes theorem eta component} at the edges (corners) of the patch, one needs information about the values of the functions in some points which lie in the coordinate system(s) of the neighbouring patch(es). A way to deal with this issue is to extend one layer of ghost cells in each direction, for each patch (refer to \S\ref{sec: connection between patches}). The field components at the ghost cells are obtained by interpolating the vectors in the neighbouring patch coordinates.

\begin{figure}
     \centering
     \includegraphics[width=0.7\textwidth]{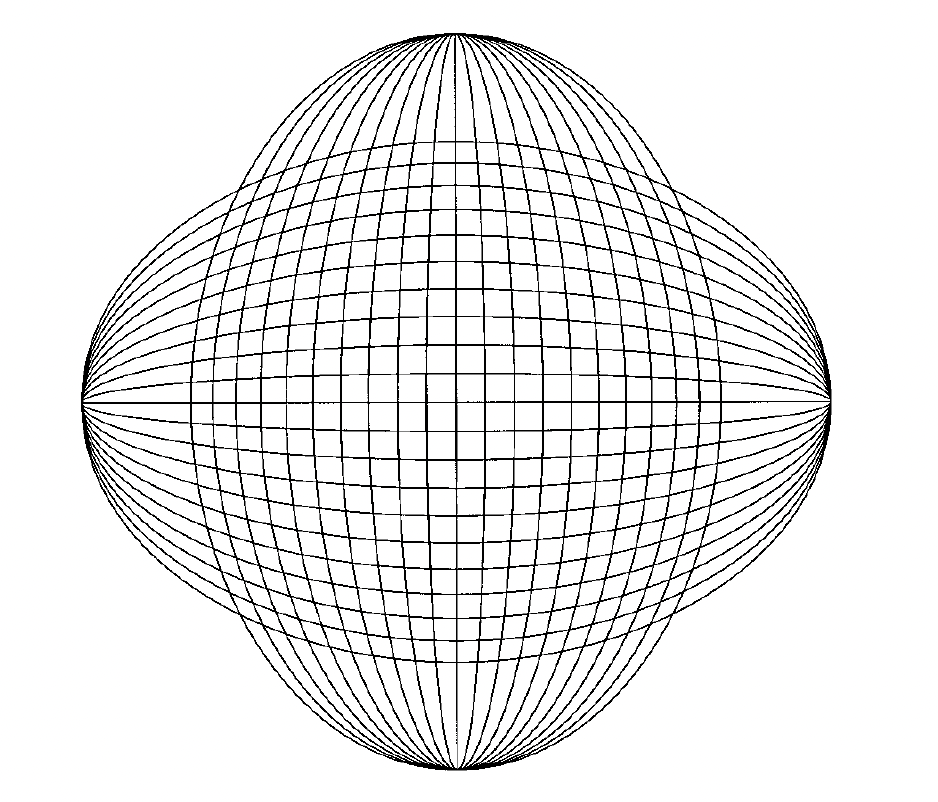}
     \caption[Construction of one of the six meshes from two sets of angularly equidistant great circles.]{Figure adopted from \cite{ronchi1996}. Construction of one of the six meshes from two sets of angularly equidistant great circles.}
     \label{fig: cubed-sphere arcs ofgreat circles}
 \end{figure}

Fig.~\ref{fig: overlap region} illustrates the mapping between two contiguous patches. By using the same regular grid size in both patches, we observe that the ghost vertical grid line in one patch (e.g., patch I in Fig.~\ref{fig: overlap region}, represented by the vertical dotted red line) coincides with the interior vertical grid line of the contiguous one (e.g., patch II in Fig.~\ref{fig: overlap region}, represented by the blue vertical line). 
This is due to the fact that the coordinate lines of the six patches which we employ to represent the entire spherical surface are arcs of great circles (as depicted in Fig.~\ref{fig: cubed-sphere arcs ofgreat circles}). Consequently, the coordinate lines of a given black join smoothly along one direction with those of the contiguous one. As a result of this choice, only a one-dimensional interpolation along the vertical $\eta$ direction will be required. It is worth noting that since $\xi$ and $\eta$ have the same grid spacing $d\xi =d\eta = \Delta$ and the same range $[-\pi/4; \pi/4]$, this idea can be applied in both the vertical and horizontal directions.

We now formalise the mapping of coordinates between two different patches. Let us consider a point close to the edge between two patches. We shall call, for each patch, $p$ the value of the point coordinate parallel to the interface, and $q$ the one pseudo-perpendicular to it (since the coordinates are not orthogonal except along the central axes of each patch). We shall use the superscript $^o$ to indicate the original coordinate system (for which we know $(p^o,q^o)$). The mapping of the point in the adjacent patch (superscript $^m$) is then given by
\begin{align}
  & q^m = {sgn}(q^{o}_{edge}) q^\circ - \frac{\pi}{2}~, \\
  & \tan (p^m) = {sgn}(q^{o}_{edge})\frac{\tan(p^o)}{\tan(q^o)}~,
\end{align}
where ${sgn}(q^{o}_{edge})$ is here used to identify the two edges in the original patch coordinate system, $q^{o}_{edge} = \pm \pi/4$, respectively. In the scenario where we are mapping ghost points to the adjacent patch they fall into, we encounter a set of points with different $\{p^o\}=\{-\pi/4 + (i/(N-1))\pi/2\}$ ($i=0,N-1$) and the same $q^o = {sgn}(q^{o}_{edge}) (\pi/4 + \Delta)$. Therefore:
\begin{align}
& q^m = - {sgn}(q^{o}_{edge}) \left(\frac{\pi}{4} - \Delta\right)~, \label{eq: mapping_edges_q} \\
& \tan (p^m) = \frac{\tan(p^o)}{\tan \left(\frac{\pi}{4} + \Delta\right)} \label{eq: mapping_edges_p}~.
\end{align}
Table~\ref{tab:edges} provides the correspondence of the direction for each edge, which involves two patches. The sign in the table indicates the direction of growth of the coordinate: if they have the same (opposite) sign, the two coordinates $p$ increase in the same (opposite) way. For a visual representation, please refer to the diagrams in section \S\ref{sec: connection between patches}.

\begin{table}
\centering
	\begin{tabular}{ | c | c | c | c | }			
		\hline
		edge & patches & $q$'s & $p$'s \\ \hline
        1  & I-II & $\xi^I=\pi/4$, $\xi^{II}=-\pi/4$ & $\eta^I$, $\eta^{II}$ \\
        2  & II-III & $\xi^{II}=\pi/4$, $\xi^{III}=-\pi/4$ & $\eta^{II}$, $\eta^{III}$ \\
        3  & III-IV & $\xi^{III}=\pi/4$, $\xi^{IV}=-\pi/4$ & $\eta^{III}$, $\eta^{IV}$ \\
        4  & IV-I & $\xi^{IV}=\pi/4$, $\xi^{I}=-\pi/4$ & $\eta^{IV}$, $\eta^{I}$ \\
        5  & I-V & $\eta^I=\pi/4$, $\eta^V=-\pi/4$ & $\xi^I$, $\xi^V$ \\
        6  & II-V & $\eta^{II}=\pi/4$, $\xi^V=\pi/4$ & $\xi^{II}$, $\eta^{V}$ \\
        7  & III-V & $\eta^{III}=\pi/4$, $\eta^V=\pi/4$ & $\xi^{III}$, $-\xi^{V}$ \\
        8  & IV-V & $\eta^{IV}=\pi/4$, $\xi^V=-\pi/4$ & $\xi^{IV}$, $-\eta^{V}$ \\
        9  & I-VI & $\eta^I=-\pi/4$, $\eta^{VI}=\pi/4$ & $\xi^I$, $\xi^{VI}$ \\
        10 & II-VI & $\eta^{II}=-\pi/4$, $\xi^{VI}=\pi/4$ & $\xi^{II}$, $-\eta^{VI}$ \\
        11 & III-VI & $\eta^{III}=-\pi/4$, $\eta^{VI}=-\pi/4$ & $\xi^{III}$, $-\xi^{VI}$ \\
        12 & IV-VI & $\eta^{IV}=-\pi/4$, $\xi^{VI}=-\pi/4$ & $\xi^{IV}$, $\eta^{VI}$ \\
        \hline
	\end{tabular}
	\caption[Coordinates at the 12 edges.]{\emph{Coordinates at the 12 edges.} Identification of coordinates at the twelve edges of the cubed sphere: pair of patch numbers, pair of values of the pseudo-perpendicular coordinate $q$ identifying the interface, parallel coordinate $p$ (the one to be mapped from one patch to the other when ghost points are defined). For a visual representation, please refer to the diagrams in section \S\ref{sec: connection between patches}. }
	\label{tab:edges}
\end{table}

Once the position of the ghost points is determined, we define a set of relative distances to the first neighbours, needed to linearly interpolate the vectors:
\begin{equation}
  W = \frac{p^m - p^o}{ \Delta} \in [0:1].
 \label{eq: interpolation weights}
\end{equation}
At the center of the edge, the distance $W$ is zero since the ghost point coincides with a point of the adjacent patch (point "O" of Fig.~\ref{fig: overlap region}). Note that the set of distances is universal, valid for any pair of patches.

The vector components at the ghost points are calculated in the coordinate system of the adjacent patch as follows:
\begin{equation}
    A_\text{gAdj}^{r, \xi, \eta}  = F1_\text{Adj}^{r, \xi, \eta}\big(1 -W \big) + F2_\text{Adj}^{r, \xi, \eta} \hspace{0.5mm}W, 
\end{equation}
where $F1_\text{Adj}^{r, \xi, \eta}$ and $F2_\text{Adj}^{r, \xi, \eta}$ are the vector components at the corresponding grid points in the adjacent patch surrounding the ghost point. Importantly, the angular components of the vector $\boldsymbol{A}$, need a change of coordinates from the adjacent to the original patch by using the Jacobians detailed in Appendix~\ref{Appendix: Jacobians}: 
\begin{align}
     A^{\xi} &= \text{JAC}(1,1) A_\text{gAdj}^{\xi} + \text{JAC}(1,2) A_\text{gAdj}^{\eta}  \nonumber\\ 
       A^{\eta} &= \text{JAC}(2,1) A_\text{gAdj}^{\xi}  + \text{JAC}(2,2) A_\text{gAdj}^{\eta}. 
\end{align}
At the edges (corners) between two (three) contiguous patches, there are two (three) coexisting coordinate systems, each one assigning slightly different values to the vector components. To guarantee identical field components at the egdes/corners between the patches, and to minimize numerical noise, we perform an averaging process after each sub-timestep and at each point along the edges. 
This averaging involves taking the values of electric currents and electric fields obtained from each patch and combining them appropriately. To achieve this correction for the angular components, an appropriate change of coordinates is required.

\section{Boundary conditions}
\label{sec: EMHD limit and Hall Induction Equation}

\subsection{Inner boundary conditions}
\label{subsec: inner BC}

For simplicity, the inner boundary conditions are imposed by demanding that the normal (radial) component of the magnetic field has to vanish at $r=R_c$. 
Under such assumptions, the Poynting flux at $r=R_c$ is zero and no energy is allowed to flow into/from the core. 

We note that, when using a second-order central difference scheme for the second derivative of a function, combined with our choice of the inner boundary conditions causes a numerical problem known as odd-even decoupling or checkerboard oscillations. This results in the numerical decoupling of two slightly different solutions, one for the odd grid points, and another one for the even grid points.
In order to relieve this, we increase the radial coupling among the nearest neighbours (found at a distance $dr$), as follows: 
\begin{align}
&& E^\xi(R_c)= \frac{1}{2}  E^\xi(R_c+dr),\nonumber\\
 && E^\eta(R_c)= \frac{1}{2}  E^\eta(R_c+dr) ,\nonumber\\
&& B^\xi(R_c-dr) = \frac{R_c}{R_c-dr} B^\xi(R_c), \nonumber\\
  && B^\eta(R_c-dr) = \frac{R_c}{R_c-dr} B^\eta(R_c). 
  \label{eq: odd-even decoupling solution inner BC}
\end{align}
In the equations above, we omit the angular dimensions for clarity.
This choice reduces the tangential current at the crust-core interface and improves the stability during the evolution. 

\begin{figure}
	\centering
	\includegraphics[width=0.5\textwidth]{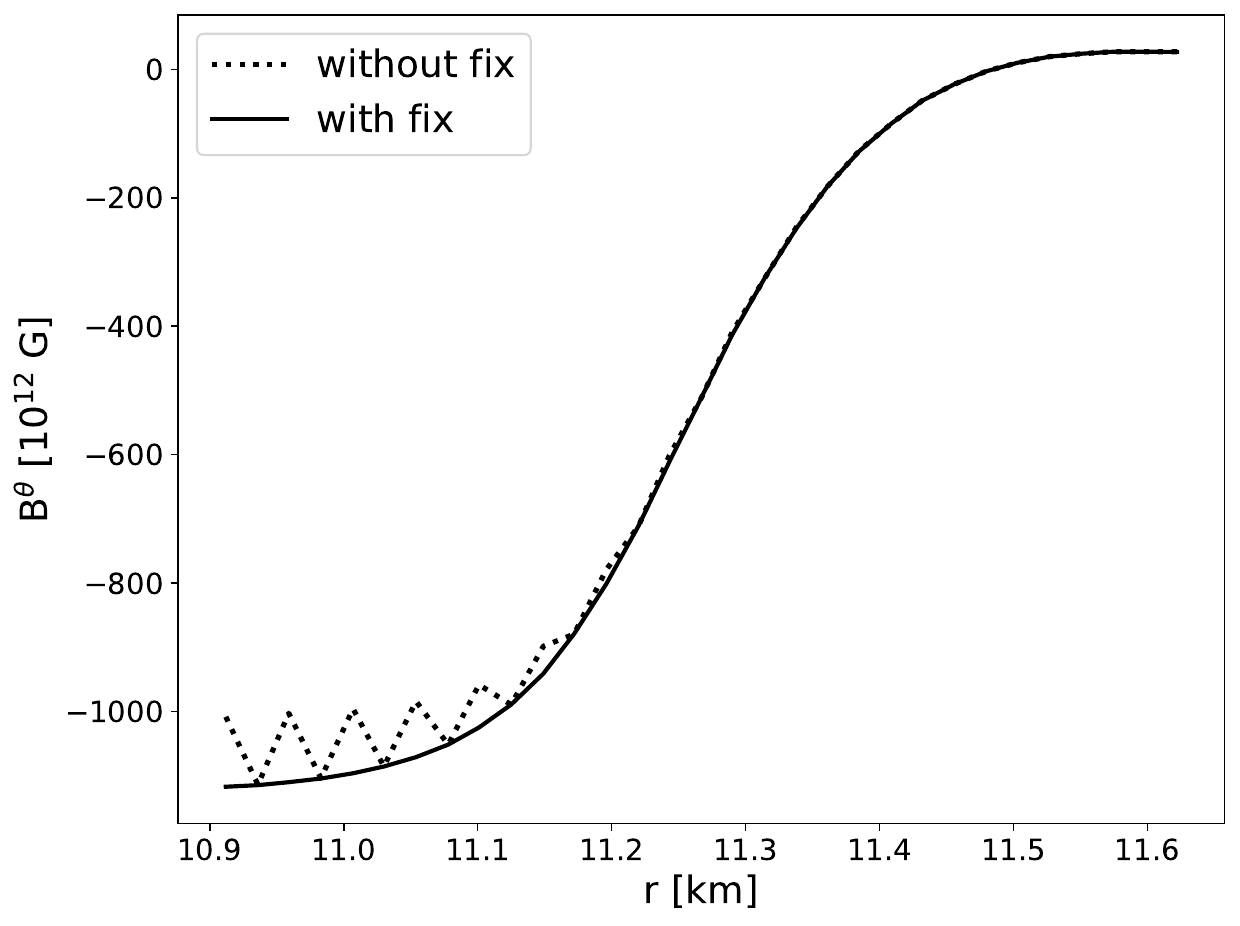}
	\caption[Difference between fixing the odd-even decoupling and not]{Difference between fixing the odd-even decoupling (eq.\,\eqref{eq: odd-even decoupling solution inner BC}, solid line) and not (dots), for a representative evolved radial profile of a magnetic field component at a given angle. As a representative example, we show $B^{\theta}$ in the upper right corner of patch II.}
	\label{fig: odd-even decoupling}
\end{figure}

In Fig.~\ref{fig: odd-even decoupling}, we illustrate a representative case of the difference in radial profile of a component with (solid line) and without the prescription above (dots).

\subsection{Outer boundary conditions: potential field }
\label{subsec: outer B.C}

We enforce an external potential (current-free) solution for the magnetic field at the surface of the star. By excluding surface current sheets, all components of the magnetic field remain continuous across the outer boundary. These boundary conditions are determined by $\nabla \times B = 0$ and $\nabla \cdot B = 0$. Consequently, the magnetic field can be expressed as the gradient of the magneto-static potential $\chi_m$, which satisfies the Laplace equation:
\begin{align}
   && \boldsymbol{B}= \boldsymbol{\nabla} \chi_m, \nonumber\\
   && \nabla^2 \chi_m = 0.
\end{align}
The spherical harmonics expansion of the scalar potential $\chi_m$ reads: 
 \begin{equation}
\chi_m = - B_0 R \sum_{l=0}^{\infty} \sum_{m=-l}^{m=+l} Y_{lm}(\theta,\phi) \bigg( b^m_l \bigg(\frac{R}{r} \bigg)^{l+1} + c^m_l\bigg(\frac{r}{R} \bigg)^l \bigg),
\label{eq: magnetostatic potential}
\end{equation}
where $B_0$ is a normalization, $b^m_{l}$ corresponds to the weight of the multipoles, and $Y_{lm}$ are the spherical harmonics. In this study, we use the $Y_{lm}$ decomposition in \S\ref{Appendix: spherical harmonics}, since we are interested in working with the real set of spherical harmonics (Laplace spherical harmonics). The latter forms an orthonormal and complete set. One can choose real functions by combining complex conjugate functions, corresponding to opposite values of $m$. 
Note that we exclude $l=0$ since it corresponds to a magnetic monopole, which violates $\boldsymbol{\nabla} \cdot \boldsymbol{B}= 0$. The dimensionless weights $b_{l}^{m}$ and $c_{l}^m$ are associated to $l$ and $m$ multipoles of two branches of solutions. The second branch, $\propto \big(r/R \big)^l$, diverges for a domain extending to $r ~ \rightarrow \infty$, like the magnetosphere, therefore we set $c_l^m=0$. 
 
The normal components of the magnetic field $B^r$ are evolving and known at the surface of the star at each timestep. However, to impose potential boundary conditions, we need to determine the angular components of the magnetic field at the surface and one cell above the surface of the star. We proceed as follows.

The continuity of $B^r$ across the surface enables us to express it in terms of the magneto-static potential as:
 \begin{equation}
     B^r= \frac{1}{e^{\lambda(r)}}\frac{\partial \chi_m}{\partial r} = \frac{B_0}{e^{\lambda(r)}} \sum_{l=0}^{\infty} \sum_{m=-l}^{m=+l} (l+1) Y_{lm}(\theta,\phi)  b^m_l \bigg(\frac{R}{r} \bigg)^{l+2}.
     \label{eq: spectral decomposition of Br}
 \end{equation}

Next, we evaluate the weights of the multipoles $b_l^m$ by applying the orthogonality properties of spherical harmonics to eq.\,\eqref{eq: spectral decomposition of Br}, resulting in:
 \begin{equation}
     b^m_l = \frac{e^{\lambda(R)}}{B_0(l+1)} \int \frac{dS^r}{r^2} B^r Y_{lm}(\theta,\phi).
     \label{eq: blm}
 \end{equation}
Using this information, we can evaluate the angular components of the magnetic field for $r\geq R$.
 \begin{equation}
    B^{\theta} = - B_0 \sum_{l=0}^{\infty} \sum_{m=-l}^{l} b^m_l \bigg(\frac{R}{r}\bigg)^{l+2} \frac{\partial Y_{lm}(\theta,\phi)}{\partial \theta},
 \end{equation}
 
\begin{equation}
    B^{\phi} =   - \frac{B_0}{sin(\theta)} \sum_{l=0}^{\infty} \sum_{m=-l}^{l} b^m_l \bigg(\frac{R}{r}\bigg)^{l+2} \frac{\partial Y_{lm}{\theta,\phi}}{\partial \phi},
\end{equation}
which are then converted into the $B^\xi$ and $B^\eta$ components in the code.

Finally, analogously to what described for the inner boundary (\S\ref{subsec: inner BC}), we prevent the radial odd-even decoupling at the surface by setting the values of the tangential components of the magnetic field as the average between the values one point above and below the surface.

A more realistic outer boundary condition can be achieved by coupling the interior crustal evolution with the magnetosphere using PINN \citep{urban2023}. While this coupling was initially performed in 2D (see Appendix~\ref{appendix: neural networks}), we are currently working on implementing it in 3D.

\section{Numerical tests}
\label{sec: numerical test}

\subsection{Diagnostics}
\label{subsec: diagnostics}

A necessary test for any numerical code is to check the instantaneous (local and global) energy balance. Any type of numerical instability usually results in the violation of the energy conservation, or any other physical constraint (the divergence condition). Therefore a careful monitoring of the energy balance is performed.
The magnetic energy balance equation for Hall-MHD can be expressed as:
\begin{equation}
    \frac{\partial }{\partial t} \bigg(e^{\zeta} \frac{B^2}{8\pi} \bigg) = - e^{2\zeta}Q_j - \boldsymbol{\nabla} \cdot \big( e^{2\zeta} \boldsymbol{S} \big),
    \label{eq: energy balance}
\end{equation}
where $Q_j= 4\pi \eta_b J^2/c^2$ is the Joule dissipation rate and $\boldsymbol{S} = c\boldsymbol{E} \times \boldsymbol{B}/4\pi$ is the Poynting vector.

Integrating eq.\,\eqref{eq: energy balance} over the whole volume of the numerical domain, we obtain the balance between the time variation of the total magnetic energy $E_{mag} = \int_V (e^{\zeta} B^2/8\pi) dV$, the Joule dissipation rate $Q_{tot} = \int_V e^{2\zeta} Q_j dV$, and the Poynting flux through the boundaries $S_{tot} = \int_S e^{2\zeta} \boldsymbol{S} \cdot \hat{n}dS $. In our case, the boundaries are the star surface and the crust-core interface, so that $S_{tot}$ is given by the integration of $S_r$ over them. Thus, the volume-integrated energy balance is
\begin{equation}
    \frac{d}{dt} E_{mag} + Q_{tot} + S_{tot} = 0.
    \label{eq: volume-integrated energy balance}
\end{equation}

In addition, we calculate the local magnetic field divergence in the cubed-sphere coordinates using Gauss' theorem:
  \begin{equation}
        \boldsymbol{\nabla} \cdot \boldsymbol{B} 
       = \frac{1}{dV}\bigg[\frac{\partial}{\partial r} (dS_r B^r ) + \frac{\partial}{\partial \xi} (dS_\xi B^\xi)  + \frac{\partial}{\partial \eta} (dS_\eta B^\eta )  \bigg].
      \label{eq: div(B)}
  \end{equation}
Starting from an initial divergence-free magnetic field (see \S\ref{appendix: initial conditions} for more details), we monitor that indeed the divergence of the magnetic field does not grow in time above some tolerable error. We achieve this using the constrained transport method implemented in our code, which ensures the conservation of the magnetic field's divergence through the finite volume scheme. To measure this, we compare the volume integral of $(\nabla\cdot\boldsymbol{B})^2$
\begin{equation}
    D_d=\int \big(\boldsymbol{\nabla} \cdot \boldsymbol{B} \big)^2 dV
    \label{eq: divB volume}
\end{equation}
to a physical quantity with the same units and scaling, e.g., the integrated values of the square of the effective current 
\begin{equation}
   D_J=\int [\boldsymbol{\nabla}\times (e^{\zeta}\boldsymbol{B})]^2 dV~,
    \label{eq: J2-star}
\end{equation}
or to $(B/<dl>)^2$, where $<dl>$ is the geometrical mean of the cell's edge lengths. 
We verify that during the evolution, the divergence of the magnetic field always remains several orders of magnitude smaller than the other quantities throughout the star. 

A detailed analysis of the spectral energy distribution is performed in this study. The explicit calculation of this quantity is done using the poloidal and toroidal decomposition of the magnetic field described (see \S\ref{appendix: Poloidal and toroidal decomposition}).
The magnetic energy content in each mode, considering the relativistic corrections, can be expressed as:
 \begin{equation}
         E_{lm} = \frac{1}{8\pi} \int  e^{\lambda+\zeta} dr ~l(l+1) \bigg[  \frac{l(l+1)}{r^2} \Phi_{lm}^2 +\big(\Phi'_{lm}\big)^2   +\Psi_{lm}^2  \bigg],
     \label{eq: spectral magnetic energy}
 \end{equation}
where $\Phi'_{lm}$ is the radial derivative of $\Phi_{lm}$, explicitly given by eq.\,\eqref{eq: radial derivative of the Phi scalar function}.
 The first two terms in eq.\,\eqref{eq: spectral magnetic energy} account for the poloidal magnetic energy and the last term accounts for the toroidal energy. The total energy density is simply $\sum_{lm} E_{lm}$. 

For a more comprehensive and quantitative analysis of the 3D magnetic evolution, we conduct a survey of the magnetic energy spectrum (Chapter~\ref{chap: 3DMT}). This allows us to observe the redistribution of magnetic energy across different spatial scales, shedding light on how the system evolves over time.

\subsection{The purely resistive benchmark}
\label{subsec: bessel test}

A classical benchmark test that enables the comparison of analytical solutions is the evolution of axisymmetric modes under Ohmic dissipation only (zero magnetic Reynolds number) and constant magnetic diffusivity, denoted as $\eta_b$. In this limit, and neglecting the general relativistic correction, the induction equation for the case of Ohmic dissipation is then given by:
\begin{equation}
    \frac{\partial \boldsymbol{B}}{\partial t} = - \eta_b \boldsymbol{\nabla} \times ( \boldsymbol{\nabla}  \times \boldsymbol{B} ).
    \label{eq: ohmic induction eq bessel}
 \end{equation}
The Ohmic eigenmodes consist of force-free solutions satisfying $\boldsymbol{\nabla} \times \boldsymbol{B} = \alpha \boldsymbol{B}$, where $\alpha$ is a constant parameter. Then, we have 
\begin{equation}
    \frac{\partial \boldsymbol{B}}{\partial t} = - \eta_b \alpha^2  \boldsymbol{B}.
    \label{eq: bessel induction eq}
\end{equation}
which shows that each component of the magnetic field decays exponentially with the same diffusion timescale $\tau_d= 1/(\eta_b \alpha^2)$.
\begin{equation}
    \boldsymbol{B}(t) = e^{-t/{\tau_d}}\boldsymbol{B}(t=0).
    \label{eq: bessel decay in time}
\end{equation}
Note that the evolution of each component is decoupled in this case.
A solution for eq.\,\eqref{eq: bessel induction eq} in spherical coordinates is described by factorized functions, where the radial parts involve the spherical Bessel functions. The regularity condition at the center selects only one branch of the spherical Bessel functions (of the first kind), which, for the $(l,m) = (1,0)$ mode, is given by:
\begin{align}
   &&  B^r =\frac{B_0 R}{r} cos\theta \bigg( \frac{sin x}{x^2}  - \frac{cos x}{x}     \bigg), \nonumber\\
    &&  B^\theta = \frac{B_0 R}{2r} sin\theta \bigg( \frac{sin x}{x^2}  - \frac{cos x}{x} - sin x    \bigg), \nonumber\\
      &&   B^\phi = \frac{\alpha R B_0 }{2} sin\theta \bigg( \frac{sin x}{x^2}  - \frac{cos x}{x} \bigg),
         \label{eq: B bessel fct}
\end{align}
where $B_0$ is the normalization factor, $x= \alpha r$ is a dimensionless quantity, and $R=10$\,km corresponds to the maximum radius of the spherical shell.

In the limit where $x$ approaches 0, we recover the solution corresponding to a homogeneous field aligned with the magnetic axis:
\begin{align}
      &&  B^r = k B_0 \frac{cos\theta}{3}, \nonumber\\
       &&  B^\theta = - k B_0 \frac{sin\theta}{3}, \nonumber\\
         && B^\phi = 0.
          \label{eq: bessel x->0}
\end{align}
For more details, we refer the reader to \S5.4 of \cite{pons2019}.

\begin{figure}
\centering
\includegraphics[width=0.45\textwidth]{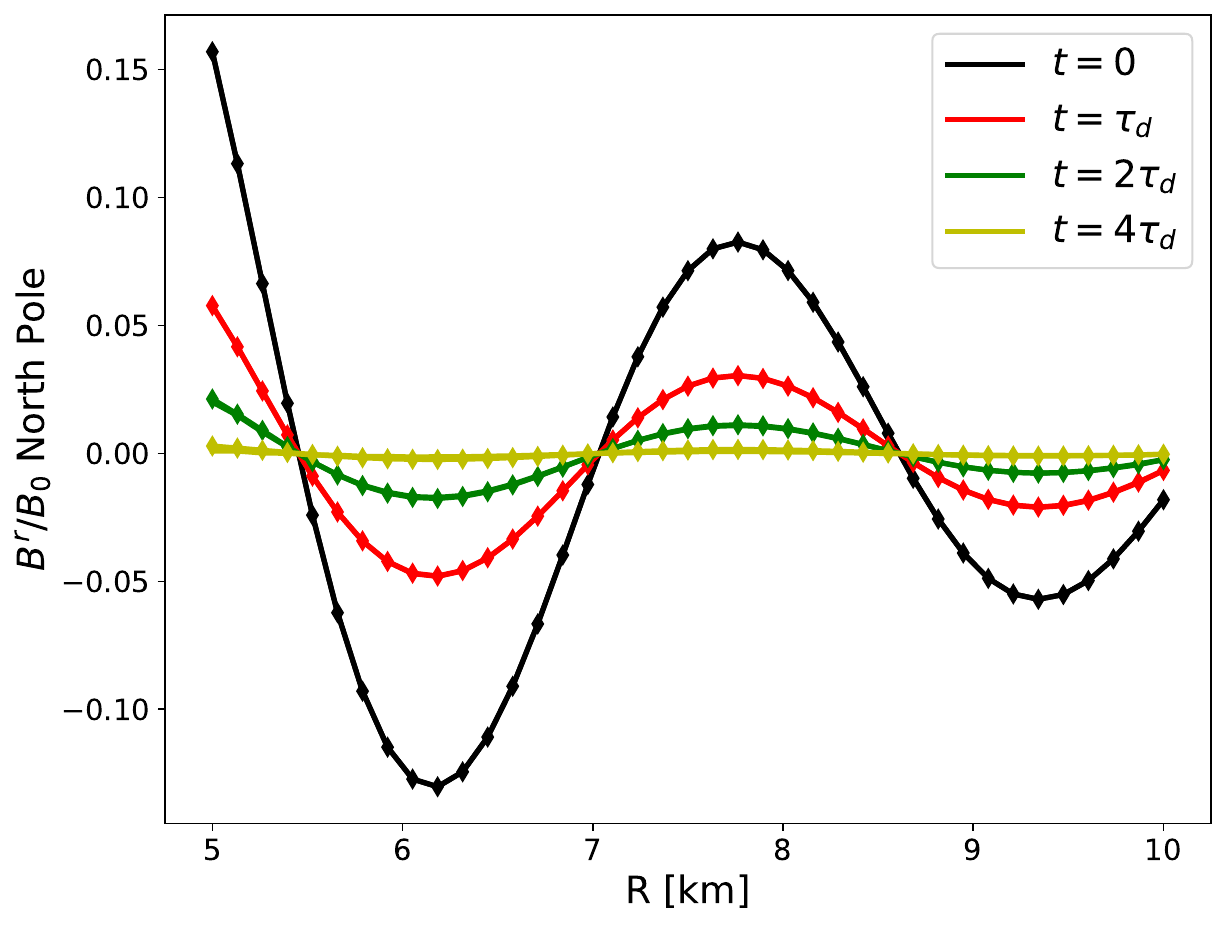}
\includegraphics[width=0.45\textwidth]{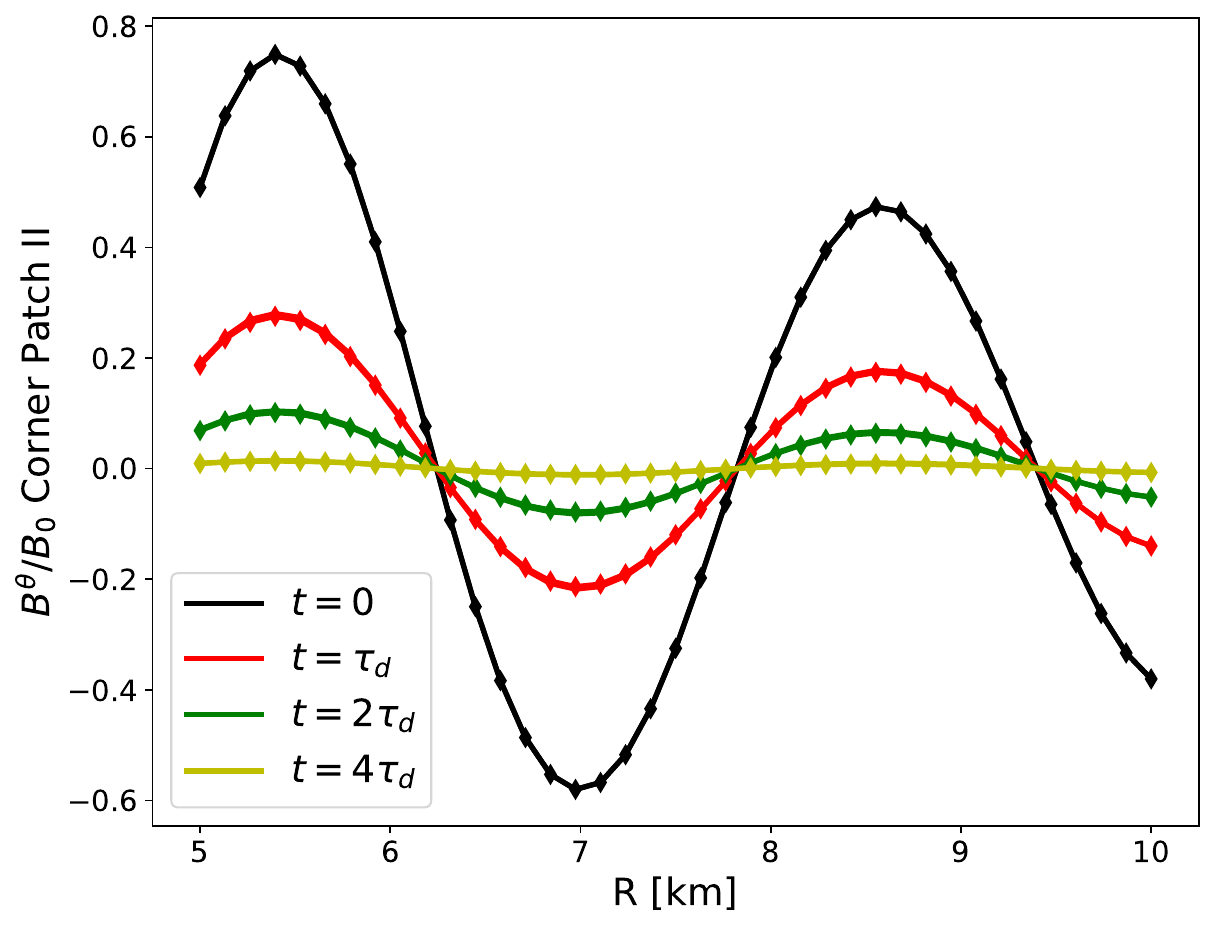} 
\includegraphics[width=0.45\textwidth]{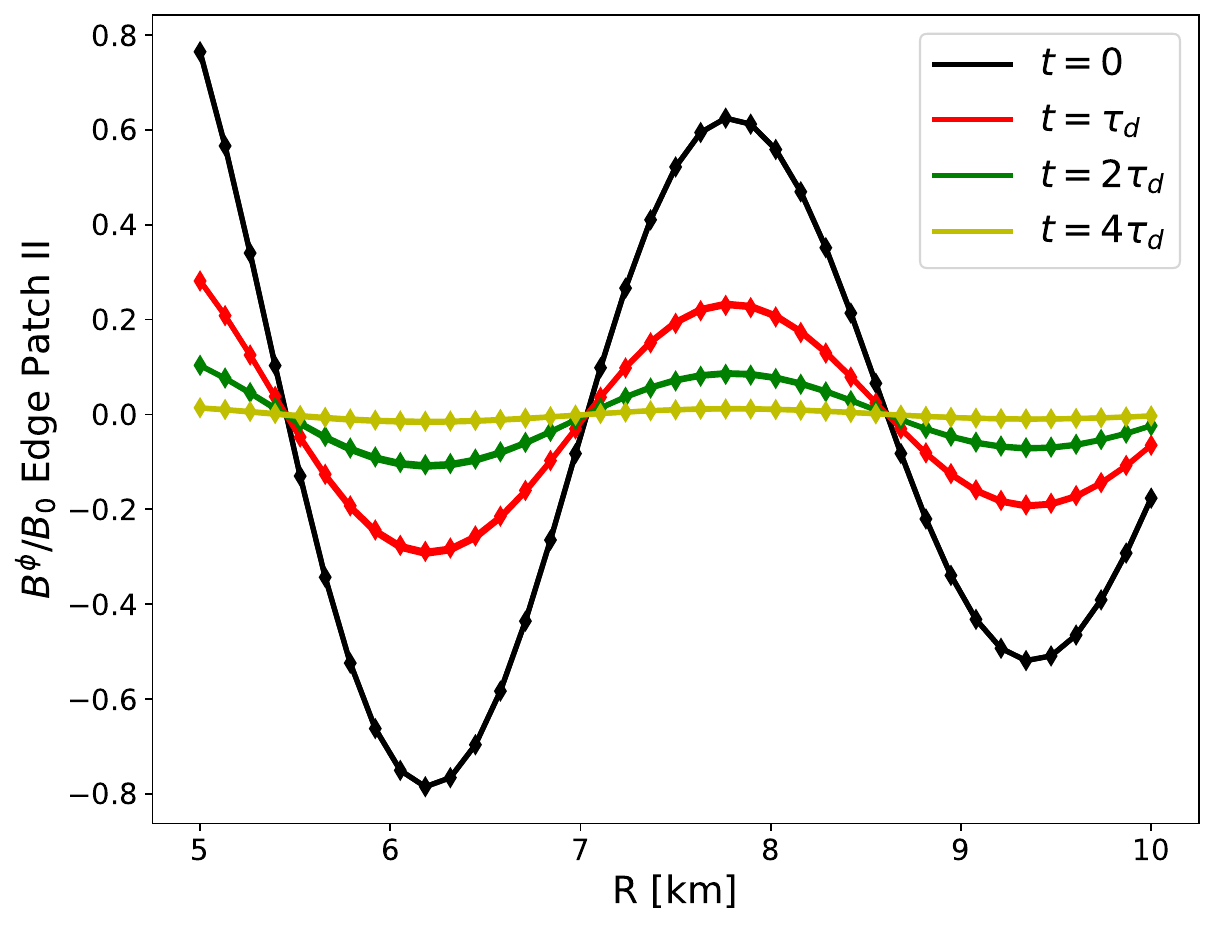}
\caption[Bessel test: Magnetic field profiles]{Bessel test: radial profiles of the magnetic field components at different Ohmic timescales, with $\tau_d=0.25$ Myr. A comparison of the numerical (solid lines) and the analytical (diamonds) solutions for a model with wave-number in unit length, $\alpha = 2$ km$^{-1}$, in a spherical shell of radius $[R_c= 5:R=10]$ km.
\emph{LHS panel:} $B^r/B_0$ radial profile at the north pole of the star. \emph{RHS panel:} $B^\theta/B_0$ radial profile at the upper right corner of patch 2 (equatorial patch). \emph{Bottom panel:} $B^\phi/B_0$ radial profile at the bottom left border of patch 2, with $B_0$ a normalization factor.}
\label{fig:Bessel test alpha=2}
\end{figure}

Considering the spherical Bessel functions as initial conditions, and imposing the analytical solutions for $B^r$, $B^\xi$, and $B^\eta$ as boundary conditions (eqs.\,\eqref{eq: B bessel fct}), we follow the evolution of the modes during several diffusion timescales.

\begin{figure}
\centering
\includegraphics[width=.5\textwidth]{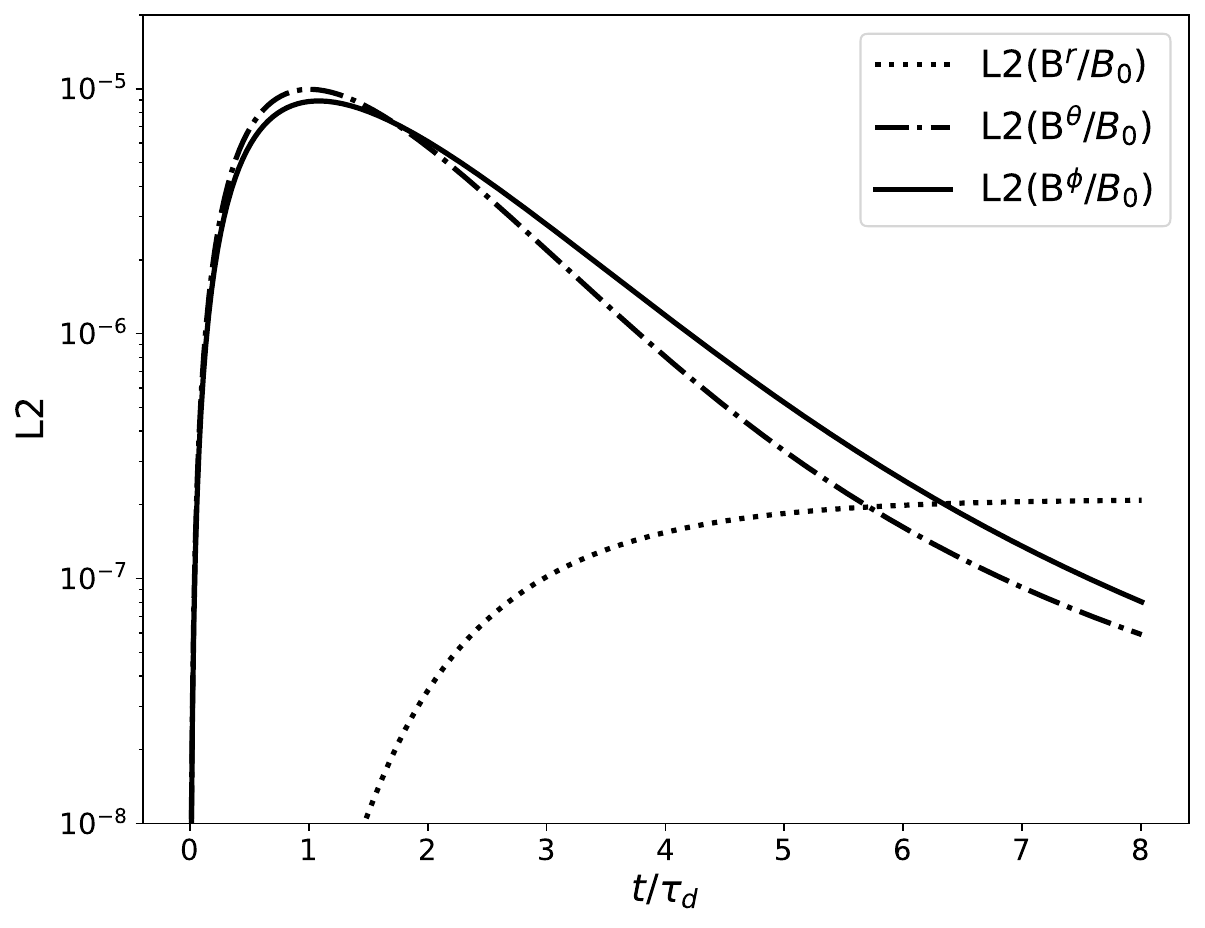}
\caption[Bessel test: Relative error]{The average absolute $L^2$ error as a function of $t/\tau_d$, 
for $B^r/B_0$ (dots), $B^\theta/B_0$ (dot-dashed lines) and $B^\phi/B_0$ (solid lines) components of the magnetic field. We have verified that the maximum absolute $L^2$ error is of the same order as the average one. The values are much smaller than the mean square of the initial magnetic field, which is of order $0.1 B_0^2$.  }
\label{fig: relative error of the Bessel test}
\end{figure}

Fig.~\ref{fig:Bessel test alpha=2} compares the numerical (solid lines) and analytical (diamonds) solutions of the magnetic field components for a magnetic field of order one, at different diffusion timescales, for a model with $\alpha = 2$\,km$^{-1}$, in a spherical shell defined by $r\in [5:10]$\,km, with a resolution of $N_r= 40$ and $N_\xi=N_\eta= 43$ points per patch in the cubed-sphere coordinates. One can notice that the magnetic field has decreased below the visible scale in the figure around $4 \tau_d$. Moreover the analytical and numerical results are indistinguishable in the graphic.

To quantify the deviation, we evaluate the average $L^2$ absolute error in terms of the deviation from the analytical solution, as shown in Fig.~\ref{fig: relative error of the Bessel test} for $B^r$ (dots), $B^\theta$ (dash-dotted lines) and $B^\phi$ (solid lines). The angular field components exhibit higher error than the radial component, primarily due to the patchy grid employed. The $L^2$ error saturates after one diffusion timescale for the two angular field components and after two diffusion timescales for the radial component. We have checked that by varying the resolution, the errors scale with $\Delta^2$, validating that the method is of second order.

\subsection{A comparison between the 2D and the 3D magnetic codes}
\label{subsec: axisymmetric - comparison 2D and 3D}

For the general case including the Hall term and with variable diffusivity and electron density, no analytical solution is available. However, since extensive results from 2D simulations are available, a detailed comparison of the 3D magnetic code presented here and the 2D code \citep{vigano2012,vigano2021}, helps to probe the validity of the results of the 3D code. 

For this comparative analysis, we use analytical fixed radial profiles for the magnetic diffusivity $\eta_b$ and the Hall prefactor $f_h$ in both codes. For the Hall prefactor $f_h$, we utilize the following fit adopted from \cite{vigano2021}
\begin{equation}
f_{h,\text{fit}} = 0.011 ~e^{k(\tilde{r}-\tilde{R_{c}})^b}
\left[ \frac{{\mathrm km}^2}{{\mathrm Myr}\, 10^{12} {\mathrm G}} \right]
    \label{eq: fh radial profile}
\end{equation}
where $k=10$, $b=1.8$, $\tilde{r}$ and $\tilde{R}_c$ are $r$ and $R_c$ given in km. This radial profile exhibits a super-exponential rise of about three orders of magnitude throughout the crust.

For the magnetic diffusivity $\eta_b$, we use the analytical radial profile 
\begin{equation}
\eta_b(r) = 6 \frac{(r-R_c)^{k_0}}{(R-R_c)^{k_0}} + 3 \frac{(R-r)^{k_1}}{(R-R_c)^{k_1}}  
\left[\mathrm{ \frac{km^2} {Myr}} \right],
    \label{eq: etab radial profile}
\end{equation}
with $k_0=3.5$ and $k_1=4$. 

The initial magnetic field is an axisymmetric crustal-confined field with a poloidal dipole of $10^{14}$\,G at the polar surface and a toroidal component consisting of a sum of a quadrupole and an octupole with a maximum value of $10^{15}$\,G. We use a grid resolution of $N_r = N_{\xi}= N_{\eta} = 30$ per patch (meaning 61 points from pole to pole and 120 along the equator). A similar resolution is used in the 2D code, e.g., $N_{\theta}=60$ and $N_r=30$.

\begin{figure}
\centering
\includegraphics[width=0.45\textwidth]{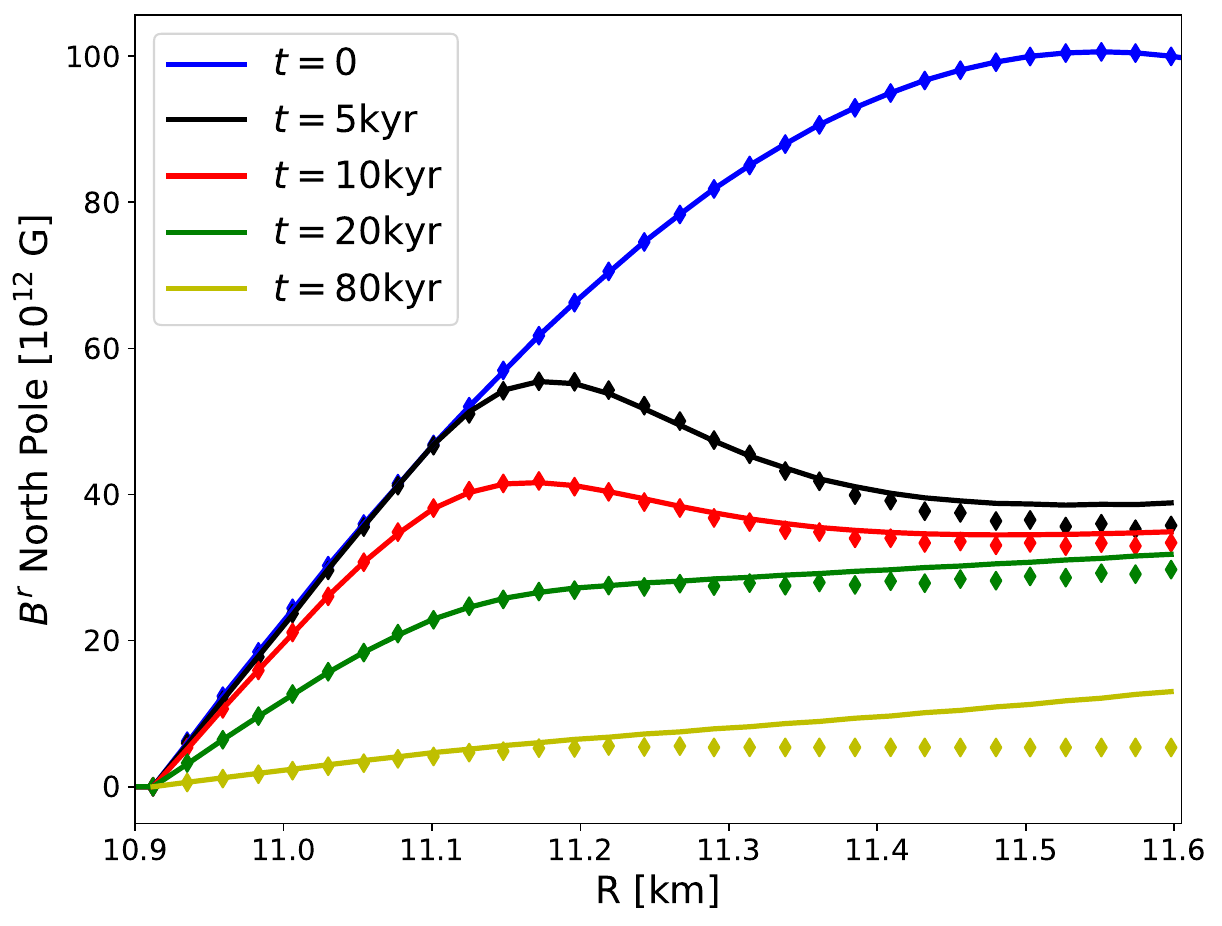}
\includegraphics[width=0.45\textwidth]{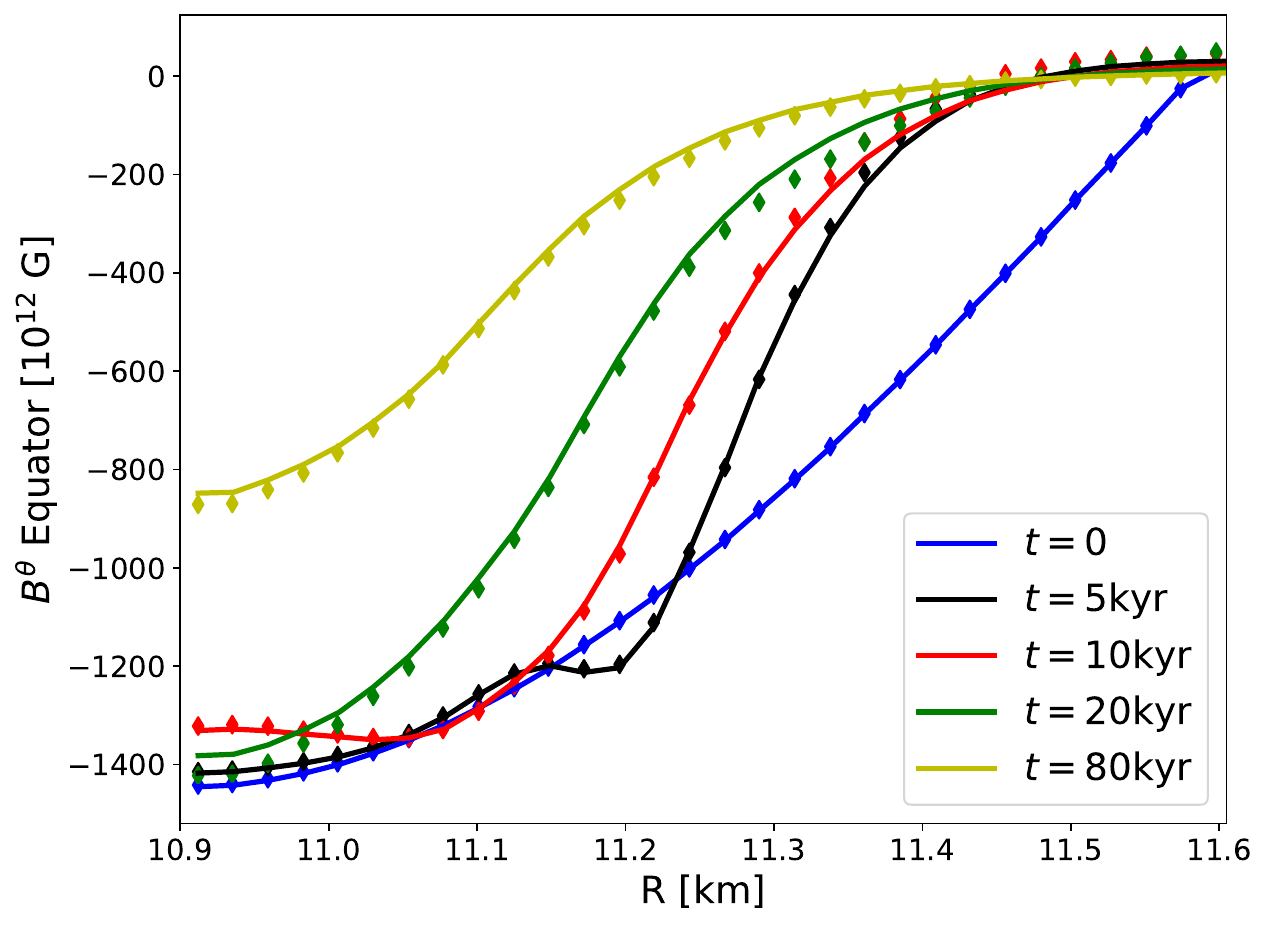} 
\includegraphics[width=0.45\textwidth]{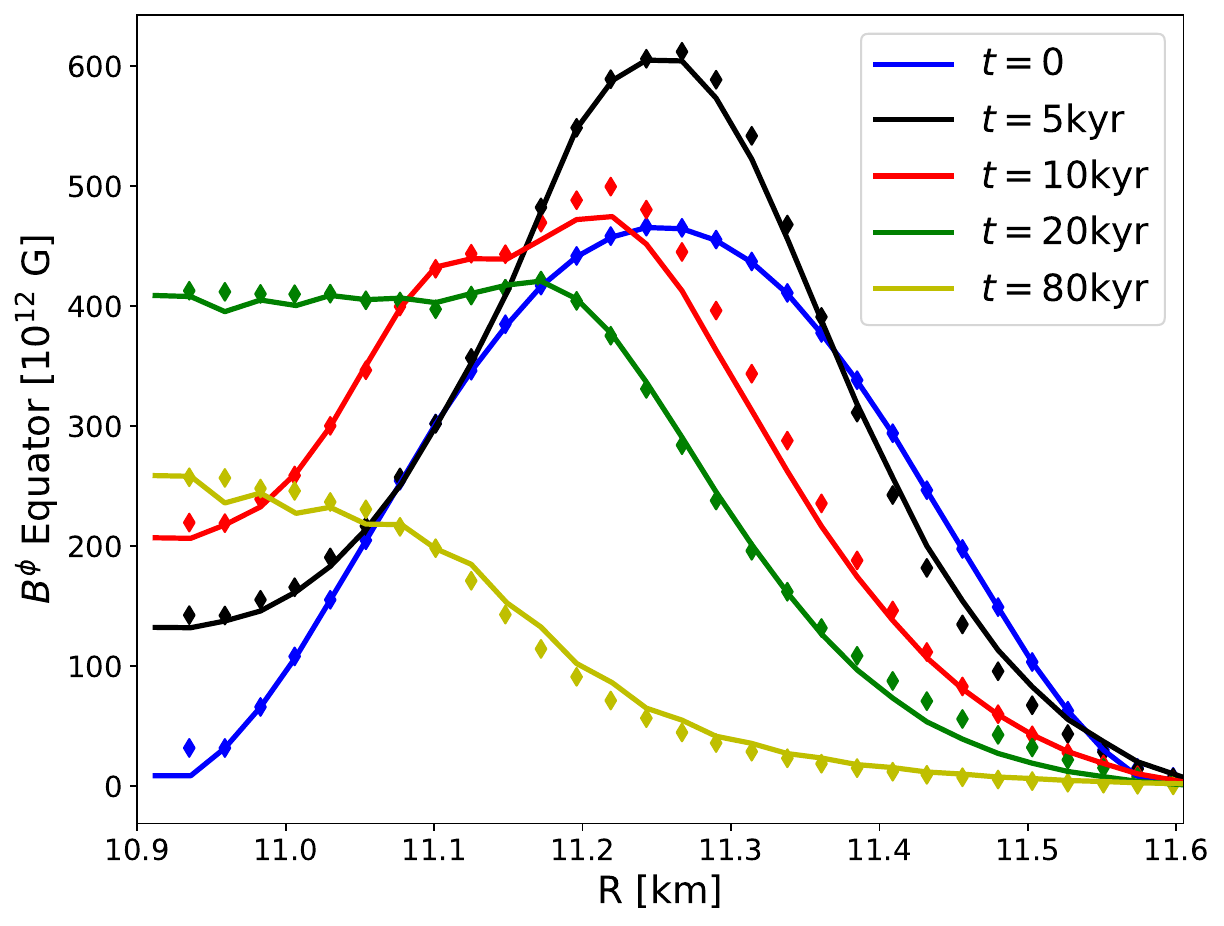}
\caption[2D vs 3D magnetic field profiles]{Radial profiles of the magnetic field components. On the left, we show the radial profile of the normal component of the magnetic field $B^r$ at the north pole of the star. In the RHS panel and the panel at the bottom, we illustrate the radial profiles of the two angular components of the magnetic field, namely $B^{\theta}$ and $B^{\phi}$, respectively. These profiles are displayed at the equator of the star. The solid lines correspond to the results of the 3D code, whereas the diamonds correspond to those of the 2D code. Different colors correspond to different evolution time.}
\label{fig: 2d-3d comparison Bfield profiles}
\end{figure}

\begin{figure}
	\centering
	\includegraphics[width=0.5\textwidth]{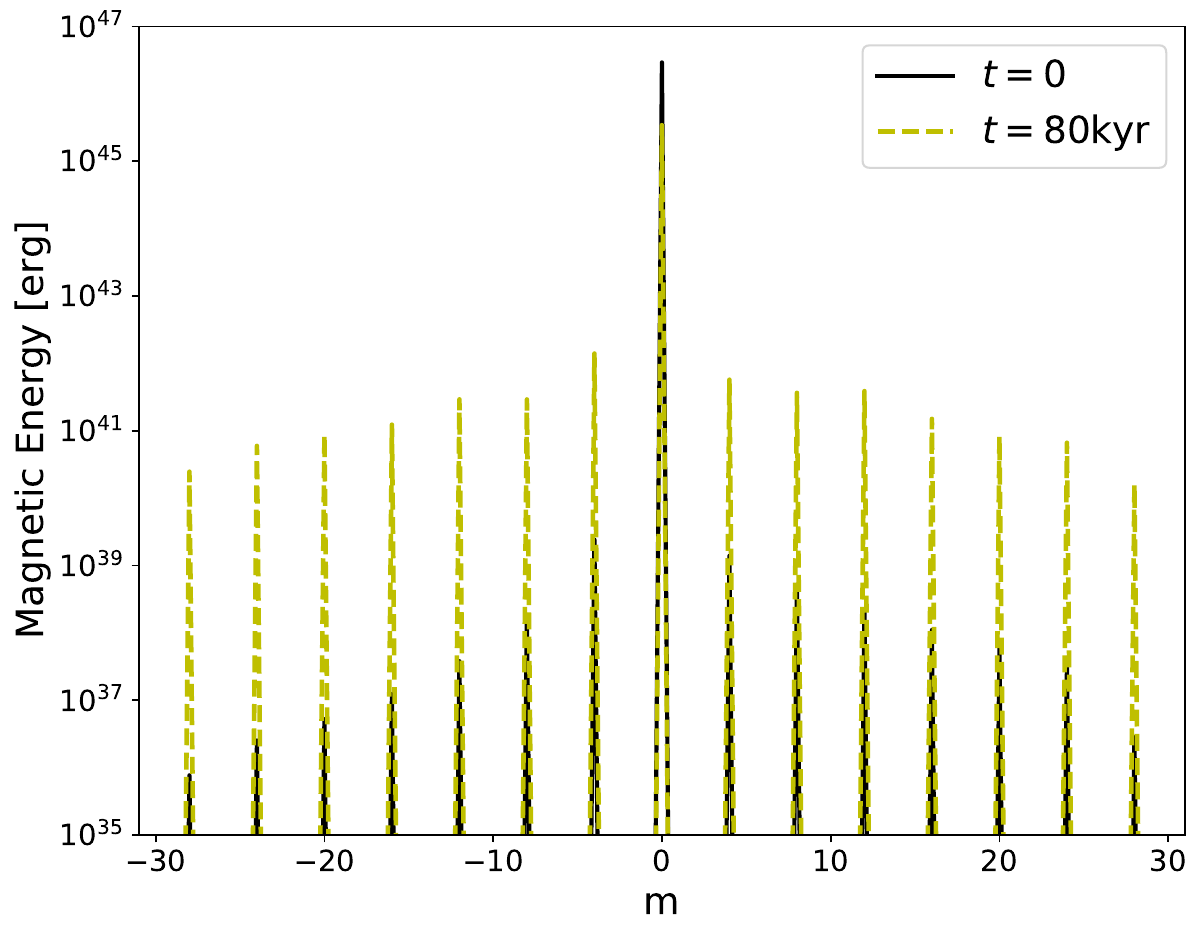}
	\caption[3D axisymmetric energy spectrum]{Evolution of the magnetic energy spectrum as a function of $m$ (summed over all $l$), for the axisymmetric case. We show $t=0$ (in black) and $t=80$\,kyr (in yellow). Throughout the evolution, the power contained in all $m \neq 0$ modes never grows above $10^{-4}$ the power in the $m=0$ mode.}
	\label{fig: 3d axisymmetric m energy spectrum}
\end{figure}

\begin{figure}
\centering
    \includegraphics[width=.45\textwidth]{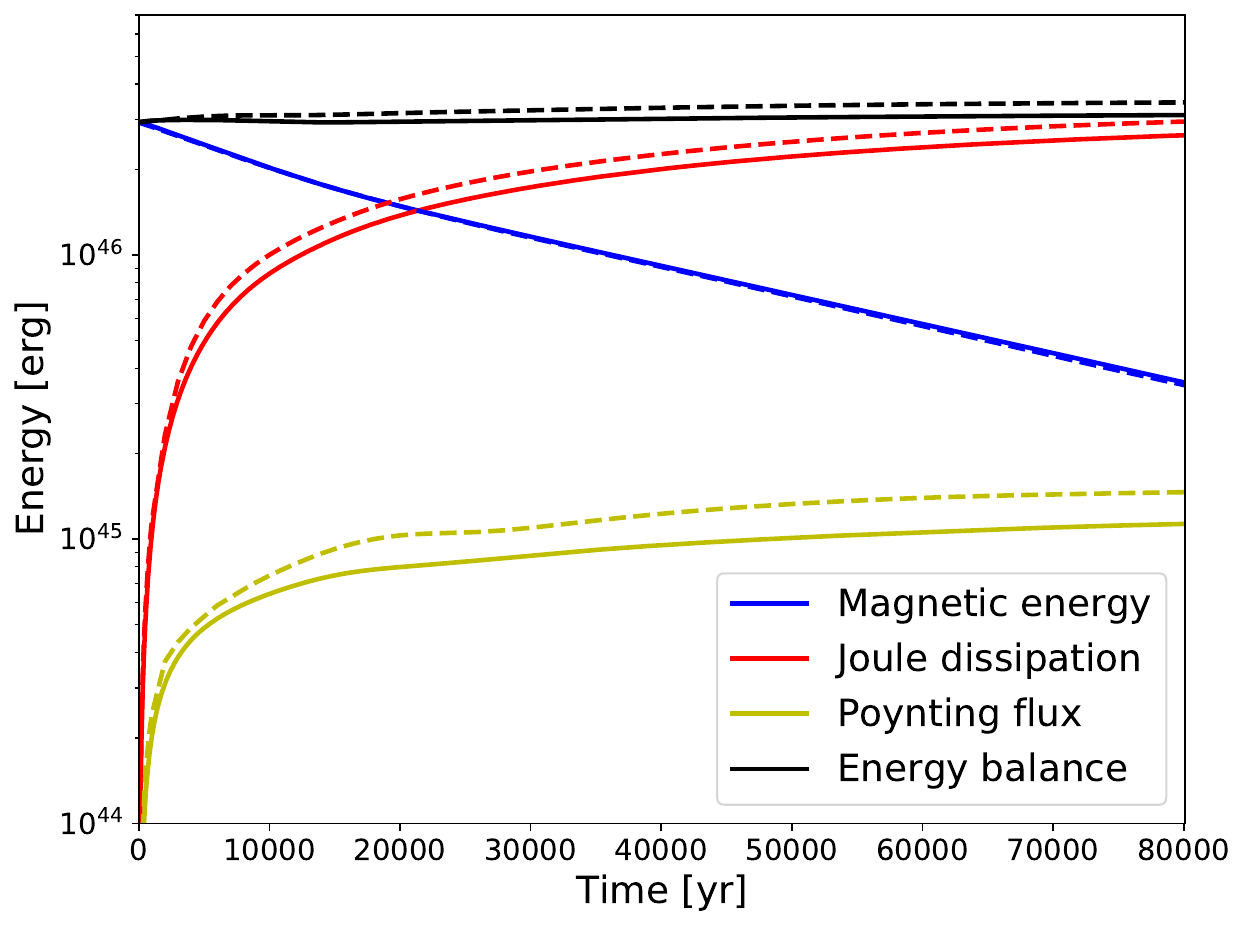}
        \includegraphics[width=0.45\textwidth]{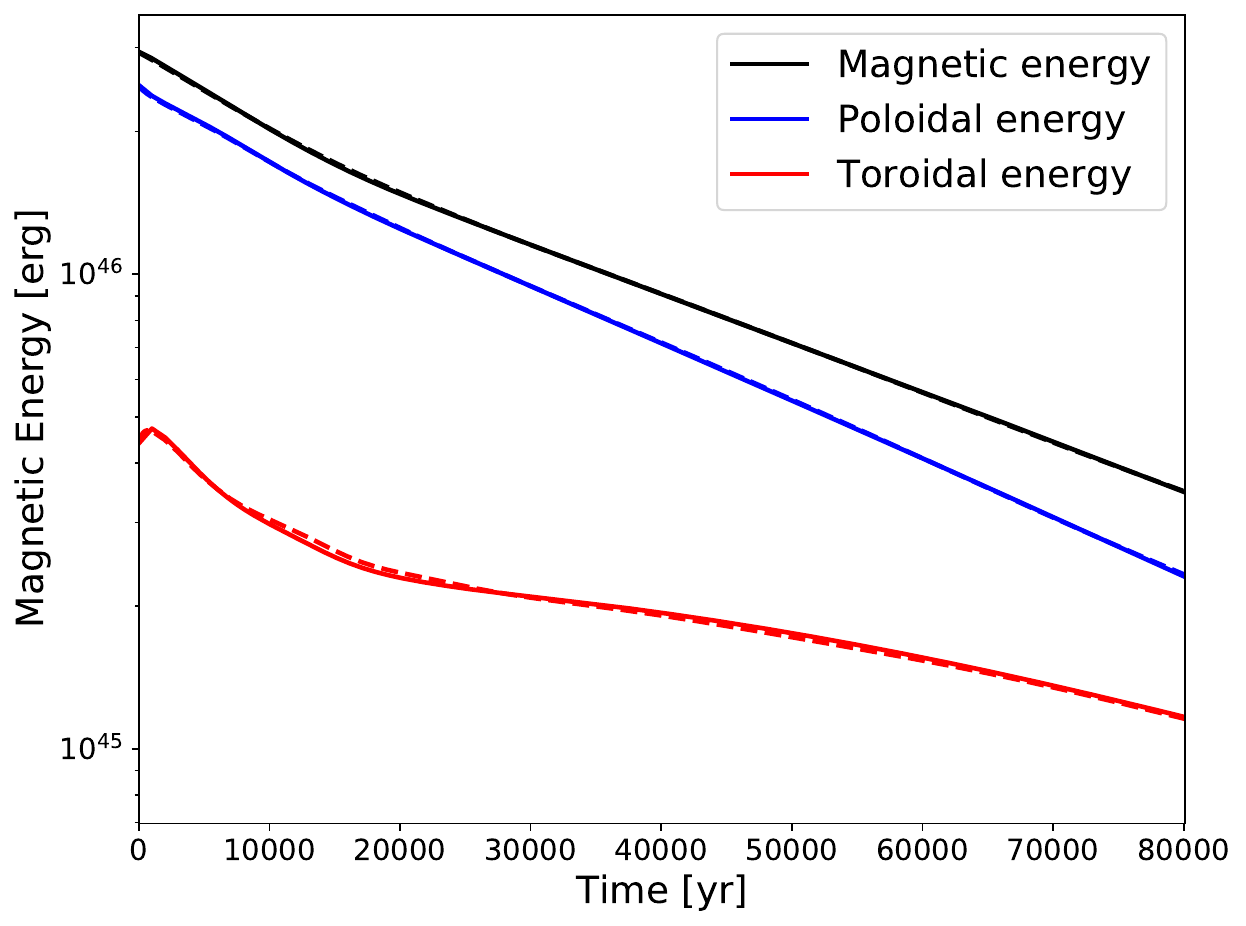}  \includegraphics[width=0.45\textwidth]{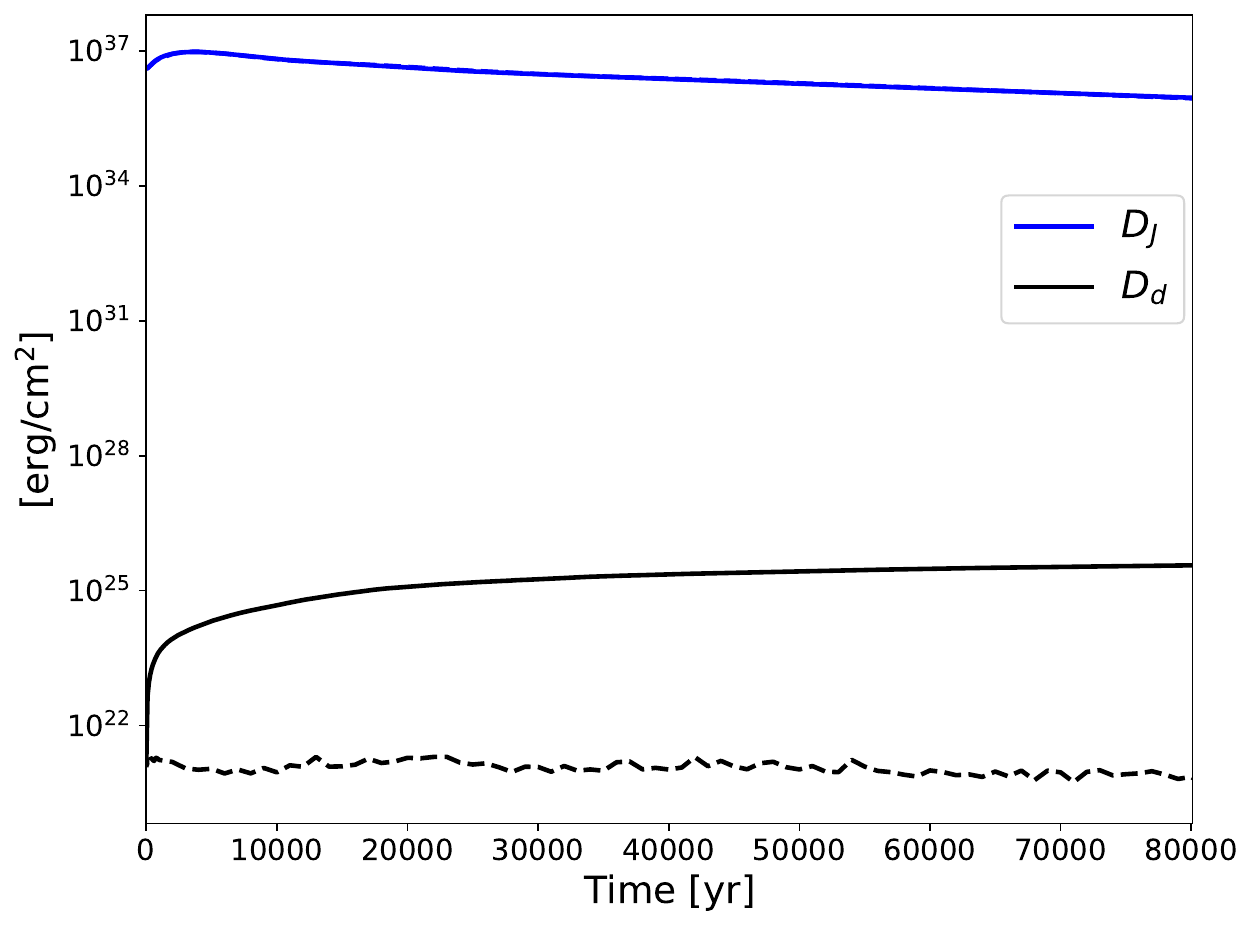}
\caption[2D vs 3D comparison]{The results of the comparison between the 2D (dashed lines) and the 3D (solid lines) codes up to $t=80$\,kyr. \emph{Left panel:} energy as a function of time. Energy balance in black, Joule dissipation in red, magnetic energy in blue and Poynting flux in yellow. \emph{Right panel:} total magnetic energy in black, poloidal magnetic energy in blue, toroidal magnetic energy in red. \emph{Bottom panel:} $D_d$ in black and $D_J$ in blue (eq.\,\eqref{eq: divB volume} and eq.\,\eqref{eq: J2-star} respectively).}
\label{fig: 2d - 3d comparison energy and divB}
\end{figure}

The results of the comparison for an evolution up to $t=80$\,kyr are displayed in Figs.~\ref{fig: 2d-3d comparison Bfield profiles}, \ref{fig: 3d axisymmetric m energy spectrum}, and \ref{fig: 2d - 3d comparison energy and divB}. The radial magnetic profiles for the three components of the magnetic field are displayed in Fig.~\ref{fig: 2d-3d comparison Bfield profiles} at $t=0$, $5$, $10$, $20$ and $80$\,kyr: $B^r$ at the north pole in the panel on the left, $B^\theta$ at the equator in the panel on the right, and $B^\phi$ at the equator in the panel at the bottom. The 3D results are represented with solid lines, whereas the diamonds correspond to the 2D results. Throughout the evolution, the maximum magnetic Reynolds number is much greater than unity, e.g., $R_m \sim 100$. Therefore, the Hall term dominates in the induction equation. The observed evolution is very similar. Local differences in the values of the components are typically less than a few percent, except for the radial component of the magnetic field at late times, which are likely due to the slightly different numerical implementation of the inner and outer boundary condition used in the two codes.

An important point is assessing to which extent the 3D numerical code preserves axial symmetry. If we start with a pure $m=0$ mode, one should expect that this symmetry is kept to some small error, during the whole evolution. 
To give a quantitative measure of possible deviations, we study the energy spectrum (eq.\,\eqref{eq: spectral magnetic energy}) by monitoring the evolution in time of each mode.

In Fig.~\ref{fig: 3d axisymmetric m energy spectrum}, we plot in logarithmic scale the energy spectrum as a function of $m$ ($E_m \equiv \sum_{l}  E_{lm}$) at $t=0$ (in black) and after $t=80$\,kyr (in yellow).
The spectral magnetic energy is concentrated at $m=0$ as expected. The rest of the modes are zero to the round-off error, except the modes with $m=\pm 4$, and higher harmonics, having anyway six or seven orders of magnitude less energy than the main one. They are caused by the discretization over the cubed-sphere grid, and in particular by the four patches that cover the tropical latitudes over the entire azimuthal direction. Their contribution to the energy spectrum is negligible. We remark that this unavoidable error introduced by the cubed-sphere grid is not increasing in time and it remains several orders of magnitude smaller than the magnetic energy contained in the $m=0$ mode after $80$\,kyr of evolution. Moreover, it decreases for higher resolution.

In Fig.~\ref{fig: 2d - 3d comparison energy and divB}, we present the different contributions to the total energy balance as a function of time in the panel on the left, the energy stored in the toroidal and poloidal components in the panel on the right, and a measure of the evolution of the divergence of the magnetic field in the panel at the bottom. The solid lines in the graph represent results from the 3D simulation, while the dashed lines correspond to the 2D simulation. We observe that the total energy is conserved more accurately in the 3D code, with a deviation of approximately $\sim 3\%$ after $80$\,kyr of evolution. In contrast, the 2D simulation exhibits a deviation of around $\sim 15\%$ during the same period. We attribute these minor differences in the energy balance to the use of spherical coordinates in the 2D simulation, which may introduce more numerical errors near the axis.

As seen in the central panel, for this model, most of the magnetic energy is stored in the poloidal field, while the toroidal field represents approximately $\sim 15 \%$ of the total magnetic energy at $t=0$ and approximately $\sim 34 \%$ at $t=80$\,kyr. The increase in the relative fraction of the toroidal energy is caused by the non-linear term, leading to some redistribution of magnetic energy between the poloidal and toroidal components.

In the right panel, we conduct a comparison between the square of the divergence of the magnetic field integrated in the star volume (eq.\,\eqref{eq: divB volume}) and the volume-integrated $J^2$ (eq.\,\eqref{eq: J2-star}). Both quantities share the same units, erg/cm$^2$, making this comparison an effective way to assess the level of conservation of the divergence constraint. The variations between the 2D and 3D values of $D_d$ are likely attributed to the different coordinate systems utilized.
However, it is important to note that despite these differences, $D_d$ remains several orders of magnitude lower than $D_J$, and it exhibits relatively constant behavior over time. Generally speaking, we can conclude that the results from both codes are in agreement, considering the expected grid/formalism-dependent numerical errors.

After presenting the \emph{MATINS} code and conducting extensive testing, as illustrated in this chapter, our focus in Chapter~\ref{chap: 3DMT} will be on performing the first 3D coupled magneto-thermal simulations using \emph{MATINS}. These simulations will incorporate the most recent temperature-dependent microphysical calculations at each point of the star, utilizing Potekhin's publicly available codes \citep{potekhin2015}. Additionally, the simulations will be based on a realistic EoS for the star structure and will include the corresponding relativistic factors in the evolution equations.
In \S\ref{eq: isothermal cooling}, we will employ isothermal cooling, and in \S\ref{sec: 3DMT}, we will conduct the first 3D coupled magneto-thermal simulation applied to a highly complex initial magnetic field topology in the crust, similar to what was recently obtained by proto-neutron star dynamo simulations.

\subsubsection*{Corresponding scientific publications}
\underline{C.~Dehman}, D.~Vigan\`o, J.A.~Pons \& N.~Rea: 2022, \textbf{3D code for MAgneto$-$Thermal evolution in Isolated Neutron Stars, MATINS: The Magnetic Field Formalism}, \emph{Mon.~Not.~Roy.~Astron.~Soc., $518$, $1222$} 
(\href{https://arxiv.org/abs/2209.12920}{\underline{arXiv:2209.12920}},\href{https://ui.adsabs.harvard.edu/abs/2023MNRAS.518.1222D/abstract}{\underline{ADS}},\href{https://doi.org/10.1093/mnras/stac2761}{\underline{DOI}}).

\clearemptydoublepage
\let\textcircled=\pgftextcircled
\chapter{3D Magneto-Thermal Simulations}
\label{chap: 3DMT}

\initial{T}he configuration of the neutron star's magnetic field at birth remains largely unknown, and understanding how to generate strong dipolar fields that can explain the timing properties of magnetars is still an open question (see e.g. \cite{igoshev21rev} for a review). Various scenarios have been discussed in the literature to explain the origin of the magnetic field in magnetars. Magnetic flux conservation can lead to the strongest magnetic fields in the case of highly magnetized progenitors that could be formed in stellar mergers \citep{ferrario06,schneider2019,makarenko21}. 
However, this may not explain the formation of millisecond magnetars, as highly magnetized progenitors tend to be slow rotators \citep{shultz2018}. An alternative scenario for achieving fast rotation and a strong magnetic field involves magnetic field amplification through a turbulent dynamo in the proto-neutron star \citep{raynaud20}. Recent MHD local and global simulations have been proposed to quantify this magnetic field amplification.

On one side, box simulations with simplified background fields have demonstrated the development of the magneto-rotational instability (MRI) \citep{balbus1991,akiyama2003,obergaulinger2014,rembiasz2017,Aloy_2021}.
On the other side, global simulations have explored various dynamo mechanisms in proto-neutron stars \citep{raynaud20,reboul2021,masada22,barrere2022}, taking into account the effects of differential and rigid rotation, as well as convection. As typically observed in dynamo simulations, the system attains an equilibrium configuration. Even though the fields are far from static, the energy distribution across scales (or multipoles) achieves a state of statistical quasi-equilibrium. This results from the balance of forces in the magnetized fluid. 

The magnetic energy within the proto-neutron star is dispersed across a broad range of spatial scales, with the non-axisymmetric and toroidal components dominating over the large-scale poloidal dipole. Although the total energy is compatible with magnetar-like magnetic fields, it is dominated by the small-scale structures in the system.

In this context, the approach and findings of these studies represent a significant advance that enhances the highly simplified depictions of static idealized equilibria known as twisted torus \citep{ciolfi2013,sur2021}, which are frequently utilized as initial field configurations. It is worth noting that these large-scale axisymmetric configurations exhibit a markedly distinct spatial distribution of magnetic energy compared to dynamo configurations. Despite this difference, they have been commonly employed as the starting point for the long-term evolution of neutron stars.

In this chapter, we present the first 3D simulations performed using \emph{MATINS}, with a background structure coming from a realistic EoS, the most recent temperature-dependent microphysical calculations, and the inclusion of the corresponding relativistic factors in the evolution equations \citep{dehman2022}. In \S\ref{eq: isothermal cooling}, we evolve the crustal temperature using a simplified treatment, adopted from \cite{yakovlev2011} and considering different non-axisymmetric initial configurations.
Then, in \S\ref{sec: 3DMT}, we perform the first 3D coupled magneto-thermal simulation using \emph{MATINS} (\cite{dehman2022} \& Ascenzi et al. in preparation) and considering a complex initial magnetic field topology similar to \cite{reboul2021} in terms of spectral distribution \citep{dehman2023c}.

\section{Isothermal cooling}
\label{sec: isothermal cooling}

\subsection{Physical setup}

In this section, we employ the analytical approximation for the temperature evolution proposed by \cite{yakovlev2011}, as follows:
\begin{equation}
    T(t) = 3.45 \times 10^8 \text{K} ~\left(1-\frac{2GM}{c^2R_\star}\right) \bigg[1+0.12 \bigg(\frac{R_\star}{10 \text{km}} \bigg)^2  \bigg] \bigg(\frac{t_c}{t} \bigg)^{1/6} e^{-\zeta}
    \label{eq: isothermal cooling}
\end{equation}
where $t_c$ is some fiducial (normalization) time-scale and it is set to the age of the Cas A supernova remnant ($330$\,yr). For our model, $M=1.4$\,M$_\odot$ and $R_\star=11.7$\,km. It has been shown that this time dependence is accurate during the neutrino cooling stage \citep{yakovlev2011}. 

The simulations consider a stratified electron number density and a temperature-dependent resistivity.
The electrical conductivity required to calculate $\eta_b$, is determined locally at each timestep, taking into account the temperature, local density, and composition.

\subsection{Initial magnetic topology}
\label{subsec: different B-field topology 3DMT}

To assess the sensitivity of our results to uncertain initial conditions, we have explored three distinct magnetic field topologies, all confined to the crust, following a similar approach to \cite{aguilera2008}. The specific radial dependence and the method for constructing a divergence-free magnetic field are outlined in detail in Appendix~\ref{appendix: initial conditions}. It is worth noting that the numerical scheme inherently preserves the local divergence up to machine error, ensuring accuracy in our simulations. 

The different models studied in this section have an average initial magnetic field of $\sim 10^{14}$\,G, corresponding to total magnetic energies of the order of $\sim 10^{45}$\,erg.
These models are summarized in Table~\ref{tab: different models}. Most of the magnetic energy is contained in the toroidal component, except for the last model. 
The variations in these models arise from the different relative weights of multipoles in their initial configurations.
In the second and third models (L5-T1e9 and L5-T2e8), we keep the temperature fixed during the evolution at $10^9$\,K and $2 \times 10^8$\,K, respectively.
It's important to note that we have deliberately chosen arbitrary combinations of a relatively small number of multipoles, rather than the expected smooth cascade over a wide range of them, as suggested by the proto-neutron star configurations mentioned earlier. The total evolution time for the first three models (L5, L5-T1e9, and L5-T2e8) is $70$\,kyr, while for the L1 model, it is $85$\,kyr, and for the L10 model, it is $100$\,kyr. However, for some models, the total evolution time is constrained by numerical instabilities that arise at late times when the temperature drops significantly below $10^8$\,K (e.g., $T(100 \,\text{kyr}) \sim 10^7$\,K from eq.\,\eqref{eq: isothermal cooling}), leading to an increase in the magnetic Reynolds number. The appearance of these instabilities also depends on the initial magnetic field strength and topology. This behavior is similar to what we have observed in our 2D magneto-thermal code \citep{vigano2021}.

For all the models displayed in Table~\ref{tab: different models}, we utilize a resolution of $N_r=40$ and $N_\xi = N_\eta =43$ per patch, which corresponds to a total of $172$ grid points around the equator and $87$ points along a meridian from pole to pole. With this resolution, we analyze up to $l=30$ for the multipole expansion.

\begin{table}
\centering
 \caption[Initial 3D magnetic field models]{Initial Models Considered. $l$ and $m$ are the initial non-zero multipoles considered in each model. $B_{avg}$ is the average initial magnetic field. $E_\text{mag}$ is the initial magnetic energy (all in the crust). $E_\text{tor}/E_\text{mag}$ is the fraction of the crustal toroidal energy. For all these models, we are confining the field lines to the crust of the star. We use the simplified cooling described in the text (eq.\,\eqref{eq: isothermal cooling}), except in two models ("deactivated").}
 \label{tab: different models}
\scalebox{0.83}{
 \begin{tabular}{|l|c|c|c|c|c|c|c|c|c|c|}
  \hline
Models & $l_\text{pol}$ & $m_\text{pol}$ &  $l_\text{tor}$ & $m_\text{tor}$ & $B_\text{avg} (t0)$  &  $B_\text{max} (t0)$  & $E_\text{mag} (t0)$&  $E_\text{tor}/E_\text{mag} $ & Simplified & $T_\text{fixed}$ \\
&  &&    & &  [G] & [G] &  [erg]& $(t0)$ & Cooling &  [K] \\
\hline
\hline
L5 & $1,2,$ &  $-1,0,1,$&  $1,2,$ &  $0,1,$& $ 2 \times 10^{14}$ & $ 7 \times 10^{14}$& $  2\times 10^{45}$ & $ 63 \%$ &   activated & -\\
& $3,5$ &  $2,3$&  $3,5$ &  $2,3$&  & &  &  &  & \\
\hline
L5-T1e9 &  $1,2,$ &  $-1,0,1,$&  $1,2,$ &  $0,1,$& $ 2 \times 10^{14}$ & $ 7 \times 10^{14}$&$ 2\times 10^{45}$ &  $ 63 \%$  & deactivated & $10^{9}$\\
&  $3,5$ &  $2,3$&  $3,5$ &  $2,3$& & &&  & & \\
\hline
L5-T2e8 &  $1,2,$ &  $-1,0,1,$&  $1,2,$ &  $0,1,$& $2 \times 10^{14}$ & $ 7 \times 10^{14}$&$ 2\times 10^{45}$ &  $ 63 \%$  & deactivated & $2 \times 10^{8}$\\
 &  $3,5$ &  $2,3$&  $3,5$ &  $2,3$& & & &   &  & \\
 \hline
  L1 & $1$ & $0$ & $1$ & $1$ & $ 3 \times 10^{14}$ & $ 6.5 \times 10^{14}$  &$ 4 \times 10^{45}$ & $ 95$\% & activated & -\\
  \hline
L10  & $1,6,$   &  $ -5,-1,0, $   & $ 1,3, $   &  $-5,0,$   & $ 10^{14}$   & $3 \times 10^{14}$ & $6 \times 10^{44}$& $ 10 \%$  & activated & -\\
& $7,10$   &  $ 1,7,8 $   & $ 7,10 $   &  $2,9$   & &  & &   & & \\
\hline
\end{tabular}}
\end{table}

\subsection{L5 model}
\label{subsec: LSCC model}

The first model, denoted as L5, has an average magnetic field strength of approximately $2 \times 10^{14}$\,G and reaches a maximum of around $7 \times 10^{14}$\,G. The initial configuration is characterized by a large-scale topology, representing a sum of multipoles up to $l=5$. Notably, this model is Hall-dominant, and throughout its evolution, it reaches a maximum magnetic Reynolds number of about $R_m \sim 200$.

\begin{figure}
\centering
\includegraphics[width=0.45\textwidth]{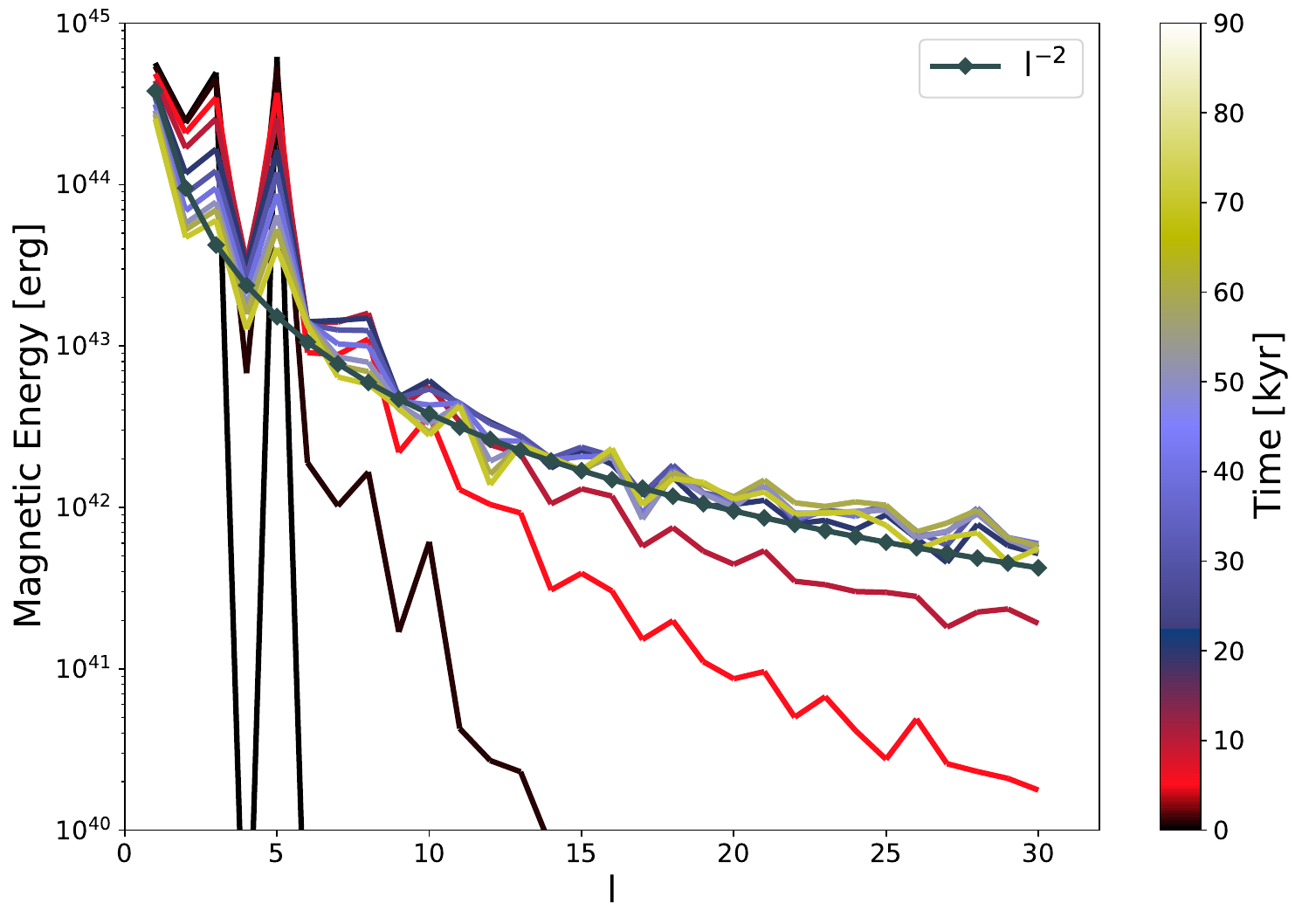}
\includegraphics[width=0.45\textwidth]{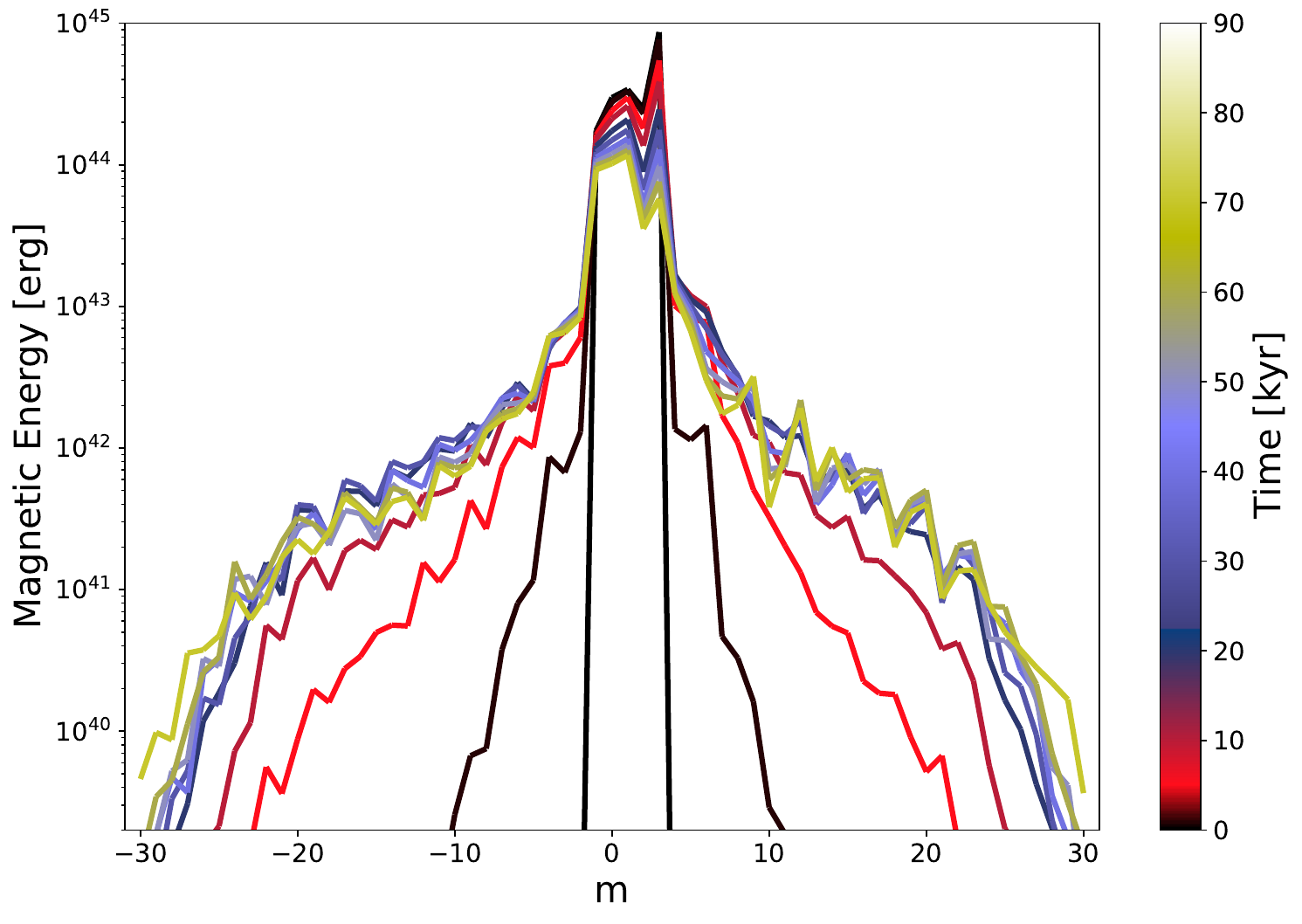}
\caption[L5 model]{\emph{L5 model}. \emph{Left panel:} $l$ energy spectrum. \emph{Right panel:} $m$ energy spectrum. The energy spectra are displayed at times $0, 1, 5, 10, 20, 30, 40, 50, 60$ and $70$\,kyr (see color bars). The $l^{-2}$ slope corresponds to the Hall cascade equilibrium distribution of magnetic energy over a quite broad range of multipoles.}
\label{fig: energy spectrum LSCC}
\end{figure}

In Fig.~\ref{fig: energy spectrum LSCC}, we examine the $l$ energy spectrum (summing eq.\,\eqref{eq: spectral magnetic energy} over all $m$'s) in the left panel and the $m$ energy spectrum (summing it over all $l$'s) in the right panel, at various evolution times. At time zero, one can clearly distinguish the multipoles imposed initially. However, as we initiate the evolution, a portion of the magnetic energy begins to transfer from the large-scale multipoles, into the smaller-scale ones. Moreover, we observe the excitation of higher-order $m$ modes in the system. By the time we reach $1$\,kyr, most of the magnetic energy is concentrated in the initially imposed multipoles, but a fraction of it has already been transferred to $l=4-10$. As we continue the evolution up to $5$ and $10$\,kyr (red curves), the transfer of energy toward small-scales persists, gradually filling the entire spectrum.
Around $20$\,kyr, the magnetic energy spectrum appears to have attained a quasi-stationary state, referred to as the Hall-saturation. Since the dissipation scales as $L^2/\eta_b$ (where $L$ represents the typical spatial scale of the field curvature), the smaller-scale structures dissipate faster compared to the larger-scale ones. Simultaneously, the larger-scales continuously feed the smaller-scale ones thanks to the Hall term in the induction equation. This process is known as the Hall cascade, leading to an equilibrium distribution of magnetic energy across a broad range of multipoles, displaying an approximate $l^{-2}$ slope \citep{goldreich1992}.

Indeed, the cascade and saturation processes require two main conditions to occur successfully: (i) Hall-dominated dynamics, which necessitates a sufficiently large magnetic field to ensure that the Hall term is dominant during the evolution of the system; (ii) an initial configuration that allows the full development of the Hall cascade. As a matter of fact, poloidal and toroidal fields exhibit asymmetric coupling, with odd multipoles of the former being more coupled to even multipoles of the latter.

In the case of axial symmetry, one can maintain a helicity-free configuration if the initial field consists only of $l=1,3,5...$ poloidal components and $l=2,4,6...$ toroidal components. Consequently, not all multipoles are excited, and only odd/even families will manifest in the spectrum.
However, in a general non-axisymmetric case with arbitrary combinations of initial multipoles, the Hall-dominated dynamics can lead to less clear relative weights of couplings between different modes. In such scenarios, the behavior of the system may be more complex, and various modes can be excited in the magnetic energy spectrum due to the interplay of different initial multipoles.

\subsection{The impact of temperature dependent microphysics}
\label{subsec: impact of microphysics}

To appreciate the role of temperature-dependent microphysics on the magnetic field evolution, we conduct a comparison up to $70$\,kyr, considering an identical magnetic field configuration in two scenarios: one with temperature evolution (L5 model) and two without temperature evolution (L5-T1e9 and L5-T2e8 models). For the L5-T1e9 model, the microphysical coefficients are computed at $T=10^9$\,K, while for the L5-T2e8 model, they are calculated at $T=2\times 10^8$\,K.
It's worth noting that $T=10^9$\,K corresponds to the temperature of a neutron star during the first years of its life, while $T=2\times 10^8$\,K corresponds to the temperature at approximately $10$\,kyr in the L5 model. This comparison allows us to understand better the influence of temperature on the magnetic field's long-term evolution under different microphysical conditions.

\begin{figure}
\centering
\includegraphics[width=.45\textwidth]{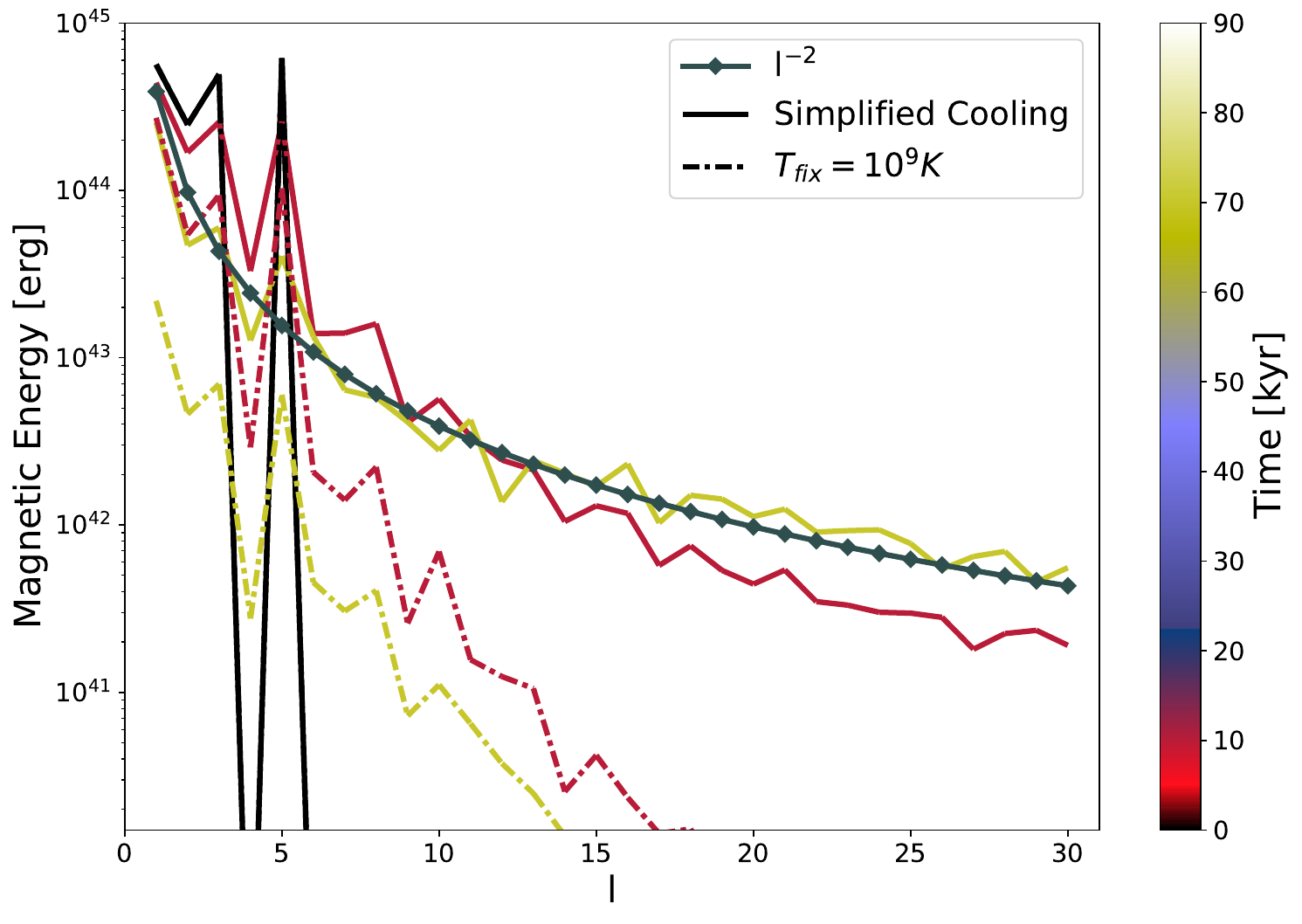}
\includegraphics[width=.45\textwidth]{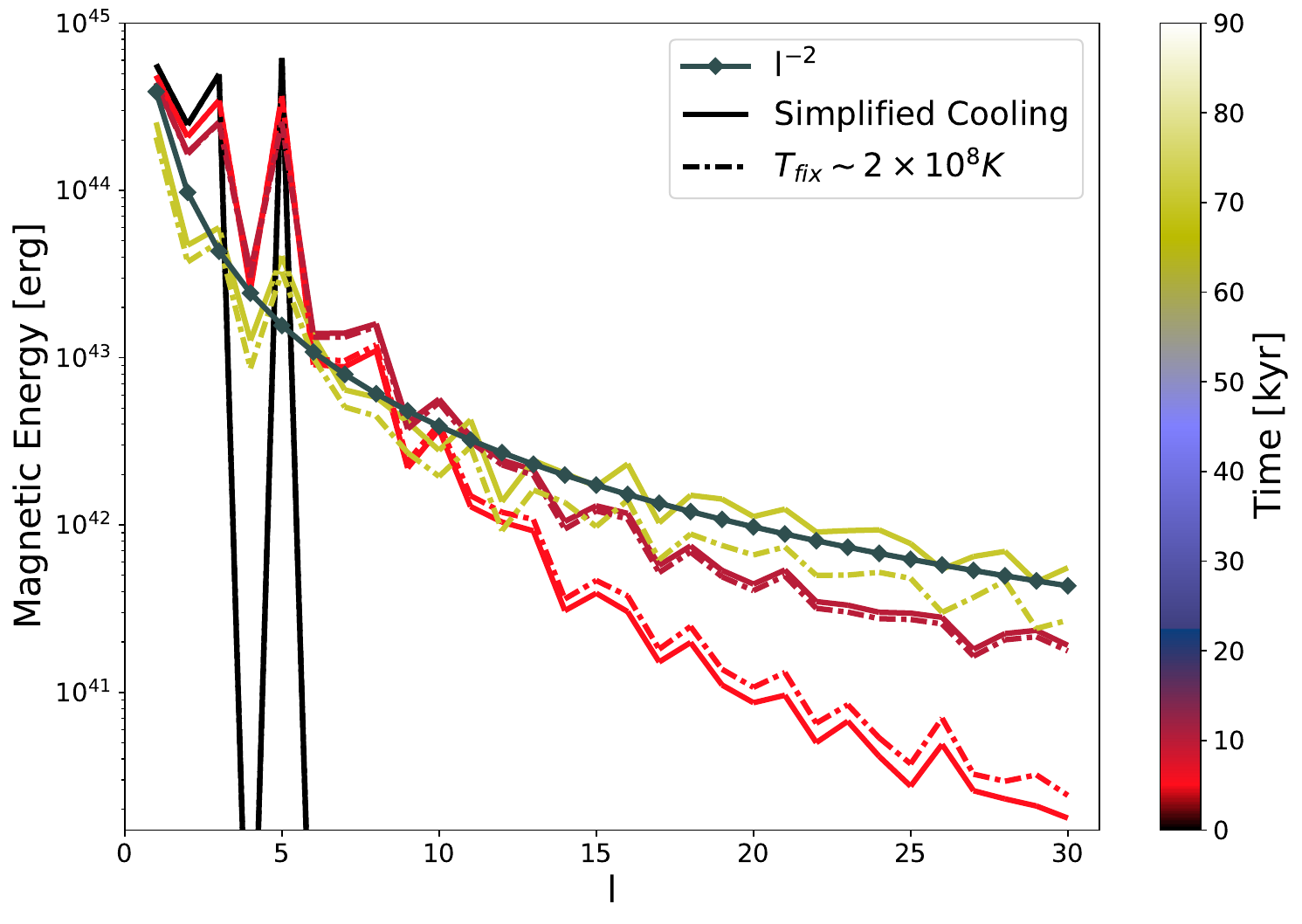}
\caption{A comparison up to $70$\,kyr of the $l$ energy spectrum for L5 model, with temperature evolution (solid lines) and without temperature evolution (dash-dotted lines). The temperature is fixed to $T=10^9$\,K in the panel on the left, and to $T=2\times 10^8$\,K in the panel on the right. In the left panel, the comparison is illustrated at $t=0$ (black), $t=10$\,kyr (dark red) and $t=70$\,kyr (yellow) and in the right panel at $t=0$ (black), $t=5$\,kyr (red), $t=10$\,kyr (dark red) and $t=70$\,kyr (yellow). The $l^{-2}$ slope corresponds to the Hall cascade equilibrium distribution of magnetic energy over a quite broad range of multipoles.}
\label{fig: Espectrum realistic vs analytical}
\end{figure}

The results at different evolution times are depicted in Fig.~\ref{fig: Espectrum realistic vs analytical}. The panel on the left illustrates the comparison between L5 (solid lines) and L5-T1e9 (dash-dotted lines) models, while the panel on the right shows the comparison between L5 (solid lines) and L5-T2e8 (dash-dotted lines) models. At the initial time, all three models overlap. Throughout the evolution, a transfer of magnetic energy towards small-scale structures is observed in all cases. However, a distinguishable behaviour emerges during the field evolution in the first case (left panel of Fig.~\ref{fig: Espectrum realistic vs analytical}). 
Model L5-T1e9 experiences significant dissipation over time with minimal redistribution of magnetic energy across different spatial scales. In other words, the $l$ energy spectrum retains the same shape at $t=10$ and $70$\,kyr for L5-T1e9. After $10$\,kyr, approximately $70\%$ of the total magnetic energy has dissipated for the L5-T1e9 model, while only $35\%$ of the total magnetic energy has dissipated for the L5 model. As a result, L5-T1e9 is predominantly Ohmic-dominant, while L5 is Hall-dominant.

In contrast, the evolution over time of the L5 and L5-T2e8 models displays a relatively comparable behavior (right panel of Fig.~\ref{fig: Espectrum realistic vs analytical}). At around $5$\,kyr, the transfer of energy is slightly more efficient in the L5-T2e8 model. This is because the temperature value considered for the L5-T2e8 model, i.e., $T=2\times10^8$\,K, is lower than the temperature value at $t\sim 5$\,kyr obtained using eq.\,\eqref{eq: isothermal cooling}. Consequently, the magnetic Reynolds number is slightly higher for the L5-T2e8 model due to the lower magnetic resistivity associated with lower temperatures.
Around $10$\,kyr, the $l$-energy spectra of the two models appear quite similar. However, at approximately $70$\,kyr, the results of the two simulations begin to diverge again. The L5-T2e8 model has experienced more dissipation compared to the L5 model, as at this stage, the magnetic Reynolds number is higher for the L5 model. Nonetheless, it's important to note that both models, L5 and L5-T2e8, remain Hall-dominant throughout the evolution.

The contrasting behaviors of the time evolution of the energy spectrum underscore the significant impact of temperature-dependent microphysics. In the first case (left panel of Fig.~\ref{fig: Espectrum realistic vs analytical}), the differences in spectra are substantial, whereas in the second case (right panel of Fig.~\ref{fig: Espectrum realistic vs analytical}), the differences are more subtle. 
For a Hall-dominated field, the specific value of the magnetic diffusivity primarily determines the resistive scale, which corresponds to the width of the inertial range where the Hall cascade is observed. 
It's important to note that in the second comparison (right panel of Fig.~\ref{fig: Espectrum realistic vs analytical}), we set the diffusivity assuming $T=2\times 10^8$\,K, which is not far from the average temperature value during the first $50$\,kyr. However, much larger differences could be obtained by manually fixing very different values of $n_e$ and $\eta_b$ (or by fixing much larger or smaller temperatures). This test demonstrates the sensitivity of the magnetic energy redistribution and dissipation behavior to temperature-dependent microphysical effects. To achieve more realistic and accurate results, a 3D magneto-thermal code coupled with realistic microphysics becomes essential, as we will present in \S\ref{sec: 3DMT}.

\subsection{Different initial multipolar topology}
\label{subsec: different initial multipoles}
 
\subsubsection{Magnetic field lines}

To investigate the influence of different magnetic field topologies, we examine three distinct models, each with different initial multipoles. In addition to the L5 model presented in \S\ref{subsec: LSCC model}, we have the L1 model, featuring a pure dipolar field with $l=1$ in both the poloidal and toroidal components. On the other hand, the L10 model encompasses a broader combination of initial multipoles, extending up to $l=10$.
Throughout the evolution, the maximum magnetic Reynolds number reaches approximately $\sim 200$ for the L5 model, $\sim 150$ for the L1 model, and $\sim 50$ for the L10 model. 

 \begin{figure}
\centering
\includegraphics[width=0.3\textwidth]{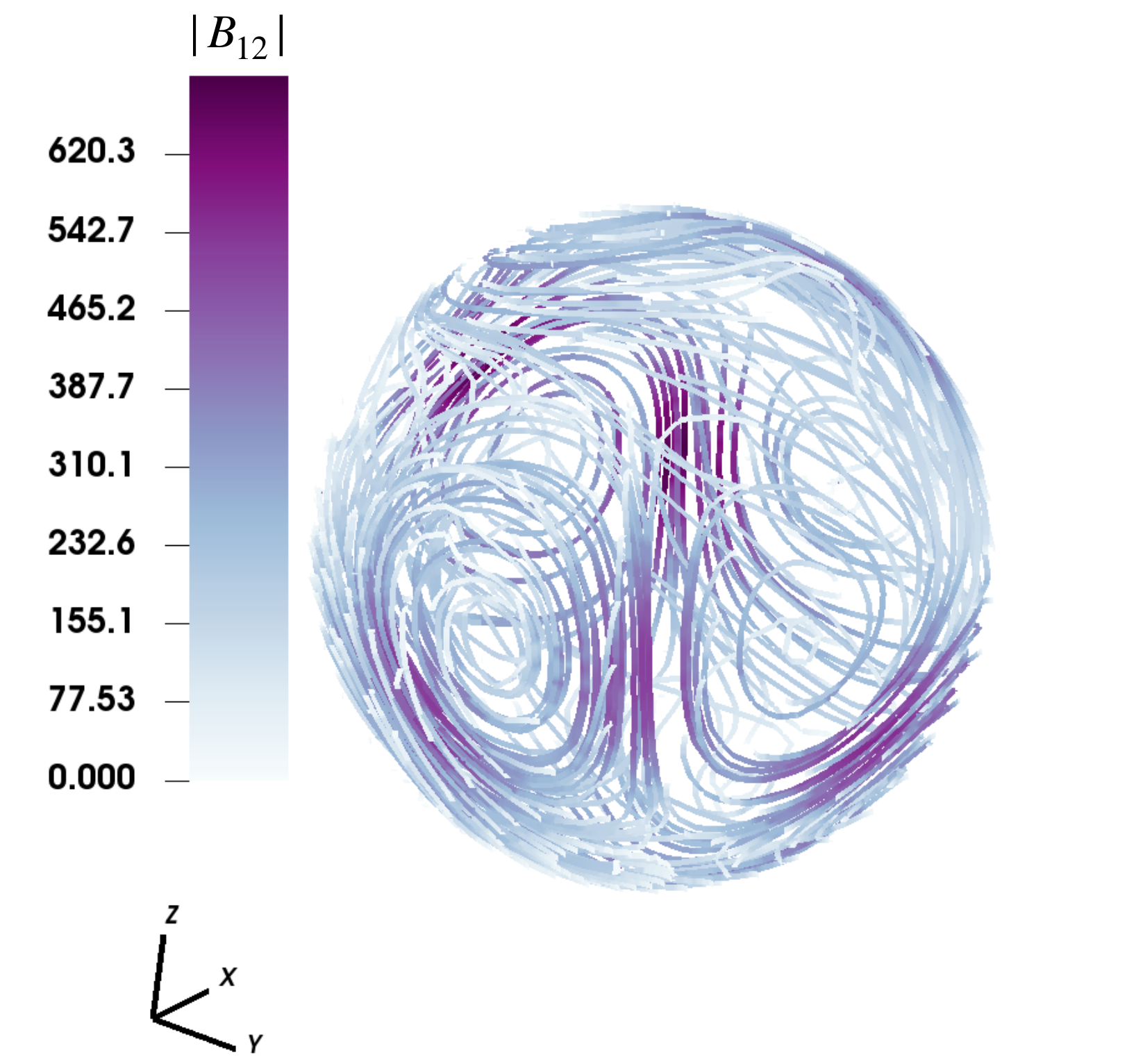}
\includegraphics[width=0.3\textwidth]{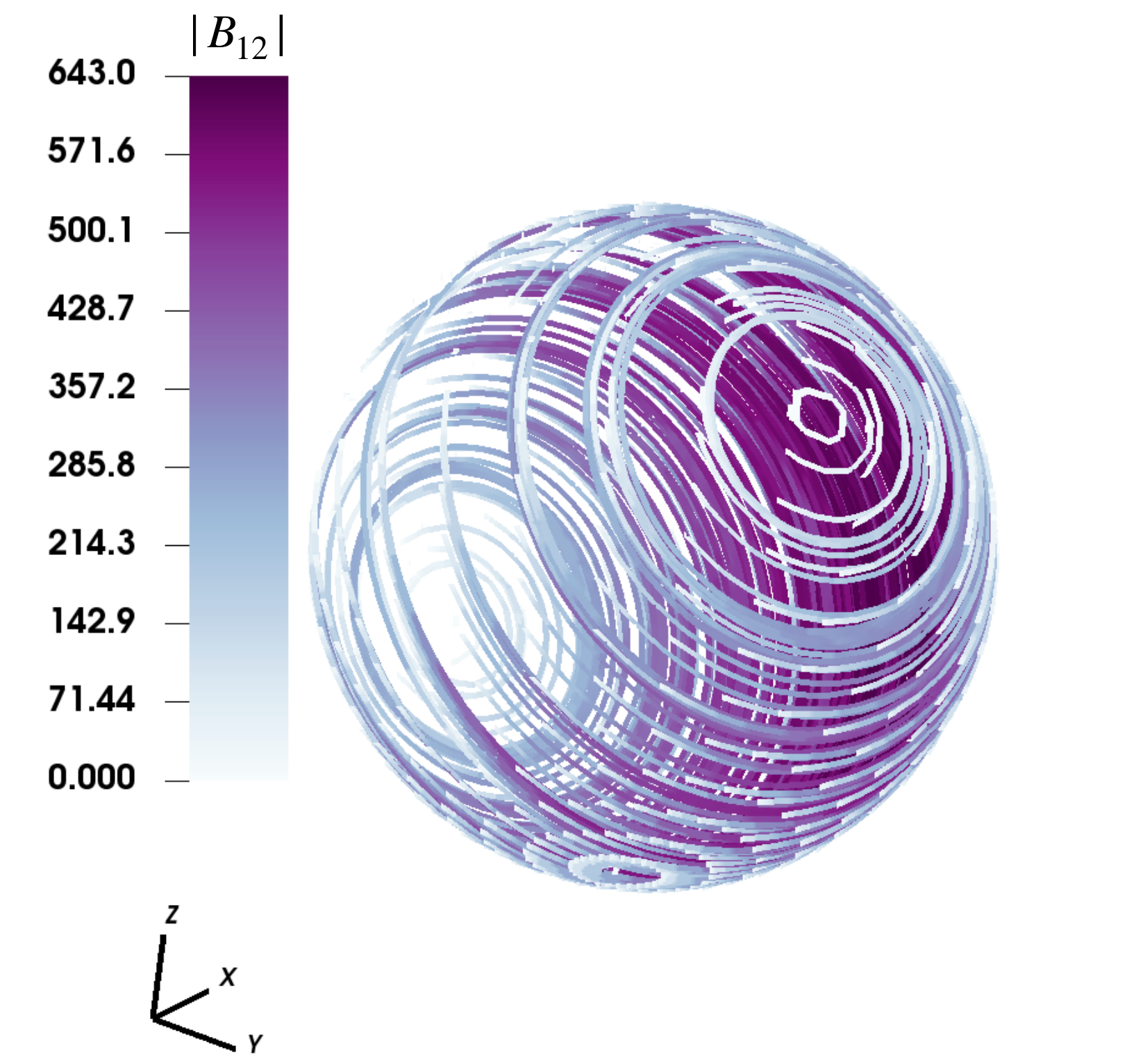}
\includegraphics[width=0.3\textwidth]{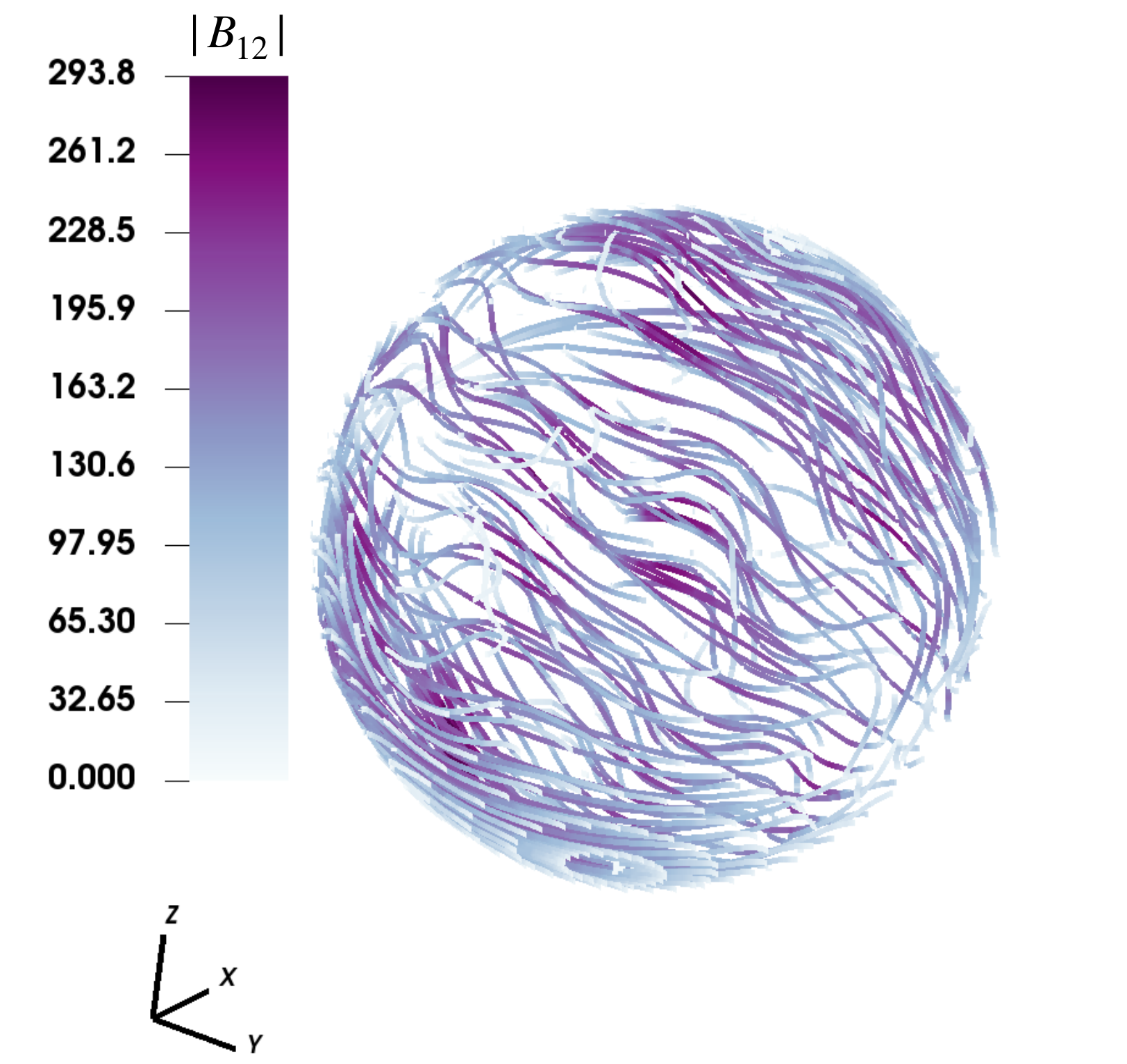} 
\includegraphics[width=0.3\textwidth]{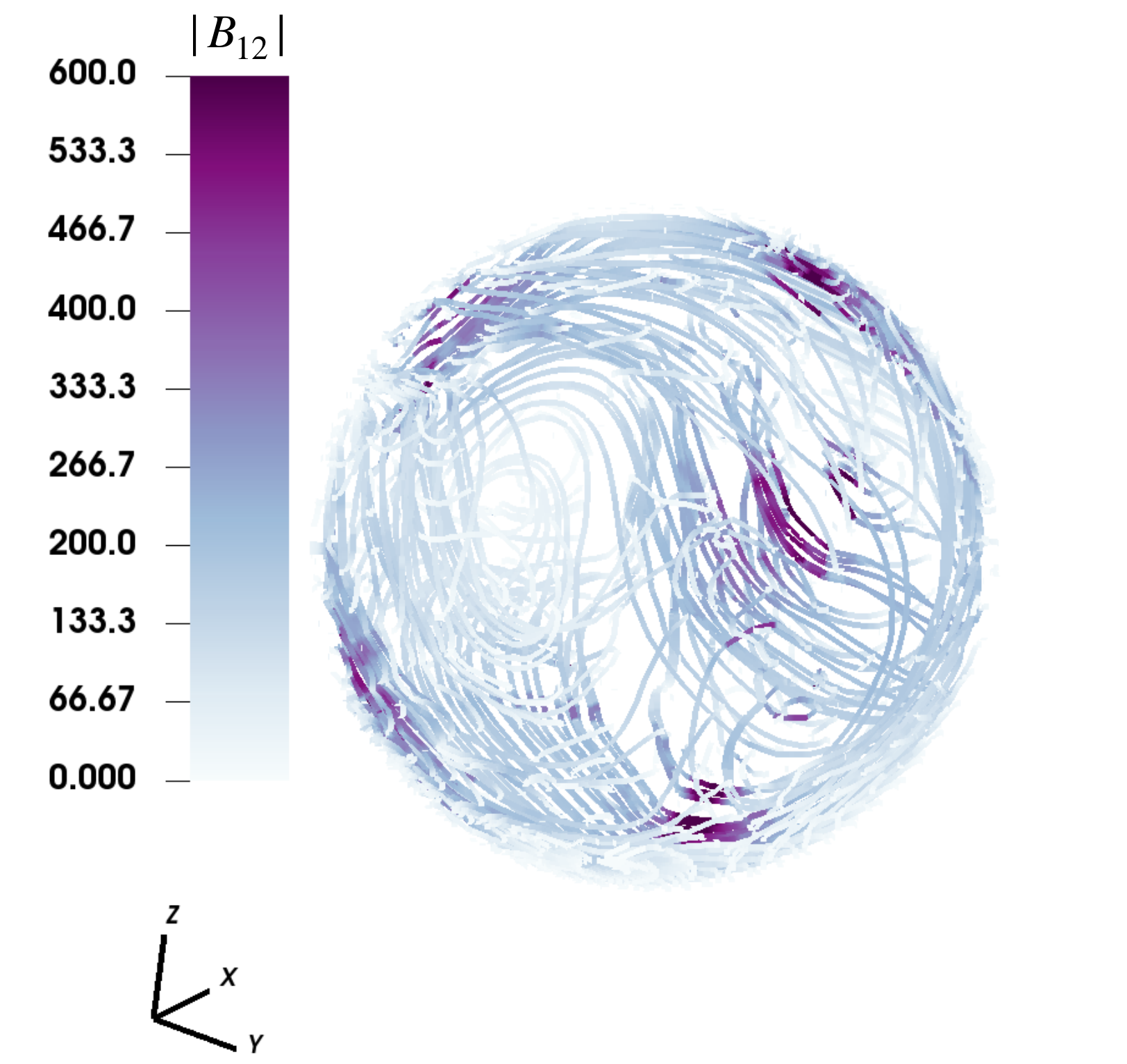}
\includegraphics[width=0.3\textwidth]{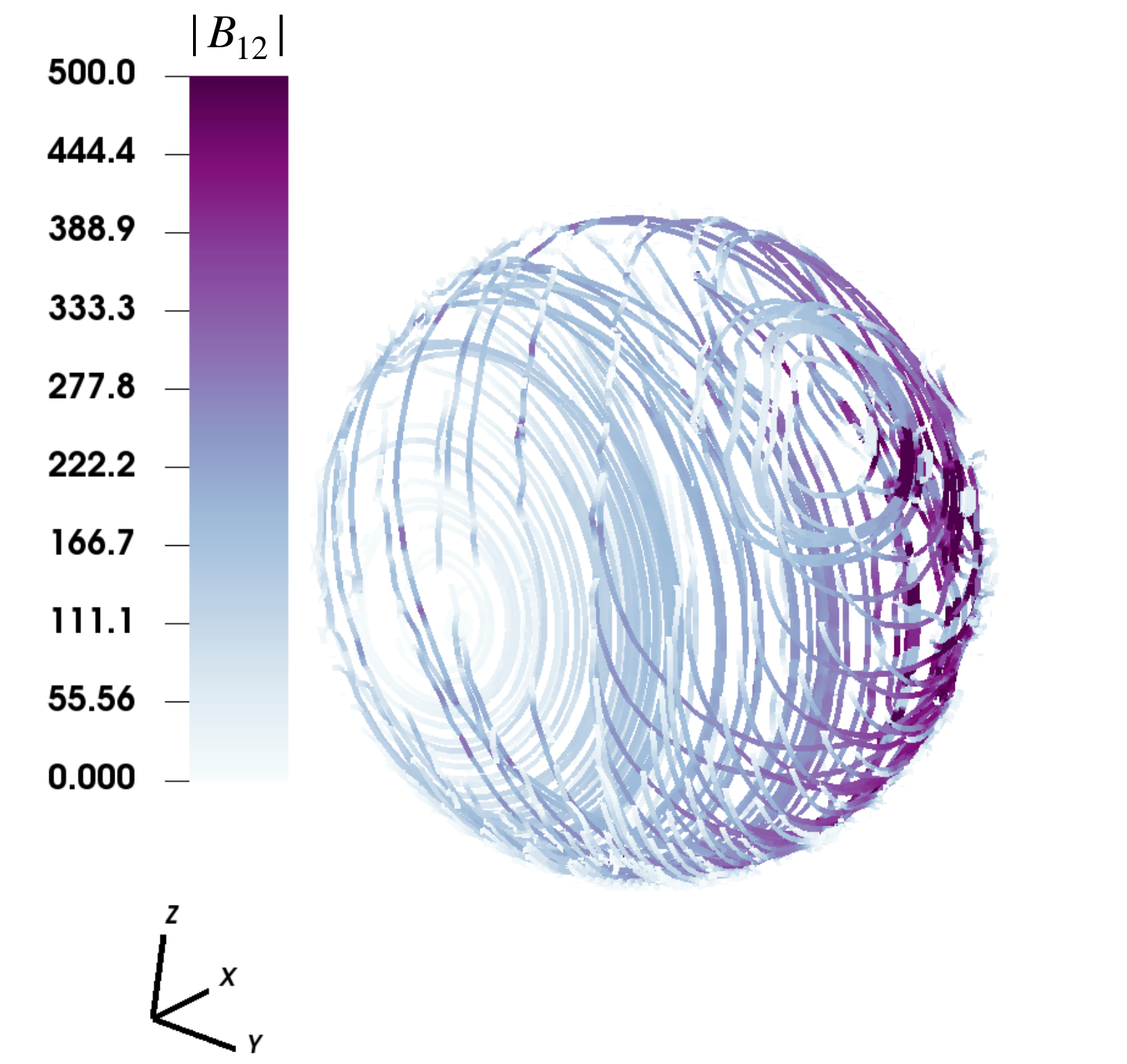}
\includegraphics[width=0.3\textwidth]{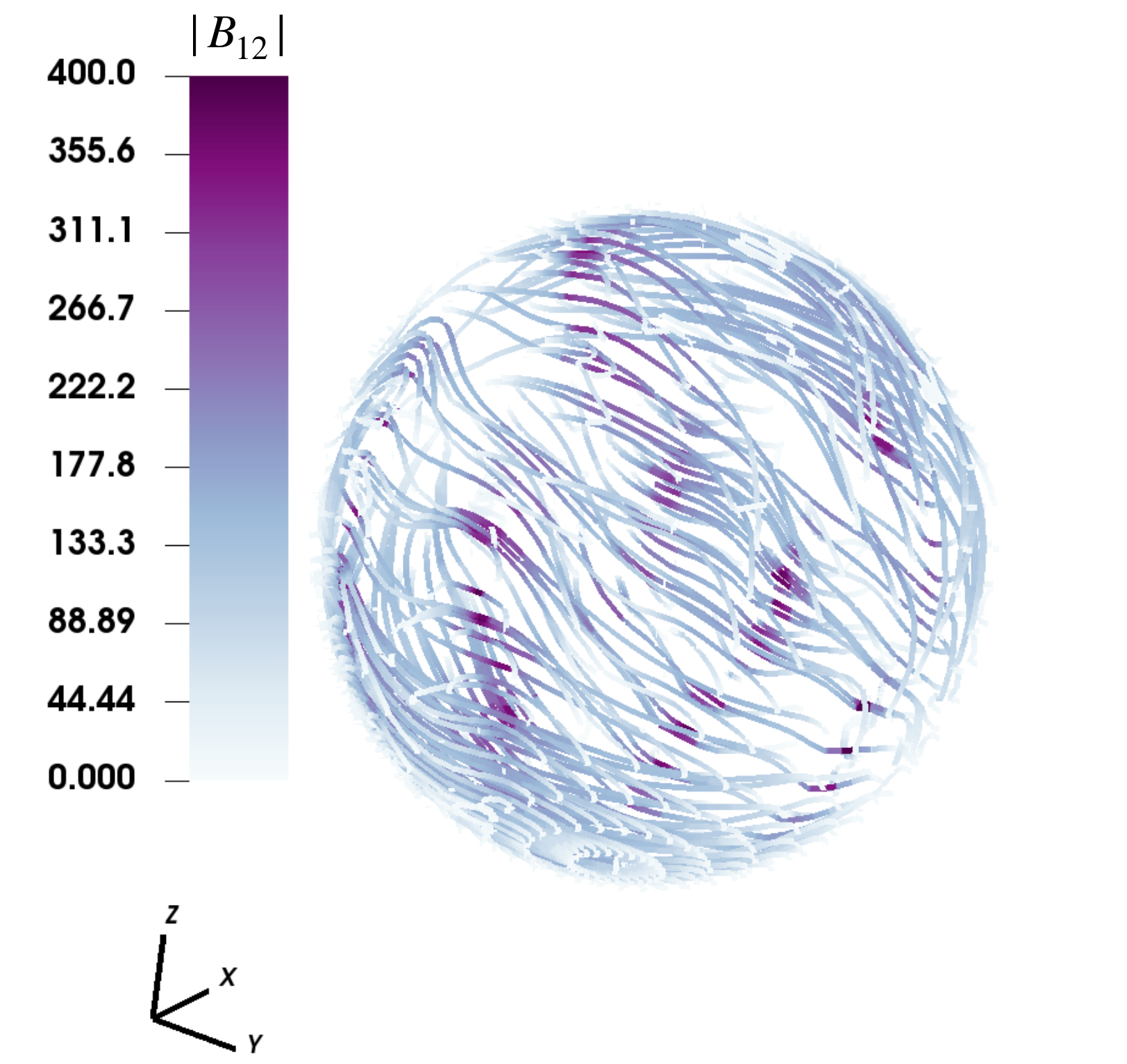}
\caption[3D Magnetic field lines]{Field lines in the crust of a neutron star, for L5 model on the left, L1 model in the center and L10 model on the right, at $t=0$ (upper panels) and after $t=50$\,kyr (bottom panels). The color scale indicates the local field intensity, in units of $10^{12}$\,G.}
\label{fig: field lines visit Dip LSCC Multi}
\end{figure}

In Fig.~\ref{fig: field lines visit Dip LSCC Multi}, we present the magnetic field lines for the L5 model (left panels), L1 model (central panels), and L10 model (right panels) at two evolution times: $t=0$ (upper row) and $t=50$\,kyr (bottom row). At $t=0$, the distinct magnetic field configurations adopted in these models are clearly discernible. However, after a few Hall timescales, approximately $50$\,kyr, the field lines become more tangled, yet one can still identify the initial magnetic field configurations. The neutron star has retained the memory of its initial large-scale topology despite the complex evolution and tangling of the magnetic field lines.

\subsubsection{Energy spectrum}

\begin{figure}
\centering
\includegraphics[width=0.45\textwidth]{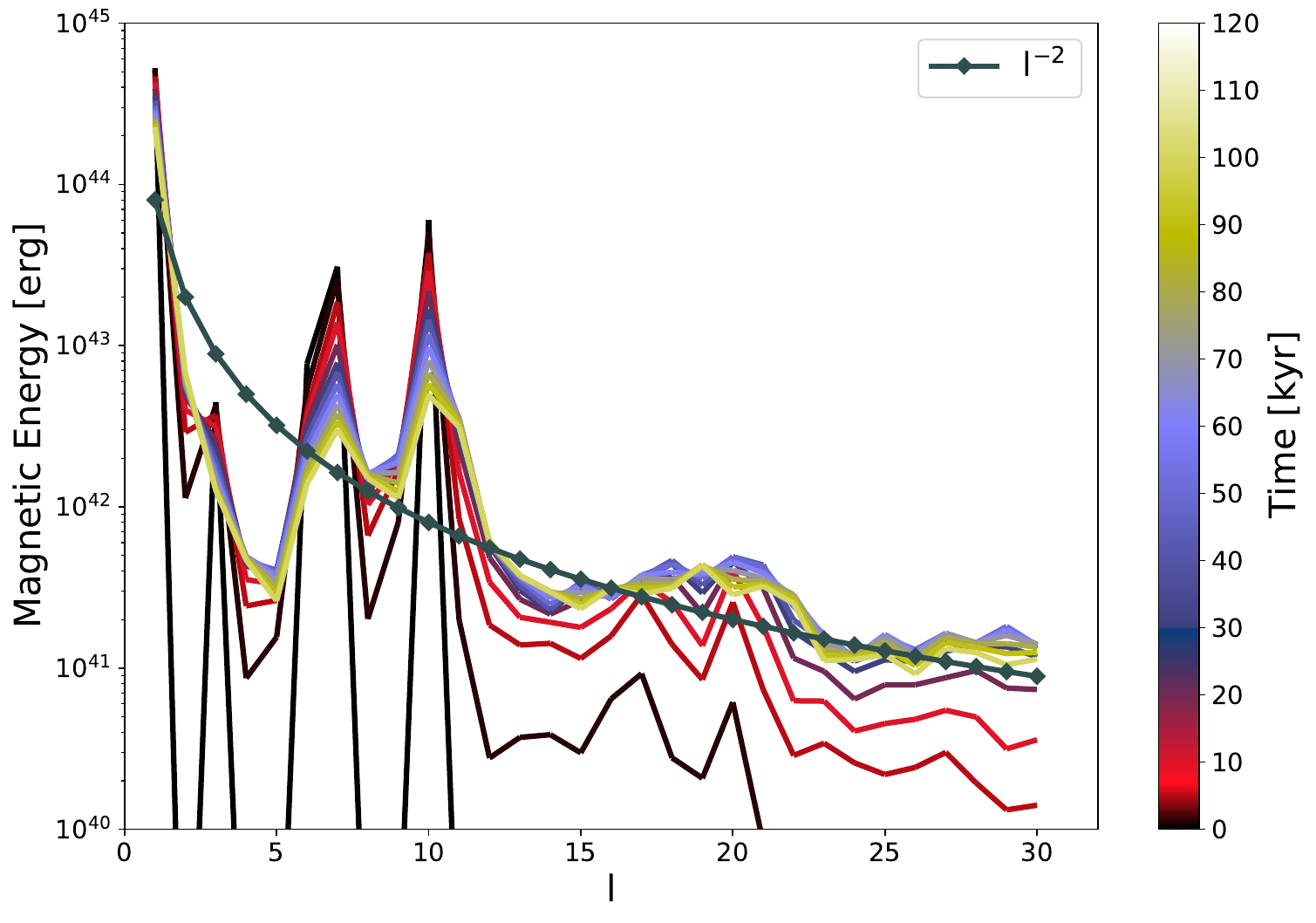}
\includegraphics[width=0.45\textwidth]{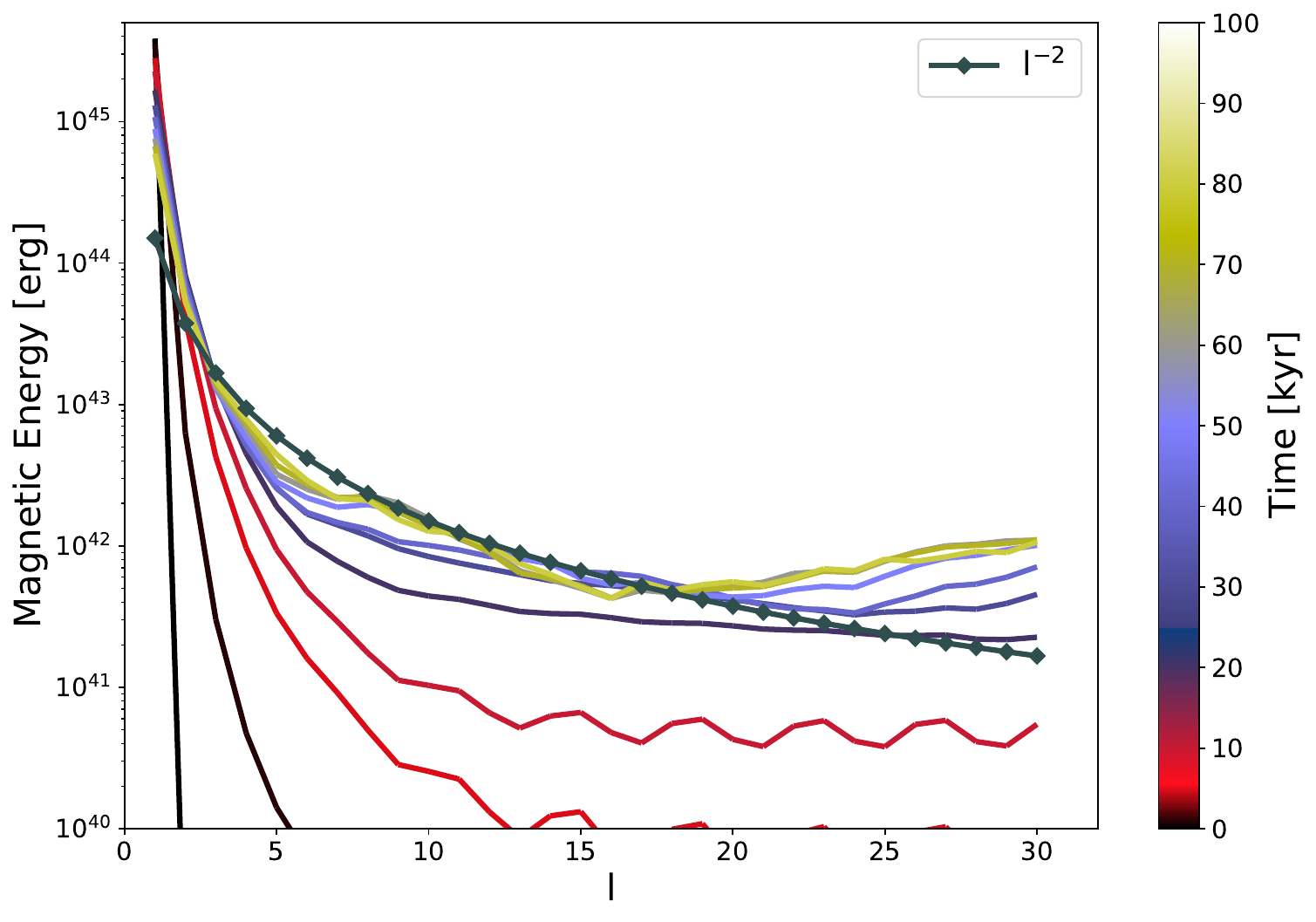}
\caption[$l$ energy spectrum for L1 and L10 models]{$l$ energy spectrum up to $100$\,kyr for L10 model (left panel) and up to $85$\,kyr for L1 model (right panel). See color bars to identify the ages. The $l^{-2}$ slope corresponds to the Hall cascade equilibrium distribution of magnetic energy over a quite broad range of multipoles.}
\label{fig: energy spectrum Dip Multi Scales models}
\end{figure}

The time evolution of the energy spectrum (as a function of $l$) is presented in Fig.~\ref{fig: energy spectrum Dip Multi Scales models}. On the left, we depict the L10 model, while on the right, we show the L1 model. For the L10 model, the weights of the initially defined multipoles can be inferred from the black energy spectrum in the left panel of Fig.~\ref{fig: energy spectrum Dip Multi Scales models}. Throughout the evolution, a redistribution of magnetic energy across different spatial scales occurs in both cases. However, the L10 model exhibits a tendency to inject magnetic energy into $l=16-18$ and $l=20$ more than other modes. Additionally, a less evident bump in the energy spectrum appears at $l=27$ and $l=29$. These bumps are more pronounced at the early stages of the evolution, e.g., up to $t=40$\,kyr.

The injected energy into these small-scale modes is relatively small compared to the initially dominant modes in the system. Nonetheless, this peculiar energy injection at small-scale structures could be indicative of the Hall instability \citep{gourgouliatos2020} that might occur in such an initial field configuration for higher magnetic Reynolds numbers.
As the evolution progresses, the magnetic energy is redistributed more uniformly across the small-scale structures, and the lower part of the spectrum, i.e., from $l=12$ up to $l=30$, follows the $l^{-2}$ slope \citep{goldreich1992}. However, the $l$-energy spectrum retains a strong memory of the initial configuration, particularly at low $l$s, throughout the entire evolution, even at $t=100$\,kyr. This is because the largest-scales have longer timescales, approximately $\sim L^2/(f_h B)$, where $L$ represents the length scale of the field (related to $l$). Consequently, it becomes challenging to transfer energy out of or into these larger-scales. In other words, the inertial range of the Hall cascade encompasses scales with sufficiently short timescales, maintaining the memory of the initial configuration over extended periods.

The transfer of magnetic energy across different spatial scales appears to be smoother for the L1 model compared to the L10 model. In the case of the L1 model, the energy spectrum is well described by an $l^{-2}$ power-law up to $l=20$. However, for smaller-scale multiples, i.e., $l > 20$, we observe an excess of energy. This injection of energy into the smallest structures grows over time and becomes more evident at around $t\sim 70-80$\,kyr, although it remains orders of magnitude lower than the dominant dipolar mode at $l=1$. Further investigation of different initial magnetic field topologies and their astrophysical implications is left for \S\ref{sec: 3DMT}, as the present section focuses on the broader understanding of the magnetic field evolution in neutron stars.

\subsubsection{Poloidal and toroidal decomposition}

\begin{figure}
\centering
\includegraphics[width=.43\textwidth]{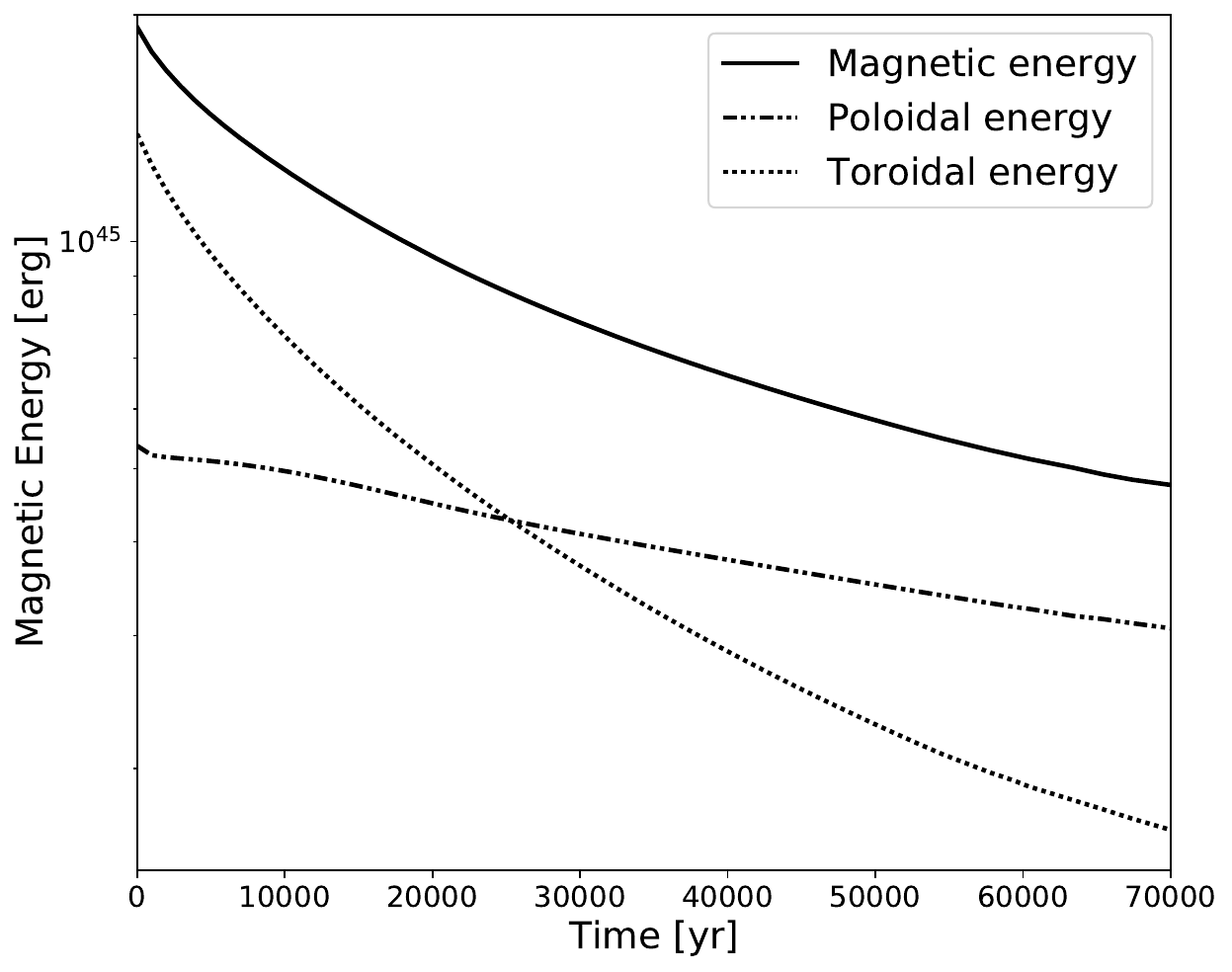}
\includegraphics[width=.5\textwidth]{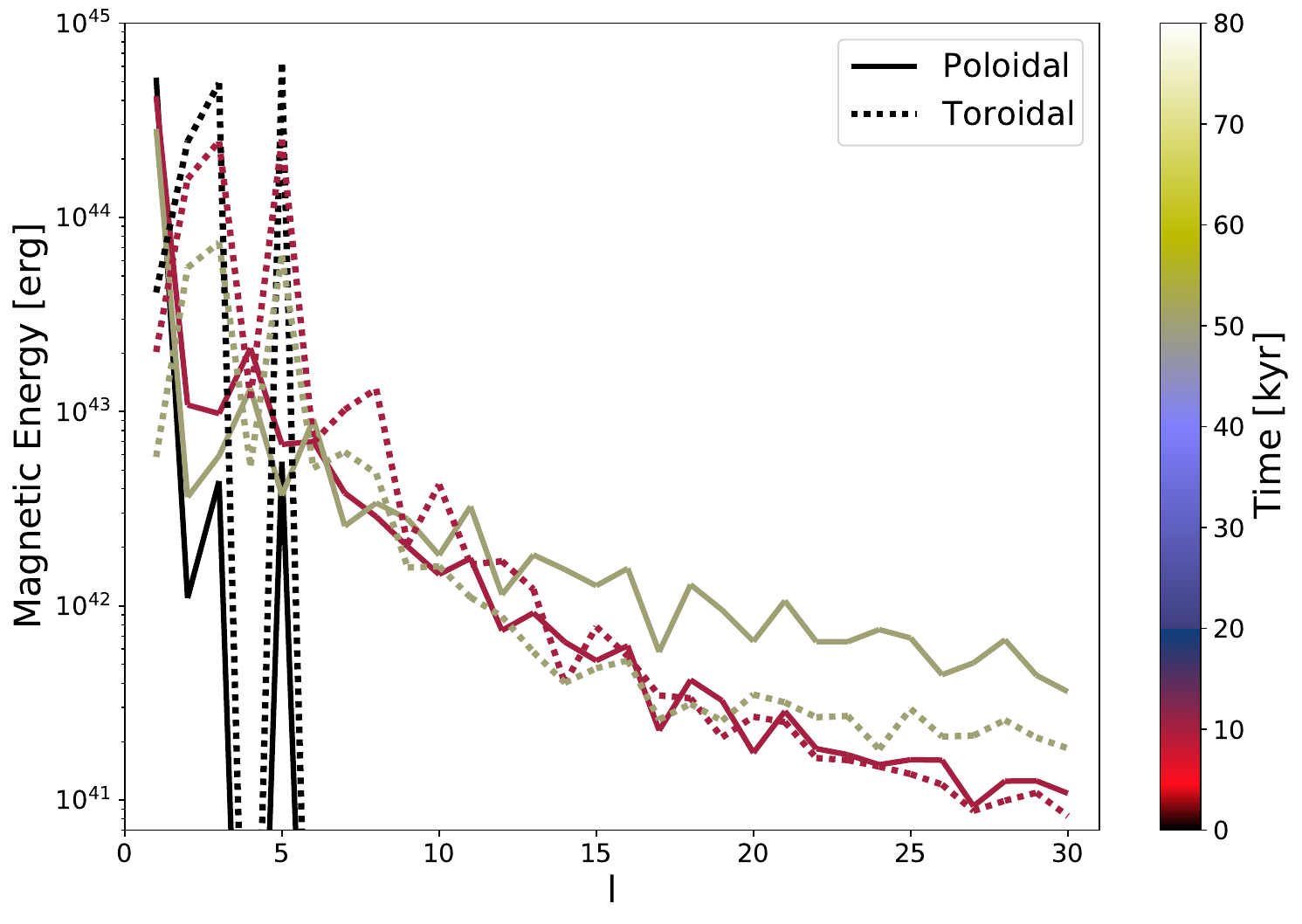}
\includegraphics[width=.43\textwidth]{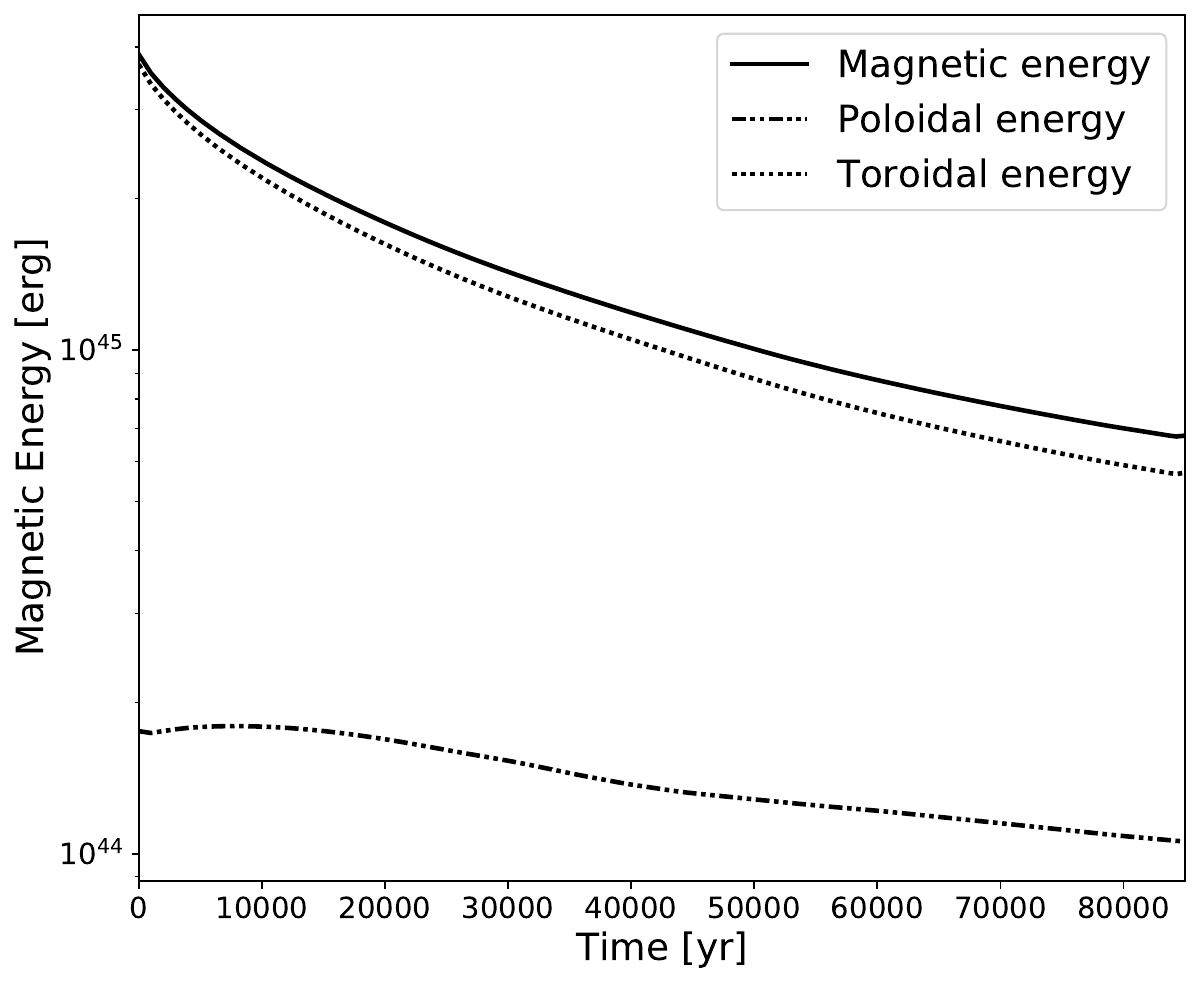}
\includegraphics[width=.5\textwidth]{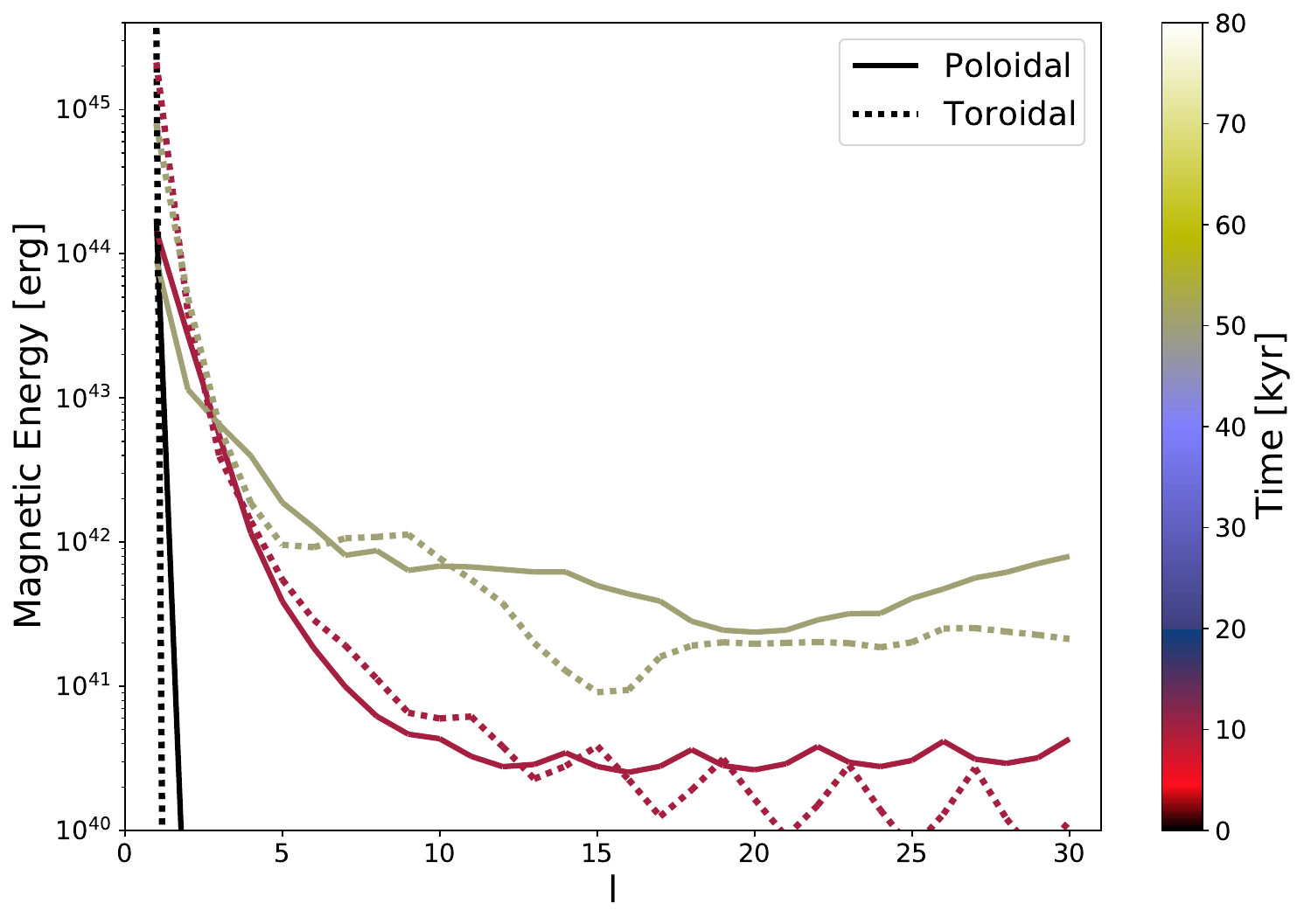}
\includegraphics[width=.43\textwidth]{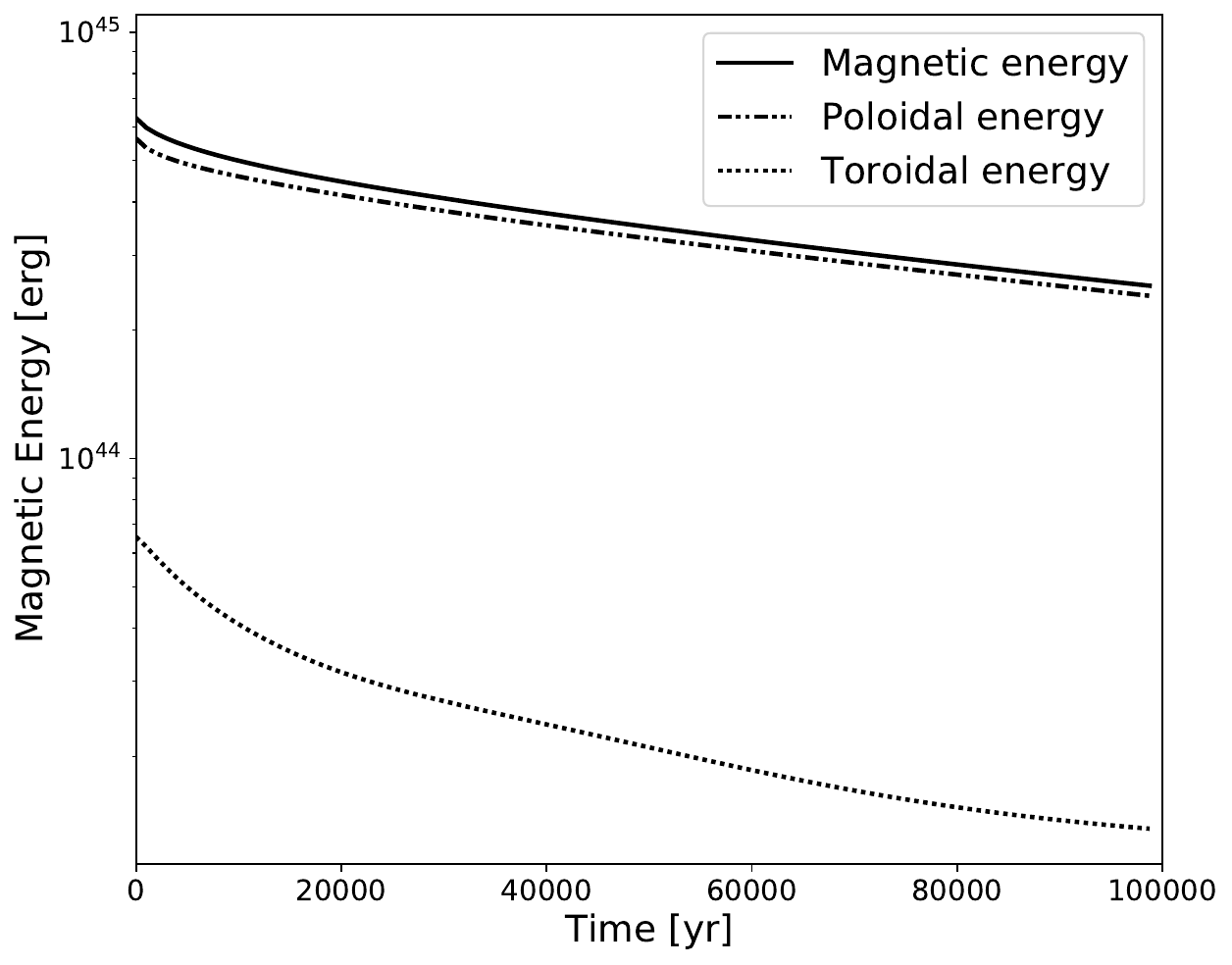}
 \includegraphics[width=.5\textwidth]{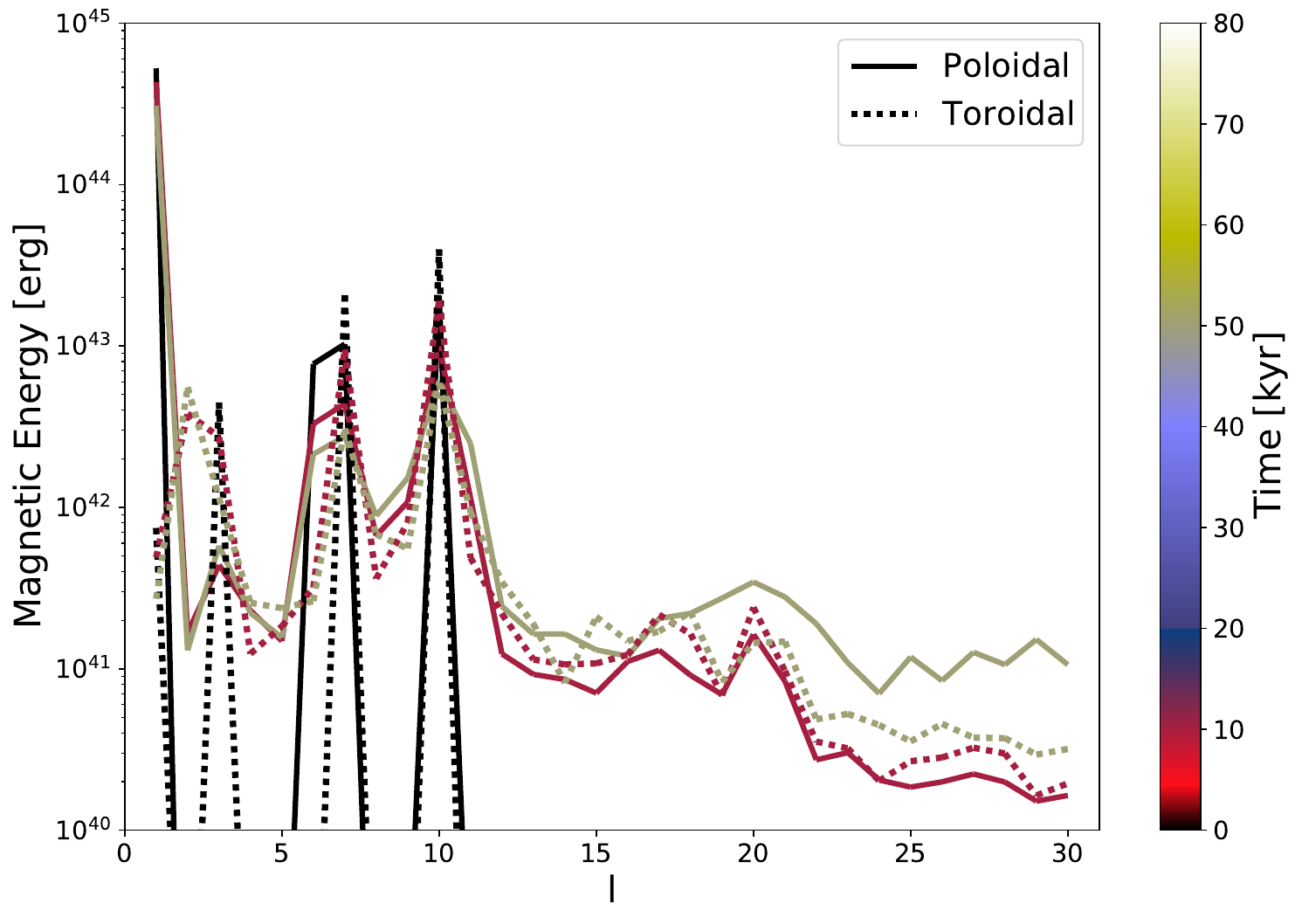}
\caption[Poloidal and toroidal energy distributions]{Decomposition of the poloidal and toroidal magnetic energy. Poloidal magnetic energy is represented with dash-dotted lines, toroidal energy with dots, and total magnetic energy with solid lines. \emph{Upper panels:} \emph{L5} model. \emph{Central panels:} \emph{L1} model. \emph{Lower panels:} \emph{L10} model. \emph{Left Column:} Poloidal and toroidal decomposition of the magnetic energy as a function of time. 
\emph{Right column:} $l$ energy spectrum at $t=0$ (black), $t=10$\,kyr (dark red), and after $50$\,kyr (olive).}
\label{fig: Toroidal- Poloidal energy Spectrum LS Dip Multi}
\end{figure}

Using the notation in eq.\,\eqref{eq: spectral magnetic energy}, we can decompose the magnetic energy into its poloidal and toroidal parts. In Fig.~\ref{fig: Toroidal- Poloidal energy Spectrum LS Dip Multi}, we present the evolution of the poloidal and toroidal magnetic energy for different models.

For the L5 model, at the early stages of evolution, the bulk of the magnetic energy is stored in the toroidal field (approximately $\sim 63\%$), with the poloidal energy accounting for about $\sim 37\%$ of the total magnetic energy. As the evolution progresses, we observe that the toroidal field tends to dissipate almost $5$ times faster than its poloidal counterpart, leading to an inversion of the poloidal-toroidal ratio. This phenomenon occurs because the toroidal energy is effectively redistributed into smaller-scale multipoles, which subsequently dissipate faster. Meanwhile, a significant portion of the poloidal energy remains in the dominant $l=1$ mode (upper right panel). However, after a few Hall timescales, the system eventually reaches a state of approximate equipartition of magnetic energy between the poloidal and toroidal energy spectra, as a result of the Hall-dominant evolution.

For the L1 model (central panels), most of the magnetic energy, approximately $\sim 90\%$, is stored in the toroidal component. In contrast, for the L10 model (bottom panels), the majority of the magnetic energy is concentrated in the poloidal component. Despite the evolution over $\sim 100$\,kyr, the magnetic energy remains predominantly stored in the dominant mode for each model (L1 in the toroidal and L10 in the poloidal). Nonetheless, approximate equipartition of magnetic energy between the poloidal and toroidal components is also achieved at approximately $\sim 10$\,kyr, but only at smaller-scales. The large-scales of the initial configuration are not easily forgotten or created and retain a significant memory throughout the evolution.

These results confirm that the system favors the redistribution of magnetic energy between the poloidal and toroidal components to stabilize the evolution. However, it's important to note that achieving this saturated configuration, known as the Hall attractor (first introduced by \cite{gourgouliatos2014}), takes tens of kyr, which coincides with the typical active timescale of magnetars. Throughout this stage, the spectra and topology still depend on the initial configuration.

\begin{figure}
\includegraphics[width=16cm, height=4.7cm]
{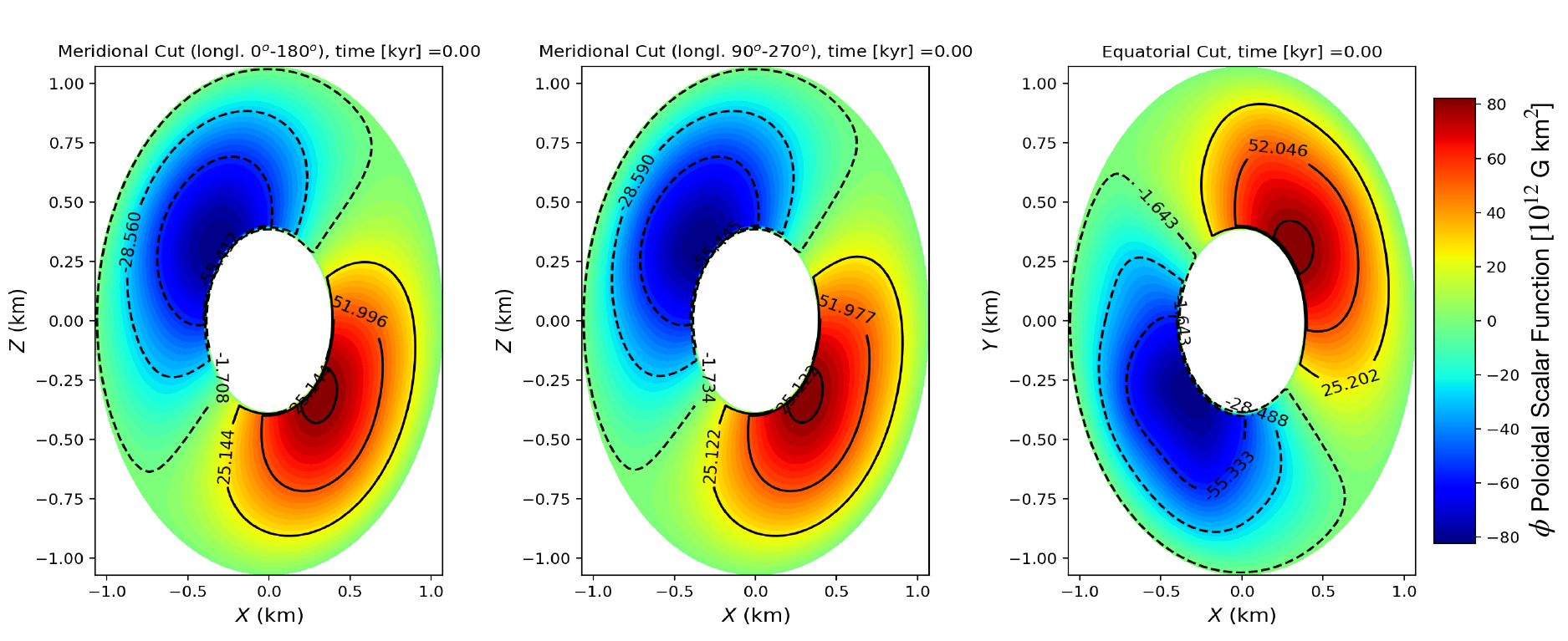}
\includegraphics[width=16cm, height=4.7cm]{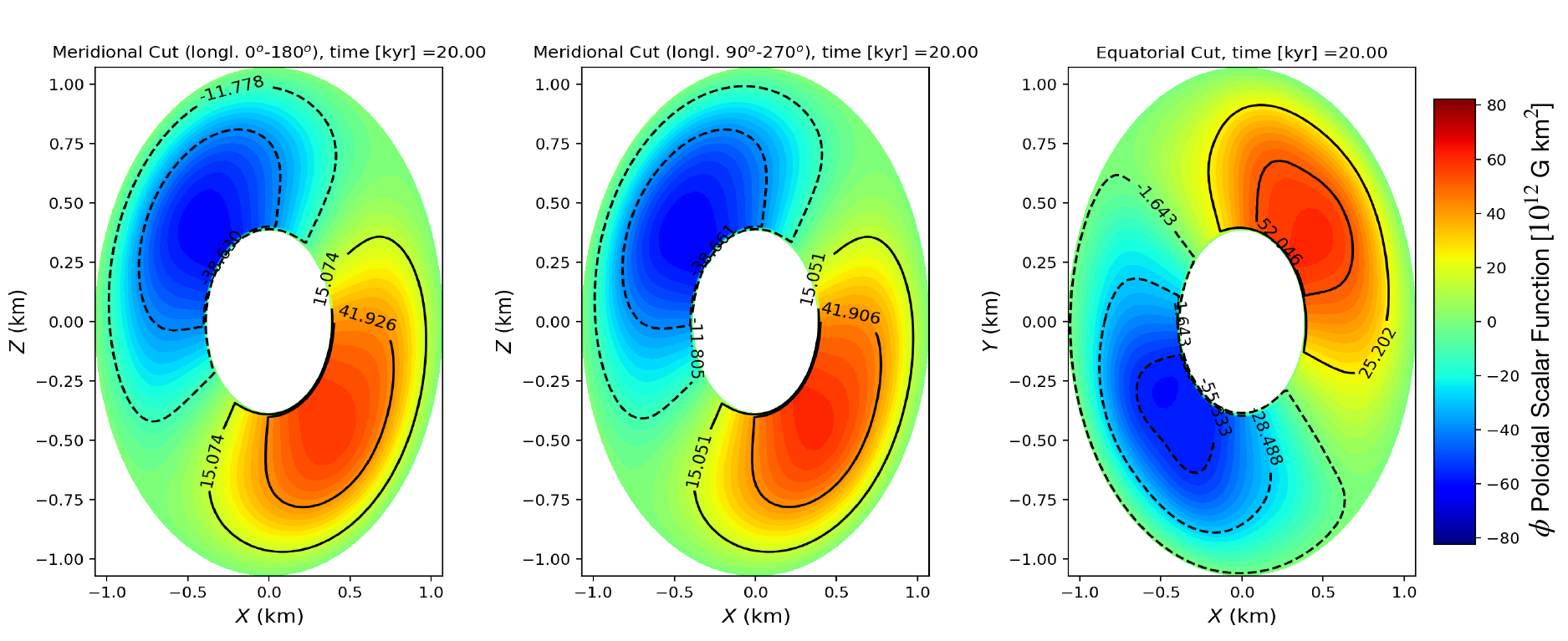}
\includegraphics[width=16cm, height=4.7cm]{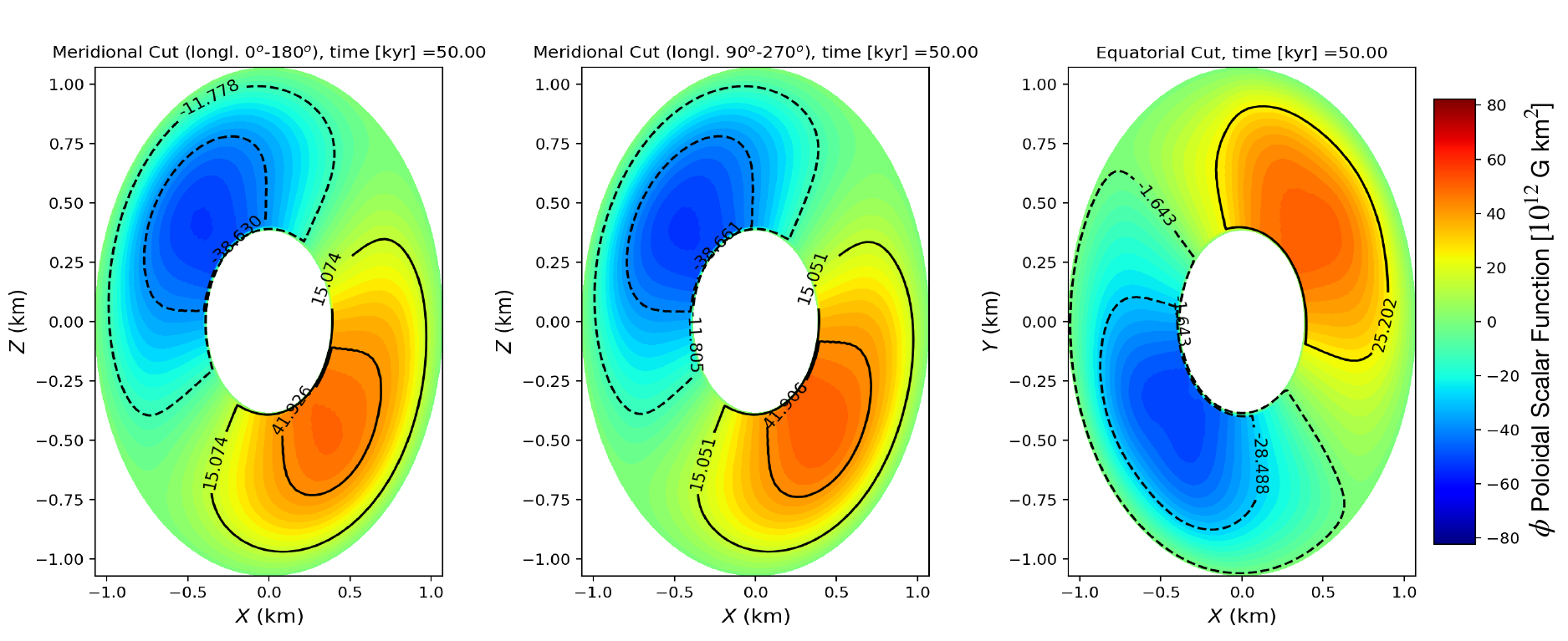}
\caption[Model L10: Poloidal scalar function]{\emph{Model L10}. Evolution of the $\Phi$ poloidal scalar function, at $0$, $20$ and $50$\,kyr (from top to bottom). In the left panels, we show the meridional cuts at the longitudes $0-180^\circ$ (cutting through the center of patches I and III). In the central panels we illustrate the meridional cut at the longitudes $90-270^\circ$ (through the center of patches II and IV). In the right panels, we represent the equatorial 2D cuts. The crust is greatly enlarged for visualisation purposes.}
\label{fig: Phi Poloidal function Multi}
\end{figure}

\begin{figure}
\includegraphics[width=16cm, height=4.7cm]{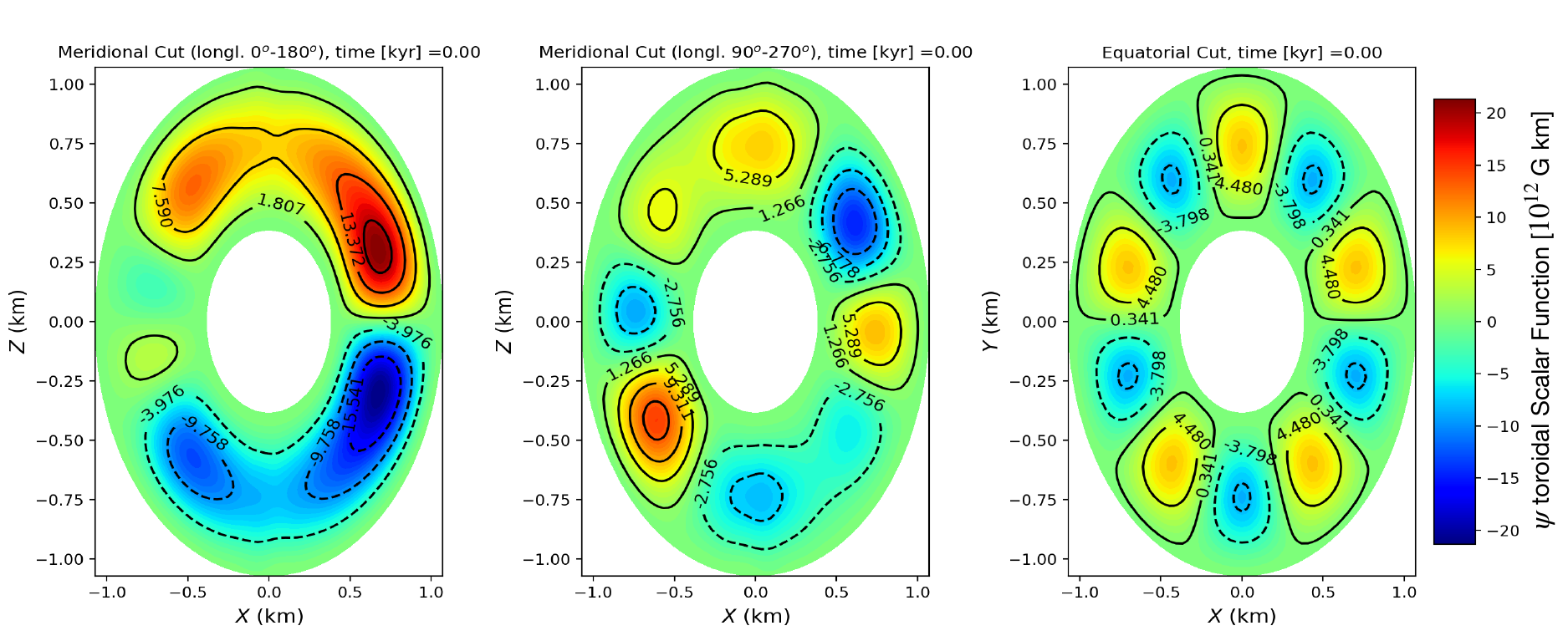}
\includegraphics[width=16cm, height=4.7cm]{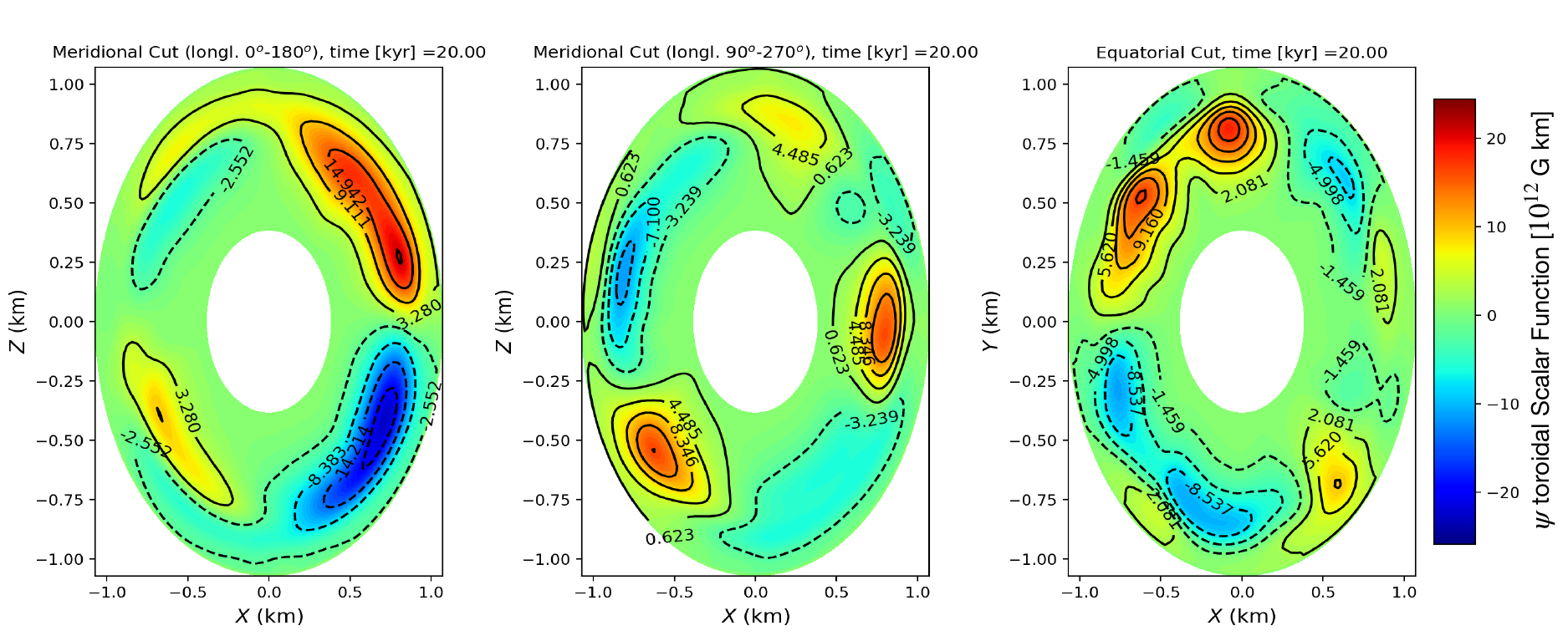}
\includegraphics[width=16cm, height=4.7cm]{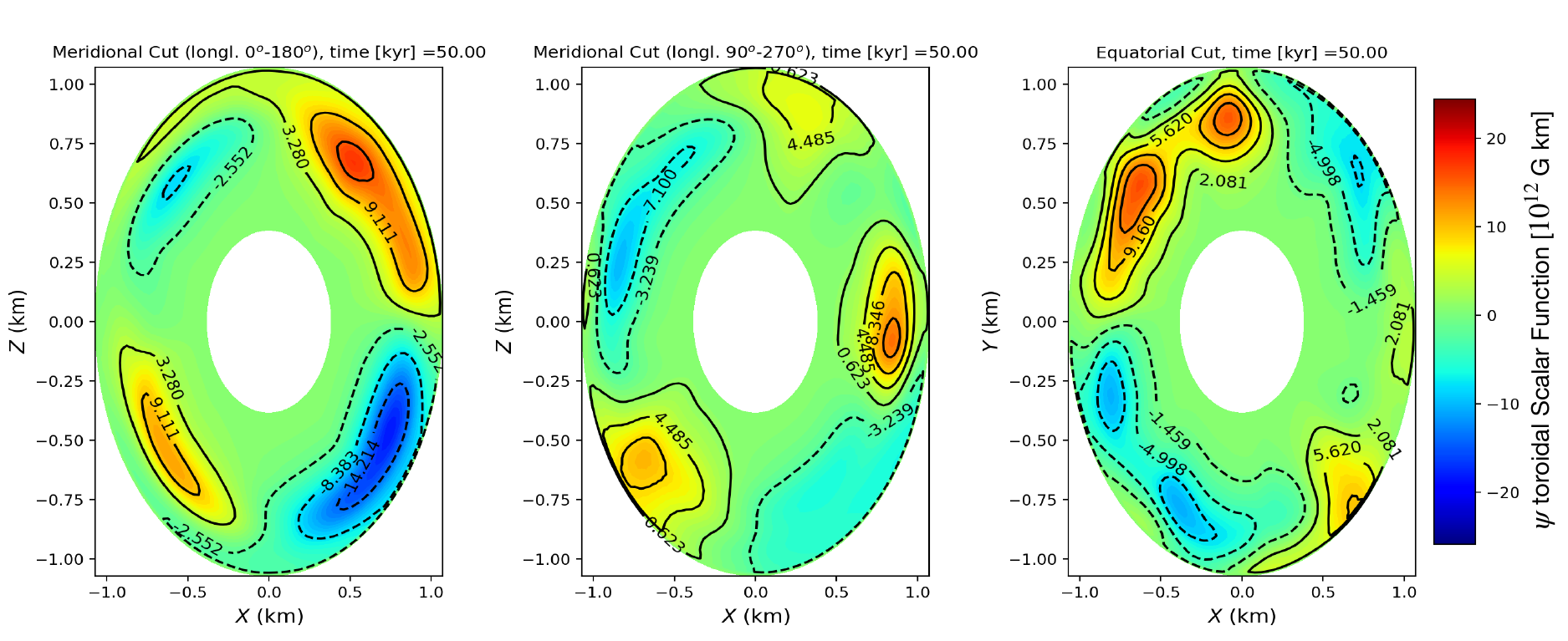}
\caption[Model L10: Toroidal scalar function]{\emph{Model L10}. The evolution of 2D cuts of the $\Psi$ toroidal scalar function. Same cuts and times as Fig.~\ref{fig: Phi Poloidal function Multi}}
\label{fig: Psi Toroidal function Multi}
\end{figure}

In Fig.~\ref{fig: Phi Poloidal function Multi} and \ref{fig: Psi Toroidal function Multi}, we illustrate the meridional cuts at longitudes $0-180^\circ$ (left panels), $90-270^\circ$ (central panels), and equatorial cuts (right panels) of the poloidal $\Phi$ and toroidal $\Psi$ scalar functions for the L10 model. The top panels represent the initial configuration at $t=0$, the central panels show the state at $20$\,kyr, and the bottom panels display the evolution at $50$\,kyr. 

Throughout the evolution, the poloidal function, initially dipole-dominated (lower right panel of Fig.~\ref{fig: Toroidal- Poloidal energy Spectrum LS Dip Multi}), undergoes only slight changes. Conversely, the initially more complex toroidal scalar function, dominated by the $l=10$ mode, experiences significant rearrangements. Furthermore, there is a drift of the toroidal scalar function toward the surface of the star, indicating the importance of coupling this code with the magnetosphere's evolution \citep{akgun2018,urban2023}.

After exploring various initial field configurations using \emph{MATINS}, our simulations have confirmed that the magnetic field's spectra and topology retain a strong memory of the initial large-scales, which are relatively resistant to restructuring or creation during the evolution. This finding highlights the importance of the large-scale configuration acquired during neutron star formation, as it significantly influences the magnetic field's topology at all stages of its evolution. Consequently, in \S\ref{sec: 3DMT}, which is based on the research conducted by \cite{dehman2023c}, we will investigate a complex initial field inspired by core-collapse supernovae simulations, utilizing for the first time the 3D coupled magneto-thermal version of the \emph{MATINS} code. This approach involves solving the heat diffusion equation as described in \S\ref{sec: neutron star cooling}, instead of adopting the simplified cooling approach mentioned in this section.
Our goal is to gain a deeper understanding of the magnetic field evolution and cooling process in neutron stars, incorporating more realistic initial conditions. 

\section{3D coupled magneto-thermal evolution}
\label{sec: 3DMT}

\begin{figure*}
    \centering
    \includegraphics[width=0.45\textwidth]{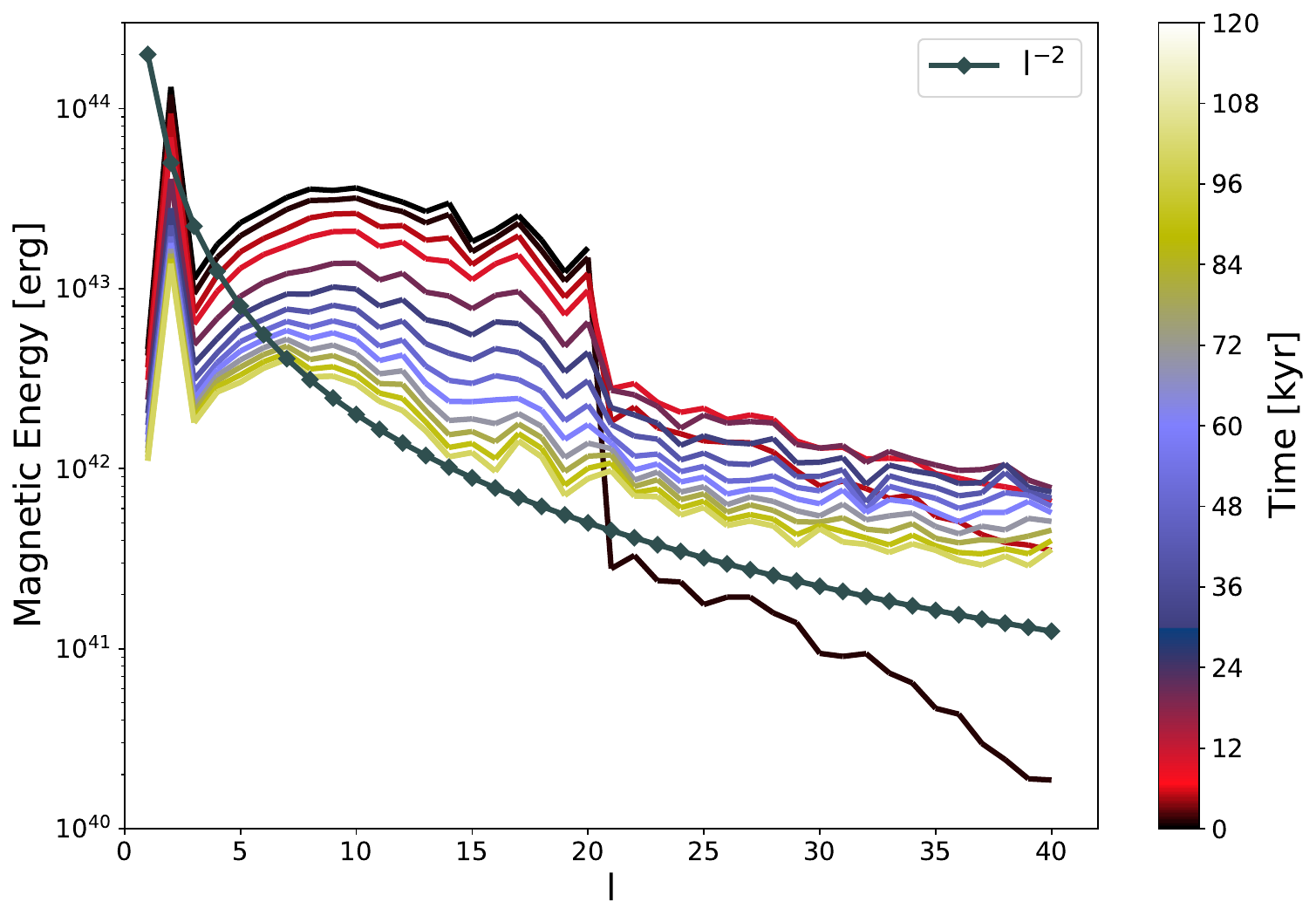}
        \includegraphics[width=0.45\textwidth]{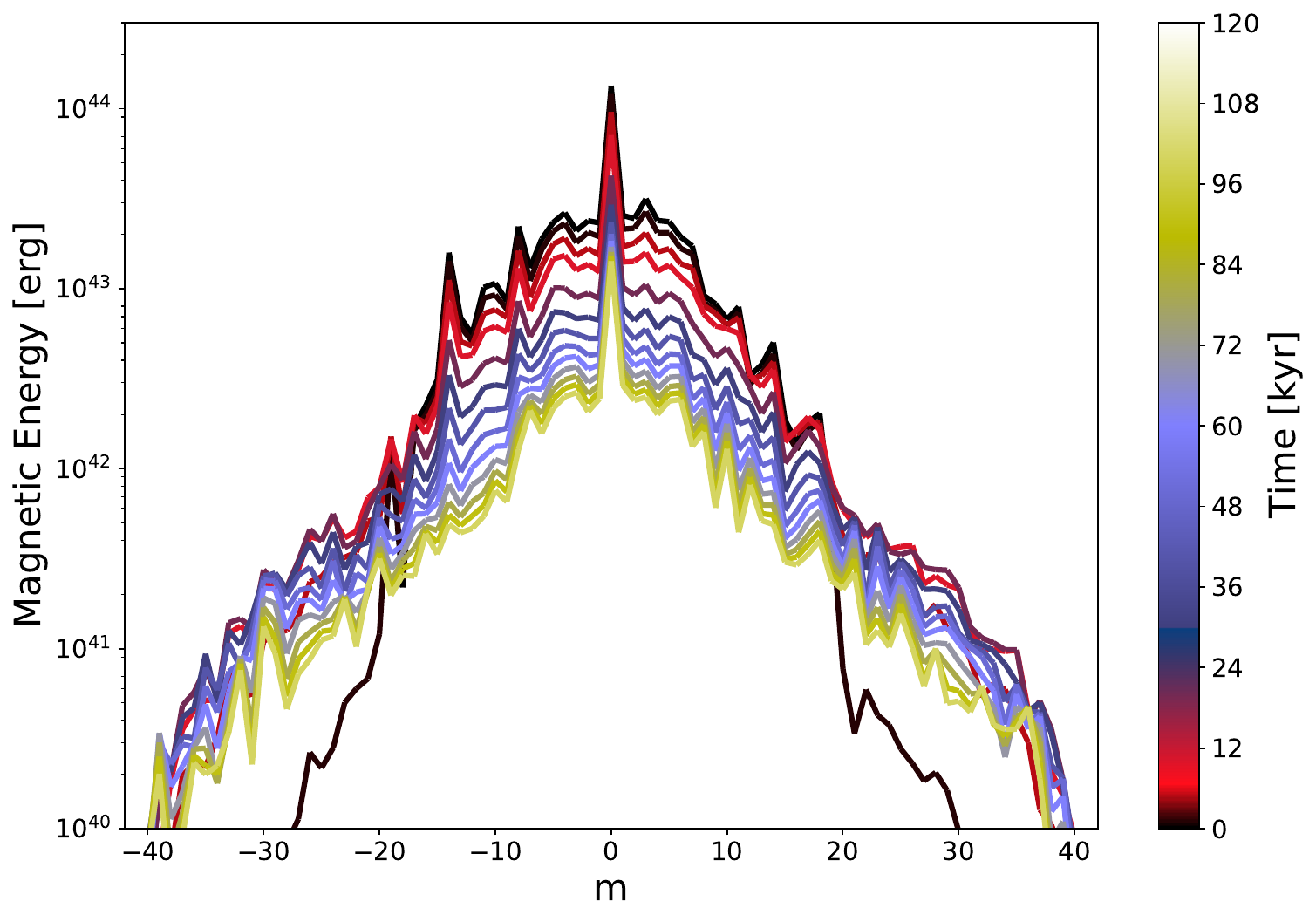}
    \caption[Spectrum of the total magnetic energy at different evolution time up to $100$\,kyr]{Spectrum of the total magnetic energy at different evolution time up to $100$\,kyr (Hall balance is reached in the system). \emph{Left panel:} $l-$energy spectrum. \emph{Right panel:} $m-$energy spectrum. }
    \label{fig: energy spectrum}
\end{figure*}

\begin{figure}
    \centering
        \includegraphics[width=0.55\textwidth]{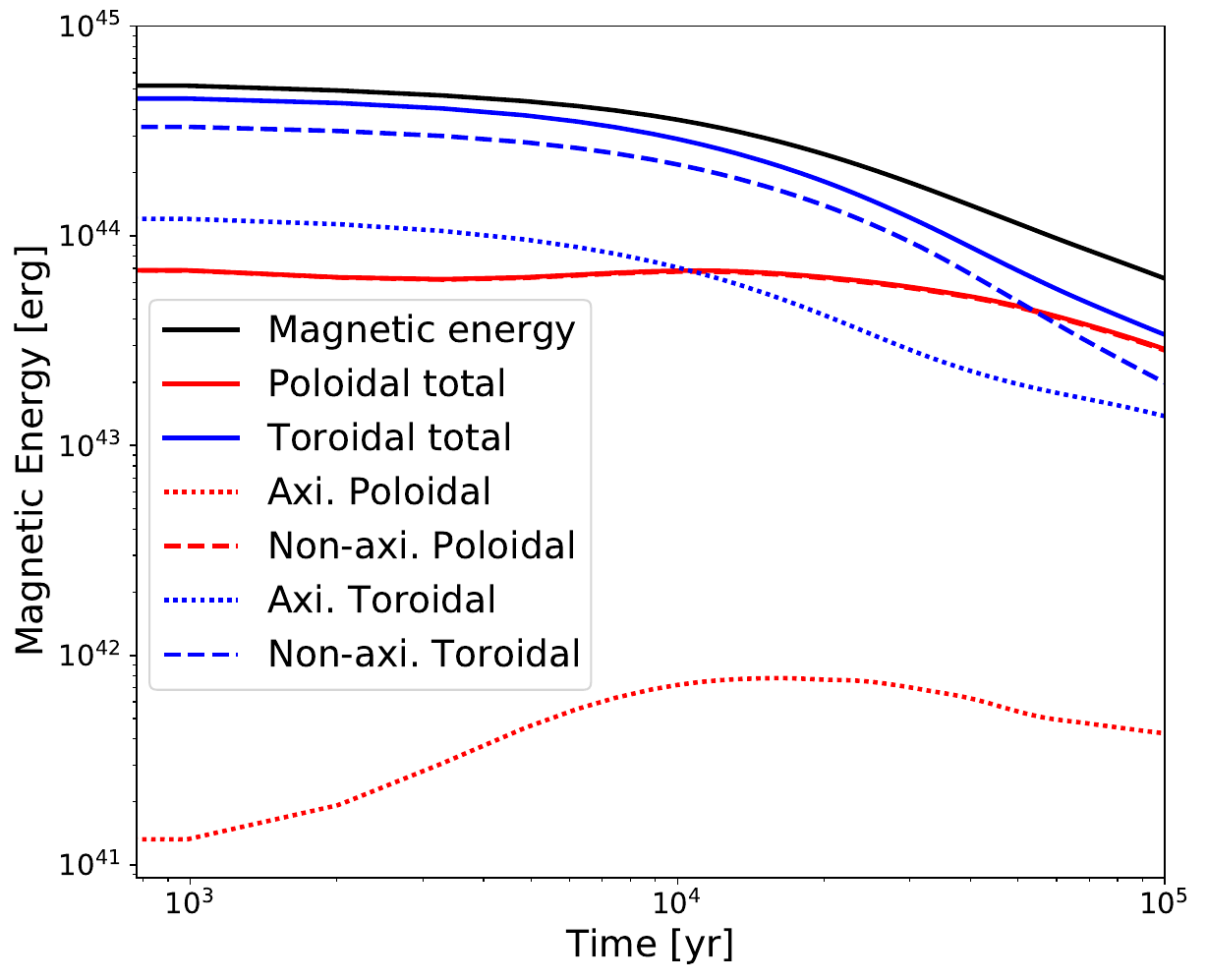}
    \caption[Decomposition of the poloidal and toroidal magnetic energy as a function of time.]{Decomposition of the poloidal and toroidal magnetic energy as a function of time. Poloidal magnetic energy is represented with red color, toroidal energy with blue, and total magnetic energy with black solid lines. The dots correspond to the axisymmetric components of these energies and the dashed lines to the non-axisymmetric components. The axisymmetric and non-axisymmetric components are taken in proto-neutron star dynamo simulations with respect to the axis of the proto-neutron star rotation. This has no role in our simulations and we define these components with respect to the magnetic axis, starting with only $m=0$ for a dipole.}
    \label{fig: Emag PT}
\end{figure}

\begin{figure*}
    \centering
    \includegraphics[width=0.47\textwidth]{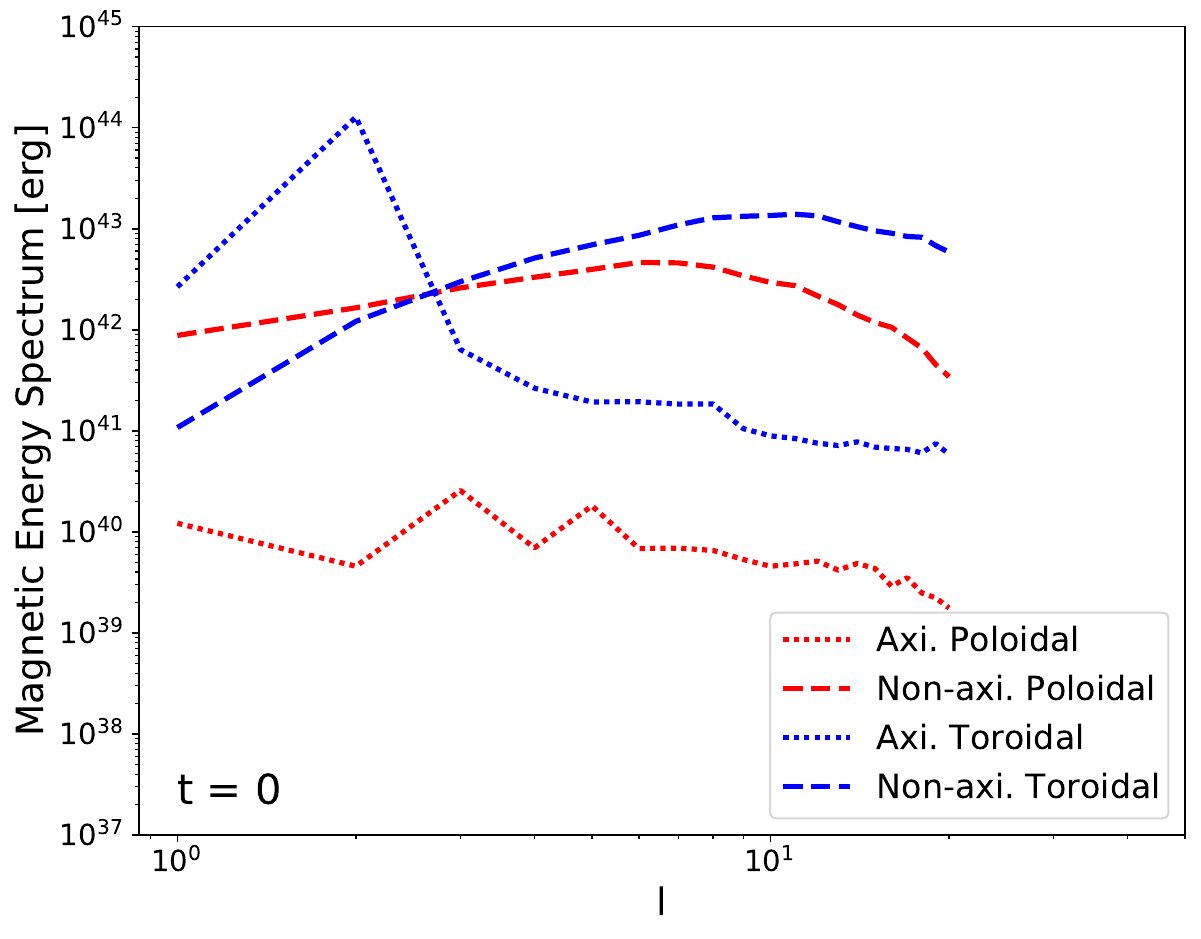}
    \includegraphics[width=0.47\textwidth]{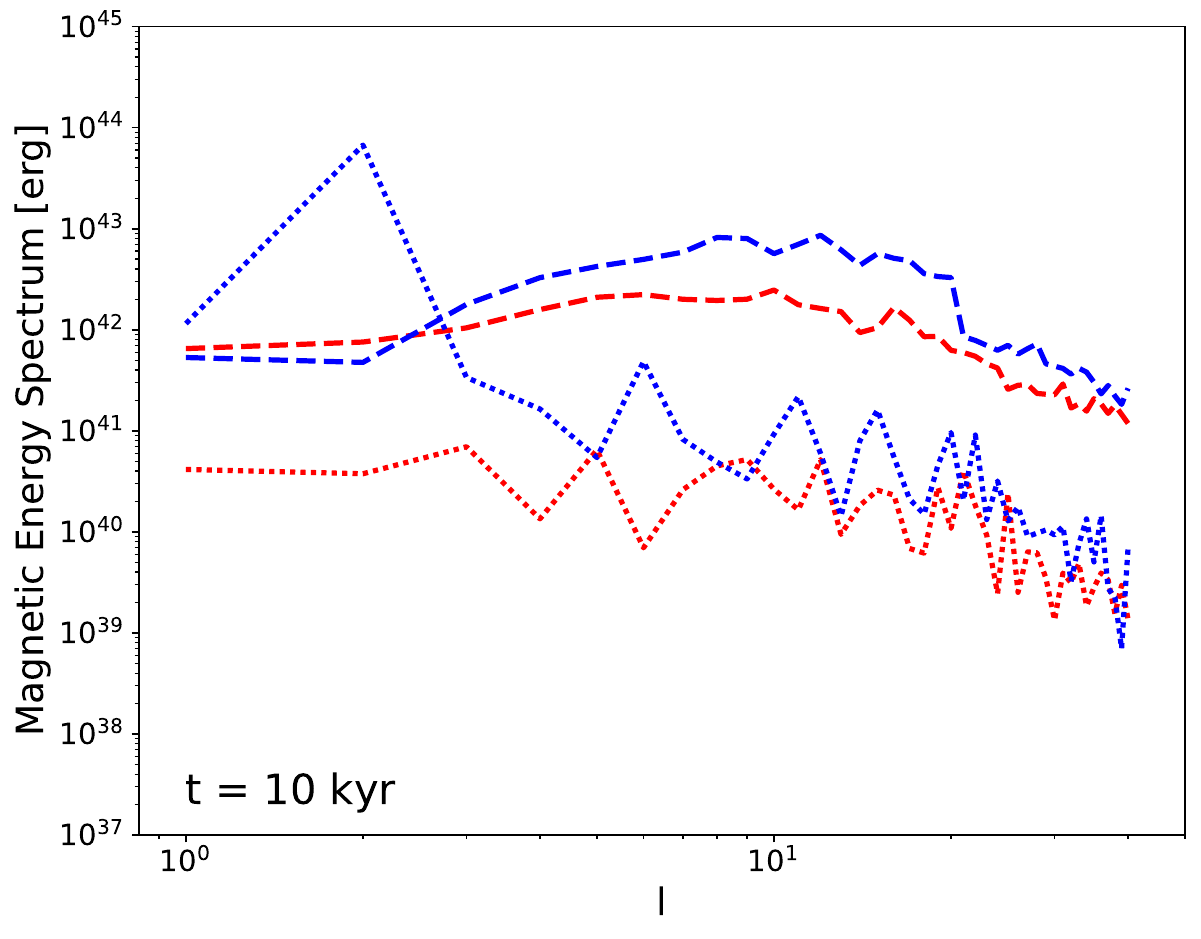}
    \includegraphics[width=0.47\textwidth]{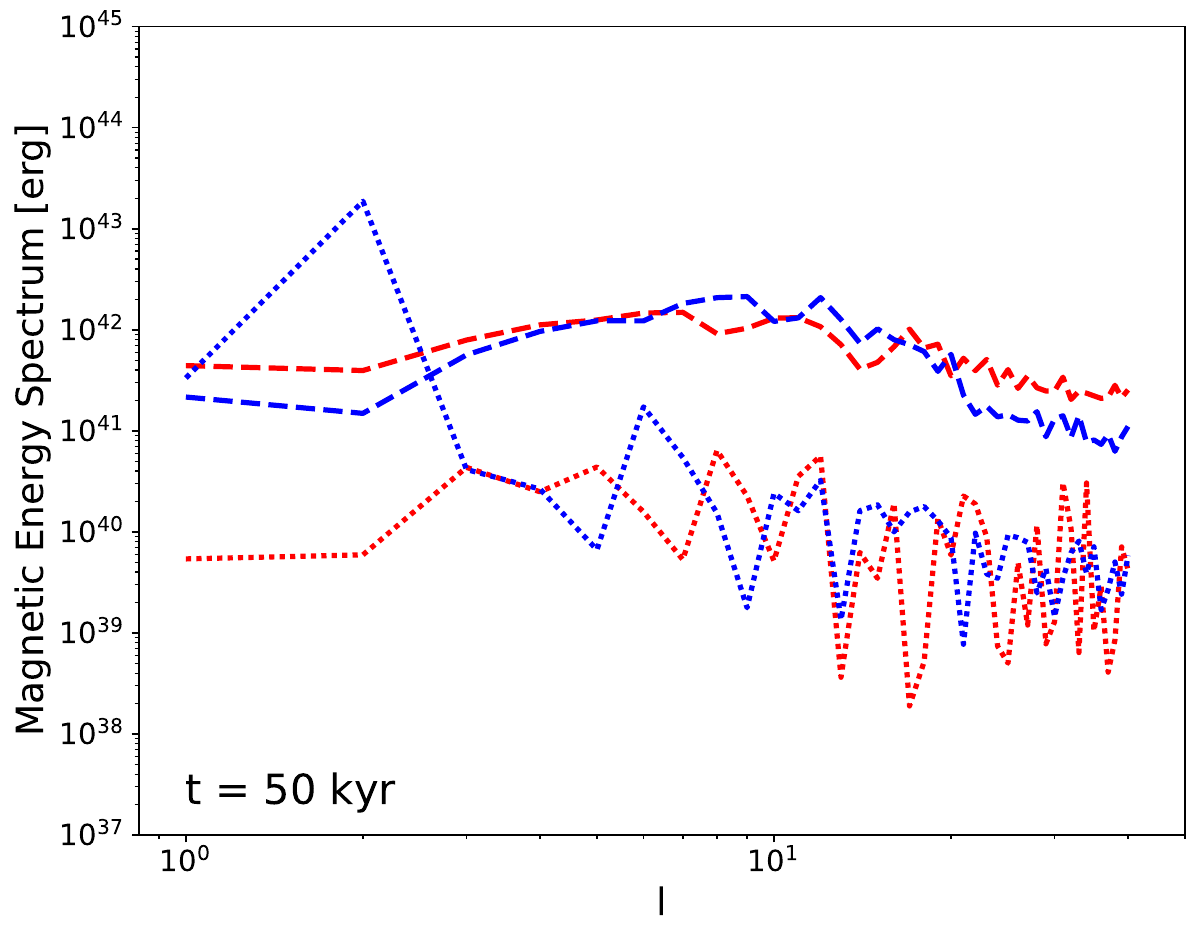}
    \includegraphics[width=0.47\textwidth]{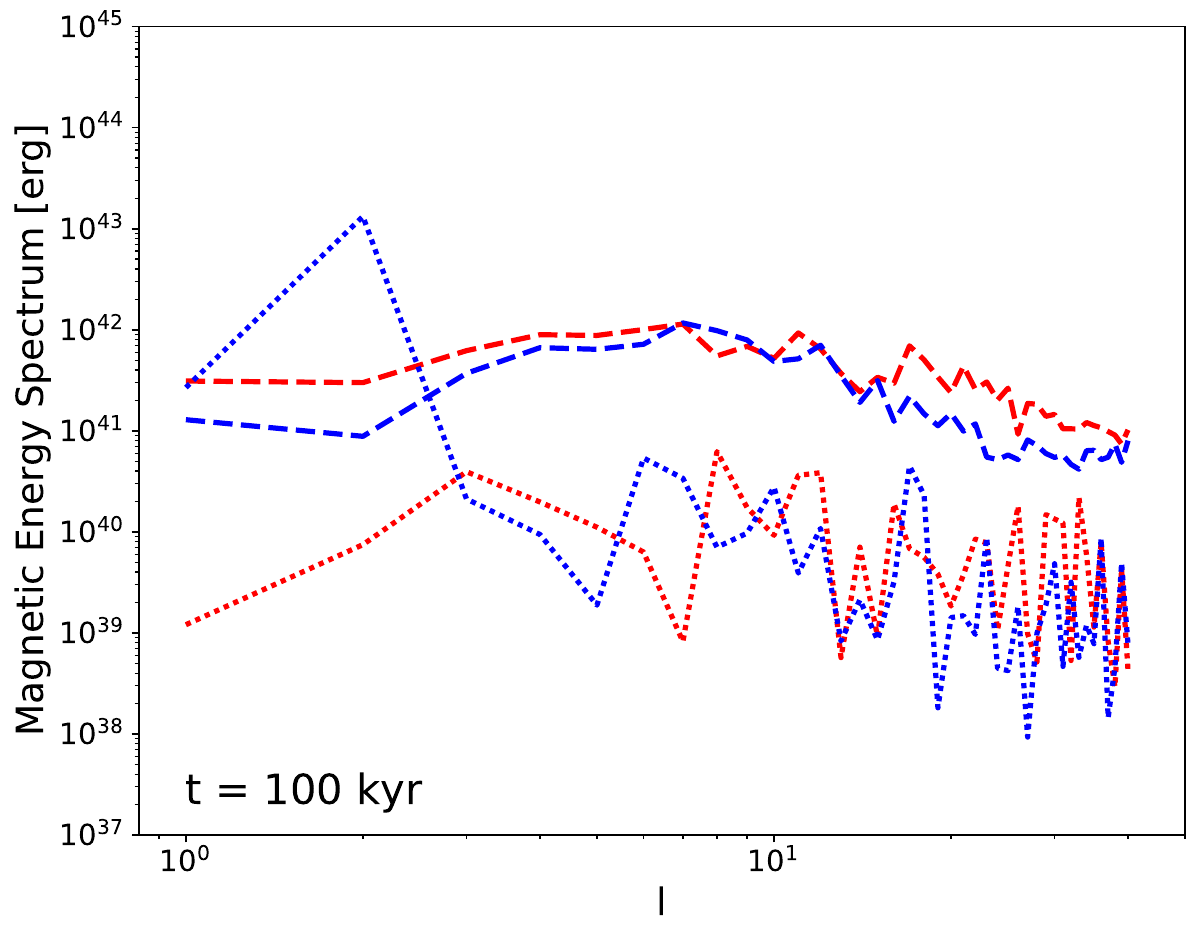}
    \caption[Spectrum of the toroidal magnetic energy (blue) and the poloidal magnetic energy (red) as a function of the spherical harmonics order $l$]{Spectrum of the toroidal magnetic energy (blue) and the poloidal magnetic energy (red) as a function of the spherical harmonics order $l$ at $t=0, 10, 50$ \& $100$\,kyr. The dots correspond to the axisymmetric components of these energies and the dashed lines to the non-axisymmetric components. The total number of multipoles in the system is $l_\text{max}=40$ and the initial number of multipoles is $l=20$. The top left panel of this figure ($t=0$) is inspired from a core-collapse generated turbulent field \citep[Fig.~7 top panel]{reboul2021}. }
    \label{fig: TP}
\end{figure*}

\subsection{Initial topology}
\label{sec: initial topology} 

In recent years, significant efforts have been dedicated to MHD simulations of core-collapse supernovae (e.g., \cite{obergaulinger2014,mosta14,mosta15,bugli20,Aloy_2021,powell22,obergaulinger22}). These simulations hold intrinsic importance in understanding the underlying explosion mechanisms and the fundamental physics of hot dense matter \citep{janka12,obergaulinger20}. Furthermore, core-collapse supernovae simulations play a crucial role in defining the characteristics of its compact remnant, a hot proto-neutron star (e.g. \cite{pons1999,barrere2022}). As described in \S\ref{sec: stellar evolution}, the highly dynamic process consists of three main stages. After the core bounce, the system has a cool central region and a hot mantle, emitting neutrinos while accreting material. In about $20$\,s, a proto-neutron star forms, becoming extremely hot and liquid with a large radius. After several minutes, it becomes transparent to neutrinos, shrinking to a final radius of $10-14$\,km, marking the birth of a neutron star. The long-term cooling is initially dominated by neutrinos and later by photon emission from the surface \citep{burrows1986,keil1995,pons1999}.

We start from the work by \cite{reboul2021}, who performed global simulations using MagIC \citep{wicht2002}. They consider a shell representative of the convective region of a proto-neutron star, and simulate the dynamo coming from the typical differential rotation profile seen during an advanced state of the core-collapse (about $1$\,s of its lifespan). At saturation, they found a very complex topology, with most of the magnetic energy contained in the toroidal axisymmetric large-scale component (especially the quadrupolar component given by winding) and in the non-axisymmetric small- or medium-scale size magnetic structures, both for the toroidal and the poloidal components. The large-scale dipolar component represents only about $\sim 5\%$ of the average magnetic field strength. 

We adopt an initial field with an angular spectral energy distribution, shown in the upper left panel of Fig.~\ref{fig: TP}, comparable to \cite[Fig.~7 top panel]{reboul2021}. Here we indicate toroidal (blue)/poloidal (red) and axisymmetric (spherical harmonics order $m=0$)/non-axisymmetric ($m\neq 0$) components as a function of the spherical harmonics degree $l$. In \emph{MATINS}, although the evolved quantities are the magnetic field components, we prescribe the initial field in terms of the toroidal and poloidal decomposition, based on the Chandrasekhar-Kendall expressions \citep{chandrasekhar1981} as detailed in \S\ref{appendix: Poloidal and toroidal decomposition}. 

The initial configuration is then a smooth cascade in $l$ (truncated for simplicity at $l=20$) and $m$, with an exception for the quadrupolar toroidal component which dominates. The average initial magnetic field of a few $10^{14}$\,G, corresponds to a total magnetic energy of the order of $\sim 6 \times 10^{44}$\,erg. Note that axial symmetry here refers to the rotational axis in the proto-neutron star phase, but the rigid rotation of the neutron star plays no role in the magnetic evolution (we don't evolve the full MHD, with the momentum equation, which includes the Coriolis force, since the background is assumed static for the solid crust). Therefore, hereafter, in our simulations, we take as the reference axis the proto-neutron star rotation axis.

Since no specific information on the radial distribution of the magnetic energy is available in \cite{reboul2021}, we adopt an arbitrary set of initial poloidal $\Phi_{lm}(r)$ and toroidal $\Psi_{lm}(r)$ radial scalar functions (see \S\ref{appendix: initial conditions} for more details). 
It's important to note that this particular choice is not based on any specific physical rationale; rather, it is solely intended to match a pure dipolar field at the surface when $t=0$.
In order to test the sensitivity of the results on the initial radial function, we also conducted a test where the radial function for the dipolar poloidal component has been applied to all the other poloidal multipoles (producing an initial current sheet at the surface, see below).

Certainly, it would be unrealistic to assume that the initial configuration of the neutron star at birth exactly matches the ones simulated during the proto-neutron star stage. In reality, there are numerous MHD timescales that can influence the evolution of the magnetic topology. On one side, proto-neutron star will experience a shrinking to the final neutron star size, which will probably imply an increase of the energy by flux conservation (here considered since we obtain similar volume-integrated energy spectra). 
On the other side, the smaller-scale will experience a fast decay as soon as the dynamo processes will stop feeding them. For this reason, we selected the first $l\leq 20$ multipoles. However, the details of the spectral slope for the small-scales are not important, since the Hall effect quickly regenerates them via direct cascade (Chapter~\ref{chap: MATINS}). 

\subsection{Results}
\label{subsec: results 3DMT}

In this section, we present the results of our simulation. The studied model has an average magnetic field of a few $10^{14}$\,G and a dominant Hall evolution (the magnetic Reynolds number is greater than unity with an average of $R_m \sim 200-300$ and a maximum reaching up to $R_m \sim 1600$ during evolution). We adopt a grid resolution of $N_r=40$ and $N_\xi=N_\eta=43$ per patch (a cubed-sphere has $6$ patches), corresponding to a total number of resolved multipoles in the system of about $l_\text{max} \sim 40$. We follow the evolution for $100$\,kyr, an age where most neutron stars have cooled down enough to be hardly detectable (bolometric thermal luminosity $L < 10^{32}$\,erg\,s$^{-1}$).
For a more quantitative analysis of the 3D magnetic evolution, we survey the magnetic energy spectrum to observe the redistribution of the magnetic energy over the different spatial scales.

\begin{figure*}
\centering
\includegraphics[width=0.9\textwidth]{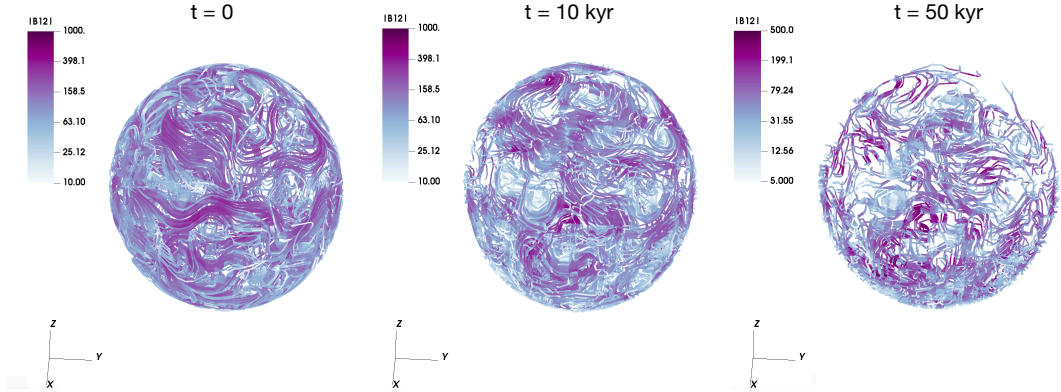}
\caption[3D field lines]{Field lines in the crust of a neutron star at $t=0$ (on the left), $t=10$\,kyr (in the center) and $t=50$\,kyr (on the right). The color scales indicates the local field intensity in units of $10^{12}$\,G.}
\label{fig: field lines visit}
\end{figure*}

We analyze the $l$ and $m$ energy spectra at different evolution times, depicted in the left and right panels of Fig.~\ref{fig: energy spectrum}, respectively. The transfer of magnetic energy occurs from large- to small-scales, where the smaller-scales are initially empty, indicating a dominance of the direct cascade. Around $20-30$\,kyr, the magnetic energy spectra reach a quasi-stationary state known as Hall-saturation.
The dissipation of small-scale structures happens more rapidly compared to large-scale ones, leading to an enhancement of Ohmic heating in the system. However, the former continuously receive energy input from the latter, facilitated by the Hall term in the induction equation. This phenomenon is referred to as the Hall cascade, which arises due to the Hall-dominated dynamics. It results in an equilibrium distribution of magnetic energy across a broad range of multipoles, approximately following an $l^{-2}$ slope, as reported by \citep{goldreich1992,dehman2022}.

Throughout $\sim100$\,kyr of evolution, the total magnetic energy drops by about half an order of magnitude as indicated in Fig.~\ref{fig: Emag PT} (solid black line). We also display in the same figure the decomposition of the total magnetic energy (black) into its poloidal (red) and toroidal (blue) components. The dots correspond to the axisymmetric components ($m=0$) and the dashed lines to the non-axisymmetric ones ($m\neq0$). The studied initial topology has a poloidal field governed by its non-axisymmetric component whereas the toroidal field has a significant contribution from both axisymmetric and non-axisymmetric pieces. For the first $\sim 10^4$\,kyr, the dissipation of magnetic energy is relatively small compared to later time. One could notice that the magnetic energy is transferred from the toroidal to the poloidal field during the evolution. The most efficient transfer in terms of relative energy increase, is for the poloidal axisymmetric field, probably because it is initially much weaker than the others. It shows a significant growth of one order of magnitude during the first $40-50$\,kyr. A saturation occurs for rest of the evolution (i.e., the system has reached the Hall balance). This quasi-constant energy trend appears for both the poloidal and the toroidal axisymmetric energy for $t \leq 50$\,kyr. At the same time, a strong transfer of magnetic energy to the poloidal non-axisymmetric component takes place. The non-axisymmetric toroidal energy tends to dissipate faster than the total magnetic energy in the system. That is because part of the toroidal non-axisymmetric is transferred to the poloidal non-axisymmetric energy. The latter dominates the axisymmetric toroidal component at $\sim 10$\,kyr and the non-axisymmetric one at $\sim 50$\,kyr. Nevertheless, the total toroidal field governs the magnetic energy at all time (solid blue line).

For a further understanding, we study in Fig.~\ref{fig: TP} the evolution in time of the spectra. We display four snapshots of the toroidal (blue)/ poloidal (red) and axisymmetric (dots)/non-axisymmetric (dashed lines) components as a function of the spherical harmonics degree $l$ at different evolution time, e.g., $t=0, 10,50$ \& $100$\,kyr. 
At $\sim10$\,kyr, a transfer of energy to the non-axisymmetric toroidal dipole (blue dashed line, $l=1$) takes place due to the inverse Hall cascade. A significant transfer of energy to the poloidal axisymmetric field happens at large- and small-scales. Instead, at late time (Hall balance is reached in the system) 
the dipolar component dissipates. For a better understanding, we discuss the behaviour of small- and large-scales independently. 
During the evolution, the small-scale modes ($ 10 \leq l \leq 40$) gain a significant fraction of the magnetic energy thanks to the Hall cascade in the system. The different behaviors are listed below
\begin{itemize}
    \item At $t=10$\,kyr, both toroidal components (i.e., axisymmetric and non-axisymmetric) decay in time, whereas the two poloidal components gain energy at small-scales. That also agrees with Fig.~\ref{fig: Emag PT} and indicates equipartition at small-scales, since isotropy is easier to achieve. As a matter of fact, the peculiar crust geometry (a thin shell of $\sim 1$\,km) and the strong stratification, play against isotropy and, therefore, equipartition between large-scale components.    
    \item At about $40-50$\,kyr, the system approaches a Hall-balance (see also Fig.~\ref{fig: energy spectrum}). At this stage, one can notice a slightly different behaviour for the axisymmetric and the non-axisymmetric components. On one hand, the axisymmetric components reach an approximate equipartition of the magnetic energy between poloidal and toroidal components at small-scales. That is in agreement to what was found using the axisymmetric 2D code \citep{pons2019}. A change of phase in the oscillations occur on a timescale of $40-50$\,kyr. On the other hand, for the non-axisymmetric modes, the slight toroidal dominance over poloidal seen at $10$\,kyr inverts at $50$ and $100$\,kyr. That is due to a transfer of energy from toroidal to poloidal field which, however, can be interpreted as equipartition of energy on the small isotropic scales. 
\end{itemize}
Minor differences appear in the spectrum at $50$ and $100$\,kyr and that is because the system has reached the Hall balance and the spectra remains stationary as it is shown in Fig.~\ref{fig: energy spectrum}. Note also how the large-scales barely evolve, compared to the others. This is another confirmation that the neutron star tends to a universal behaviour (Hall cascade) for intermediate and small-scales, but it has a strong memory of the large-scale magnetic topology at birth \citep{dehman2022}. This has important implications to relate current observables to the formation process (proto-neutron star stage).

The field lines in the crust of a neutron star at $t=0$, $10$ and $50$\,kyr are displayed in Fig.~\ref{fig: field lines visit}. The magnetic field lines are highly multipolar and many small-scale structures cover the surface. The field lines are very tangled throughout evolution, making it difficult to discern any clear dominant component.  

\begin{figure}
    \centering
      \includegraphics[width=0.65\textwidth]{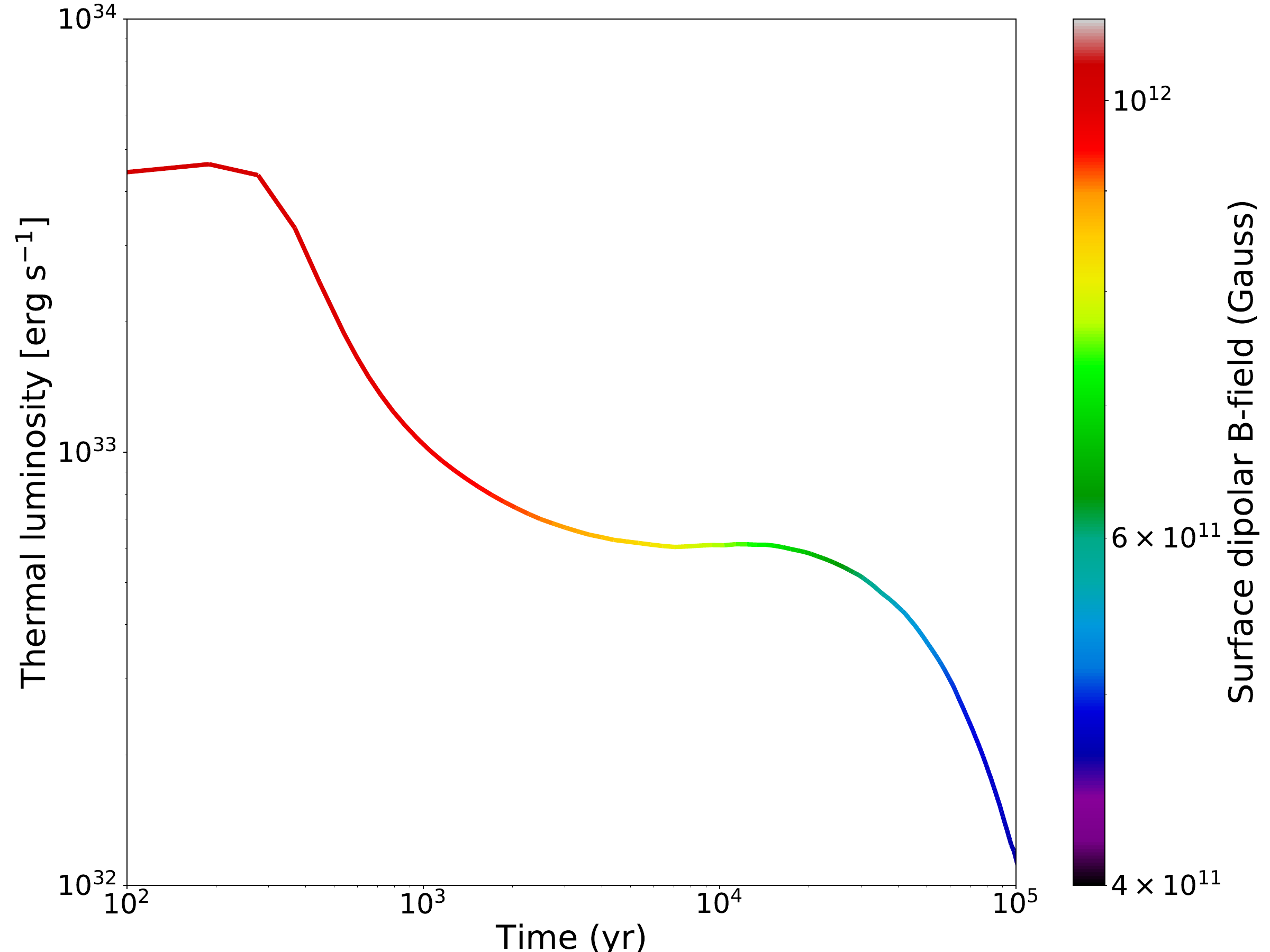}
    \caption[Luminosity curve as a function of time.]{Luminosity curve as a function of time. The colorbar indicates the evolution of the dipolar component of the poloidal magnetic field at the surface of the star (value at the magnetic pole). }
    \label{fig: luminosity bdip}
\end{figure}

In Fig.~\ref{fig: luminosity bdip}, we show the evolution of the thermal luminosity and (in color scale) of $B_p$, the dipolar component of the poloidal magnetic field at the surface of the star (value at the magnetic pole). 
For such an initial configuration, the luminosity ranges from $5\times10^{32}-10^{33}$\,erg/s in the neutrino cooling era, soon after, in the photon cooling era (t $\sim 10^5$\,yr), the luminosity drops sharply below $10^{32}$\,erg/s. The abrupt change in the slope of the cooling curves at ages $\sim 300$\,yr is due to the choice of the superfluid gap \citep{ho2015}. 
The rapid cooling during the photon cooling era is also caused by the low core heat capacity, which in turn depends on the assumed pairing details. A comprehensive revision of the microphysics embedded in magneto-thermal models can be found in \S\ref{sec: microphysics}. On the other hand, $B_p$ drops from $\sim 10^{12}$\,G to $\sim 4\times 10^{11}$\,G without any noticeable increase. 

At the surface of the star, we define the average field strength in each multipole $l$ as follows 
\begin{align}
 \bar{B}^\text{surf}_l &=  \bigg[ \frac{1}{4\pi} \int d\Omega \big((B^r)^2+ (B^\theta)^2 + (B^\phi)^2\big) \bigg]^{0.5}              \nonumber\\ 
& =\bigg[ \frac{B_0^2}{4\pi} \sum_m (b^m_l)^2  ~ \bigg( e^{-2\lambda(R)} (l+1)^2  + l(l+1)\bigg)\bigg]^{0.5},~ 
    \label{eq: B surface}
 \end{align}
where $e^{-2\lambda(R)}$ is the relativistic metric correction at the surface, $b_l^m$ are the dimensionless weights of the multipoles entering in the spherical harmonics decomposition of the radial magnetic field (eq.\,\eqref{eq: blm}), and $B_0$ is the normalization used in the code.

\begin{figure}
\centering
\includegraphics[width=0.49\textwidth]{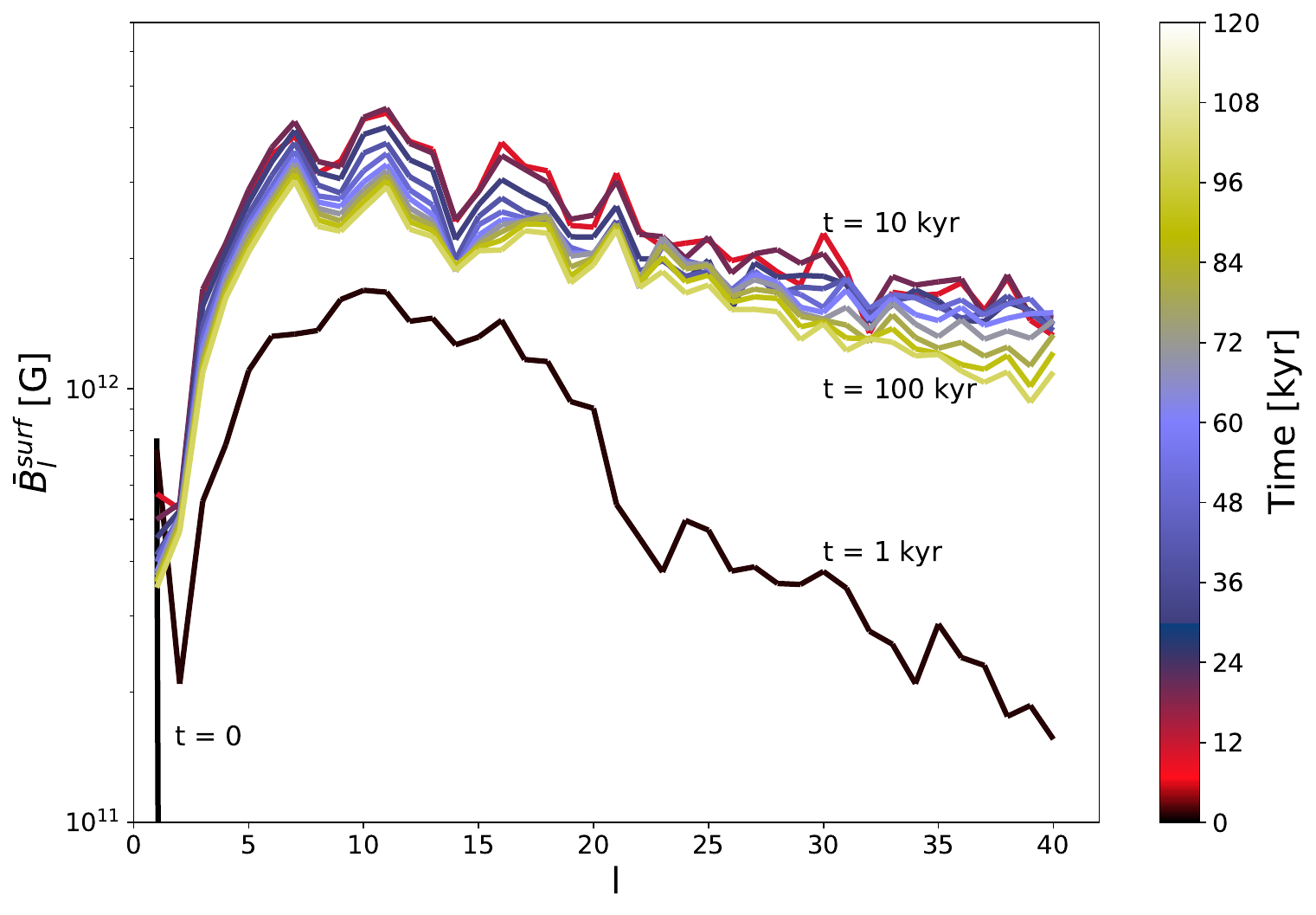}
\includegraphics[width=0.49\textwidth]{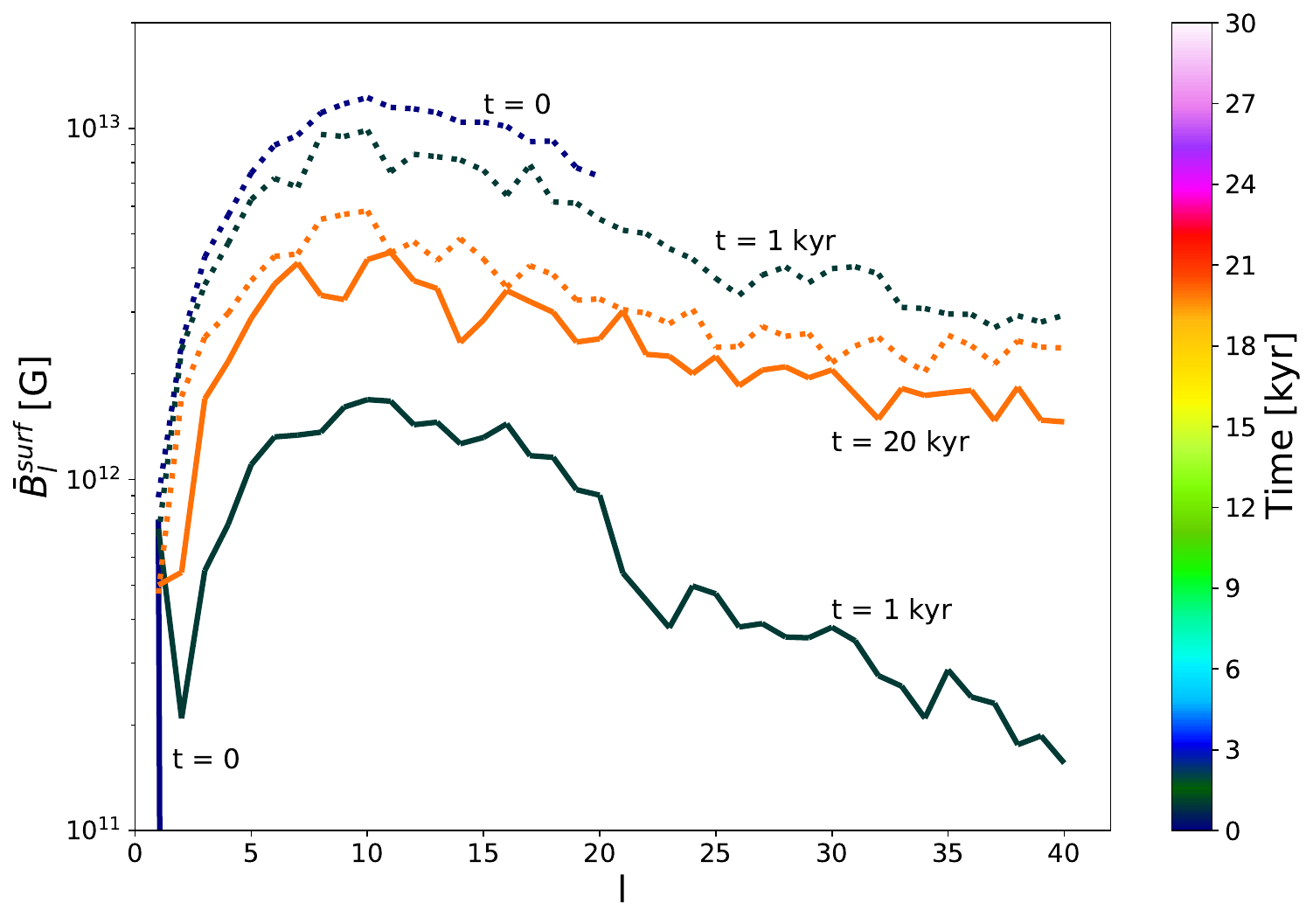}
\caption[Time evolution of the surface average field strength as a function of $l$]{Time evolution of the surface average field strength as a function of $l$ (eq.\,\eqref{eq: B surface}). 
\emph{Left Panel:} evolution up to $100$\,kyr. The colorbar reflects the evolution in time. \emph{Right Panel:} Comparison of the evolution for different radial functions at $t=0$ (black), $1$\,kyr (red) and $20$\,kyr (yellow). The solid lines correspond to the radial function that fits the potential boundary conditions (\S\ref{appendix: initial conditions} of Appendix~\ref{appendix: Magnetic field formalism}) and results in an initial purely dipolar field at surface (black solid lines), whereas the dots correspond to \citep[eq.\,(8)]{aguilera2008} applied to both toroidal and poloidal field.
The latter allows an initial distribution of small- and large-scale multipoles on the star's surface.}
\label{fig: blm surface}
\end{figure}

In the left panel of Fig.~\ref{fig: blm surface}, we show the time evolution of $\bar{B}^\text{surf}_l$ as a function of $l$.  
At t=0, the neutron star surface is dominated by the dipolar mode due to the specific radial function that we assume for each multipole in this case. However, as soon as we start the evolution the small-scale modes become dominant at about $1$\,kyr. Then the system reaches some sort of balance throughout the evolution. The balanced configuration is dominated by small-scale structures only ($l\geq 10$). This finding implies that during the evolution the $B_p$ value (the colorbar of Fig.~\ref{fig: luminosity bdip}) is much weaker than the one inferred for most magnetars ($10^{14}-10^{15}$\,G), when the classical spin-down formula is used to interpret the timing properties.

So far we have focused only on different choices of the tangential distribution of magnetic energy (the initial multipole weights). However, the initial radial profile of the topology (i.e., the set of radial functions of each multipole $(l,m)$) is also an important parameter that potentially affect our results (also connected to the outer and inner boundary conditions). For this reason, in the right panel of Fig.~\ref{fig: blm surface}, we compare the evolution for two different initial sets of radial functions (both equally arbitrary), which result in very different initial shapes of $\bar{B}^\text{surf}_l$. The solid lines correspond to a set of radial functions that allow a smooth matching with a pure dipole outside (see \S\ref{appendix: initial conditions}), confining inside all the other multipoles. The dots correspond to another set of radial functions that allows an initial distribution of small- and large-scale multipoles on the star's surface, with an initial non-zero tangential current that is quickly dissipated by the re-arrangement of the field. 
Both simulations are quickly dominated by small-scale structures (around $\sim 1$\,kyr, shown in red) and converge to a similar spectral distribution of $\bar{B}^\text{surf}_l$ after approximately $20$\,kyr, signifying the attainment of Hall balance.
From the comparison of these two simple choices, it seems that, independently from the initial radial distribution of the magnetic field inside the neutron star's crust, the star surface is anyway dominated by small-scale structures. 

\subsection{Discussion}
\label{subsec: discussion 3DMT}

Based on our analysis of the results (\S\ref{subsec: results 3DMT}), we observed that the surface dipolar component experiences no significant growth over this timescale, regardless of the initial radial distribution of the magnetic field in the neutron star crust (see right panel of Fig.~\ref{fig: blm surface}). We argue that a dominant surface dipolar magnetic field could only arise if the initial magnetic energy distribution is primarily concentrated in the dipolar component alone (either in the crust or in the core). Similar to our previous work and PARODY-based studies \citep{gourgouliatos2020, igoshev21}, we find a wide range of spatial scales in which the magnetic energy is distributed, and the large-scale components are sub-dominant. However, it remains unclear how core collapse could lead to a neutron star with such an almost pure dipolar configuration. In fact, several observations suggest that internal non-dipolar components (such as toroidal and multipolar components) are dominant in various astrophysical sources, including low field magnetars \citep{rea2013, tan2023}, high-B pulsars \citep{zhu2011}, CCOs with outbursts \citep{rea2016}, spectral features in magnetars \citep{tiengo2013}, and XDINS \citep{borghese2017}. Lastly, in \cite{dehman2020}, the 2D simulations demonstrated that the dipolar poloidal field ($B_p$) played a minor role in determining magnetars' bursting activity, and the magnetic energy stored in the star's crust served as a better indicator.

These features, already partially explored in \cite{gourgouliatos2020,igoshev21} (where they started with only small-scale multipoles), could be suitable for describing CCOs, X-ray sources with luminosity ranging between $10^{32}-10^{34}$\,erg/s, located at the centres of supernova remnants, with an estimated surface dipole magnetic field in the range $10^{10}-10^{11}$\,G. The dissipation of such weak large-scale component alone cannot provide sufficient thermal energy to power their observed X-ray luminosity. It is believed instead that CCOs have a hidden strong magnetic field due to the fall-back accretion \citep{ho2011,vigano2012}. This magnetic field dissipates in the star interior to provide the bright thermal luminosity of CCOs (Fig.~\ref{fig: luminosity bdip}). 

It is plausible that certain physical effects may come into play during the collapse or in the early life of the neutron star, resulting in the presence of large-scale dipole magnetic fields at the star's surface, with intensities exceeding $10^{13}$\,G. One potential mechanism could involve a strong inverse Hall cascade triggered by helical magnetic fields (e.g., \cite{brandenburg2020} for box simulations). However, such behavior has not been observed in global simulations thus far, indicating that there might be some crucial dynamics that we are currently overlooking, which could lead to the organization of sustaining currents on a large scale. 

On one hand, the proto-neutron star dynamo simulations are intrinsically influenced by the chosen simplistic boundary conditions (see, e.g., \cite{raynaud20} for a comparison between perfect conductor and potential configurations), which are, in any case, not realistic for the dense, hot, and plasma-filled environment of the proto-neutron star. On the other hand, the magnetic field sustained by currents in the core might be significant, particularly in the long-term, given that its evolution timescales are presumably longer (although see \cite{gusakov2020}). Nevertheless, it remains uncertain how to generate a strong poloidal dipolar field, supported by robust toroidal and organized currents, during the proto-neutron star or early neutron star stages.

A radically different alternative, compatible with the absence of strong dipolar fields, is conceivable. It is possible that the values of $B_p$ are consistently lower if the electromagnetic torque is primarily dominated by magnetospheric effects, such as particle winds and the presence of strong and extended loops charged with particles. This concept finds support in observations of the Sun \citep{yeates2018} and Zeeman-Doppler studies of main sequence stars \citep{neiner2009}, which have inferred magnetospheric topologies displaying field lines stretched by the wind and exhibiting complex configurations. The presence of strong magnetospheric structures could also align with the existence of small hotspots (with radii less than $1$\,km), as commonly inferred from magnetars' X-ray thermal spectra during quiescence or outbursts. Additionally, these structures could be associated with the more extensively studied resonant Compton scattering, which gives rise to the observed non-thermal tails in spectra. However, quantifying the supplementary contributions of the magnetosphere to the spin-down remains limited by the models proposed by \cite{tong2013}, warranting further investigation to support this scenario. In this regard, the coupling of the interior evolution with the magnetosphere plays a crucial role in enabling the flow of currents and, ultimately, leading to the presence of a larger surface dipolar magnetic field (Appendix~\ref{appendix: neural networks}).

Finally, this study marks a groundbreaking achievement as we present, for the very first time, a fully 3D coupled magneto-thermal evolution applied to an initial complex magnetic field topology, while incorporating a realistic star structure, the microphysics, and the general relativistic corrections. Our simulations complement and extend recent efforts in describing crust-confined 3D magnetic field evolution from \emph{PARODY}-based published works \citep{wood2015, gourgouliatos2016, gourgouliatos2020, degrandis2020, degrandis2021, igoshev21a, igoshev21}, which had limitations in predicting thermal luminosity. Moreover, our simulations break free from axial symmetry restrictions, employing the cubed-sphere formalism to properly treat the axis, distinguishing them from the conventional 2D magneto-thermal code \citep{vigano2012, vigano2021} and box simulations \citep{brandenburg2020}.

\subsubsection*{Corresponding scientific publications}
\underline{C.~Dehman}, D.~Vigan\`o, J.A.~Pons \& N.~Rea: 2022, \textbf{3D code for MAgneto$-$Thermal evolution in Isolated Neutron Stars, MATINS: The Magnetic Field Formalism}, \emph{Mon.~Not.~Roy.~Astron.~Soc., $518$, $1222$} 
(\href{https://arxiv.org/abs/2209.12920}{\underline{arXiv:2209.12920}},\href{https://ui.adsabs.harvard.edu/abs/2023MNRAS.518.1222D/abstract}{\underline{ADS}},\href{https://doi.org/10.1093/mnras/stac2761}{\underline{DOI}}). \\ \\ 
\underline{C.~Dehman}, D.~Vigan\`o, S. Ascenzi, J.A.~Pons \&  N. Rea: 2023, \textbf{3D evolution of neutron star magnetic-fields from a realistic core-collapse turbulent topology}, 
\emph{Mon.~Not.~Roy.~Astron.~Soc. $523$, $5198$} (\href{https://arxiv.org/abs/2305.06342}{\underline{arXiv:2305.06342}},\href{https://ui.adsabs.harvard.edu/abs/2023arXiv230506342D/abstract}{\underline{ADS}},\href{https://doi.org/10.1093/mnras/stad1773}{\underline{DOI}}).

\clearemptydoublepage
\let\textcircled=\pgftextcircled
\chapter{Conclusions}
\label{chap: conclusions}

\initial{I}n this thesis, we have presented a comprehensive study of the magneto-thermal evolution in isolated neutron stars. Our investigations have shed light on the physics of neutron star interiors, which operate under extreme conditions within these compact celestial objects. Throughout our research, we have conducted numerous astrophysical applications, covering both theoretical and observational aspects. Furthermore, we developed a groundbreaking 3D code called \emph{MATINS}, which accurately models thermal luminosity, timing properties, and magnetic field evolution, while considering realistic star structures and microphysics. Notably, our simulation tackled the highly complex initial magnetic field topology in the crust, similar to that recently obtained by proto-neutron star dynamo simulations \citep{dehman2023c}. As a result, \emph{MATINS} stands as the most realistic 3D coupled magneto-thermal code to date.

Several key conclusions can be drawn from the diverse studies performed during this doctorate. To begin with, a comprehensive understanding of the cooling process in young and middle-aged magnetars is crucial for interpreting observational data. In particular, we reevaluated neutron star cooling curves, taking into account envelope models and extreme magnetic field topologies. Our research revealed that the surface temperature ($T_s$) is highly sensitive to the magnetic field strength. Specifically, the state-of-the-art envelope model \citep{potekhin2015} predicted significantly higher $T_s$ for strong magnetic fields compared to older models, leading to substantial changes in photon luminosity: from high luminosity epoch (during the neutrino cooling era) to a very low luminosity phase (soon after entering the photon cooling era). The impact of this trend is most evident in models where the magnetic field penetrates the star's core, with electric currents circulating there. In contrast, in crustal-confined models, the additional energy released by Joule heating close to the star's surface is very effective and governs the energy balance equation, which, in turn, counterbalances the effect. As a consequence, the location of bulk electrical currents within the star determines whether middle-aged magnetars appear relatively bright or exhibit very low luminosities, making their persistent emission undetectable as X-ray sources.

Our results also contribute to a better understanding of different isolated neutron star classes based on observational data. For instance, bright middle-aged magnetars like 1E 2259+586 can be explained by crustal-field with magnetized light envelopes, while all XDINSs require a strong component of crustal-field. However, the envelope model can vary, being either light or heavy, and either magnetic or non-magnetic. Additionally, the middle-aged faint pulsars, such as PSR B0656+14, may be explained by the presence of a light element envelope. This becomes particularly relevant for neutron stars with low magnetic fields, where light envelopes result in cooler neutron stars compared to heavy envelopes as they age beyond $10^4$\,yrs, irrespective of the field configuration. Furthermore, the existence of strongly magnetized neutron stars with detectable thermal emission at later times strongly supports the presence of a crustal magnetic field.

This study emphasizes the crucial consideration of all factors in the complex theory of neutron star cooling. Neglecting the envelope's role or using non-magnetized envelopes can lead to discrepancies up to one order of magnitude compared to observational data. Conversely, accurately estimating surface luminosity is vital for constraining source properties like surface magnetic field strength or age. These insights have significant implications for population synthesis studies of pulsars and magnetars. The inability to detect certain sources introduces observational biases that can greatly influence birth rate and field distribution predictions.

In Chapter~\ref{chap: Comparison with observations}, a thorough comparison was conducted between observations and theoretical simulations. To comprehend the behavior of cold and young observed pulsars, machine learning techniques were employed for statistical analysis. The results revealed that EoSs capable of triggering fast-cooling processes within the first $10^3$\,yrs of their lifespan can account for these cold and young sources. This study contributes significantly to constraining nuclear EoS, as a suitable EoS should elucidate both exceptionally bright objects like magnetars and extremely faint objects at young ages. Taking into account a simplified meta-modeling approach \citep{margueron2017}, the excluded range is estimated to encompass approximately 75\% of the proposed EoSs.

Then in Chapter~\ref{chap: outburst}, we focused on exploring the magnetic-related activity and crustal failures in young magnetars. Through simulations, we made a remarkable discovery: the dipolar component, which most FRB-magnetar models currently rely on, actually plays a minor role in determining the bursting activity of the magnetosphere when triggered by crustal failures. Our investigations revealed that the crustal failure rate changes significantly by several orders of magnitude when varying the magnetic field configuration from a large-scale, smooth core-threaded field to a multipolar field confined in the crust, even with the same dipolar poloidal value ($B_p$). However, we observed that the crustal magnetic energy serves as a reliable tracer of the expected number of events, regardless of the specific field topology. To better constrain the detectable fraction of events, further observations and a deeper understanding of disturbance propagation and burst emission mechanisms are essential.

The core of this thesis lies in the development and application of the \emph{MATINS} 3D code. The \emph{MATINS} code will soon be available to the public, providing a valuable resource for the high-energy astrophysics community focused on exploring various astrophysical aspects related to highly magnetized neutron stars. The code can be accessed at the following URL:
\begin{center}
\href{https://github.com/csic-ice-magnesia/MATINS}{https://github.com/csic-ice-magnesia/MATINS.}
\end{center}
We have intentions to consistently enhance and expand this open-access code at regular intervals.

\emph{MATINS} deals with realistic nuclear EoS allowing to obtain a realistic star's structure and microphysics. The code is based on finite volume scheme applied to the cubed-sphere formalism, and it is second-order accurate in space and fourth-order accurate in time. The cubed-sphere formalism is a peculiar gridding technique widely used in different fields of physics, and it allows to solve partial differential equations in spherical geometry avoiding the axis singularity problem: a common problem that emerges when adopting finite volume/difference scheme in spherical coordinates.

We have shown that \emph{MATINS} is stable and can follow to late times the evolution of the internal coupled magneto-thermal evolution in the crust of neutron stars. It conserves the total energy contained in the system and the divergence-free condition of the magnetic field. Moreover, it has been extensively tested, against analytical solutions, e.g., the purely resistive test (\S\ref{subsec: bessel test}), and numerical axisymmetric solutions replicable by our 2D code (\S\ref{subsec: axisymmetric - comparison 2D and 3D}). 

The simulations performed using the \emph{MATINS} code are described in Chapter~\ref{chap: 3DMT}. Initially, we adopted an isothermal cooling profile based on \cite{yakovlev2011}. Different initial field configurations were explored (\S\ref{appendix: initial conditions}) using this code. Our simulations (see \S\ref{sec: isothermal cooling}) confirm that a strong magnetic field in the range of $10^{14}-10^{15}$\,G, results in the Hall cascade redistributing energy across a wide range of scales, exhibiting a slope of approximately $\sim l^{-2}$. Moreover, at small scales, there is an approximate equipartition of energy between the poloidal and toroidal components. However, reaching this saturated configuration, known as the Hall attractor, takes tens of kyr, which coincides with the typical activity timescale of magnetars. During this stage, the spectra and topology retain a strong memory of the initial large scales, which are considerably more resistant to restructuring or creation. This underscores the significance of the large-scale configuration during neutron star formation in determining the magnetic field topology at any age during its evolution.

Finally, we tackled the heat diffusion equation in 3D and conducted simulations considering an exceptionally complex initial magnetic field topology derived from proto-neutron star dynamo simulations \citep{dehman2023c}. This marks a pioneering achievement as we present, for the first time, the most realistic 3D simulation to date. It includes fully 3D coupled magneto-thermal evolution (\S\ref{sec: 3DMT}), a realistic star's structure, temperature-dependent microphysics, solves the axis singularity problem thanks to the cubed-sphere formalism, accounts for relativistic corrections, and utilizes the state-of-the-art envelope model existing in the literature, along with an initial magnetic field topology derived from proto-neutron star dynamo simulations. With this initial setup, we found that the simulation does not account for the X-ray luminosity observed in Magnetars. Instead, it can explain the luminosity observed in low-field magnetars and CCOs. Furthermore, the surface dipolar component, responsible for the dominant electromagnetic spin-down torque, does not exhibit any increase over time when initiated from this initial topology.

The work presented here is not without limitations. The potential boundary conditions assumed in the simulations, where the current is not allowed to circulate outside the star, may affect our results. However, we have recently conducted a study (Appendix~\ref{appendix: neural networks}) where we coupled the interior evolution with the magnetosphere in 2D, using the PINN approach. This study highlighted the significance of such coupling on the interior evolution. The PINN approach has made this investigation more feasible in three dimensions, allowing for more comprehensive insights into the system. Currently, we are actively working on extending these results to the 3D domain. 

Furthermore, we anticipate that future research will delve into understanding the 3D magnetic field evolution in the core of a neutron star. Additionally, our finite volume code, utilizing cubed-sphere coordinates, could be extended and generalized to include MHD simulations and turbulence, eliminating the need to cut the axis. This enhancement would enable a fully spherical three-dimensional finite volume/difference code capable of studying turbulence.

In conclusion, the work presented here has made a significant contribution to the field of neutron star research, offering crucial insights into their complex magneto-thermal evolution. These findings effectively bridge the gap between theoretical models and observational data, providing valuable understanding of the behavior of these enigmatic cosmic entities.
Moreover, the versatility of these simulations extends to various astrophysical studies, encompassing investigations into crustal failures, outbursts, population synthesis, and more. We are hopeful that this research will establish a robust foundation for future investigations into the physics of neutron stars and their astrophysical significance.

\clearemptydoublepage

%
\appendix
\let\textcircled=\pgftextcircled
\chapter{The Cubed-Sphere Formalism}
\label{appendix: cubed-sphere formalism}

\section{Coordinates transformations}
\label{app:coordinates}

All evolution calculations represented in Chapter~\ref{chap: MATINS} and \ref{chap: 3DMT} are performed in the cubed sphere coordinates, with a few exceptions. The potential boundary conditions are imposed in spherical coordinates (\S\ref{subsec: outer B.C}).
Thus, a transformation from spherical to cubed-sphere coordinates is needed at each magnetic timestep. This transformation has also been used in some other cases, in particular when defining the initial magnetic field in spherical coordinates, e.g., the Bessel test (\S\ref{subsec: bessel test}). Instead, the transformation from cubed-sphere to spherical coordinates is applied to generate the output files.

We follow the same approach as in \cite{ronchi1996}, but using Schwarzschild interior metric solution of the TOV equation (\S\ref{sec: TOV}). We introduce the auxiliary variables that will be used in our formalism
\begin{align}
   X &\equiv \tan(\xi), \nonumber\\    
   Y &\equiv \tan(\eta), 
    \nonumber\\  
   \delta &\equiv 1+ X^2 + Y^2,
     \nonumber\\
     C &\equiv (1+ X^2)^{1/2} \equiv  \frac{1}{\cos (\xi)},
 \nonumber\\
 D &\equiv (1+ Y^2)^{1/2} \equiv  \frac{1}{\cos (\eta)}.
   \label{eq: cubed sphere variables app}
\end{align}

The metric tensor exhibits the same functional dependence on the auxiliary variables across all patches. In the unit vector basis, it can be expressed as follows:
\begin{equation} 
g_{ij}=
\begin{pmatrix}
1 &0  & 0\\
0 & 1 &  - \frac{XY}{CD} \\
0 &  - \frac{XY}{CD} & 1
\end{pmatrix}
\label{eq: metric tensor app}
\end{equation}
The inverse of the tensor metric reads
\begin{equation}
g_{ij}^{-1}=
\begin{pmatrix}
1 &0  & 0\\
0 & \frac{C^2 D^2}{\delta} &   \frac{CDXY}{\delta} \\
0 &  \frac{CDXY}{\delta} & \frac{C^2 D^2}{\delta}
\end{pmatrix}
\label{eq: inverse metric tensor app}
\end{equation}
In all patches, the radial versor $\boldsymbol{e}_r$ is orthogonal to the plane formed by $\boldsymbol{e}_\xi$ and $\boldsymbol{e}_\eta$ unit vectors, which are not in general orthogonal to each other.

The covariant components (lower indexes) of a vector are obtained by contracting the metric tensor with its contravariant components, as follows:  
\begin{equation}
V_i = g_{ij} V^j.
\label{eq: contravariant to covariant}
\end{equation}
Conversely, the contravariant components (upper indices) of a vector are obtained by contracting the inverse of the metric with the covariant components of a vector.
\begin{equation}
V^i= g_{ij}^{-1} V_j.
\label{eq: covariant to contravariant}
\end{equation}

\begin{figure}
	\centering
	\includegraphics[width=0.7\textwidth]{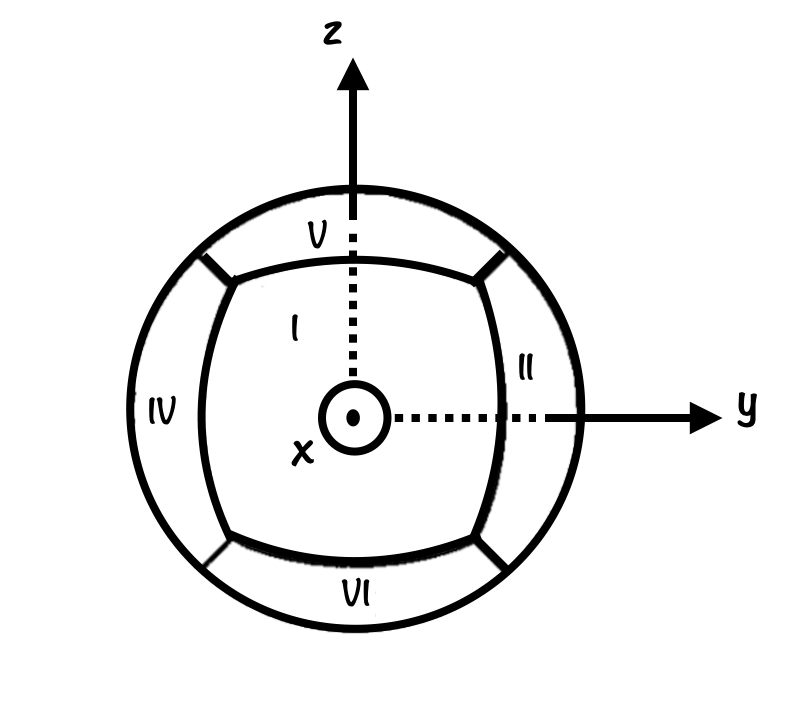}
	\caption[The cubed sphere represented with respect to Cartesian coordinates.]{The cubed sphere represented with respect to Cartesian coordinates. The x-axis is directed outwards piercing the centers of patch I ($\boldsymbol{e}_x$) and III ($-\boldsymbol{e}_x$, behind, not visible), the y-axis goes to the right, piercing the centers of patch II ($\boldsymbol{e}_y$) and IV ($-\boldsymbol{e}_y$), the-z axis is directed upwards, piercing the centers of patch V ($\boldsymbol{e}_z$) and VI ($-\boldsymbol{e}_z$).}
	\label{fig: Cubed-sphere in cartesian axis}
\end{figure}

The spherical coordinates consist, as usual, of: $r\in [R_c,R_\star]$, representing the distance to the origin between the crust-core interface and the surface; $\theta \in [0,\pi]$, known as the co-latitude, or polar angle, which measures the angle with respect to the North pole ($x=y=0$, positive $z$ in Cartesian coordinates); $\phi \in [0,2\pi]$, denoting the azimuth angle defined in the $x-y$ plane, starting from the $x$-axis. Each patch of the unit sphere is centered around a Cartesian axis, as illustrated in Fig.~\ref{fig: Cubed-sphere in cartesian axis}.
The transformations between these different coordinate systems are identical to those described in \cite{ronchi1996,lehner2005}.

The coordinate directions of the patches are indicated in the exploded view of Fig.~\ref{fig:full_grid} and can be described qualitatively as follows:\footnote{The direction is denoted for brevity using the symbol ``$\rightarrow$'', and is exact only at the center of each patch. As the points approach the edges, the indicated direction becomes less straightforward, and the angular deviation increases up to a maximum of $\pi/4$ (notably at the three-patch common corners).}: 
\begin{itemize}
	\item Patch I: center in $x=1$ ($\theta=\frac{\pi}{2}, \phi=0$); $\boldsymbol{e}_\xi \rightarrow \boldsymbol{e}_y$, $\boldsymbol{e}_\eta \rightarrow \boldsymbol{e}_z$.
	\item Patch II: center in $y=1$ ($\theta=\frac{\pi}{2}, \phi=\frac{\pi}{2}$); $\boldsymbol{e}_\xi \rightarrow -\boldsymbol{e}_x$, $\boldsymbol{e}_\eta \rightarrow \boldsymbol{e}_z$.
	\item Patch III: center in $x=-1$ ($\theta=\frac{\pi}{2}, \phi=\pi$); $\boldsymbol{e}_\xi \rightarrow -\boldsymbol{e}_y$, $\boldsymbol{e}_\eta \rightarrow \boldsymbol{e}_z$.
	\item Patch IV: center in $y=-1$ ($\theta=\frac{\pi}{2}, \phi=\frac{3\pi}{2}$); $\boldsymbol{e}_\xi \rightarrow \boldsymbol{e}_x$, $\boldsymbol{e}_\eta \rightarrow \boldsymbol{e}_z$.
	\item Patch V: center in $z=1$ ($\theta=0$, $\phi$ undefined); $\boldsymbol{e}_\xi \rightarrow \boldsymbol{e}_y$, $\boldsymbol{e}_\eta \rightarrow -\boldsymbol{e}_x$.
	\item Patch VI: center in $z=-1$ ($\theta=\pi$, $\phi$ undefined); $\boldsymbol{e}_\xi \rightarrow \boldsymbol{e}_y$, $\boldsymbol{e}_\eta \rightarrow \boldsymbol{e}_x$.
\end{itemize}

With this notation, we have the following relations between cubed sphere, spherical and Cartesian coordinates, for each patch:\footnote{The transformation from cubed sphere to Cartesian coordinates is taken from \S4.1.3 of \cite{lehner2005}. They use a grid that is equally spaced in $b\equiv X$ and $a\equiv Y$, rather than being equally spaced in $\xi$ and $\eta$. Note also that their patches 0-5 correspond to I-VI here, in the same order. The $\theta(\xi,\eta)$ and $\phi(\xi,\eta)$ are derived in this work with simple trigonometric relations, starting from the definitions $X(\theta,\phi)$ and $Y(\theta,\phi)$. For the polar patches, we have employed the identities $\sin(\arctan\beta) = \pm  \mid \beta\mid /\sqrt{1+\beta^2}$  (i.e., $"+"$ if $\beta>0$ and $"-"$ if $\beta<0$) and $\cos(\arctan\beta)=1/\sqrt{1+\beta^2}$.}

\begin{itemize}
  \item Patch I (Equator)
  \begin{align}
  X&= y/x=\tan \phi \nonumber\\
  Y&= z/x=1/\tan \theta \cos \phi\nonumber\\
  x&= r/ \sqrt{\delta}, \hspace{1mm}y= rX/ \sqrt{\delta}, \hspace{1mm}z= rY/ \sqrt{\delta} \nonumber\\
  \theta &= \arctan[(\cos\xi\tan\eta)^{-1}] = \arctan(C/Y) \nonumber\\
  \phi &= \xi
  \end{align}
  
  \item Patch II (Equator)
  \begin{align}
  X&=  - x/y=-1/ \tan \phi \nonumber\\
  Y&= z/y=1/\tan \theta \sin \phi \nonumber\\
  x&= - r X/ \sqrt{\delta}, \hspace{1mm}y= r/ \sqrt{\delta}, \hspace{1mm}z= rY/ \sqrt{\delta} 
  \nonumber\\
  \theta &=  \arctan(C/Y) \nonumber\\
  \phi &= \xi + \frac{\pi}{2}
  \end{align}
  
  \item Patch III (Equator)
  \begin{align}
  X&=  y/x=\tan \phi \nonumber\\
  Y&= -z/x=- 1/\tan \theta \cos \phi \nonumber\\
  x&= - r/ \sqrt{\delta}, \hspace{1mm}y= - rX/ \sqrt{\delta}, \hspace{1mm}z= rY/ \sqrt{\delta}\nonumber\\
  \theta &=  \arctan(C/Y) \nonumber\\
  \phi &= \xi + \pi
  \end{align}

  \item Patch IV (Equator)
  \begin{align}
  X&= - x/y=-1/ \tan \phi \nonumber\\
  Y&= - z/y= - 1/\tan \theta \sin \phi\nonumber\\
  x&= rX/ \sqrt{\delta}, \hspace{1mm}y= - r/ \sqrt{\delta}, \hspace{1mm}z= rY/ \sqrt{\delta}\nonumber\\
  \theta &=  \arctan(C/Y) \nonumber\\
  \phi &= \xi + \frac{3\pi}{2}
  \end{align}

  \item Patch V (North)
  \begin{align}
  X&= y/z= \tan \theta \sin \phi \nonumber\\
  Y&= -x/z= - \tan \theta \cos \phi\nonumber\\
  x&= - rY/ \sqrt{\delta}, \hspace{1mm}y=  rX/ \sqrt{\delta}, \hspace{1mm}z= r/ \sqrt{\delta}\nonumber\\
  \theta &=   \arctan\sqrt{ \delta - 1} \nonumber\\
  \text{if } \hspace{1mm} (X>0,  Y<0) \hspace{2mm}
  \phi &= - \arctan(X/Y) \hspace{2mm} (\text{region}  \hspace{1mm}\alpha)
  \nonumber\\
  \text{if } \hspace{1mm} (X>0,  Y>0) \hspace{2mm}
  \phi &= \pi - \arctan(X/Y) \hspace{2mm} (\text{region}  \hspace{1mm}\beta)
    \nonumber\\
  \text{if } \hspace{1mm} (X<0,  Y>0) \hspace{2mm}
  \phi &= \pi - \arctan(X/Y) \hspace{2mm} (\text{region}  \hspace{1mm}\gamma)
    \nonumber\\
  \text{if } \hspace{1mm} (X<0,  Y<0) \hspace{2mm}
  \phi &= 2\pi - \arctan(X/Y) \hspace{2mm} (\text{region}  \hspace{1mm}\delta)   
  \end{align}
   
  \item Patch VI (South)
  \begin{align}
  X&= - y/z= - \tan \theta \sin \phi \nonumber\\
  Y&= -x/z= - \tan \theta \cos \phi \nonumber\\
 x&= rY/ \sqrt{\delta}, \hspace{1mm}y=  rX/ \sqrt{\delta}, \hspace{1mm}z= -r/ \sqrt{\delta}\nonumber \\
   \theta &=   \pi - \arctan(\sqrt{\delta}-1)  
   \nonumber\\
  \text{if } \hspace{1mm} (X>0,  Y<0) \hspace{2mm}
  \phi &= \pi + \arctan(X/Y) \hspace{2mm} (\text{region}  \hspace{1mm}\alpha)
  \nonumber\\
  \text{if } \hspace{1mm} (X>0,  Y>0) \hspace{2mm}
  \phi &= \arctan(X/Y) \hspace{2mm} (\text{region}  \hspace{1mm}\beta)
    \nonumber\\
  \text{if } \hspace{1mm} (X<0,  Y>0) \hspace{2mm}
  \phi &= 2\pi + \arctan(X/Y) \hspace{2mm} (\text{region}  \hspace{1mm}\gamma)
    \nonumber\\
  \text{if} \hspace{1mm} (X<0,  Y<0) \hspace{2mm}
  \phi &= \pi + \arctan(X/Y) \hspace{2mm} (\text{region}  \hspace{1mm}\delta)
  \end{align}
\end{itemize}
Please note that along the equatorial-centered patches I-II-III-IV, the $\xi$ coordinate coincides with the $\phi$ coordinate in spherical coordinates, with a phase shift of $(0, \pi/2, \pi, 3\pi/2)$ respectively. Additionally, the transformation into the $\theta$ coordinate remains the same in all four patches, as they cover the same co-latitude.

For the polar patches, the transformation is less trivial. It's important to remember that the $\arctan$ function tends to $\pi/2$ (i.e., patches I-IV and $X>0$ in patches V and VI) and $3\pi/2$ (i.e., $X<0$ in patches V and VI) if the argument tends to $\pm\infty$ (i.e., when the denominator $Y$ of the ratios $X/Y$ and $C/Y$ approaches zero). This behavior is essential to consider when dealing with the coordinate transformations in these polar regions.

\begin{figure}
	\centering
	\includegraphics[width=0.9\textwidth]{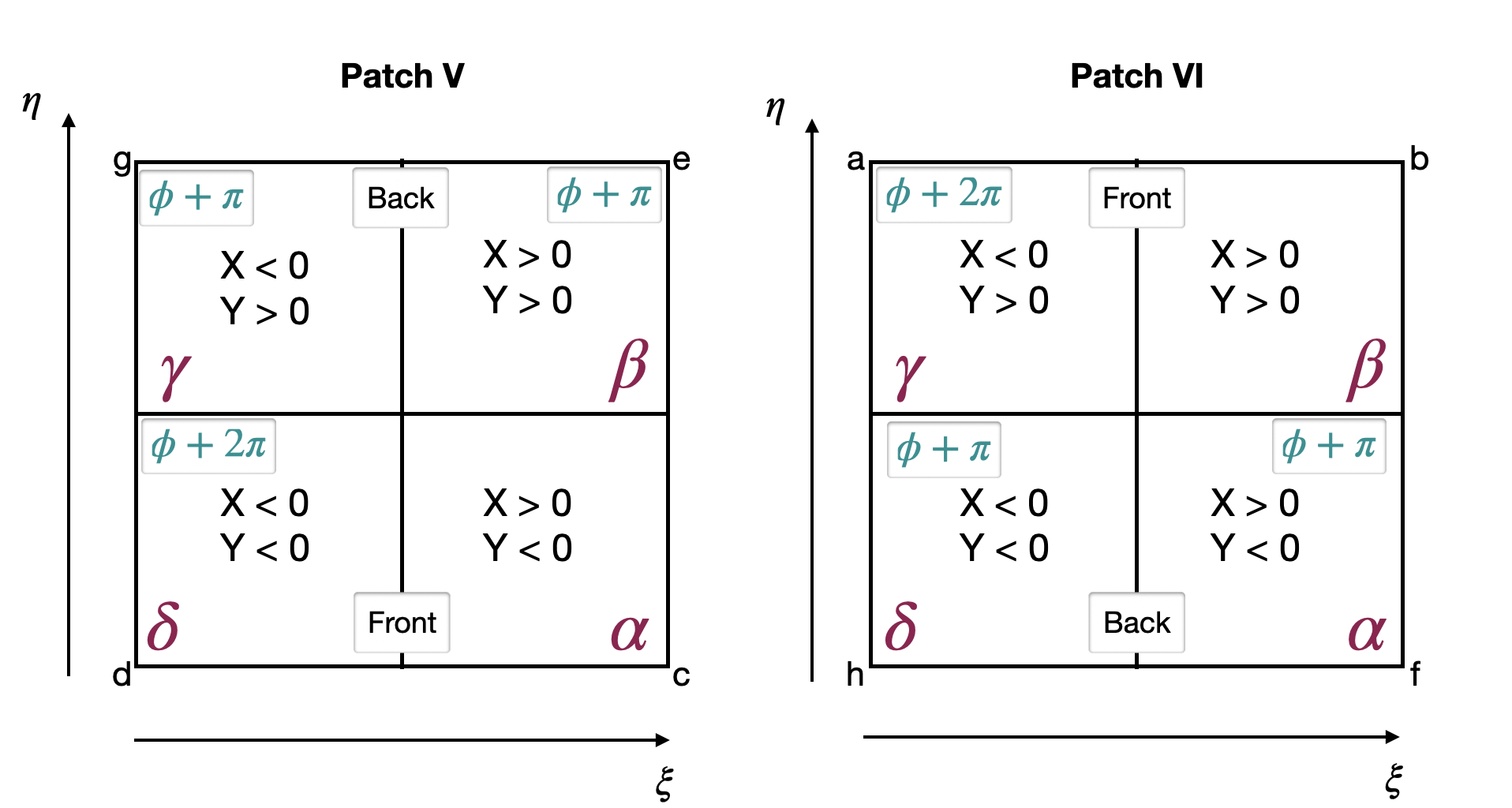}
	\caption[Subdivision of patch V and patch VI.]{Subdivision of patch V and patch VI. Each of these patches is divided into four sub-regions, and each of these sub-regions has a different sign of the $X/Y$ ratio. This subdivision is crucial to properly define $\phi$ in the range $[0:2\pi]$.
	} 
	\label{fig: subregions patch V and patch VI}
\end{figure}

To ensure that $\phi$ is defined in the range $[0;2\pi]$ and $\theta$ in the range $[0;\pi]$ within patch V and patch VI, a subdivision of each of these patches is required. This subdivision becomes crucial as the sign of the $X/Y$ ratio changes within these subregions, as shown in Fig.~\ref{fig: subregions patch V and patch VI}. Consequently, in order to guarantee that $\phi$ spans the interval $[0;2\pi]$, a "$+\pi$" adjustment (i.e., subregions $\beta$ and $\gamma$ of patch V and subregions $\alpha$ and $\delta$ of patch VI) or a "$+2\pi$" adjustment (i.e., subregion $\delta$ of patch V and subregion $\gamma$ of patch VI) must be added to the expression of $\phi$. This ensures proper continuity and coverage of the full range for the azimuthal angle.

\section{Jacobians}
\label{Appendix: Jacobians}

In order to transform vectors from spherical coordinates to cubed sphere coordinates, we need the Jacobian matrices. 
Hereafter we indicate only the $2\times 2$ Jacobian relating the transformation of the tangential components, since the radial coordinate is the same.

\begin{itemize}
	\item Patch I-IV (Equator)
	\begin{equation}
	\begin{pmatrix}
	A^\xi \\ A^\eta  
	\end{pmatrix} =
	\begin{pmatrix}
	0 & CD/\delta^{1/2}\\
	-1 & XY/\delta^{1/2}\\
	\end{pmatrix}
	\begin{pmatrix}
	A^\theta \\ A^\phi 
	\end{pmatrix}
	\label{eq: spherical to equator}
	\end{equation}
	
	\begin{equation}
	\begin{pmatrix}
	A^\theta \\ A^\phi 
	\end{pmatrix}
	=
	\begin{pmatrix}
	XY/CD & -1 \\
	\delta^{1/2}/CD & 0  \\
	\end{pmatrix}
	\begin{pmatrix}
	A^\xi \\ A^\eta  
	\end{pmatrix}
	\label{eq: equator to spherical}
	\end{equation}
	
	\item Patch V (North)
	\begin{equation}
	\begin{pmatrix}
	A^\xi \\ A^\eta  
	\end{pmatrix} =\frac{1}{\big(\delta-1 \big)^{1/2}} \begin{pmatrix}
	DX & -DY/\delta^{1/2}\\
	CY & CX/\delta^{1/2}  \\
	\end{pmatrix}
	\begin{pmatrix}
	A^\theta \\ A^\phi 
	\end{pmatrix} 
	\label{eq: spherical to patch 5}
	\end{equation}
	
	\begin{equation}
	\begin{pmatrix}
	A^\theta \\ A^\phi 
	\end{pmatrix}
	= \frac{1}{(\delta-1)^{1/2}}
	\begin{pmatrix}
	X/D & Y/C \\
	-Y \delta^{1/2} /D   & X \delta^{1/2} /C\\
	\end{pmatrix}
	\begin{pmatrix}
	A^\xi \\ A^\eta  
	\end{pmatrix}
	\label{eq: patch 5 to spherical}
	\end{equation}
	
	\item Patch VI (South)
	\begin{equation}
	\begin{pmatrix}
	A^\xi \\ A^\eta  
	\end{pmatrix} =\frac{1}{\big(\delta-1 \big)^{1/2}} \begin{pmatrix}
	- DX & DY/\delta^{1/2}\\
	- CY & -CX/\delta^{1/2}  \\
	\end{pmatrix}
	\begin{pmatrix}
	A^\theta \\ A^\phi 
	\end{pmatrix} 
	\label{eq: spherical to patch 6}
	\end{equation}
	
	\begin{equation}
	\begin{pmatrix}
	A^\theta \\ A^\phi 
	\end{pmatrix}
	=  \frac{1}{(\delta-1)^{1/2}}
	\begin{pmatrix}
	-X/D & - Y/C \\
	Y\delta^{1/2}/D   & - X\delta^{1/2}/C\\
	\end{pmatrix}  \begin{pmatrix}
	A^\xi \\ A^\eta  
	\end{pmatrix}
	\label{eq: patch 6 to spherical}
	\end{equation}
	
\end{itemize}
Keep in mind that the quantities $X$, $Y$, $D$, $C$, and $\delta$ (eqs.\,\eqref{eq: cubed sphere variables app}) are functions of $\xi$ and $\eta$, which means that the Jacobian varies depending on the location within the patch. Additionally, it's worth noting that in the equatorial patches, vectors transform in a consistent manner due to the symmetrical construction of the four patches. In these equatorial patches, $\xi$ and $\eta$ are directed in the same way, resulting in their mutual interfaces aligning along the $\eta$ direction. However, this symmetry does not apply to the polar patches, where the orientation of $\xi$ and $\eta$ is different, making the vector transformations more complex in these regions.

On the axis, the angular components of vectors in spherical coordinates and the corresponding Jacobians are ill-defined, and as a result, they are not used. Therefore, when a transformation from spherical to cubed-sphere coordinates is required (e.g., for boundary conditions or when the initial field is given in spherical coordinates), the angular components in the cubed-sphere coordinates are averaged. This averaging process involves considering the 8 closest neighbors in the tangential direction surrounding the axis point at a specific radial layer.

At each patch edge, to transition from the coordinate system of the adjacent patch to that of the original patch, we utilize a Jacobian matrix to compute the vectors at the ghost cells and at the interface. The Jacobian is constructed by passing through spherical coordinates. For instance, to transition from the north patch to an equatorial patch, $\text{JAC}$, is obtained by multiplying the Jacobian required to go from the north patch to spherical coordinates (eq.\,\ref{eq: patch 5 to spherical}) and the Jacobian needed to go from spherical coordinates to an equatorial patch in cubed-sphere coordinates (eq.\,\ref{eq: spherical to equator}). On the other hand, to move from an equatorial patch to Patch VI, $\text{JAC}$ results from the multiplication of eq.\,\eqref{eq: equator to spherical} and eq.\,\eqref{eq: spherical to patch 6}.

\subsection{Dot product}
Considering the metric tensor defined in eq.\,\eqref{eq: metric tensor}, the dot product is given by:
\begin{align}
      \boldsymbol{a}\cdot\boldsymbol{b}&= \begin{pmatrix}
a^r &a^\xi  & a^\eta 
\end{pmatrix} \begin{pmatrix}
1 &0  & 0\\
 0 & 1 &  - \frac{XY}{CD} \\
0 &  - \frac{XY}{CD} & 1
\end{pmatrix} \begin{pmatrix}
b^r \\ b^\xi \\b^\eta \end{pmatrix} 
      \nonumber\\ 
   &=    a^{r} b^{r}  + a^{\xi} b^{\xi}+ a^{\eta} b^{\eta} - \frac{XY}{CD} \big(a^{\xi} b^{\eta}+ a^{\eta} b^{\xi} \big).\nonumber\\ 
   \label{eq: dot product}
\end{align}
The presence of the mixing term $- \frac{XY}{CD} \big(a^{\xi} b^{\eta}+ a^{\eta} b^{\xi} \big)$ arises from the fact that $\boldsymbol{e}_\xi \cdot \boldsymbol{e}_\eta \neq 0$ since these two unit vectors are non-orthogonal. Therefore, the off-diagonal terms of the metric tensor are non-zero due to the non-orthogonality of these basis vectors.

\subsection{Cross product}

\subsubsection{Contravarient component}

The contravariant component of the cross product is given by
\begin{equation}
    (\boldsymbol{a} \times \boldsymbol{b})^{l} = a^{i} b^{j} g^{kl} \tilde{\varepsilon}_{ijk},
          \label{eq: contravariant cross product}
    \end{equation}
where $g^{kl}$ is the inverse of the metric, $\tilde{\varepsilon}_{ijk}= \sqrt{g} \hspace{0.5mm}\varepsilon_{ijk} $ is the covariant Levi-Civita tensor, $\varepsilon_{ijk}$ is the usual Levi-Civita symbol, and $\sqrt{g}$ is the square root of the determinant of the metric. The latter is given by 
\begin{equation} \text{det}[g]=
    \begin{vmatrix}
1 & 0 & 0\\
0 & 1 & -\frac{XY}{CD} \\
0 & -\frac{XY}{CD}& 1
\end{vmatrix}
=1 - \frac{X^2Y^2}{C^2D^2}=   \frac{\delta}{C^2D^2},
\end{equation}
with $\sqrt{g} = \delta^{1/2}/CD$.

The contravariant component of the cross product is then written as follows:
\begin{align}
 \big(  \boldsymbol{a}\times \boldsymbol{b}\big)^l
      &= 
         \frac{\delta^{1/2}}{ CD} \big( a^{\xi} b^{\eta} - a^{\eta} b^{\xi} \big) \boldsymbol{e}_r
              \nonumber\\
      & +
      \frac{1}{\delta^{1/2}} \bigg(CD \big( a^{\eta} b^{r} - a^{r} b^{\eta} \big)+ XY  \big( a^{r} b^{\xi} - a^{\xi} b^{r} \big) \bigg) \boldsymbol{e}_\xi
      \nonumber\\
      & + \frac{1}{\delta^{1/2}} \bigg( XY \big( a^{\eta} b^{r} - a^{r} b^{\eta} \big)+ CD  \big( a^{r} b^{\xi} - a^{\xi} b^{r} \big)  \bigg) \boldsymbol{e}_\eta. 
      \label{eq: contravariant cross product our metric}
\end{align}

\subsubsection{Covariant component}
\label{app: covariant cross product}
The covariant components of the cross product are given by:
\begin{equation}
    (\boldsymbol{a} \times \boldsymbol{b})_{l} = a_{i} b_{j} g_{kl}. \tilde{\varepsilon}^{ijk},
\end{equation}
Here, $g_{kl}$ is the metric, $\tilde{\varepsilon}^{ijk}$ is the contravariant Levi-Civita tensor, and it is given by:
\begin{equation}
    \tilde{\varepsilon}^{ijk} = \frac{\text{sgn}(\text{det}[g])}{\sqrt{g}} \varepsilon^{ijk},   
\end{equation}
where $\text{sgn}(\text{det}[g]) = (-1)^q$. If the metric signature contains an odd number of negatives ($q$), then the sign of the components of this tensor differs from the standard Levi-Civita symbol. However, to simplify our work and avoid dealing with the contravariant Levi-Civita tensor, we can write
\begin{align}
    (\boldsymbol{a} \times \boldsymbol{b})_{l} &= a_{i} b_{j} g_{kl} \tilde{\varepsilon}^{ijk} 
    \nonumber\\
    &= g_{kl} g_{iu} g_{jv} \tilde{\varepsilon}^{ijk} a^{u} b^{j}.
\end{align}
The covariant Levi-Civita tensor can be expressed in terms of the contravariant Levi-Civita tensor as follows:
\begin{equation}
\tilde{\varepsilon}_{uvl}  = g_{kl} g_{iu} g_{jv}  \tilde{\varepsilon}^{ijk}.
\end{equation}
Therefore, the covariant components of the cross product are given by: 
\begin{equation}
   (\boldsymbol{a} \times \boldsymbol{b})_{l} =  \tilde{\varepsilon}_{uvl} a^{u} b^{v},
    \label{eq: covariant cross product}  
    \end{equation}
where, with the help of the metric tensor from eq.\,\eqref{eq: metric tensor}, they can be expressed as follows:
 \begin{align}
    \big(\boldsymbol{a}\times \boldsymbol{b}\big)_l &= \frac{\delta^{1/2}}{CD}  \big(a^\xi b^\eta - a^\eta b^\xi \big) \boldsymbol{e}^r
        \nonumber\\
    & + \frac{\delta^{1/2}}{CD} \big(a^\eta b^r - a^r b^\eta \big)  \boldsymbol{e}^\xi
    \nonumber\\
    & + \frac{\delta^{1/2}}{CD} \big(a^r b^\xi - a^\xi b^r \big) \boldsymbol{e}^\eta.
    \label{eq: covariant cross product metric} 
 \end{align}
In our work, we utilize the covariant components of the cross product to compute the covariant surface components (as discussed \S\ref{subsec: Metric}). These components are then employed in the curl operator within the context of finite volume schemes (as detailed in \S\ref{subsec: finite Volume schemes}).

\section{Infinitesimal geometrical elements}
\label{app: geometrical elements}

\subsection{Infinitesimal lengths: contravariant components}

The contravariant components of the infinitesimal length elements of a sphere with radius $r$ are:
 \begin{equation}
     dl^r = e^{\lambda(r)}dr,
     \label{eq: length r contravariant}
 \end{equation}
 \begin{equation}
     dl^\xi = \frac{2rD}{\delta} \frac{d\xi}{cos^2\xi}  =  \frac{2rC^2D}{\delta} d\xi ,
          \label{eq: length xi contravariant}
 \end{equation}
  \begin{equation}
    dl^\eta = \frac{2rC}{\delta} \frac{d\eta}{cos^2\eta}  = \frac{2rCD^2}{\delta} d\eta ,
          \label{eq: length eta contravariant}
 \end{equation}

\subsection{Infinitesimal lengths: covariant components}

By utilizing eq.\,\eqref{eq: contravariant to covariant}, we can calculate the covariant components of the length element as follows:
\begin{align}
       dl_r&= g_{rr}dl^{r} +g_{r\xi}dl^{\xi} + g_{r\eta}dl^{\eta}
        \nonumber\\
        &= dl^r = e^{\lambda(r)}dr.
             \label{eq: length r covariant}
\end{align}
\begin{align}
       dl_\xi&= g_{\xi r} dl^{r} +g_{\xi\xi}dl^{\xi} + g_{\xi\eta}dl^{\eta} 
        \nonumber\\
        &= dl^{\xi} -\frac{XY}{CD} dl^{\eta} \nonumber\\
        &= \frac{2rD}{\delta} (C^2 d\xi  - XYd\eta  ) .
         \label{eq: length xi covariant}
\end{align}

\begin{align}
       dl_\eta &= g_{\eta r} dl^{r} +g_{\eta\xi} dl^{\xi} + g_{\eta\eta} dl^{\eta} 
        \nonumber\\
        &= dl^{\eta} -\frac{XY}{CD} dl^{\xi} \nonumber\\
        &= \frac{2rC}{\delta} (D^2 d\eta  - XYd\xi  ) . 
         \label{eq: length eta covariant}
\end{align}

\subsection{Infinitesimal areas: contravariant components}
The area elements are determined by the magnitude of the cross product of the various length elements. 

The contravariant components of the infinitesimal radial surface element are as follows:
 \begin{equation}
     \boldsymbol{dS^r} = \boldsymbol{dl^\xi} \times \boldsymbol{dl^\eta}.
 \end{equation}
 Using eq.\,\eqref{eq: contravariant cross product}, we get
 \begin{align}
     \boldsymbol{dS^r} &= \frac{r^2CD}{\delta^2} \frac{4d\xi d\eta}{cos^2\xi cos^2 \eta}  \frac{e^{-\lambda} \delta^{1/2}}{C D} \boldsymbol{e}_r
     \nonumber\\
     &= \frac{4r^2e^{-\lambda}C^2D^2}{\delta^{3/2}} d\xi d\eta\, \boldsymbol{e}_r.
      \label{eq: surface r contravariant}
 \end{align}
The contravariant component of the infinitesimal surface element $\xi$ is determined by
 \begin{align}
     \boldsymbol{dS^\xi} &= \boldsymbol{dl^\eta} \times \boldsymbol{dl^r}
     \nonumber\\
     &= \frac{2r C D^2e^{\lambda(r)}  dr d\eta} {\delta^{3/2}} (CD\boldsymbol{e}_\xi + XY\boldsymbol{e}_\eta).
           \label{eq: surface xi contravariant}
 \end{align}
The contravariant component of the infinitesimal surface element $\eta$ is determined by \begin{align}
     \boldsymbol{dS^\eta} &= \boldsymbol{dl^r} \times \boldsymbol{dl^\xi}
       \nonumber\\
     &= \frac{2r D C^2 e^{\lambda(r)}dr d\xi}{\delta^{3/2}}(XY \boldsymbol{e}_\xi + CD\boldsymbol{e}_\eta).
     \label{eq: surface eta contravariant}
 \end{align}

\subsection{Infinitesimal areas: covariant components} 
Following eq.\,\eqref{eq: covariant cross product}, we can define the covariant components of the surface elements in terms of the contravariant length elements.

The covariant component of the radial surface element is defined as
\begin{align}
         dS_r&= \mid\boldsymbol{dl_\xi} \times \boldsymbol{dl_\eta} \mid
         \nonumber\\
         &= dl^\xi dl^\eta \tilde{\varepsilon}_{\xi \eta r}
         \nonumber\\
         &=  dl^\xi dl^\eta \sqrt{g} \hspace{0.5mm} \varepsilon_{\xi \eta r}
         \nonumber\\
         &=  \frac{4r^2}{\delta^{3/2}} C^2D^2 d\eta d\xi
         \nonumber\\
         &= dS^r,
\label{eq: surface r covariant}
 \end{align}
as expected from eq.\,\eqref{eq: contravariant to covariant}.
The covariant component of the infinitesimal surface element $\xi$ is defined as: 
 \begin{align}
        dS_\xi&=\mid \boldsymbol{dl_\eta} \times \boldsymbol{dl_r} \mid
         \nonumber\\
         &= dl^\eta dl^r \tilde{\varepsilon}_{\eta r \xi}
         \nonumber\\
         &=   \frac{2re^\lambda D }{\delta^{1/2}}dr d\eta.
         \label{eq: surface xi covariant}
 \end{align}
Whereas, the covariant component of the infinitesimal surface element $\eta$ is defined as:
  \begin{align}
         dS_\eta&=\mid \boldsymbol{dl_r} \times \boldsymbol{dl_\xi} \mid
         \nonumber\\
         &=  dl^r dl^\xi  \tilde{\varepsilon}_{r \xi \eta}
         \nonumber\\
         &=   \frac{2re^{\lambda(r)}C }{\delta^{1/2}} dr d\xi.
         \label{eq: surface eta covariant}
 \end{align}
 
\subsection{Infinitesimal volume element}

The infinitesimal volume element is obtained by calculating the mixed product of the three geometrical lengths.
 \begin{align}
    dV &= \big(\boldsymbol{dl}^\xi \times \boldsymbol{dl}^\eta \big) \cdot \boldsymbol{dl}^r = \big(\boldsymbol{dl}_\xi \times \boldsymbol{dl}_\eta \big) \cdot \boldsymbol{dl}_r = \boldsymbol{dS}_r \boldsymbol{dl}^r
  \nonumber\\
    &= \big(\boldsymbol{dl}^\eta \times \boldsymbol{dl}^r \big) \cdot \boldsymbol{dl}^\xi = \big(\boldsymbol{dl}_\eta \times \boldsymbol{dl}_r \big) \cdot \boldsymbol{dl}_\xi= \boldsymbol{dS}_\xi \boldsymbol{dl}^\xi
\nonumber\\       
    &= \big(\boldsymbol{dl}^r \times \boldsymbol{dl}^\xi \big) \cdot \boldsymbol{dl}^\eta=  \big(\boldsymbol{dl}_r \times \boldsymbol{dl}_\xi \big) \cdot \boldsymbol{dl}_\eta  =  \boldsymbol{dS}_\eta \boldsymbol{dl}^\eta
    \nonumber\\
    &= e^{\lambda(r)} \frac{4r^2 C^2D^2}{\delta^{3/2}} dr d\eta d\xi.
     \label{eq: mixed product app }
 \end{align}

\section{Treatment of the ghost cells}
\label{sec: connection between patches}

When computing the curl operator introduced in eqs.\,\eqref{eq: stokes theorem radial component}-\eqref{eq: stokes theorem eta component} at the edges (corners) of the patch, one needs information about the values of the functions in some points which lie in the coordinate system(s) of the neighbouring patch(es). A way to deal with this issue is to extend one layer of ghost cells in each direction, for each patch. The handling of ghost cells adds another layer of complexity to the cubed-sphere coordinates. This complexity arises due to the six distinct coordinate systems composing the cubed-sphere coordinates. 

\begin{figure}[ht]
    \centering
    \includegraphics[width=\textwidth]{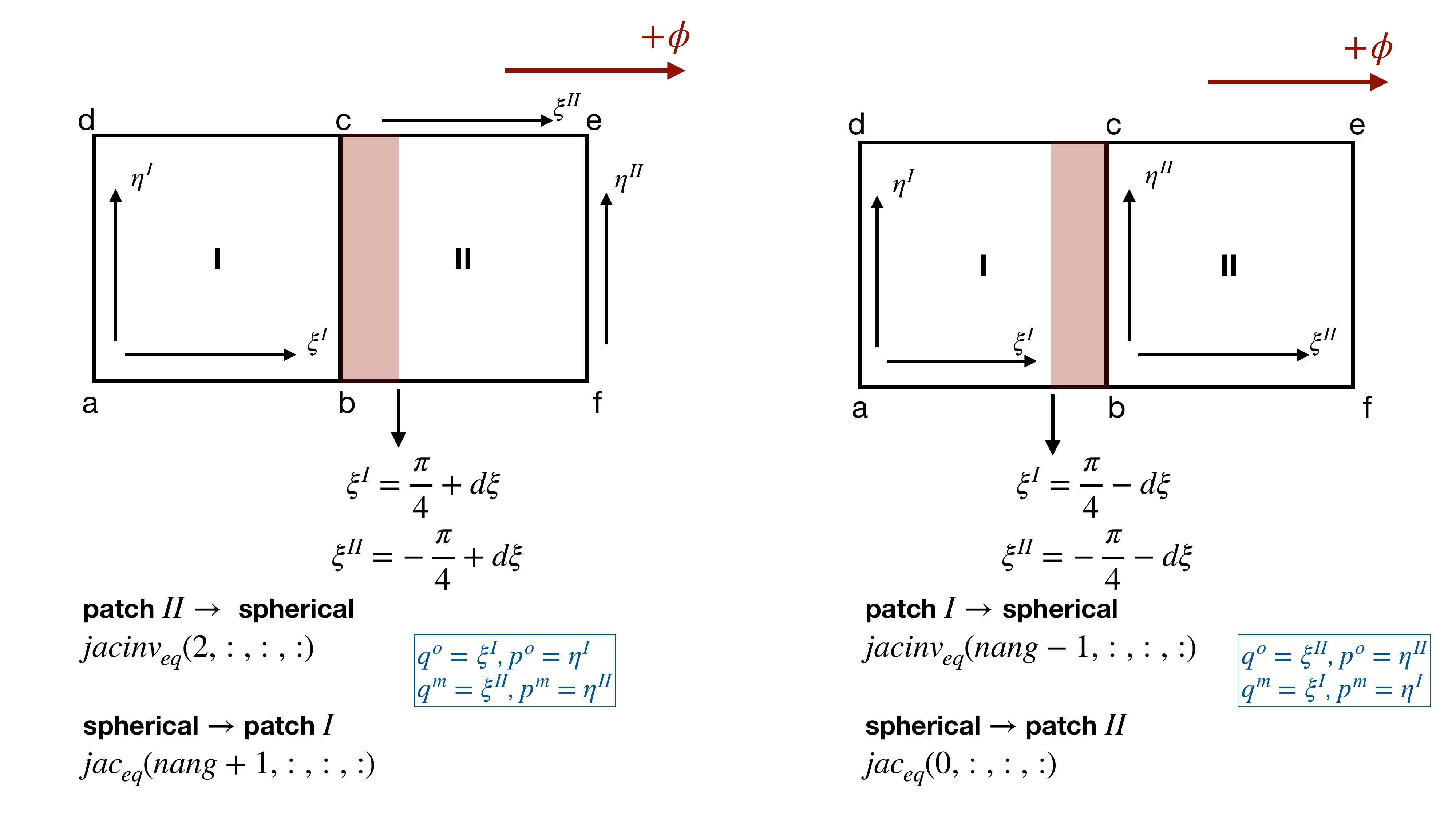}
    \caption{Schematic depiction of two adjacent equatorial blocks, e.g., patch I and patch II, is provided. The LHS and RHS panels illustrate the ghost cells of patch I and patch II in pink, respectively. The orientation of the axes $\xi^{I}$, $\eta^I$, $\xi^{II}$, and $\eta^{II}$ is depicted in each diagram. The $\phi$ axis represents the orientation of the spherical $\phi$ coordinate. Here, $p$ denotes the value of the point coordinate parallel to the interface, and $q$ represents the coordinate pseudo-perpendicular to it. The original coordinate system (for which we know $(p^o,q^o)$) is indicated by the superscript $^o$, while the mapping of the point in the adjacent patch is denoted with the superscript $^m$. The Jacobian $jacinv_{eq}$ corresponds to eq.\,\eqref{eq: equator to spherical}, while $jac_{eq}$ corresponds to eq.\,\eqref{eq: spherical to equator}. This schematic representation applies to every pair of contiguous equatorial blocks, including patch II-III, patch III-IV, and patch IV-I.}
    \label{fig: equatorial patches}
\end{figure}

In this section, we provide a series of schematic depictions illustrating pairs of adjacent patches. These depictions significantly aid in understanding the arrangement of ghost cells, particularly when it involves connecting one of the equatorial patches with either the north (patch V) or south patch (patch VI). In the figures below, each panel illustrates the ghost cells of the original patch in pink. The orientation of the axes within each patch, i.e., $\xi$ and $\eta$, is displayed in each diagram. The $\phi$ axis corresponds to the orientation of the spherical $\phi$ coordinate. In this context, $p$ represents the value of the point coordinate parallel to the interface, while $q$ signifies the coordinate pseudo-perpendicular to it. It is important to note that the coordinates are non-orthogonal except along the central axes of each patch. The original coordinate system, denoted by $(p^o, q^o)$, is indicated using the superscript $^o$, whereas the mapping of a point in the adjacent patch is marked with the superscript $^m$. For additional details, please refer to \S\ref{subsec: edges of the patches} and Table~\ref{tab:edges}. The Jacobian $jacinv$ corresponds to the transformation from a specific patch of the cubed-sphere coordinates to spherical coordinates, while the Jacobian $jac$ corresponds to the transformation from spherical coordinates to one of the cubed-sphere coordinates. The schematic representation displayed in Fig.~\ref{fig: equatorial patches} is applicable to every pair of contiguous equatorial blocks, encompassing patch II-III, patch III-IV, and patch IV-I. However, the complexity increases when dealing with the interconnection of an equatorial patch with either the northern patch or the southern patch.

\begin{figure}
    \centering
    \includegraphics[width=\textwidth]{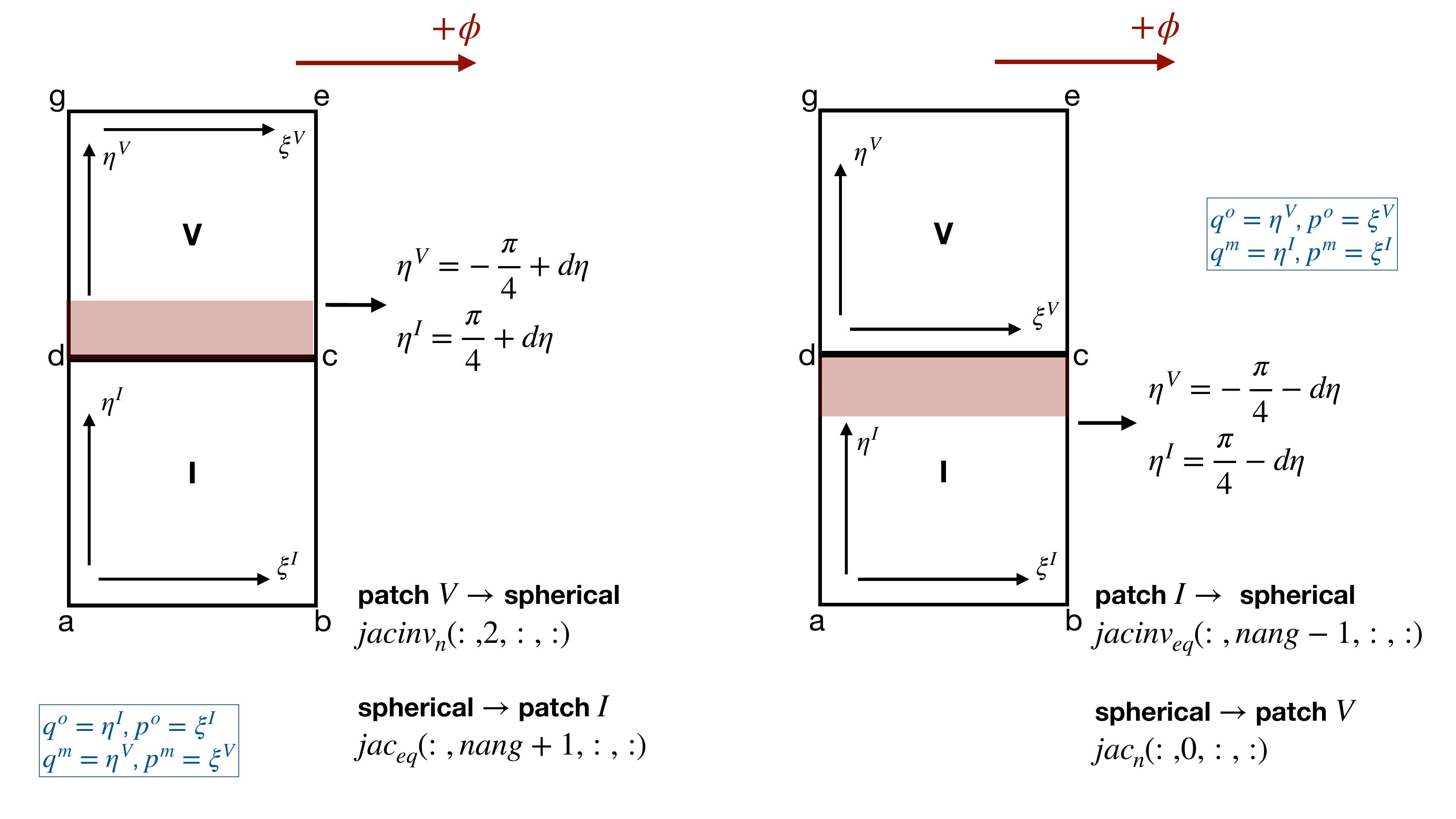}
    \caption{Same as Fig.~\ref{fig: equatorial patches}. The Jacobian $jacinv_{n}$ corresponds to eq.\,\eqref{eq: patch 5 to spherical}, while $jac_{n}$ corresponds to eq.\,\eqref{eq: spherical to patch 5}. The LHS and RHS panels illustrate the ghost cells of patch I and patch V in pink, respectively.}
    \label{fig: I-V}
\end{figure}

\begin{figure}
    \centering
    \includegraphics[width=\textwidth]{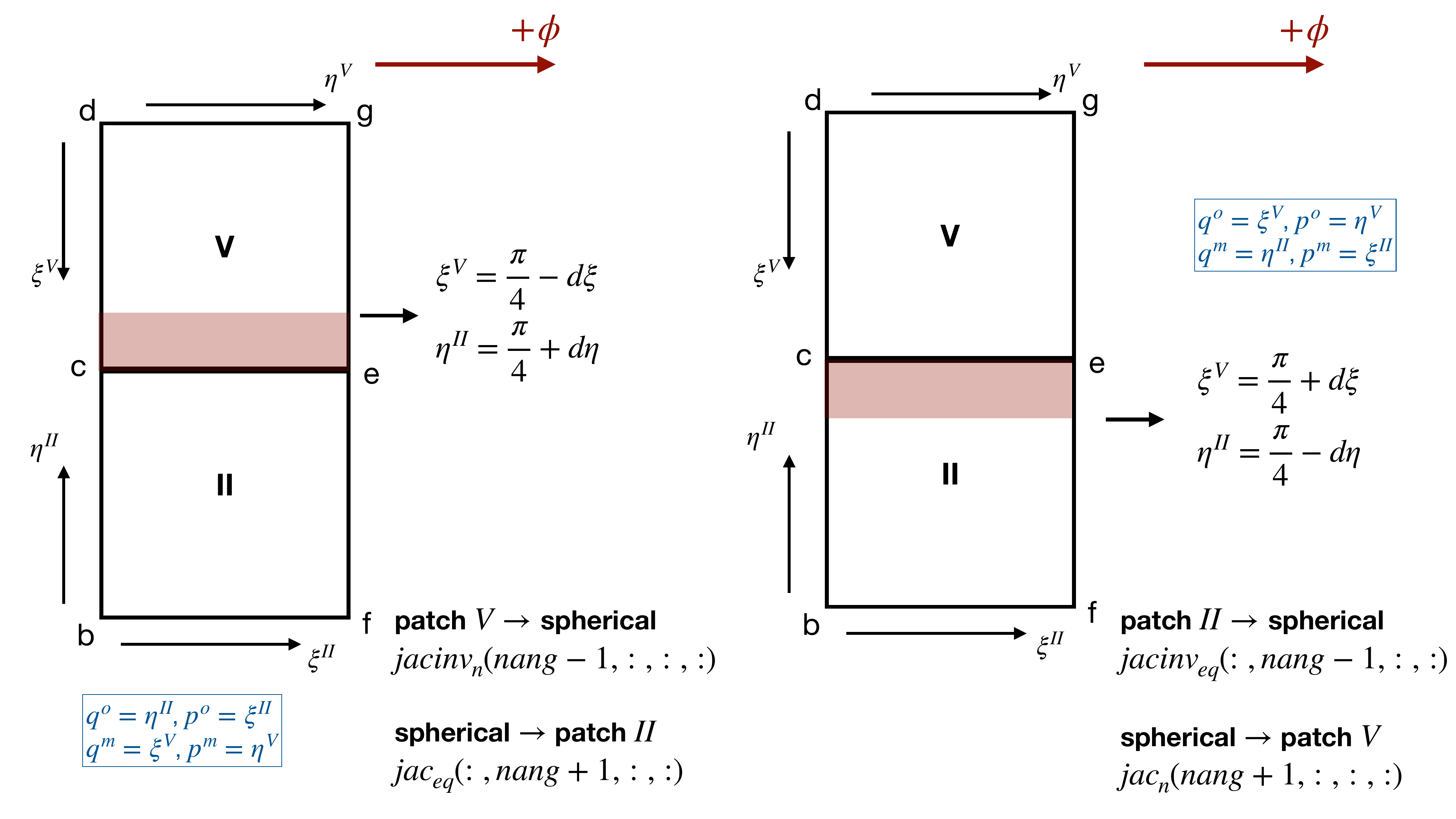}
    \caption{Same as Fig.~\ref{fig: equatorial patches} and Fig.~\ref{fig: I-V}. The LHS and RHS panels illustrate the ghost cells of patch II and patch V in pink, respectively.}
    \label{fig: II-V}
\end{figure}

\begin{figure}
    \centering
    \includegraphics[width=\textwidth]{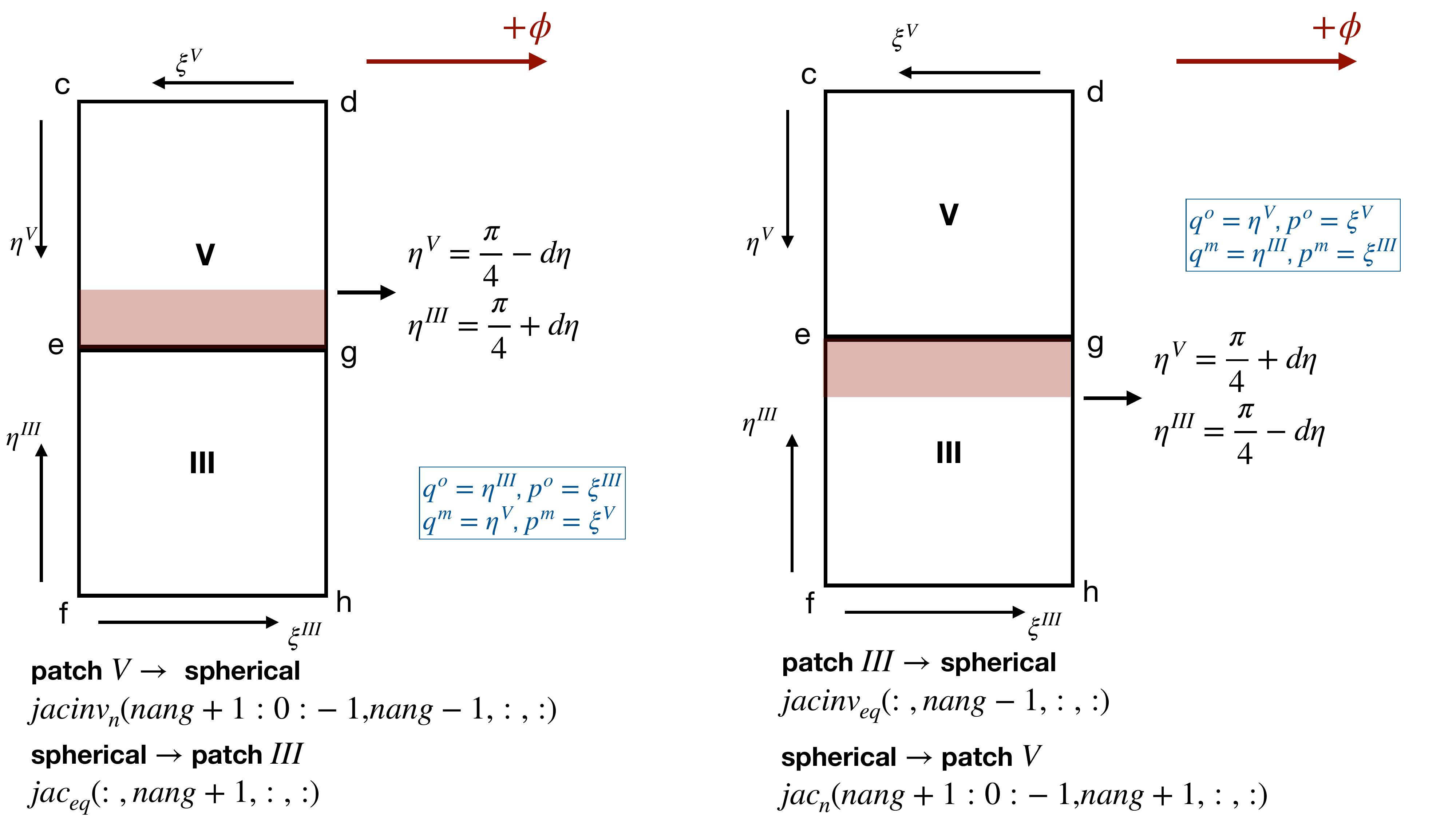}
    \caption{Same as Fig.~\ref{fig: equatorial patches} and Fig.~\ref{fig: I-V}. The LHS and RHS panels illustrate the ghost cells of patch III and patch V in pink, respectively.}
    \label{fig: III-V}
\end{figure}

\begin{figure}
    \centering
    \includegraphics[width=\textwidth]{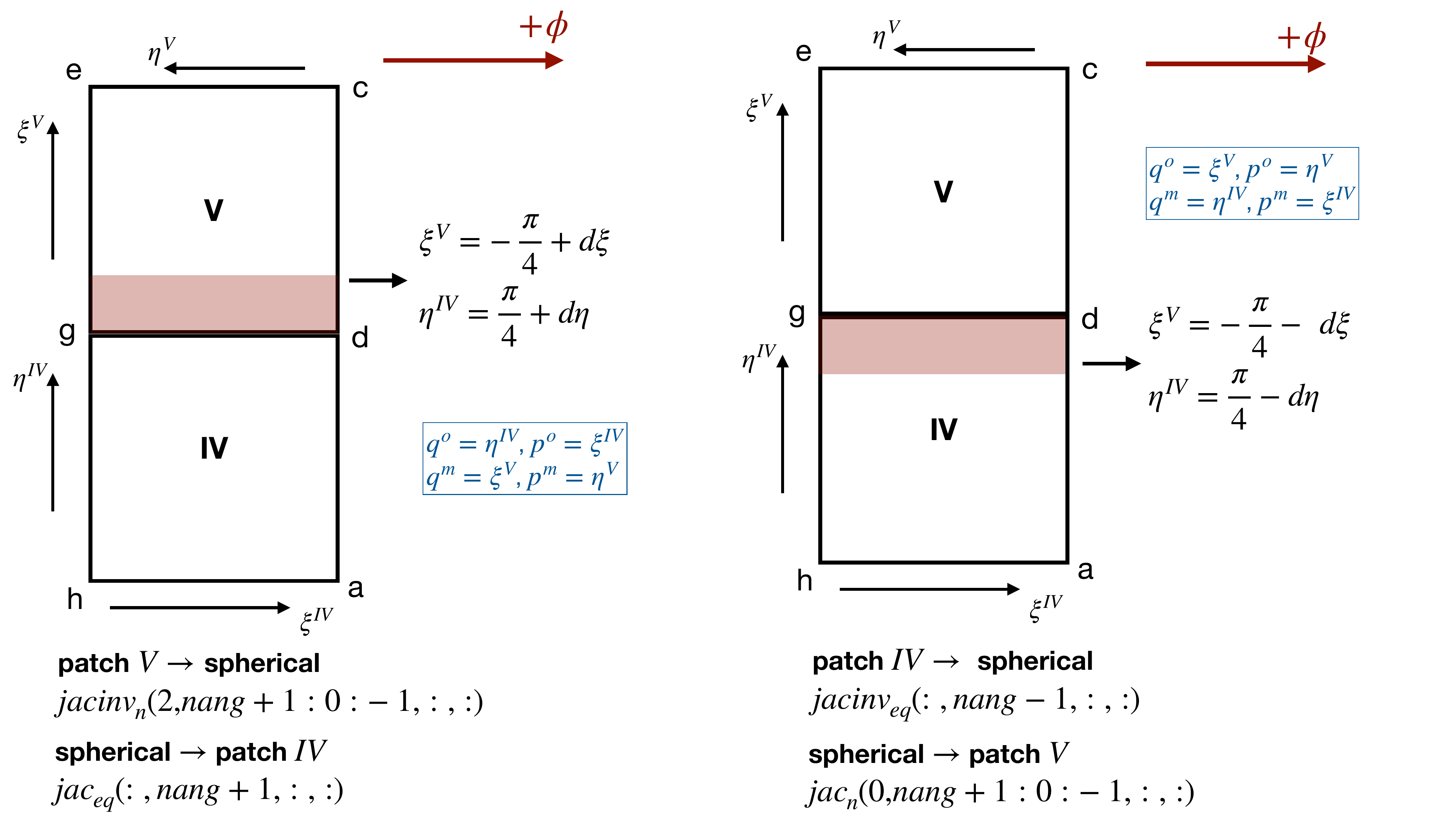}
    \caption{Same as Fig.~\ref{fig: equatorial patches} and Fig.~\ref{fig: I-V}. The LHS and RHS panels illustrate the ghost cells of patch IV and patch V in pink, respectively.}
    \label{fig: IV-V}
\end{figure}

\begin{figure}
    \centering
    \includegraphics[width=\textwidth]{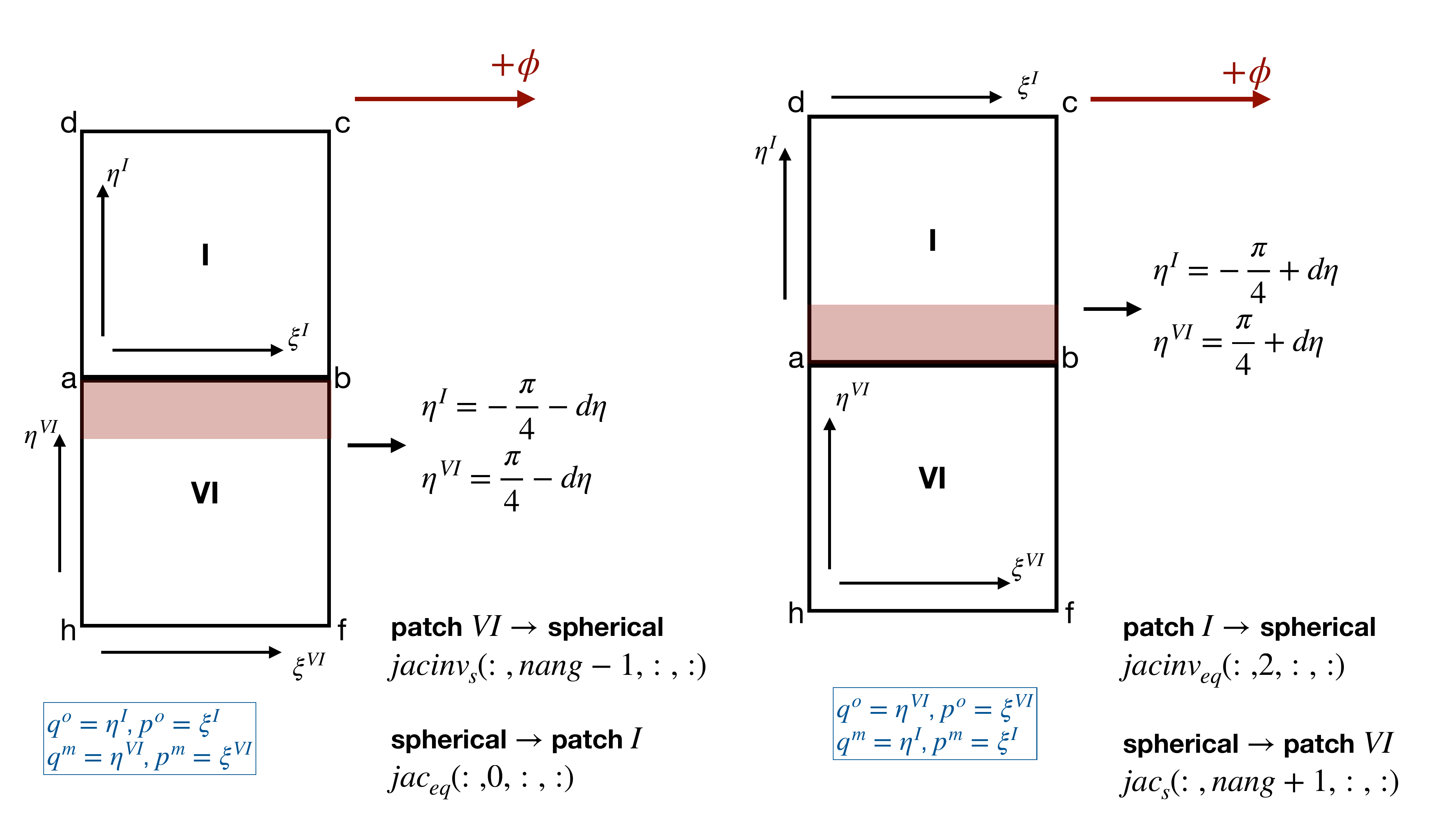}
    \caption{Same as Fig.~\ref{fig: equatorial patches}. The Jacobian $jacinv_{s}$ corresponds to eq.\,\eqref{eq: patch 6 to spherical}, while $jac_{s}$ corresponds to eq.\,\eqref{eq: spherical to patch 6}. The LHS and RHS panels illustrate the ghost cells of patch I and patch VI in pink, respectively.}
    \label{fig: I-VI}
\end{figure}

\begin{figure}
    \centering
    \includegraphics[width=\textwidth]{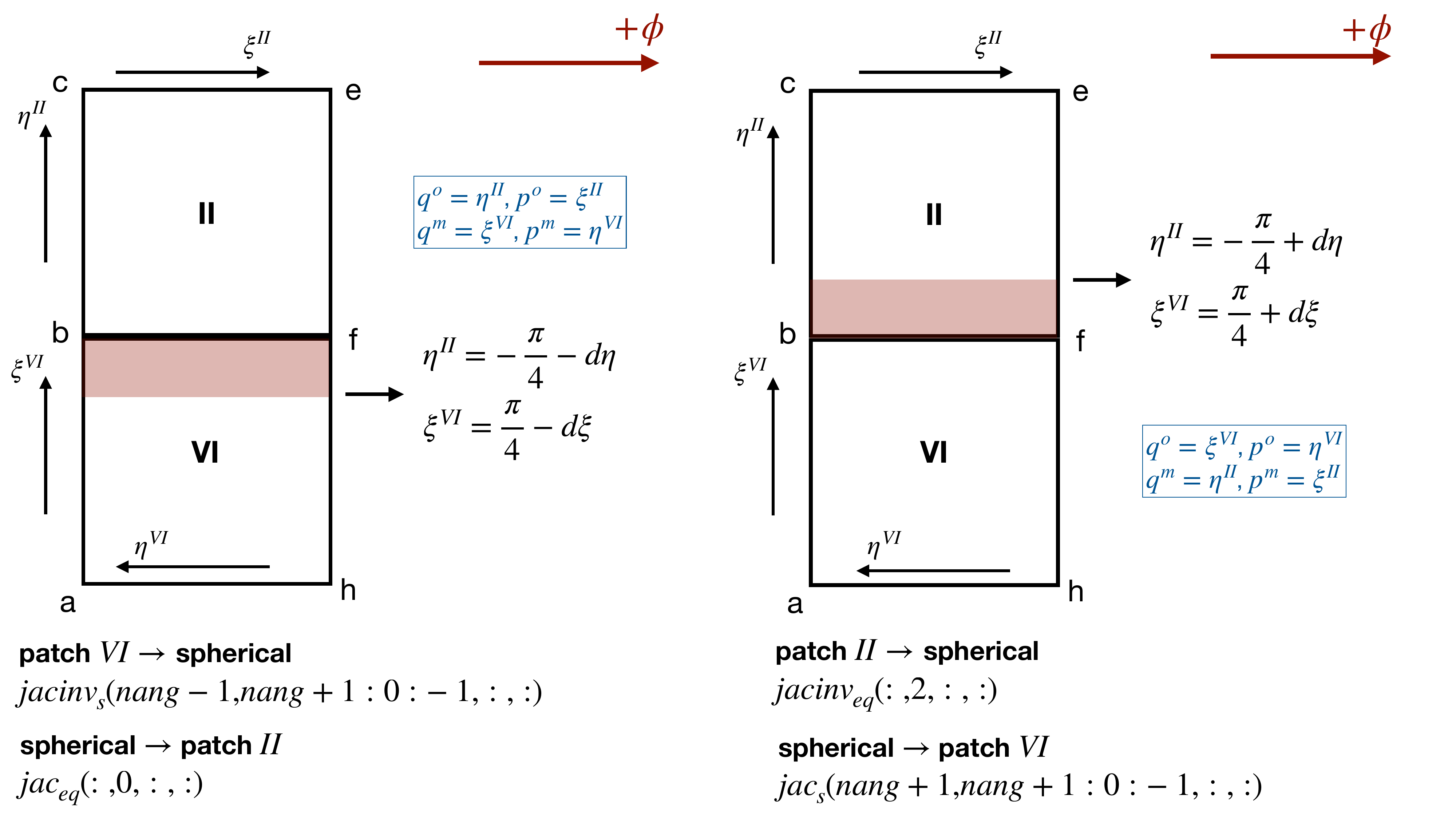}
    \caption{Same as Fig.~\ref{fig: equatorial patches} and Fig.~\ref{fig: I-VI}. The LHS and RHS panels illustrate the ghost cells of patch II and patch VI in pink, respectively.}
    \label{fig: II-VI}
\end{figure}

\begin{figure}
    \centering
    \includegraphics[width=\textwidth]{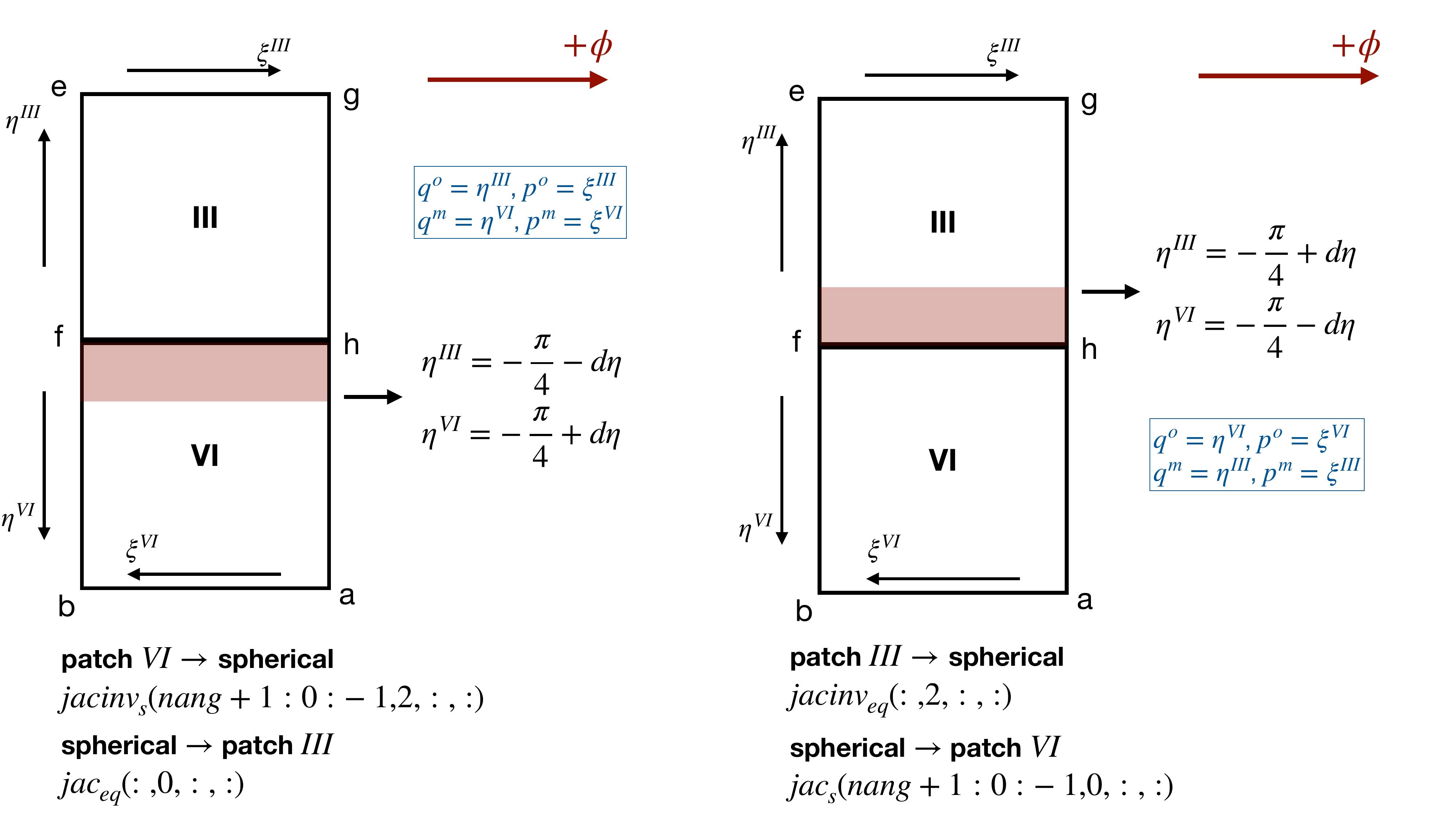}
    \caption{Same as Fig.~\ref{fig: equatorial patches} and Fig.~\ref{fig: I-VI}. The LHS and RHS panels illustrate the ghost cells of patch III and patch VI in pink, respectively.}
    \label{fig: III-VI}
\end{figure}

\begin{figure}
    \centering
    \includegraphics[width=\textwidth]{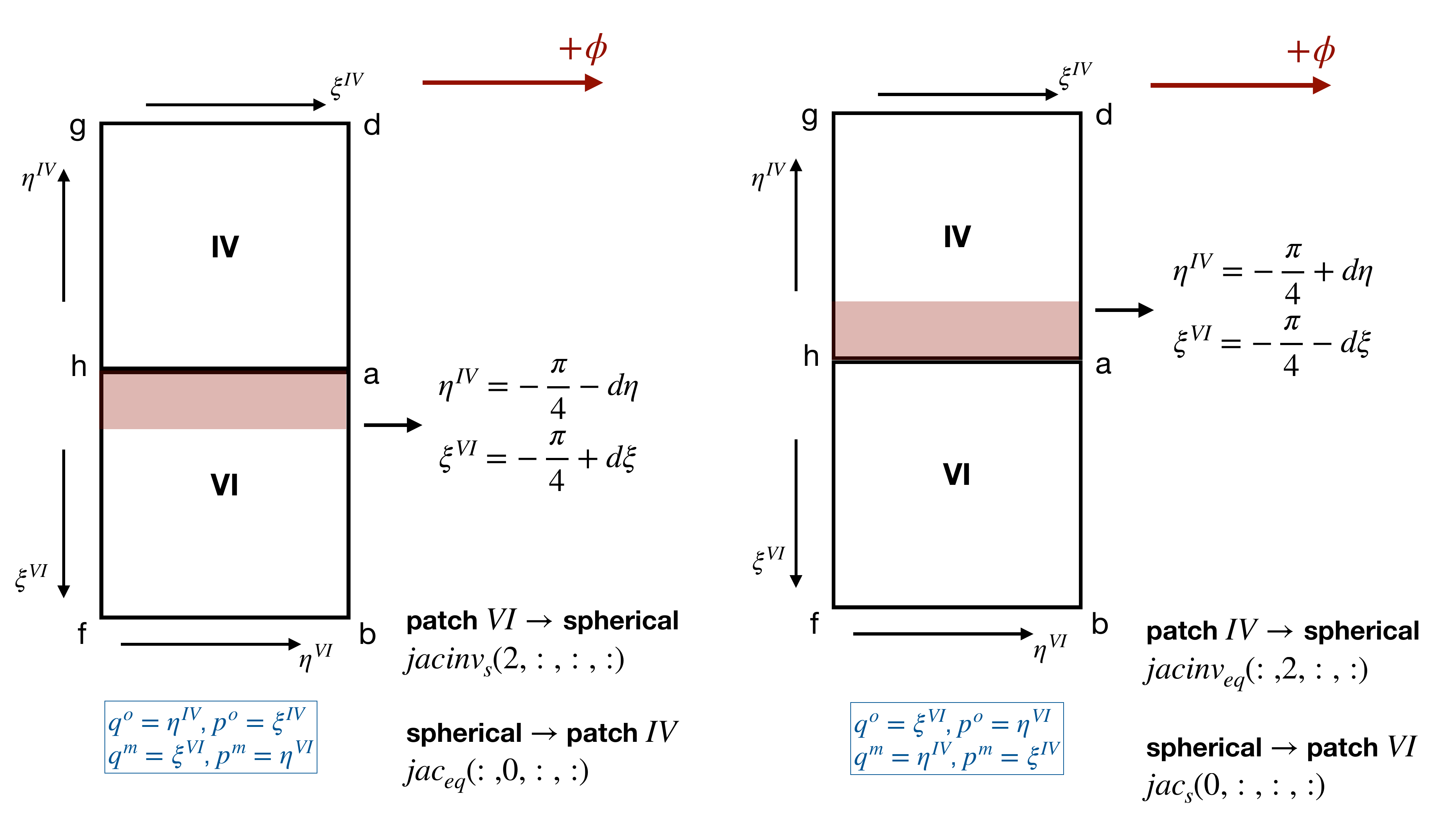}
    \caption{Same as Fig.~\ref{fig: equatorial patches} and Fig.~\ref{fig: I-VI}. The LHS and RHS panels illustrate the ghost cells of patch IV and patch VI in pink, respectively.}
    \label{fig: IV-VI}
\end{figure}

\clearemptydoublepage
\let\textcircled=\pgftextcircled
\chapter{Three-Dimensional Magnetic Field Formalism}
\label{appendix: Magnetic field formalism}

\section{Poloidal and toroidal decomposition}
\label{appendix: Poloidal and toroidal decomposition}

In MHD, various formalisms can be used to describe the magnetic field. In this context, we present the most common notations found in the literature. For any three-dimensional, solenoidal vector field $\boldsymbol{B}$, representing the magnetic field, we can always introduce the vector potential $\boldsymbol{A}$ such that:
\begin{equation}
    \boldsymbol{B} = \boldsymbol{\nabla} \times \boldsymbol{A}~. 
\end{equation}
The magnetic field $\boldsymbol{B}$ can be represented using two scalar functions, $\Phi(x)$ and $\Psi(x)$, which define its poloidal and toroidal components, respectively, based on the Chandrasekhar-Kendall expressions \citep{chandrasekhar1957}:
  \begin{align}
      \boldsymbol{B}_\text{pol} &=  
    \boldsymbol{\nabla} \times \big( \boldsymbol{\nabla} \times \Phi \boldsymbol{k}   \big) =  - \boldsymbol{r} \nabla^2 \Phi + \nabla \Bigg(\frac{\partial (r \Phi)}{\partial r} \Bigg)~,
      \nonumber\\
        \boldsymbol{B}_\text{tor} &= 
        \boldsymbol{\nabla}\times \Psi \boldsymbol{k}~.
    \label{eq: poloidal toroidal field components}
  \end{align}
Here, $\boldsymbol{k}$ is an arbitrary vector. This decomposition proves particularly useful in problems where $\boldsymbol{k}$ can be taken to be normal to the physical boundaries, and the boundary conditions in the toroidal direction are periodic. Consequently, when dealing with a spherical domain and using the cubed-sphere coordinates $(r, \xi, \eta)$, the suitable choice for $\boldsymbol{k}$ is $\boldsymbol{r}$.

In accordance with the notation of \cite{geppert1991}, the fundamental concept involves expanding the poloidal function $\Phi$ and the toroidal function $\Psi$ as series of spherical harmonics at time zero to establish the initial conditions. Expressing the two scalar functions, $\Phi$ and $\Psi$, as series of spherical harmonics, we get:
  \begin{equation}
        \Phi(t,r,\theta,\phi) = \frac{1}{r}\sum_{l,m} \Phi_{lm}(r,t) Y_{lm}(\theta,\phi) ,
        \label{eq: phi scalar functions}
  \end{equation}
  \begin{equation}
      \Psi(t,r,\theta,\phi)  = \frac{1}{r}\sum_{l,m} \Psi_{lm}(r,t) Y_{lm}(\theta,\phi).
       \label{eq: Psi scalar functions} 
  \end{equation}
In this context, the expansion involves considering the degree $l = 1, ..., l_{max}$ and the order $m = -l, ..., l$ of the multipole. It is important to note that in 3D, the toroidal field is a combination of the two tangential components of the magnetic field, while the poloidal field is a mixture of all three components. This presents a more intricate scenario compared to the 2D case, where the toroidal part is solely comprised of the azimuthal component, and the poloidal part consists of the other two components of the magnetic field.

Combining the poloidal and toroidal components of the magnetic field, one can express the three components of the magnetic field in spherical coordinates as follows:
\begin{align}
 B^{r} &= \frac{1}{r^2} \sum_{lm} l(l+1)  \Phi_{lm}(r) Y_{lm}(\theta,\phi),   \nonumber\\
 B^{\theta} &=   \frac{1}{r} \sum_{lm} \Phi'_{lm}(r) \frac{\partial Y_{lm}(\theta,\phi)}{\partial \theta} +  \frac{1 }{r sin\theta} \sum_{lm} \Psi_{lm}(r) \frac{\partial Y_{lm}(\theta,\phi)}{ \partial \phi},
 \nonumber\\
  B^{\phi} &=    - \frac{1}{r}\sum_{lm} \Psi_{lm}(r) \frac{\partial Y_{lm}(\theta,\phi)}{ \partial \theta}  + \frac{1}{r sin\theta}  \sum_{lm}  \Phi'_{lm}(r) \frac{\partial Y_{lm}(\theta,\phi)}{\partial \phi},
 \label{eq: magnetic field components spectral}
\end{align}
where $\Phi'_{lm}$ represents the radial derivative of $\Phi_{lm}$ and is given by: 
 \begin{equation}
\Phi'_{lm}(r)= e^{-\lambda(r)} \frac{\partial \Phi_{lm}(r)}{\partial r} + \frac{1 - e^{-\lambda(r)}}{r} \Phi_{lm}(r). 
\label{eq: radial derivative of the Phi scalar function}
 \end{equation}
Refer to \S\ref{sec: Mathematical derivation of Bfield} for a thorough explanation of the derivations.

\section{Spectral magnetic energy}
\label{app: spectral magnetic energy}
To achieve a comprehensive quantitative analysis and gain insights into the results, as well as to investigate the distribution of magnetic energy across various spatial scales and multipoles within the defined system, we examine the energy spectrum. Our goal is to track the evolution of the $l$ and $m$ energy spectrum over time within the volume of a neutron star. To compute the spectral energy, we utilize the poloidal and toroidal decomposition of the magnetic field as defined in eq.\,\eqref{eq: magnetic field components spectral}. The magnetic energy content in each mode, accounting for relativistic corrections, can be expressed as follows:
 \begin{equation}
   E_{lm} = \frac{1}{8\pi} \int  e^{\lambda+\zeta} dr ~l(l+1) \bigg[  \frac{l(l+1)}{r^2} \Phi_{lm}^2 +\big(\Phi'_{lm}\big)^2   +\Psi_{lm}^2  \bigg].
     \label{eq: spectral magnetic energy app}
 \end{equation}
The $e^\lambda$ and $e^\zeta$ terms are seen in \S\ref{sec: TOV}.

The first two terms in eq.\,\eqref{eq: spectral magnetic energy app} account for the poloidal magnetic energy and the last term accounts for the toroidal energy. The total energy density can be obtained by summing over all $l$ and $m$ modes, i.e., $\sum_{lm} E_{lm}$.

To determine the spectral energy distribution as given in eq.\,\eqref{eq: spectral magnetic energy app}, one needs to reconstruct the three radial scalar functions, namely $\Phi_{lm}$, $\Phi'_{lm}$, and $\Psi_{lm}$, which are defined as follows:
\begin{equation}
     \Phi_{lm}(r) = \frac{1}{l(l+1) } \int dS_r B^r Y_{lm} (\theta,\phi),
\end{equation}

\begin{equation}
\Phi'_{lm}(r) =  \frac{1}{l(l+1)} \int \frac{dS_r}{r} \bigg(B^{\theta}\frac{\partial Y_{lm}}{\partial \theta}  +  \frac{B^{\phi}}{sin\theta} \frac{\partial Y_{lm}}{\partial \phi}  \bigg), 
\end{equation}
and 
  \begin{equation}
      \Psi_{lm}(r) =   \frac{1}{l(l+1)} \int \frac{dS_r}{r} \bigg( \frac{B^{\theta}}{sin\theta}\frac{\partial Y_{lm}}{\partial \phi}  -  B^{\phi} \frac{\partial Y_{lm}}{\partial \theta}  \bigg). 
\end{equation}
For a detailed explanation of the derivations, refer to \S\ref{sec: Mathematical derivation of Espectral}.

For $r\geq R$, the spectral magnetic energy in c.g.s. units, including the relativistic correction, can be expressed as:
  \begin{equation}
      E^{r\geq R}_{lm} = \frac{1}{8\pi}\int e^{\lambda(r) + \zeta(r)} r^2 dr \sum_{lm}   b_{lm}^2 \Bigg( e^{-2\lambda(r)} (l+1)^2 + l(l+1) \Bigg) \Bigg( \frac{R}{r} \Bigg)^{(2l+4)} . 
    \label{eq: surface energy}
  \end{equation}
Refer to \S\ref{sec: Mathematical derivation of Esurf} for a comprehensive derivation.

\section{Initial models}
\label{appendix: initial conditions}

The initial topology of the magnetic field can be constructed by selecting a specific set of spherical harmonics that define the desired configuration. For example, for a dipole, we use $Y_{l=1, m}$, for a quadrupole, we use $Y_{l=2, m}$, and for a multipolar topology, we can create it by combining several spherical harmonics, expressed as $\sum_{lm} Y_{lm}(\theta,\phi)$.

The choice of the set of spherical harmonics determines the angular part of the magnetic field topology. This selection offers flexibility, allowing researchers to tailor the topology as per their requirements. In our study, we enforce potential magnetic boundary conditions and use a set of radial scalar functions $\Phi_{lm}(r)$ and $\Psi_{lm}(r)$ that ensure a smooth matching with the potential boundary conditions. This approach enables us to construct a magnetic field configuration that satisfies the desired properties at the boundaries.

We adopt the radial profile of the dipolar poloidal scalar function, $\Phi_{l=1,m}(r)$, as described in eq.\,(8) of \cite{aguilera2008}. This profile smoothly connects to a pure dipolar field outside, neglecting relativistic corrections. It ensures continuity in the function and its first derivative. The expression for $\Phi_{l=1,m}(r)$ is given by:
\begin{equation}
    \Phi_{l=1,m}(r) = \Phi_0 \mu r (a + tan(\mu R_\star) b )
    \label{eq: funa function}
\end{equation}
where $\Phi_0$ is the normalization and 
\begin{equation}
    a= \frac{sin(\mu r)}{(\mu r)^2} - \frac{cos(\mu r)}{\mu r}, \hspace{5mm} b= -  \frac{cos(\mu r)}{(\mu r)^2} - \frac{sin(\mu r)}{(\mu r )},
\end{equation}
$\mu$ is a parameter related to the magnetic field curvature, and its value needs to be determined for a specific surface radius $R_\star$.

For higher-order multipoles of the poloidal scalar function $(l>1)$ and all the toroidal scalar function contributions $\Psi_{lm}(r)$, we restrict their presence to the crust of a neutron star as follows: 
\begin{equation}
     \Phi_{l>1,m}(r),~ \Psi_{lm}(r) \propto - \big(R -r \big)^2 \big( r-R_c\big)^2 ,
\end{equation}
where the proportionality means that every multipole can have a different normalization (i.e., its initial weight). Note that this particular choice has no specific physical ground, further than matching at $t=0$ with a pure dipolar field at the surface. 

From $\Phi_{lm}(r)$ and $\Psi_{lm}(r)$, we construct the magnetic field components defined by eqs.\,\eqref{eq: poloidal toroidal field components}. We utilize the curl operator in cubed-sphere coordinates, given by eqs.\,\eqref{eq: stokes theorem radial component}-\eqref{eq: stokes theorem eta component}. This construction ensures that the initial topology of the magnetic field is divergence-free up to machine error, and effectively eliminates any axis-singularity issues.

\section{Spherical harmonics}
\label{Appendix: spherical harmonics}

The spherical harmonics are characterized by forming an orthonormal and complete set, as expressed by the following integral equation:
\begin{equation}
     \int_0^{2\pi} d\phi \int_0^{\pi}   Y^{m}_l(\theta,\phi)  Y^{m' *}_{l'}(\theta,\phi) sin\theta d\theta = \delta_{ll'} \delta_{mm'},
\end{equation}
where $\delta_{ll'}$ and $\delta_{mm'}$ are Kronecker delta functions. This property is essential for various mathematical and physical applications involving real spherical harmonics on the surface of the sphere. $\delta$ equals $1$ when $l=l'$ and $m=m'$; otherwise, $\delta$ is equal to $0$. The notation $Y^{m'*}_{l'}(\theta,\phi)$ represents the conjugate of the spherical harmonics.

We are interested in utilizing the real set of spherical harmonics, also known as Laplace spherical harmonics. To work with real functions, one can construct them by combining complex conjugate functions corresponding to opposite values of $m$. In this manner, the real spherical harmonics $Y_{lm}(\theta,\phi)$ are defined as follows: 
\begin{equation*}
\begin{cases}
\frac{(-1)^m}{\sqrt{2}}\big(Y_l^m +Y_l^{m*}  \big) = \sqrt{\frac{2l+1}{2\pi} \frac{(l-m)!}{(l+m)!}} P^m_l(cos\theta) cos(m\phi)  & \quad m >0 , \\ \\
 \sqrt{\frac{2l+1}{4\pi}}  P^0_l(cos\theta)  & \quad m=0, \\ \\
\frac{(-1)^m}{i\sqrt{2}}\big(Y_l^{|m|} - Y_l^{|m|*}  \big) = \sqrt{\frac{2l+1}{2\pi} \frac{(l-|m|)!}{(l+|m|)!}} P^{|m|}_l(cos\theta) sin(|m|\phi)  & \quad m <0.
\end{cases}
\label{eq: real spherical harmonics}
\end{equation*}

Indeed, the real set of spherical harmonics also constitutes an orthonormal and complete set, leading to the following expression for their orthonormality condition:
\begin{equation}
  \int_0^{2\pi} d\phi \int_0^{\pi}   Y_{lm}(\theta,\phi)  Y_{l'm'}(\theta,\phi) sin\theta d\theta = \delta_{ll'} \delta_{mm'}, 
\end{equation}
where the solid angle $d\Omega =\sin\theta d\theta  d\phi  = dS_r/r^2$. This relationship allows us to express the orthonormality condition in the cubed-sphere formalism as:
\begin{equation}
  \int \frac{dS_r}{r^2}   Y_{lm}(\xi,\eta)  Y_{l'm'}(\xi,\eta)= \delta_{ll'} \delta_{mm'}.
  \label{eq: orthogonality of real spherical harmonics}
\end{equation}
Here, $\xi$ and $\eta$ are dependent on the specific patch to which they are associated. Other spherical harmonics properties relevant to our formalism are:
\begin{equation}
    \int d\Omega \frac{\partial Y_{lm}}{\partial\theta}\frac{\partial Y_{l'm'}}{\partial\theta} = \bigg(l(l+1) -\frac{(2l+1)m}{2}\bigg)\delta_{ll'} \delta_{mm'},
\label{eq: orthogonality of square of dYlm/dtheta }
\end{equation}
and 
\begin{equation}
    \int \frac{d\Omega}{sin^2\theta} \frac{\partial Y_{lm}}{\partial\phi}\frac{\partial Y_{l'm'}}{\partial\phi} = \frac{(2l+1)m}{2}\delta_{ll'} \delta_{mm'}.
\label{eq: orthogonality of square of (dYlm/dphi/sin theta)}
\end{equation}
The Laplace transform of the spherical harmonics is defined as:
\begin{equation}
    \frac{1}{Y}  \frac{\partial^2 Y}{\partial \theta^2} + \frac{cos\theta}{ Y sin\theta} \frac{\partial Y}{\partial \theta} + \frac{1}{Y sin^2 \theta } \frac{\partial^2 Y}{\partial \phi^2}  = -l (l+1). 
    \label{eq: laplace spherical harmonics}
\end{equation}

To determine the spherical harmonics $Y_{lm}$ and their derivatives, $\frac{\partial Y_{lm}}{\partial\theta}$ and $\frac{\partial Y_{lm}}{\partial\phi}$, needed in our formalism, we express them in terms of the associated Legendre polynomials, which constitute a non-orthogonal set of functions defined as:
\begin{equation}
    P^m_l(x) = (-1)^m (1-x^2)^{m/2} \frac{d^m P_l(x)}{dx^m}.
\end{equation}
Here, $P_l(x)$ represents the Legendre polynomial. To compute the associated Legendre polynomials for different values of $l$ and $m$, we use one of the many recurrence relations available, such as the following:
 \begin{equation*}
P^m_l(x) = \begin{cases}
\frac{1}{l-m}\bigg(x(2l-1)P^m_{l-1}(x) - (l+m-1) P^m_{l-2}(x) \bigg)  & \quad m \geq 0 , \\ \\
 (-1)^l (2l-1)!! (1-x^2)^{l/2} & \quad m=l.
\end{cases}
\end{equation*}
Note that the associated Legendre polynomials $P^m_l$ for $m<0$ are not required in our formalism. Once we have the set of associated Legendre polynomials, we can easily find the spherical harmonics using the definition of the real spherical harmonics, $Y_{lm}$.

The partial derivative $\partial Y_{lm}(\theta,\phi)/\partial \phi$, is given by:
\begin{equation}
\frac{\partial Y_{lm}(\theta,\phi)}{\partial \phi} = \begin{cases}
-m \sqrt{\frac{2l+1}{2\pi} \frac{(l-m)!}{(l+m)!}} P^m_l(cos\theta) sin(m\phi)  & \quad m > 0 , \\ \\
 0& \quad m=0, \\ \\
|m| \sqrt{\frac{2l+1}{2\pi} \frac{(l-|m|)!}{(l+|m|)!}} P^{|m|}_l(cos\theta) cos(|m|\phi) & \quad m <0.
\end{cases}
\end{equation}
For the partial derivative $\partial Y_{lm}(\theta,\phi)/\partial \theta$, we have:
\begin{equation*}
\frac{\partial Y_{lm}(\theta,\phi)}{\partial \theta} = \begin{cases}
 \sqrt{\frac{2l+1}{2\pi} \frac{(l-m)!}{(l+m)!}}  cos(m\phi) \frac{\partial P^m_l(cos\theta)}{\partial \theta} & \quad m >0 , \\ \\
 \sqrt{\frac{2l+1}{4\pi}}  \frac{\partial P^0_l(cos\theta)}{\partial \theta}  & \quad m=0, \\ \\
 \sqrt{\frac{2l+1}{2\pi} \frac{(l-|m|)!}{(l+|m|)!}}  sin(|m|\phi) \frac{\partial P^{|m|}_l(cos\theta)}{\partial \theta}  & \quad m <0.
\end{cases}
\end{equation*}
The expression for $\partial P^m_l(\cos\theta)/\partial \theta$
for $m \geq 0$ is non-zero when $l>0$ and is given by:
  \begin{equation*}
\begin{cases}
- \frac{1}{2} \bigg((l+m)(l-m+1) P^{m-1}_l(cos\theta) - P^{m+1}_l(cos\theta)    \bigg) & \quad 0 < m \leq l-1 , \\ 
P^1_l(cos\theta) & \quad m=0, \\ 
- l P^{l-1}_l(cos\theta) & \quad m =l .
\end{cases}
\end{equation*}
The choice of the recurrence relation for $m>0$ ensures that $B^{\theta}$ goes to zero at the poles of the sphere.

\section{Mathematical derivation of the spectral components of the magnetic field}
\label{sec: Mathematical derivation of Bfield}

In this section, we provide a comprehensive derivation of eq.\,\eqref{eq: magnetic field components spectral}. This derivation encompasses the incorporation of general relativistic corrections within a three-dimensional framework. The toroidal component of the magnetic field is given by  
 \begin{align}
    \boldsymbol{B}_\text{tor} &=  - \boldsymbol{r}\times \boldsymbol{\nabla} \Psi~, \nonumber\\
    &= \frac{\boldsymbol{e}_{\theta} }{ sin\theta} \frac{\partial\Psi}{\partial\phi} - \boldsymbol{e}_\phi \frac{\partial\Psi}{\partial\theta}~.
        \label{eq: Btor}
 \end{align}
$\Psi$ is defined in eq.\,\eqref{eq: Psi scalar functions}, thus
 \begin{equation}
      \boldsymbol{B}_\text{tor} = \frac{\boldsymbol{e}_{\theta} }{r sin\theta} \sum_{lm} \Psi_{lm}(r) \frac{\partial Y_{lm}(\theta,\phi)}{ \partial \phi} - \frac{\boldsymbol{e}_\phi}{r}\sum_{lm} \Psi_{lm}(r) \frac{\partial Y_{lm}(\theta,\phi)}{ \partial \theta}~.
      \label{eq: Btor derived}
 \end{equation}
The poloidal component of the magnetic field can be derived from the toroidal vector potential. The latter is expressed as follows: 
\begin{equation}
     \boldsymbol{A}_\text{tor} =  - \boldsymbol{r}\times \boldsymbol{\nabla} \Phi ~.
       \label{eq: Ator}
\end{equation}
$\Phi$ is defined in eq.\,\eqref{eq: phi scalar functions}, thus
  \begin{equation}
    \boldsymbol{A}_\text{tor} = \frac{\boldsymbol{e}_{\theta} }{r sin\theta} \sum_{lm} \Phi_{lm}(r) \frac{\partial Y_{lm}(\theta,\phi)}{ \partial \phi} - \frac{\boldsymbol{e}_\phi}{r}\sum_{lm} \Phi_{lm}(r) \frac{\partial Y_{lm}(\theta,\phi)}{ \partial \theta}. 
         \label{eq: Ator decomposed}
 \end{equation}
The poloidal component of the magnetic field is then given by:
 \begin{align}
      \boldsymbol{B}_\text{pol} &=  \boldsymbol{\nabla} \times \boldsymbol{A}_\text{tor}~,  \nonumber\\
        &= \boldsymbol{e}_{r} \bigg(\frac{1}{r} \frac{\partial A^{\phi}}{\partial\theta} - \frac{1}{r sin\theta} \frac{\partial A^{\theta}}{\partial \phi} + \frac{A^{\phi} cos\theta}{r sin\theta}  \bigg)               \nonumber\\
        &- \boldsymbol{e}_{\theta}  \bigg(\frac{\partial A^{\phi}}{\partial r} + \frac{A^{\phi}}{r}  \bigg)      \nonumber\\
        &+ \boldsymbol{e}_{\phi} \bigg(\frac{\partial A^{\theta}}{\partial r} + \frac{A^{\theta}}{r}  \bigg).
        \label{eq: Bpol as a function of Ator}
  \end{align}
Using eq.\,\eqref{eq: Ator decomposed}, one can express the radial component of the magnetic field, $B^r$, as follows:
\begin{equation}
    B^r = - \frac{1}{r^2} \sum_{lm} \Phi_{lm}(r) \bigg[\frac{\partial^2 Y_{lm} (\theta,\phi)}{\partial \theta^2} + \frac{1}{sin^2\theta} \frac{\partial^2 Y_{lm} (\theta,\phi)}{\partial \phi^2} + \frac{cos\theta}{sin\theta} \frac{\partial Y_{lm} (\theta,\phi)}{\partial \theta}    \bigg].
\end{equation}
By utilizing the property of Laplace's spherical harmonics defined in eq.\,\eqref{eq: laplace spherical harmonics}, $B^r$ can be expressed as: 
\begin{equation}
    B^r = \frac{1}{r^2} \sum_{lm} l(l+1) \Phi_{lm}(r) Y_{lm}(\theta,\phi). 
\end{equation}
The angular derivatives of the poloidal component of the magnetic field, $B^\theta_\text{pol}$ and $B^\phi_\text{pol}$, including the relativistic correction, are expressed as follows:
\begin{equation}
B^\theta_\text{pol} = \frac{1}{r} \sum_{lm}  \frac{\partial Y_{lm}(\theta,\phi)}{\partial \theta} \Bigg[ e^{-\lambda(r)} \frac{\partial \Phi_{lm}(r)}{\partial r} + \frac{(1 - e^{-\lambda(r)})}{r} \Phi_{lm}(r) \Bigg],
   \label{eq: Bth pol} 
\end{equation}
\begin{equation}
B^\phi_\text{pol} = \frac{1}{r\, sin\theta} \sum_{lm}  \frac{\partial Y_{lm}(\theta,\phi)}{\partial \phi} \Bigg[ e^{-\lambda(r)} \frac{\partial \Phi_{lm}(r)}{\partial r} + \frac{(1 - e^{-\lambda(r)})}{r} \Phi_{lm}(r) \Bigg].
   \label{eq: Bphi pol} 
\end{equation}
Using eq.\,\eqref{eq: radial derivative of the Phi scalar function}, we can express the poloidal component of the magnetic field as follows:
\begin{align}
     B_\text{pol} &= \boldsymbol{e}_{r}  \frac{1}{r^2} \sum_{lm}  l(l+1)   \Phi_{lm}(r) Y_{lm}(\theta,\phi)   \nonumber\\
     &+ \boldsymbol{e}_{\theta} \frac{1}{r} \sum_{lm} \Phi'_{lm}(r) \frac{\partial Y_{lm}(\theta,\phi)}{\partial \theta}
     \nonumber\\ &+ \boldsymbol{e}_{\phi} \frac{1}{r sin\theta}  \sum_{lm} \Phi'_{lm}(r) \frac{\partial Y_{lm}(\theta,\phi)}{\partial \phi}~. 
     \label{eq: Bpol}
\end{align}
Here, the term $\Phi'_{lm}(r)$ incorporates the relativistic corrections (eq.~\eqref{eq: radial derivative of the Phi scalar function}).

Combining the poloidal and toroidal components of the magnetic field (eq.\,\eqref{eq: Btor derived} and eq.\,\eqref{eq: Bpol}), the three components of the magnetic field in 3D spherical coordinates in spectral space are given by:
\begin{equation}
   B^{r} =  \frac{1}{r^2} \sum_{lm} l(l+1) \Phi_{lm}(r) Y_{lm}(\theta,\phi) ~, 
   \label{eq: Br spherical harmonic}
\end{equation}
\begin{equation}
    B^{\theta}=   \frac{1}{r} \sum_{lm} \Phi'_{lm}(r) \frac{\partial Y_{lm}(\theta,\phi)}{\partial \theta} +  \frac{1 }{r\, sin\theta} \sum_{lm} \Psi_{lm}(r) \frac{\partial Y_{lm}(\theta,\phi)}{ \partial \phi}~,
       \label{eq: Bth spherical harmonic}
\end{equation}
\begin{equation}
     B^{\phi} =   - \frac{1}{r}\sum_{lm} \Psi_{lm}(r) \frac{\partial Y_{lm}(\theta,\phi)}{ \partial \theta}  + \frac{1}{r\, sin\theta}  \sum_{lm}\Phi'_{lm}(r) \frac{\partial Y_{lm}(\theta,\phi)}{\partial \phi}~.
 \label{eq: Bphi spherical harmonic}
\end{equation}

\section{Mathematical derivation of the spectral magnetic energy}
\label{sec: Mathematical derivation of Espectral}

To compute the spectral magnetic energy, one needs to determine three radial scalar functions, $\Phi_{lm}(r),\, \Phi'_{lm}(r)$,
and $\Psi_{lm}(r)$. To compute $\Phi_{lm}(r)$, we utilize the expression for $B^r$ of the magnetic field (eq.\,\eqref{eq: Br spherical harmonic}) and the orthogonality property of spherical harmonics (see \S\ref{Appendix: spherical harmonics}). This allows us to express $\Phi_{lm}(r)$ as follows:
\begin{align}
    r^2\int d\Omega B^r \sum_{lm} Y_{lm}(\theta,\phi) &= \sum_{lm}  l(l+1) \Phi_{lm}(r)\int d\Omega Y_{lm} Y_{l'm'}     \nonumber\\
&= \sum_{lm} \Phi_{lm}(r)  l(l+1)~, 
\end{align}
thus 
\begin{equation}
   \sum_{lm} \Phi_{lm}(r) =  \sum_{lm} \frac{1}{l(l+1)  } \int dS_r B^r Y_{lm} (\theta,\phi). 
\end{equation}
The angular components of the magnetic field contain mixed toroidal and poloidal contributions. Therefore, to calculate $\Phi'_{lm}(r)$, we first multiply eq.\,\eqref{eq: Bth spherical harmonic} by $ \frac{\partial Y_{lm}}{ \partial \theta}$ and integrate over the solid angle:
  \begin{equation}
    r \int d\Omega \, B^{\theta} \,  \frac{\partial Y_{lm}}{\partial\theta}  = \sum_{lm}   \Phi'_{lm}(r) \int d\Omega \Bigg(\frac{\partial Y_{lm}}{\partial\theta}\Bigg)^2 + \sum_{lm}  \Psi_{lm}(r)   \int \frac{d\Omega}{sin\theta} \frac{\partial Y_{lm}}{\partial\theta} \frac{\partial Y_{lm}}{\partial\phi}.
    \label{eq: r bth spec Clm}
  \end{equation}
Then, we multiply eq.\,\eqref{eq: Bphi spherical harmonic} by $\frac{1}{\sin\theta}\frac{\partial Y_{lm}}{\partial\phi}$ and integrate over the solid angle:
\begin{equation}
 r \int    \frac{d\Omega \,  B^{\phi}}{sin\theta}  \frac{\partial Y_{lm}}{\partial\phi} = - \sum_{lm}  \Psi_{lm}(r)   \int \frac{d\Omega}{sin\theta} \frac{\partial Y_{lm}}{\partial\theta} \frac{\partial Y_{lm}}{\partial\phi} + \sum_{lm} \Phi'_{lm}(r)  \int \frac{d\Omega}{sin^2\theta}
    \Bigg(\frac{\partial Y_{lm}}{\partial\phi}\Bigg)^2.
    \label{eq: r bphi spec Clm}
  \end{equation}
Summing eq.\,\eqref{eq: r bth spec Clm} and eq.\,\eqref{eq: r bphi spec Clm}, and applying the orthogonality properties of spherical harmonics  (eqs.\,\eqref{eq: orthogonality of square of dYlm/dtheta } and \eqref{eq: orthogonality of square of (dYlm/dphi/sin theta)}), we obtain:
\begin{equation}
   \sum_{lm} \Phi'_{lm}(r) =  \sum_{lm} \frac{1}{l(l+1)} \int \frac{dS_r}{r} \Bigg(B^{\theta}\frac{\partial Y_{lm}}{\partial \theta}  +  \frac{B^{\phi}}{sin\theta} \frac{\partial Y_{lm}}{\partial \phi}  \Bigg). 
\end{equation}
To calculate $\Psi_{lm}(r)$, we multiply eq.\,\eqref{eq: Bth spherical harmonic} by $\frac{1}{\sin\theta}\frac{\partial Y_{lm}}{\partial\phi}$ and integrate over the solid angle:
   \begin{equation}
    r \int \frac{d\Omega\, B^{\theta} }{sin\theta}\frac{\partial Y_{lm}}{\partial\phi}  = \sum_{lm}    \Phi'_{lm}(r) \int \frac{d\Omega}{sin\theta} \frac{\partial Y_{lm}}{\partial\theta} \frac{\partial Y_{lm}}{\partial\phi} + \sum_{lm}  \Psi_{lm}(r)   \int \frac{d\Omega}{sin^2\theta} \Bigg( \frac{\partial Y_{lm}}{\partial\phi}\Bigg)^2.
    \label{eq: r bth spec dlm}
  \end{equation}
Then, we multiply eq.\,\eqref{eq: Bphi spherical harmonic} by $\frac{\partial Y_{lm}}{\partial\theta}$ and integrate over the solid angle:
   \begin{equation}
 r \int    d\Omega \, B^{\phi}  \,  \frac{\partial Y_{lm}}{\partial\theta} = - \sum_{lm}  \Psi_{lm}(r)   \int d\Omega \Bigg( \frac{\partial Y_{lm}}{\partial\theta} \Bigg)^2 + \sum_{lm}   \Phi'_{lm}(r) \int \frac{d\Omega}{sin\theta}
    \frac{\partial Y_{lm}}{\partial\phi}  \frac{\partial Y_{lm}}{\partial\theta}.
    \label{eq: r bphi spec dlm}
  \end{equation}
  Subtracting eq.\,\eqref{eq: r bth spec dlm} from eq.\,\eqref{eq: r bphi spec dlm} and applying the orthogonality properties provided in eqs.\,\eqref{eq: orthogonality of square of dYlm/dtheta } and \eqref{eq: orthogonality of square of (dYlm/dphi/sin theta)}, we obtain:
  \begin{equation}
   \sum_{lm} \Psi_{lm}(r) =  \sum_{lm} \frac{1}{l(l+1)} \int \frac{dS_r}{r} \Bigg( \frac{B^{\theta}}{sin\theta}\frac{\partial Y_{lm}}{\partial \phi}  -  B^{\phi} \frac{\partial Y_{lm}}{\partial \theta}  \Bigg). 
\end{equation}
The spectral magnetic energy is expressed as follows:  \begin{equation}
      E_\text{mag} = \frac{1}{2} \int e^{\lambda(r)} r^2 dr \int d\Omega \big( (B^r)^2 + (B^{\theta})^2 + (B^{\phi})^2    \big). 
  \end{equation}
The radial magnetic energy is defined as:  \begin{align}
   E^\text{rad}_{lm}&= \int e^{\lambda(r)} r^2 dr \int d\Omega\,  (B^r)^2 \nonumber\\
   &= \sum_{lm} \int e^{\lambda(r)} \frac{dr}{r^2}  \int  d\Omega \, \Phi_{lm}(r) \, \Phi_{l'm'}(r) \,Y_{lm}(\theta,\phi) Y_{l'm'}(\theta,\phi) ll'(l+1) (l'+1).
     \end{align}
By applying the orthogonality property of spherical harmonics (eq.\,\eqref{eq: orthogonality of real spherical harmonics}), we can derive:
\begin{equation}
     E^\text{rad}_{lm} =  \sum_{lm}  \int e^{\lambda(r)} \frac{dr}{r^2}    l^2(l+1)^2  \Phi_{lm}^2(r). 
\end{equation}
The angular magnetic energy is expressed as follows (assuming $l=l'$ and $m=m'$):
\begin{align}
   E^\text{ang}_{lm} &= \int e^{\lambda(r)} r^2 dr \int d\Omega \bigg((B^{\theta})^2 + (B^{\phi})^2   \bigg) \nonumber\\
   &= \int e^{\lambda(r)} r^2 dr \int d\Omega \Bigg[ \frac{1}{r^2} \sum_{lm} \bigg(\Phi'_{lm}(r)\bigg)^2 \Bigg(\frac{\partial Y_{lm}(\theta,\phi)}{\partial \theta} \Bigg)^2 \nonumber\\
   &+ \frac{1}{r^2 sin^2 \theta} \sum_{lm} \Psi_{lm}^2(r) \Bigg(\frac{\partial Y_{lm}(\theta,\phi)}{\partial \phi} \Bigg)^2 \nonumber\\
   &+ \frac{2}{r^2 sin \theta} \sum_{lm}\Phi'_{lm}(r)\Psi_{lm}(r) \frac{\partial Y_{lm}(\theta,\phi)}{\partial \theta}   \frac{\partial Y_{lm}(\theta,\phi)}{\partial \phi} \nonumber\\
   &+ \frac{1}{r^2} \sum_{lm} \Psi_{lm}^2(r) \Bigg(\frac{\partial Y_{lm}(\theta,\phi)}{\partial \theta} \Bigg)^2 \nonumber\\
     &+ \frac{1}{r^2 sin^2\theta} \sum_{lm} \bigg(\Phi'_{lm}(r)\bigg)^2 \Bigg(\frac{\partial Y_{lm}(\theta,\phi)}{\partial \phi} \Bigg)^2 
     \nonumber\\
     &- \frac{2}{r^2 sin \theta} \sum_{lm}\Phi'_{lm}(r)\Psi_{lm}(r) \frac{\partial Y_{lm}(\theta,\phi)}{\partial \theta}   \frac{\partial Y_{lm}(\theta,\phi)}{\partial \phi}
   \Bigg].
\end{align}
Thus,
\begin{align}
   E^\text{ang}_{lm} = \int e^{\lambda(r)} dr  \sum_{lm} \Bigg[ \bigg(\Phi'_{lm}(r) \bigg)^2 + \Psi_{lm}^2(r)\Bigg] \int d\Omega \Bigg[ \Bigg(\frac{\partial Y_{lm}(\theta,\phi)}{\partial \theta} \Bigg)^2  + \frac{1}{sin^2\theta} \Bigg(\frac{\partial Y_{lm}(\theta,\phi)}{\partial \phi} \Bigg)^2   \Bigg]. \nonumber\\
\end{align}
Utilizing the orthogonality properties of spherical harmonics as defined in eqs.\,\eqref{eq: orthogonality of square of dYlm/dtheta } and \eqref{eq: orthogonality of square of (dYlm/dphi/sin theta)}, we find:
\begin{equation}
     E^\text{ang}_{lm} =  \int e^{\lambda(r)} dr  \sum_{lm} l(l+1) \bigg[ \bigg(\Phi'_{lm}(r) \bigg)^2 + \Psi_{lm}^2(r)\bigg].   
\end{equation}
The spectral magnetic energy in c.g.s. units, including the general relativistic correction, is expressed as:
 \begin{equation}
         E_{lm} = \frac{1}{8\pi} \int  e^{\lambda(r)+\zeta(r)} dr ~l(l+1) \bigg[  \frac{l(l+1)}{r^2} \Phi_{lm}^2(r) +\big(\Phi'_{lm}(r)\big)^2   +\Psi_{lm}^2(r)  \bigg].
 \end{equation}

\section[Mathematical derivation of the spectral energy for r >= R]{Mathematical derivation of the spectral energy for $r\geq R$}
\label{sec: Mathematical derivation of Esurf}

At the surface of the star, the radial scalar function is known (see \S\ref{subsec: outer B.C} for more details). This is a result of considering potential boundary conditions ($\nabla^2 \chi_m = 0$). Therefore, for $r \geq R$, the three components of the magnetic field are defined as $\boldsymbol{B} = \boldsymbol{\nabla}\chi_m$:
\begin{align}
   B^r &= e^{-\lambda(r)}\sum_{lm} Y_{lm} b_{lm} (l+1) \bigg( \frac{R}{r} \bigg)^{(l+2)}, \nonumber\\
   B^\theta &= -  \sum_{lm} b_{lm} \frac{\partial Y_{lm}}{\partial \theta} \Bigg( \frac{R}{r} \Bigg)^{(l+2)}, \nonumber\\ 
   B^\phi &= - \frac{1}{sin\theta} \sum_{lm} b_{lm} \frac{\partial Y_{lm}}{\partial \phi} \Bigg( \frac{R}{r} \Bigg)^{(l+2)}. 
\end{align}
The radial spectral energy $E^{r\geq R}_{r}$ is expressed as follows: \begin{align}
     E^{r\geq R}_r &= \int e^{\lambda(r)} r^2 dr \int d \Omega (B^r)^2 ~,\nonumber\\
 &= \int e^{-\lambda(r)} r^2 dr \sum_{lm} (l+1)^2 b_{lm}^2 \Bigg( \frac{R}{r} \Bigg)^{(2l+4)}~.  
 \end{align}
The angular component of the magnetic energy $E^{r\geq R}_{\theta}$, is defined as:
  \begin{equation}
      E^{r\geq R}_{\theta} = \int e^{\lambda(r)} r^2 dr \sum_{lm} \sum_{l'm'} b_{lm} b_{l'm'} \Bigg( \frac{R}{r} \Bigg)^{(l+2)} \Bigg( \frac{R}{r} \Bigg)^{(l'+2)} \int d\Omega \frac{\partial Y_{lm}}{\partial \theta} \frac{\partial Y_{l'm'}}{\partial \theta}~.
  \end{equation}
Applying the orthogonality property as described in eq.\,\eqref{eq: orthogonality of square of dYlm/dtheta }, we obtain:
   \begin{equation}
      E^{r\geq R}_{\theta} = \int e^{\lambda(r)} r^2 dr\sum_{lm} b_{lm}^2\Bigg( \frac{R}{r} \Bigg)^{(2l+4)} \Bigg( l(l+1) - \frac{(2l+1)m}{2} \Bigg)~.
      \label{eq: Eth surf}
      \end{equation}
The angular component of the magnetic energy $E^{r\geq R}_{\phi}$, is expressed as:
\begin{equation}
         E^{r\geq R}_{\phi} = \int e^{\lambda(r)} r^2 dr \sum_{lm} \sum_{l'm'} b_{lm} b_{l'm'} \Bigg( \frac{R}{r} \Bigg)^{(l+2)} \Bigg( \frac{R}{r} \Bigg)^{(l'+2)} \int \frac{d\Omega}{sin^2 \theta} \frac{\partial Y_{lm}}{\partial \phi} \frac{\partial Y_{l'm'}}{\partial \phi}~.
\end{equation}
Applying the orthogonality property as described in eq.\,\eqref{eq: orthogonality of square of (dYlm/dphi/sin theta)}, we obtain: 
 \begin{equation}
      E^{r\geq R}_{\phi} =  \int e^{\lambda(r)} r^2 dr\sum_{lm} b_{lm}^2\Bigg( \frac{R}{r} \Bigg)^{(2l+4)}  \frac{(2l+1)m}{2}~ . 
      \label{eq: Ephi surf}
      \end{equation}
Note that eq.\,\eqref{eq: Ephi surf} cancels with the second term of eq.\,\eqref{eq: Eth surf}. Therefore, the spectral magnetic energy for $r\geq R$, in c.g.s. units and including the relativistic correction, is expressed as:
  \begin{equation}
      E^{r\geq R}_{lm} = \frac{1}{8\pi}\int e^{\lambda(r) + \zeta(r)} r^2 dr \sum_{lm}   b_{lm}^2 \Bigg( e^{-2\lambda(r)} (l+1)^2 + l(l+1) \Bigg) \Bigg( \frac{R}{r} \Bigg)^{(2l+4)} . 
  \end{equation}

\clearemptydoublepage
\let\textcircled=\pgftextcircled
\chapter{Crust–Magnetosphere Coupling Using Physics-Informed Neural Networks}
\label{appendix: neural networks}

\initial{T}he magnetosphere of a neutron star plays a crucial role in explaining various observational properties \citep{beloborodov2009, akgun2017}. Although our work does not delve into the 3D connection between the star's interior magnetic evolution and its magnetosphere, we present the results of a novel approach using physics-informed neural networks (PINN). This approach allows us to examine the astrophysical scenario and investigate the impact of the force-free magnetosphere on the neutron star's magnetic field evolution in 2D.

Conducting a global simulation for the entire domain is computationally expensive because it requires an elliptic solver for the exterior solution \citep{akgun2016,akgun2017,Stefanou23}. To tackle this challenge, \cite{urban2023} explored the applicability of a PINN solver for elliptic problems, with a specific focus on modeling the force-free magnetospheres of non-rotating, axisymmetric neutron stars.

In this appendix, we provide a concise explanation of what a PINN is in \S\ref{sec: NN}. Moving forward to \S\ref{sec: axisymmetric magnetic field formalism}, we offer a brief reminder of the axisymmetric magnetic field formalism in terms of poloidal and toroidal stream functions, denoted as $\cal{P}$ and $\cal{T}$, respectively. In \S\ref{subsec: potential PINN}, we utilize PINN to recover the 2D potential boundary conditions, allowing us to assess its performance. Subsequently, in \S\ref{subsec: FF BC}, we couple the interior 2D magneto-thermal evolution with the force-free magnetosphere, providing a brief evaluation of the impact of the new boundary conditions on the final outcomes.

\section{Physics-Informed Neural Networks}
\label{sec: NN}

\begin{figure}
    \centering
    \includegraphics[width=0.7\columnwidth]{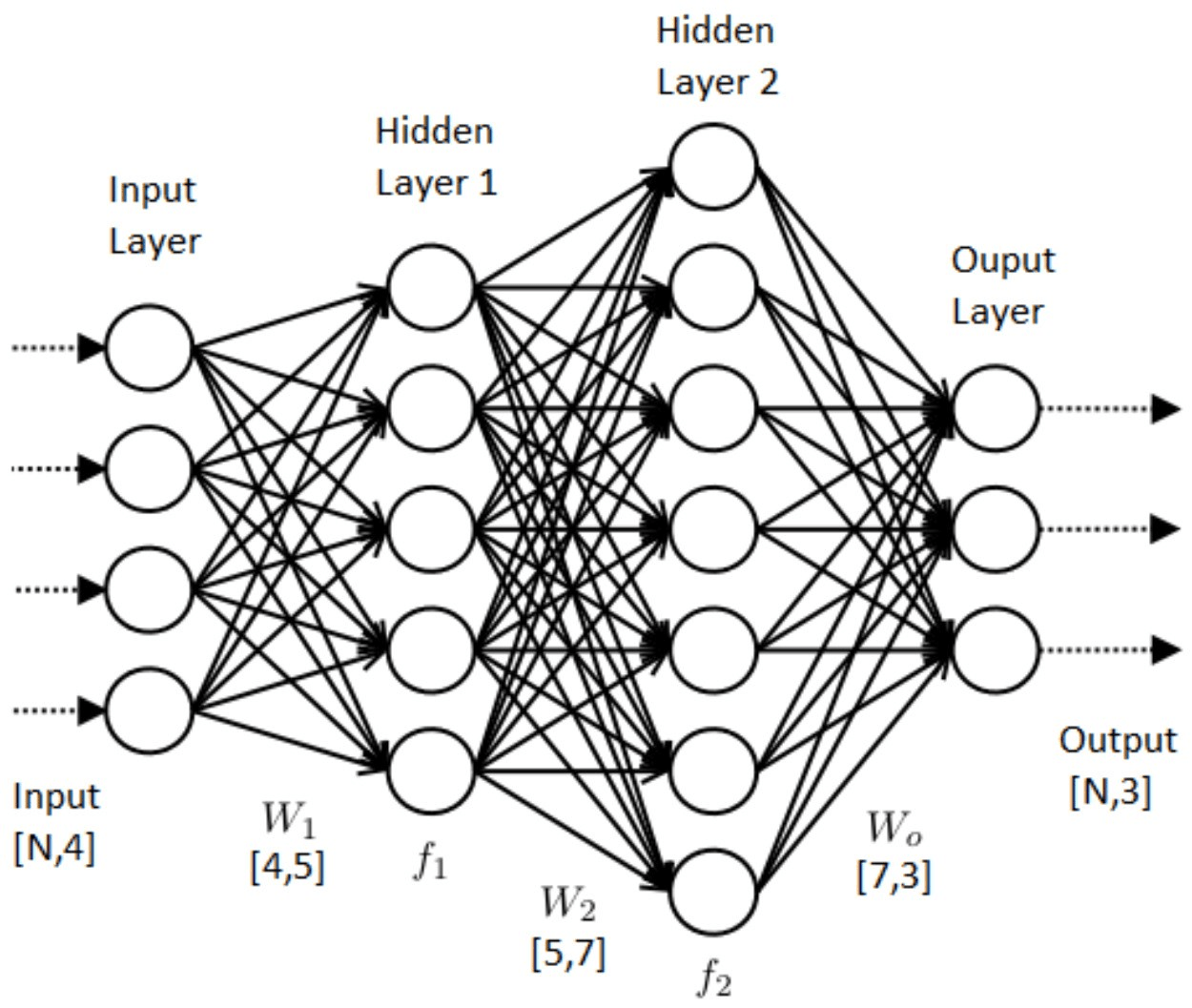}
    \caption{A schematic representation of a deep fully-connected neural network.}
    \label{fig:fcnn_sketch}
\end{figure}

Neural networks (NNs) serve as powerful tools for approximating complex mathematical functions \citep{Hornik_Stinchcombe_White_1989}. They consist of interconnected layers of neurons, with each neuron performing transformations on the input data. The specific arrangement of these interconnected layers defines the NN's architecture. The most common type is the Fully Connected Neural Network (FCNN), where neurons are linked to all neurons in the preceding and succeeding layers, as depicted in Fig.~\ref{fig:fcnn_sketch}. Training a NN involves adjusting its parameters, namely weights and biases, to minimize the disparity between the predicted output and the desired output. This optimization process is achieved through backpropagation, which utilizes gradient descent to update the parameters.

The standard approach to training a NN involves using data with true solution values at specific points in the input domain. However, this method requires a large number of training examples to establish a reliable relationship between inputs and outputs. For astrophysical systems, this would mean relying on data from an extensive set of observations.

The key innovation in PINN is the integration of physical laws into the training process. Instead of focusing on the difference between predictions and real data or exact solutions, PINN minimizes the residual of the governing partial differential equations (PDEs). Although using data in the loss function can be helpful for optimization, it is not essential in practice. The loss function used in training aims to make the NN's prediction as close to an exact solution as possible, incorporating the physical constraints directly \citep{urban2023}.

One of the notable advantages of PINNs is their ability to handle various boundary conditions and source terms, making them versatile PDE solvers capable of approximating solutions for different operators and boundary conditions. By intelligently encoding the boundary conditions as part of the network's input, PINNs generalize well to provide solutions for various points and boundary conditions in the domain. For a more detailed exploration of the PINN solver for elliptic problems, we suggest referring to \cite{urban2023}.

\section{2D magnetic field formalism}
\label{sec: axisymmetric magnetic field formalism}

In general, any magnetic field (or any divergence-less, i.e. solenoidal field) can be expressed as the sum of a poloidal and a toroidal field \citep{chandrasekhar1957,chandrasekhar1981}, as described in Appendix~\ref{appendix: Magnetic field formalism}. Specifically, in the case of axisymmetry, we define the poloidal and toroidal stream functions as $\cal{P}$ and $\cal{T}$, respectively. The magnetic field can then be represented as:
\begin{equation}
    \boldsymbol{B} = \boldsymbol{\nabla} \cal{P} \times  \boldsymbol{\nabla} \phi + \cal{T}  \boldsymbol{\nabla} \phi,
\end{equation}
where $\boldsymbol{\nabla} \phi= \frac{\boldsymbol{e_\phi}}{r \sin\theta}$, with $\boldsymbol{e_{\phi}}$ being the azimuthal unit vector, $r$ representing the radius, and $\theta$ indicating the polar angle in spherical coordinates ($r, \theta, \phi$). The poloidal and toroidal stream functions can be expressed in terms of the vector potential $A_\phi$ and $B_{\phi}$, respectively:
\begin{equation}
    \cal P  = r \, sin\,\theta ~
    \,A_\phi,
    \label{eq: P poloidal flux function}
\end{equation}
\begin{equation}
    \cal T  = r \, sin\, \theta ~
    \,B_\phi.
    \label{eq: T toroidal flux function}
\end{equation}

\subsection{Potential boundary conditions}
\label{subsec: potential PINN}

We initiate by testing the PINN's performance considering a crustal-confined magnetic field topology and implementing vacuum boundary conditions, which imply no electrical currents circulating in the envelope and across the surface. To enforce these boundary conditions, we utilize a multipole expansion of the radial magnetic field at the surface, following the method described in \cite{pons2009,pons2019}.

The coefficients of the multipole expansion in 2D can be computed from the radial component of the magnetic field at the surface of the star as follows: 
\begin{equation}
    b_l = \frac{2l+1}{2(l+1)} \int_0^\pi B^r(R,\theta) P_l(\cos\theta) \sin\theta d\theta,
    \label{eq: bl}
\end{equation}
where $P_l$ are the Legendre polynomials. 

During the evolution, at each time step, we calculate the $b_l$ coefficients using eq.\,\eqref{eq: bl}. In the classical approach, we explicitly reconstruct the values of $B^r$ and $B^\theta$ in the external ghost cells as follows:
\begin{align}
    B^r &= \sum_{l=1}^{l_\text{max}}{b_l\left(l+1\right)P_l\left(\cos \theta \right)\left(\frac{R}{r}\right)^{l+2}} \label{Br_vac}, \\ 
    B^\theta &= -\sin \theta \sum_{l=1}^{l_\text{max}}{b_l P_l'\left(\cos \theta \right)\left(\frac{R}{r}\right)^{l+2}}, \label{Bth_vac}
\end{align}
where $R$ represents the radius of the neutron star. For brevity, we refer to this procedure using the nomenclature "OLD". 

In the new PINNs approach, we utilize the $b_l$ coefficients obtained from the Legendre decomposition as inputs to the PINN. The PINN, in turn, provides values of the poloidal flux function ${\cal P}$ (eq.\,\eqref{eq: P poloidal flux function}) or any required component of the magnetic field by taking derivatives. Admittedly, in this case (vacuum boundary conditions), the PINN approach may not offer any apparent advantage since we already know how to construct the analytical solution. However, our intention is to ensure that the simulations produce results without any undesirable effects before proceeding to more complex cases.

\begin{figure}
    \centering
    \includegraphics[width=.47\textwidth]{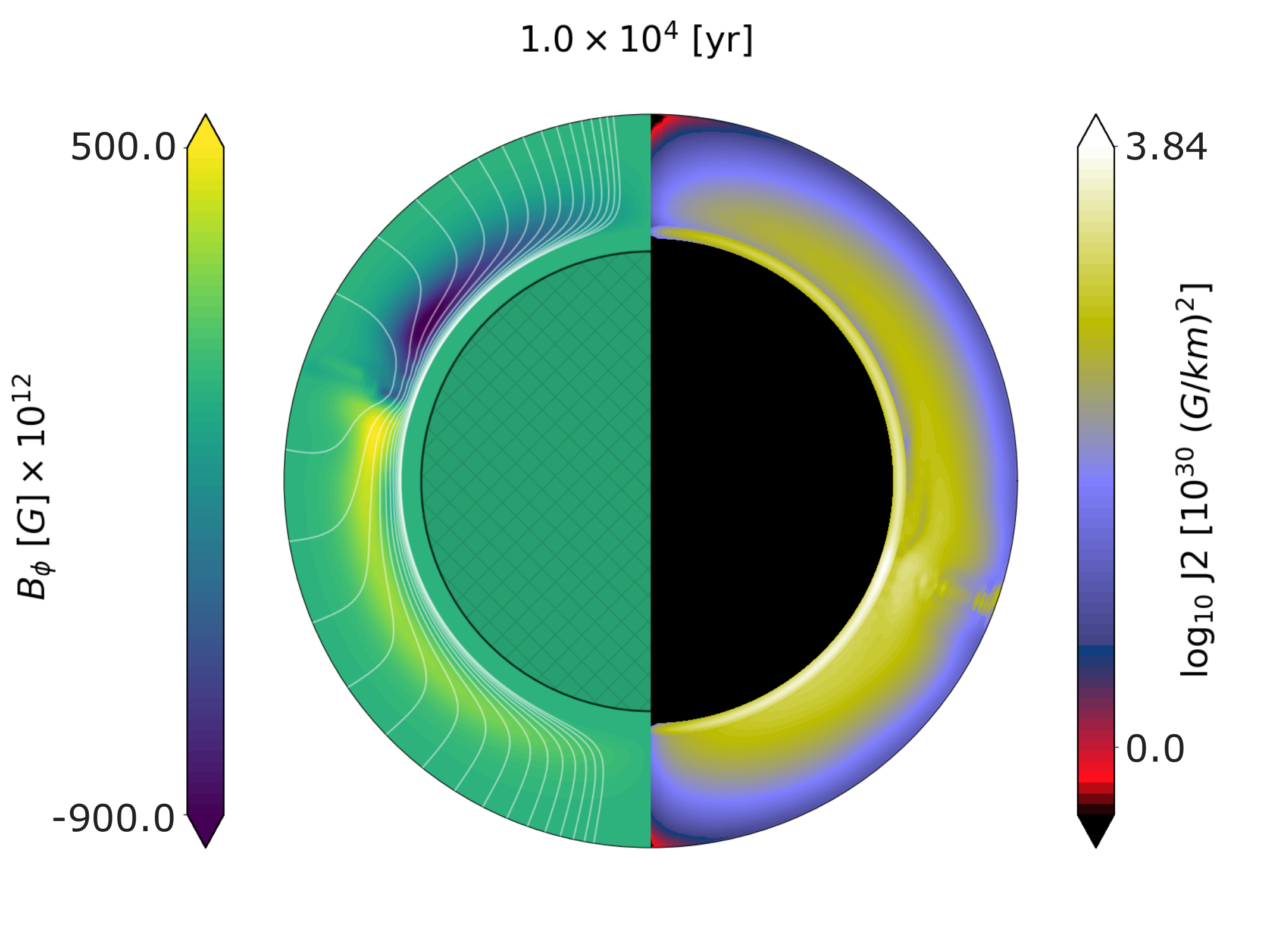}
    \includegraphics[width=.47\textwidth]{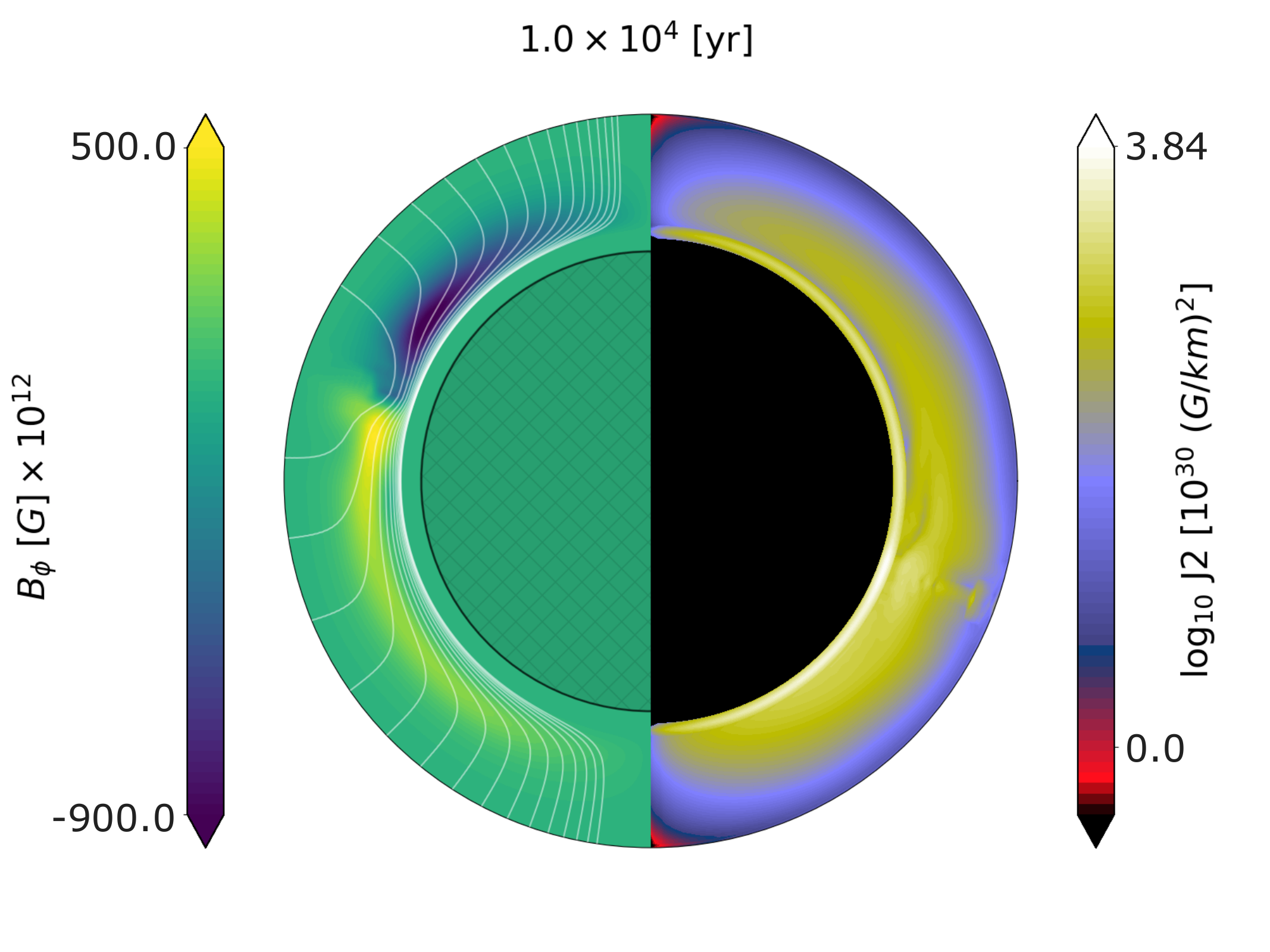}
    \caption[A snapshot of the magnetic field evolution and the electric current at $10$\,kyr, obtained using OLD and PINN.]{A snapshot of the magnetic field evolution and the electric current at $10$\,kyr, obtained using OLD (left panel) and PINN (right panel). In the left hemisphere, we show the meridional projection of the magnetic field lines (white lines) and the toroidal field (colors). In the right hemishphere, we display the square of the modulus of the electric current, i.e., $|J|^2$ (note the $\log$ scale). The crust has been enlarged by a factor of 8 for visualization purposes.}
    \label{fig: B field OLD PINN}
\end{figure}

To compare the different techniques employed, as described above, we conducted axisymmetric crustal-confined magnetic field simulations using the 2D magneto-thermal code \citep{vigano2021}. The simulations were performed on a grid with $99$ angular points (from pole to pole) and $200$ radial points. The initial magnetic field consists of a poloidal component of $10^{14}$\,G (value at the pole) and a sum of a dipole ($b_1=1$), a quadrupole ($b_2=0.6$), and an octupole ($b_3=0.3$). Additionally, the initial toroidal quadrupolar component has also a maximum initial value of $10^{14}$\,G.

We set the maximum number of multipoles to $l_{\text{max}}=7$ for the PINN and $l_{\text{max}}=50$ for the classical approach (OLD). The reason for using different $l_{\text{max}}$ values is to assess the impact of truncating the multipole number when employing the PINN. Training the PINN with a high number of multipoles requires additional hidden layers, making the training process computationally expensive. However, our primary objective in this study is to test the effectiveness of the PINN rather than achieving extremely accurate results.

The results of the comparison at $t=10$\,kyr are depicted in Fig.~\ref{fig: B field OLD PINN}. On the left (right) side, we present the magnetic field profiles obtained with OLD (PINN) boundary conditions. The overall evolution of the magnetic field and electric current is very similar between the two approaches. However, slight differences arise due to the multipolar truncation in the PINN case. It is important to note that if we set the same maximum number of multipoles for both systems, we obtain nearly identical results.

\subsection{Force-free boundary conditions}
\label{subsec: FF BC}

To couple the internal field evolution with a force-free magnetosphere using the PINN solver, we need to extend the vacuum case (\S\ref{subsec: potential PINN}) with additional steps. In our magneto-thermal evolution code, we impose external boundary conditions by providing the values of the magnetic field components in two radial ghost cells for every angular cell. In the vacuum case, once the multipolar decomposition of the radial field over the surface is known, the solution in the ghost cells can be built analytically. However, in the general case, one must solve an elliptic equation on a different grid, which must extend far from the surface to properly capture the asymptotic behavior at long distances \citep{akgun2016,Stefanou23}.

This process is indeed computationally intensive since it needs to be repeated tens of thousands of times as the interior field evolves. However, once we have trained the PINN, it becomes a valuable tool that enables us to rapidly obtain the necessary values of the solution in the ghost cells. We proceed as follows. At each evolution time step, we must know the toroidal function $\cal{T(P)}$. For the sake of simplicity in this application, we use a quadratic function:
\begin{equation}
    \cal{T (P)} = s_1 \cal{P} + s_2 \cal{P}^2, 
    \label{eq:T-of-P-quadratic}
\end{equation}
where $s_1$ and $s_2$ are parameters controlling the relative strength of the toroidal and poloidal components, the region where the toroidal field is non-zero, and the non-linearity of the model.

Then, at each time step, we perform a fit to the values obtained from the internal evolution just one cell below the surface. This fitting process yields the coefficients $s_1$ and $s_2$ required for the quadratic interpolation defined in eq.\,\eqref{eq:T-of-P-quadratic}.

Next, we include $s_1$ and $s_2$ as additional input parameters during the forward pass of the PINN \citep{urban2023}. Subsequently, the PINN provides us with the poloidal flux function $\cal{P}$ and the magnetic field components necessary for the ghost cells in the magneto-thermal evolution code. The toroidal flux function $\cal{T}$ can be constructed from $\cal{P}$ (eq.\,\eqref{eq:T-of-P-quadratic}), which is then used to determine the $B_{\phi}$ component of the magnetic field using eq.\,\eqref{eq: T toroidal flux function}. Armed with this information, the internal evolution can proceed to the next time step.

\begin{figure}
    \centering
    \includegraphics[width=.47\textwidth]{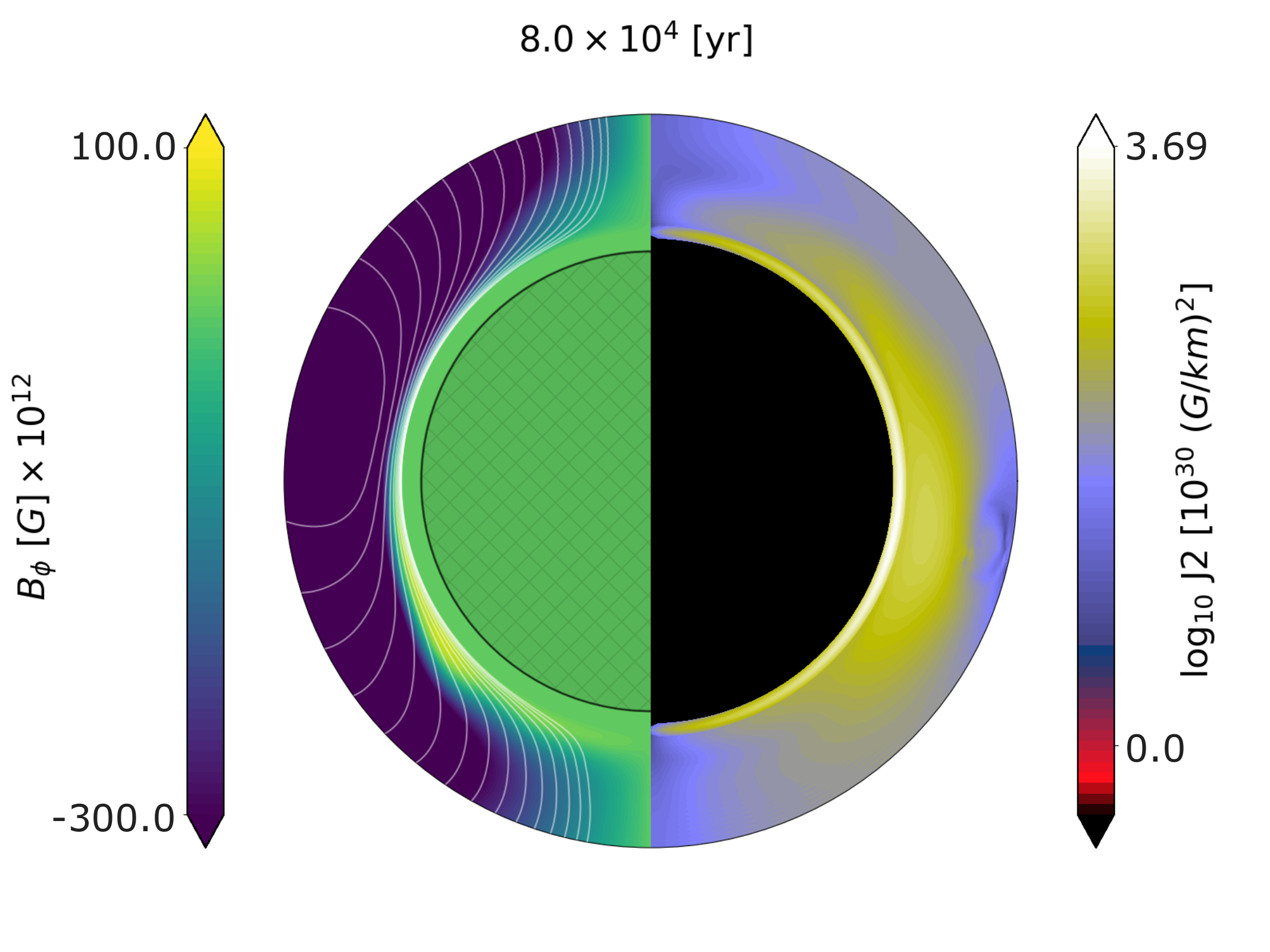}
    \includegraphics[width=.47\textwidth]{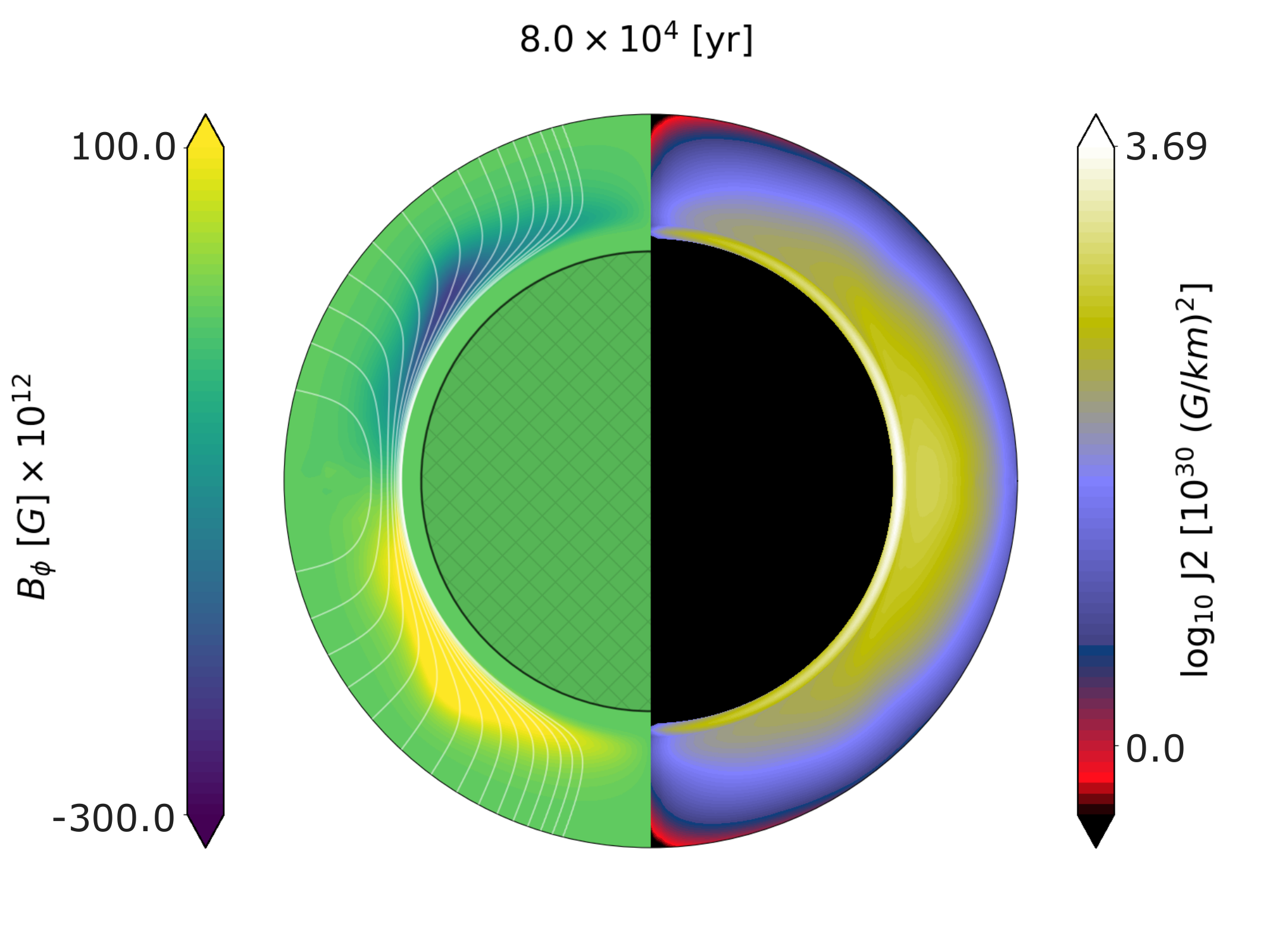}
    \caption[A snapshot of the magnetic field evolution and the electric current at $80$\,kyr, obtained using force-free boundary conditions and vacuum boundary conditions.]{Same as Fig.~\ref{fig: B field OLD PINN}.
    A snapshot of the magnetic field evolution and the electric current at $80$\,kyr. Left panel: force-free boundary conditions. Right panel: vacuum boundary conditions. }  
    \label{fig: B field PINN FF vs VAC}
\end{figure}

We assume an initial force-free magnetic field with a poloidal component of $3\times 10^{14}$\,G at the polar surface and a maximum toroidal field of $3\times 10^{14}$\,G. To investigate the impact of different boundary conditions, we consider two cases: force-free boundary conditions (left panel of Fig.~\ref{fig: B field PINN FF vs VAC}) and vacuum boundary conditions (right panel of Fig.~\ref{fig: B field PINN FF vs VAC}).

The results of the comparison are shown in two snapshots at $t=80$\,kyr of the evolution, using the same initial model. We observe distinct magnetic field evolutions depending on the applied boundary conditions. With force-free boundary conditions (left panel), a stronger toroidal dipole is present near the surface, slightly displaced towards the north. This stronger toroidal component compresses the poloidal field lines closer to the poles. On the other hand, when using vacuum boundary conditions (right panel), the poloidal field lines retain a certain symmetry with respect to the equator, and the dominant toroidal component becomes quadrupolar and concentrated at the crust/core interface.

Moreover, the distribution of the electric current in the stellar crust differs between the two cases. Enforced by the vacuum boundary conditions, the current tends to vanish around the poles and close to the surface, similar to what was observed in Fig.~\ref{fig: B field OLD PINN}, despite the initial field topology being different. However, with force-free boundary conditions, currents near the surface are not forced to vanish. In the left panel, the slightly more yellowish region in the northern hemisphere and mid-latitudes indicates that significant current flows into the magnetosphere. These differences in current configurations have significant implications for the observed temperature distribution, as discussed in \cite{akgun2018}.

This study highlights the importance of coupling the evolution of the interior magnetic field with the magnetospheric field. Interestingly, our results demonstrate that the computational cost of using a PINN is more than an order of magnitude lower than that of a finite difference scheme, with only a slight compromise on accuracy. These promising findings pave the way for future extensions to 3D problems, which would otherwise be prohibitively expensive with generalized boundary conditions. Currently, we are actively working on extending these results to the 3D domain.

\clearemptydoublepage
\let\textcircled=\pgftextcircled
\chapter{Observational Data}
\label{appendix: observational data}

\initial{W}ithin this appendix, we present a comprehensive overview of the data reduction processes and analytical methodologies utilized in our study of the selected sample of isolated neutron stars, as outlined in Chapter~\ref{chap: Comparison with observations}. The timing properties and age estimates of these stars are detailed in Table~\ref{tab:timing-1}, whereas Table~\ref{tab:spectral} delves into the emission properties attributed to the thermally emitting neutron stars.

\section{On the data reduction process}
\label{subs:data-red}

For all the chosen sources, we examined the available observations within the \xmm\ and/or \cxo\ archives, selecting only one or a few observations per target. In particular, our selection adhered to the subsequent criteria:\\
(i) For each source, we gave preference to \xmm\ observations over \cxo\ when exposure times were comparable, owing to the former's larger collecting area. We favored \cxo\ for young neutron stars accompanied by notably bright nebulae, where \cxo's excellent spatial resolution played a crucial role in discerning the thermal emission from the source. \\
(ii) When multiple observations were available, preference was given to those with the longest exposure. Additionally, we selected multiple observations conducted very close in time, such as those separated by approximately a few days.\\
(iii) In the case of variable sources, like magnetars, our search was limited to observations carried out during quiescence, and we again selected the one with the lengthiest exposure. \\
After identifying and downloading the final dataset, we proceeded with the data reduction procedures. The subsequent paragraphs detail these procedures for both instruments.

In this study, we will exclusively incorporate data from the EPIC-pn detector \citep{Struder2001} aboard \xmm. Data reprocessing was conducted utilizing \xmm\, Science Analysis Software (SAS) v. 20.0.0. To enhance the quality of the data, we initially filtered the event files to exclude periods of elevated background activity. We systematically examined potential pile-up effects using the \textsc{epatplot} tool.
For source count extraction, a circle with a 30-arcsec radius, centered on the source coordinates, was typically employed. Nevertheless, exceptions were made for sources enveloped by a pulsar wind nebula, for which a 15-arcsec region was chosen to minimize contamination from neighboring non-thermal emissions. Conversely, a region of the same size and shape, positioned at a sufficient distance from the source on the same charge-coupled device (CCD), was selected for background estimation.

All \cxo\ observations utilized in this study were conducted using the Advanced CCD Imaging Spectrometer \cite[ACIS;][]{Garmire2003}. The data underwent processing, reduction, and spectral extraction within \textsc{CIAO} v. 4.14, employing standard pipelines. The most up-to-date \textsc{CALDB} calibration files were consistently applied. 
For the source region extraction, a circular region with a 3-pixel radius was employed. This size was chosen to minimize potential contamination from any surrounding nebula (if present). Similarly, a region of the same dimensions, positioned far from both the source and the nebula, was designated as the background region.
Science products, including spectra, redistribution matrix files, and ancillary response files, were generated using the \texttt{specextract} task. In a few cases, the measured count rate reached a level where pile-up effects became significant. In these cases, a \textsc{pileup} component was incorporated into the spectral model used to describe the data in \textsc{Xspec}, allowing for accurate incorporation of pile-up-induced alterations in the spectra.

\section{Spectral analysis}
\label{sec: spectral analysis}

The spectral analysis was carried out using \textsc{Xspec} v.12.12.1. To ensure reliable statistical results, all spectra were binned to contain a minimum of 20 counts per bin, allowing for the use of $\chi^2$ statistics. However, for a few sources, the spectral statistics were insufficient for this grouping strategy and the corresponding $\chi^2$ statistics. In these cases, a grouping of 10 counts per bin was employed, and the Cash-statistics were adopted instead.

In all data models employed to describe the data, we included the \textsc{tbabs} component to account for interstellar absorption effects. The photoelectric cross-sections were set to the values provided by \cite{Verner1996}, while the elemental abundances were set to the standard tables by \cite{Wilms2000}.

Various models were tested for each source, including a simple blackbody component (BB) described using \textsc{bbodyrad} within \textsc{Xspec}, a double \textsc{bbodyrad} model (2BB), a combination of blackbody and power-law components (BB+PL), and a double blackbody plus power-law model (2BB+PL). To estimate the thermal luminosity for each source, the flux associated with the blackbody component(s) was calculated using the \texttt{flux} command in \textsc{Xspec}. Finally, the known distances of the sources were utilized to convert these fluxes into thermal luminosities ($L_{th}$). To account for uncertainties in the estimated distance and/or the flux estimation process, a conservative 20\% relative error was assigned to each $L_{th}$ value. Detailed information on estimated $L_{th}$ values, along with selected parameters for the best-fitting model, can be found in Table~\ref{tab:spectral}.

\begingroup 
\setlength\tabcolsep{4pt}
\footnotesize

\setlength\LTcapwidth{\textwidth} 

\setlength\LTleft{0pt}            
\setlength\LTright{0pt}           

\begin{longtable}{l l c c c c c c l l }
\caption{\small Timing properties and age estimates of our sample of isolated neutron stars. Here $\dot{E}_{rot}=3.9\times 10^{46}\dot{P}/P^3$\,erg/s is the rotational energy loss and $B_p=6.4\times 10^{19}(P\dot{P})^{1/2}$\,G is the magnetic field strength at the pole, assuming that rotational energy losses are dominated by dipolar magnetic torques.
Sources with multiple/variable $\dot{P}$ values in the literature are labelled with a $^*$. For references, see the ATNF pulsar catalog$^2$, the McGill magnetar catalog$^3$, and our online catalog$^1$. We report here in the "method" column a reference for the method used in archival work to identify $\tau_r$, i.e. $^1$: Supernova Remnant (SNR), $^2$: Proper Motion (PM). True age references: A15=\cite{Allen2015}, B18=\cite{Borkowski2018},
B20=\cite{Borkowski2020}, B19=\cite{Braun2019}, C99=\cite{Corbel1999}, F06=\cite{Fesen2006}, L07=\cite{Leahy2007}, K13=\cite{Kothes2013}, K19=\cite{Knies2018}, M09=\cite{Motch2009}, M20=\cite{Mayer2020}, N07=\cite{Ng2007}, P12=\cite{Park2012}, R88=\cite{Roger1988}, S06=\cite{Smith2006}, S11=\cite{Sushch2011}, S13=\cite{Sasaki2013}, S18=\cite{Sasaki2018}, T08=\cite{Tian2008}, T11=\cite{Tetzlaff2011}, T12=\cite{Tetzlaff2012}, T13=\cite{Tendulkar2013},  W97=\cite{Wang1997}, Y04=\cite{YarUyaniker2004}, Z16=\cite{Zhou2016}.
}
\label{tab:timing-1}
\\
\hline
\hline
Source & assoc./nick &  $P$ & $\log(\dot{P})$ & $\log(\dot{E}_{rot})$ & $\log(B_p)$ & $\log(\tau_c)$ & $\log(\tau_r)$ & method & Ref. \\
& & [s] & & [erg/s] & [G] &  [yr] &  [yr] & \\
\hline
        \hline
        \multicolumn{9}{c}{Magnetars} \\
        \hline
        SGR 1627-41         & SNR G337.0-0.1      & 2.59  & -10.7 & 34.6 & 14.6 & 3.3 & 3.7 & SNR$^1$ & C99 \\ 
        1E 2259+586         & SNR CTB109   & 6.98  & -12.3 & 32.1 & 14.1 & 5.4 & 4.0 & SNR & S13 \\ 
        XTE J1810-197       & -           & 5.54  & -11.5 & 32.8 & 14.4 & 4.5 & - \\ 
        SGR 1806-20         & W31 open cluster  & 7.75  & -10.1 & 33.2 & 15.2 & 3.2 & 2.9 & PM$^2$ & T13\\ 
        CXOU J1647-4552     & Westerlund 1 & 10.61 & -12.0 & 31.5 & 14.3 & 5.2 & - \\ 
        SGR J0501+4516      & SNR HB9        & 5.76  & -11.2 & 33.1 & 14.6 & 4.2 & 4.0 & SNR & L07 \\ 
        1E 1547-5408        & SNR G327.24-0.1  & 4.5 & 2.07  & -10.3 & 35.3 & 14.8 & 2.8 & SNR  & \\ 
        SGR J0418+5729      & -           & 9.08  & -14.4 & 29.3 & 13.1 & 7.6 & - \\ 
        SGR J1833-0832      & -           & 7.57  & -11.5 & 32.5 & 14.5 & 4.5 & - \\ 
        Swift J1822.3-1606  & -           & 8.44  & -12.9 & 30.9 & 13.8 & 6.0 & - \\ 
        Swift J1834.9-0846  & -           & 2.48  & -11.1 & 34.3 & 14.5 & 3.7 & - \\ 
        1E 1048.1-5937      & -           & 6.46  & -10.7 & 33.5 & 14.9 & 3.7 & - \\ 
        SGR J1745-2900      & Galactic Center   & 3.76  & -10.5 & 34.3 & 14.8 & 3.3 & - \\ 
        SGR J1935+2154      & FRB         & 3.24  & -10.8 & 34.2 & 14.6 & 3.6 & - \\ 
        1E 1841-045         & SNR Kes73 & 11.79 & -10.4 & 33.0 & 15.1 & 3.7 & 2.8 & SNR & T08 \\ 
        SGR 1900+14         & open cluster & 5.2   & -10.0 & 34.4 & 15.1 & 3.0 & 3.7 & PM & T13 \\ 
        4U 0142+614         & -           & 8.69  & -11.7 & 32.1 & 14.4 & 4.8 & - \\ 
        1RXS J170849.0-400910 & -           & 11.01 & -10.7 & 32.8 & 15.0 & 4.0 & - \\ 
        CXOU J010043.1-721134  & SMC         & 8.02  & -10.7 & 33.1 & 14.9 & 3.8 & - \\ 
        CXOU J171405.7-381031 & SNR CTB37B      & 3.83  & -10.2 & 34.6 & 15.0 & 3.0 & SNR  \\ 
        SGR 0526-66         & SNR N49 - LMC   & 8.05  & -10.4 & 33.5 & 15.0 & 3.5 & SNR & P12 \\ 
        PSR J1119-6127 & SNR  G292.2-0.5 &  0.41 & -11.4 & 36.4 & 13.9 & 3.2 &    3.6-3.9 &      SNR\\
        PSR J1846-0258      & -           & 0.33  & -11.1 & 37.0 & 14.0 & 2.9 & SNR \\ 
        PSR J1622-4950      & -           & 4.33  & -10.8 & 33.9 & 14.7 & 3.6 & - \\ 
        Swift J1818.0-1607  & -           & 1.36  & -10.3 & 35.9 & 14.7 & 2.7 & - \\ 
        SGR J1830-0645      & -           & 10.42 & -11.1 & 32.4 & 14.8 & 4.3 & - \\ 
        3XMM J185246.6+003317     & -           & 11.56 & -12.9 & 30.6 & 13.9 & 6.1 & - \\ 
        Swift J1555.2-5402  & -           & 3.86  & -10.7 & 34.2 & 14.8 & 3.4 & - \\ 
        \hline
        \multicolumn{9}{c}{Central Compact Objects (CCOs)} \\
        \hline
        CXOU J185238.6+0040 & SNR Kes79 & 0.11  & -17.1 & 32.5 & 10.8 & 8.3 & 3.8 & SNR & Z16 \\ 
        1E 1207.4-5209      & SNR G296.5+10.0   & 0.42  & -16.7 & 31.1 & 11.3 & 8.5 & 3.8 & SNR & R88 \\ 
        RX J0822.0-4300     & SNR PuppisA        & 0.11  & -17.0 & 32.3 & 10.8 & 8.4 & 3.6 & SNR & M20\\ 
        1E 161348-5055      & SNR RCW103      & 2.40  & $<$-9.1  & -    & -    & -   & 3.3 & SNR & B19\\ 
      CXOU J085201.4-461753 & SNR Vela Jr       & -     & -     & -    & -    & -   & 3.3 & SNR & A15 \\ 
       CXOU J232327.9+584842 & SNR CasA       & -     & -     & -    & -    & -   & 2.5 & SNR & F06\\ 
            1E 0102.2-7219 & B0102-72.3  & -     & -     & -    & -    & -   & 3.3 \\ 
      2XMM J104608.7-594306 & Homunculus  & -     & -     & -    & -    & -   & 4.0 & SNR & S06 \\ 
     CXOU J160103.1-513353 & G330.2+1.0  & -     & -     & -    & -    & -   & 2.9 & SNR & B18 \\ 
        1WGA J1713.4-3949 & G347.3-0.5     & -     & -     & -    & -    & -   & 3.2 & SNR & W97 \\ 
      XMMU J172054.5-372652 & G350.1-0.3     & -     & -     & -    & -    & -   & 2.8 & SNR & B20 \\ 
      XMMU J173203.3-344518 & HESS J1731-347 & -     & -     & -    & -    & -   & 3.6 & SNR & T08 \\ 
      CXOU J181852.0-150213 & G015.9+0.002   & -     & -     & -    & -    & -   & 3.5 & SNR & S18 \\ 
        \hline
       \multicolumn{9}{c}{X-ray Dim Isolated Neutron Stars (XDINSs)} \\
       \hline
       RXJ0420.0-5022 & - & 3.45 & -13.56 & 31.4 & 13.3 & 6.3 & - & - \\ 
       RXJ1856.5-3754 & - & 7.06 & -13.52 & 30.5 & 13.5 & 6.6 & 5.6 & PM & T11\\ 
       RXJ2143.0+0654 & - & 9.43 & -13.39 & 30.3 & 13.6 & 6.6 & - & - \\ 
       RXJ0720.4-3125 & - & 8.39 & -13.16 & 30.7 & 13.7 & 6.3 & 5.9 & PM & T11 \\ 
       RXJ0806.4-4123 & - & 11.37 & -13.26 & 30.2 & 13.7 & - & - \\ 
       RXJ1308.6+2127 & - & 10.31 & -12.96 & 30.6 & 13.8 & 6.2 & 6.0 & PM & M09 \\ 
       RXJ1605.3+3249 & - & 3.39 & -11.80 & 33.2 & 14.2 & 4.5 & 5.7  & PM & T12 \\ 

       \hline
       \multicolumn{9}{c}{Rotation-Powered Pulsars (RPPs)} \\
       \hline
         PSR J0538+2817 & SNR S147       &  0.14 & -14.4 & 34.7 &  12.2 & 5.8 & $\sim$4.6 &        SNR & N07 \\
               PSR B1055-52 & -              &  0.20 & -14.2 & 34.5 & 12.3 & 5.7 &         -  &          -\\
            PSR J0633+1746 & aka Geminga    &  0.24 & -14.0 & 34.5 & 12.5 & 5.5 &         -  &          -\\
               PSR B1706-44 & -              &  0.10 & -13.0 & 36.5 & 12.8 & 4.2 &         -  &          -\\
              PSR B0833-45 & SNR Vela       &  0.09 & -12.9 & 36.8 & 12.8 & 4.1 &    3.7-4.2 &        SNR & S11 \\
              PSR B0656+14 & SNR Monogem    &  0.39 & -13.3 & 34.6 & 13.0 & 5.0 &  $\sim$4.9 &        SNR & K18 \\
               PSR B2334+61 & SNR G114.3+0.3 & 0.49 & -12.7 & 34.8 & 13.3 & 4.6 &  $\sim$4.0 &        SNR & Y04 \\
           PSR J1740+1000 & -               & 0.15 & -11.7 & 37.4 & 13.6 & 3.1 &         -  &          -\\          
            PSR J0726-2612 & -               &  3.44 & -12.5 & 32.4 & 13.8 & 5.3 &         -  &          -\\
          PSR J1819-1458 &     RRAT        &  4.26 & -12.2 & 32.5 & 14.0 & 5.1 &         -  &          -\\
           PSR J1718-3718 & -               &  3.38 & -11.8 & 33.2 & 14.2 & 4.5 &         -  &          -\\
           1RXS J1412+7922 &         Calvera &  0.06 & -14.5 & 35.8 & 11.9 & 5.5 &         -  &           -\\ 
           PSR J1357-6429 &  HESS J1356-645 &  0.17 & -12.4 & 36.5 & 13.2 & 3.9 &         -  &           -\\
            PSR J1741-2054 & -               &  0.41 & -13.8 & 34.0 & 12.7 & 5.6 &         -  &           -\\
              PSR B1822-09 & -               &  0.77 & -13.3 & 33.7 & 13.1 & 5.4 &         -  &           -\\
            PSR J1836+5925 &    Next Geminga &  0.17 & -14.8 & 34.0 & 12.1 & 6.3 &         -  &           -\\
            PSR J1957+5033 & -               &  0.38 & -14.2 & 33.7 & 12.5 & 5.9 &         -  &           -\\
            PSR J0205+6449 &            3C58 &  0.07 & -12.7 & 37.4 & 12.9 & 3.7 &        2.9 &       SNR & K13\\
\hline
\hline
\end{longtable}

\begin{longtable}{l c c c c c c c c}
\caption{\small Emission properties of the thermally emitting neutron
  stars. Whenever we used multiple black bodies to model the spectra, we report here only the parameters corresponding to the hottest component. $^\dagger$: due to the paucity of counts, spectral fits could not be performed and only upper limits on the thermal luminosity could be posed. $^a$: since the distance of the source is unknown, we fixed it to the arbitrary value of 10 kpc. References for distances:
  A15=\cite{Allen2015}; B03=\cite{Brisken2003}; B08=\cite{Bibby2008}; B14=\cite{Bower2014}; B21=\cite{Bailes2021}; C02=\cite{Clark2002}: C11=\cite{Chang2012}; C75=\cite{Caswell1975}; D03=\cite{Dodson2003}; D06=\cite{Durant2006}; D09=\cite{Davies2009}; D20=\cite{Ding2020}; F03=\cite{Fukui2003}; F07=\cite{Faherty2007}; F89=\cite{Frail1989}; G00=\cite{Giacani2000}; G08=\cite{Gaensler2008}; G99=\cite{Gaensler1999}; H13=\cite{Halpern2013}; H99=\cite{Hurley1999}; K05=\cite{Kaspi2005}; K06=\cite{Keller2006}; K08=\cite{Kumar2008}; K12=\cite{Kothes2012}; K95=\cite{Koribalski1995}; L08=\cite{Leahy2008}; L11=\cite{Lin2011}; M01=\cite{McClureGriffiths2001}; M06=\cite{Macri06}; MG06=\cite{McGowan2006}; M13=\cite{Miller2013}; M14=\cite{Marelli2014}; N07=\cite{Ng2007}; P07=\cite{Posselt2007}; P15=\cite{Posselt2015}; R22=\cite{Rigoselli2022}; R95=\cite{Reynoso1995}; RH95=\cite{Reed1995}; S06=\cite{Smith2006}; S11=\cite{Speagle2011}; S12=\cite{Scholz2012}; T08a=\cite{Tian2008}; T08b=\cite{Tian2008_1731}; TN10=\cite{Tetzlaff2010}; TV10=\cite{Tiengo2010}; T11=\cite{Tetzlaff2011}; T12=\cite{Tian2012}; T22=\cite{Tanashkin2022}; V10=\cite{VanderHorst2010}; W10=\cite{Walter2010}; Z05=\cite{Zou2005}; Z14=\cite{Zhou2014}; Z21=\cite{Zyuzin2021}. 
}
\label{tab:spectral}
\\
\hline
\hline
Source & $d$  & Ref & $kT_{bb}$ & $R_{bb}$ & $\log(L)$ & best fit & $kT_{cool}$ & $\log(L_{cool})$ \\
 & [kpc] & & [eV] & [km] &  [erg/s] & model & [eV] & [erg/s] \\
\hline
  \hline
       \multicolumn{9}{c}{Magnetars} \\       
       \hline
                  SGR 1627-41   & 11.0$\pm$0.3 & H11 & 1000$^{+300}_{-200}$ & $<$0.22 & 33.08 & BB & $<$300 &  $<$34.9 \\
                  1E 2259+586 &  3.2$\pm 0.2$ & K12 & 630$^{+40}_{-50}$ & 1.1$\pm$0.7 & 34.79 & 2BB & $<$120 &  $<$33.4\\
                  XTE J1810-197 & 3.6 & D20 & 240$\pm$30 & 2.0$^{+2.0}_{-1.5}$ & 34.56 &  2BB & - & - \\ 
                  SGR 1806-20   & 8.6 & B08 & 790$\pm$10 & 0.9$^{+0.8}_{-0.6}$ & 35.15 & 2BB &  $<$250 &  $<$34.7 \\
                  CXOU J1647-4552   & 4.0$^{+ 1.5}_{- 1.0}$ & C02 & 530$^{+30}_{-50}$ & 0.5$^{+0.4}_{-0.3}$ & 33.54 & 2BB & $<$120 & $<$33.4\\
                  SGR J0501+4516 & 1.5$^{+ 1.0}_{- 0.5}$ & L11 & 570$\pm$110 & 0.20$\pm$0.04 & 33.11 & RCS & $<$100 &  $<$32.9\\
                  1E 1547-5408   &  4.5$\pm 0.5$ & T10 & 620$^{+650}_{-130}$ & $<$0.9 & 33.36 & 2BB & $<$150 &  $<$33.8\\
                  SGR J0418+5729   & 2.0$\pm$0.3 & V10 & 320$\pm$30 & 0.07$^{+0.07}_{-0.05}$ & 30.90 & BB & $<$40 &  $<$31.5\\
                  SGR J1833-0832$^\dagger$   & 10.0$^a$ & - & - & - & $<$33.90 & - & - & - \\
                  Swift J1822.3-1606   & 1.6$\pm 0.3$ & S12 & 490$^{+200}_{-110}$ & 0.06$^{+0.11}_{-0.05}$ & 32.48 & 2BB & $<$70 & $<$32.5\\
                  Swift J1834.9-0846$^\dagger$  & 4.2$\pm$0.3 & L08 & - & - & $<$32.30 & - & - & - \\
                  1E 1048.1-5937   & 2.7$\pm 1.0$  & D06 & 820$^{+40}_{-35}$ & 0.8$\pm$0.4 & 35.66 & 2BB & $<$100 & $<$33.1 \\
                  SGR J1745-2900   & 8.3$\pm$0.3 & B14 & 710$\pm$25 & 0.6$\pm$0.2 & 34.02 & BB & - & -  \\ 
                  SGR J1935+2154   & 9.0$\pm$1.7 & B21 & 490$^{+50}_{-40}$ & 1.5$^{+1.3}_{-0.9}$ & 33.60 & 2BB \\ 
                  1E 1841-045   & 8.5$^{+1.3}_{-1.0}$ & T08a & 430$^{+30}_{-40}$ & 6.6$^{+4.0}_{-3.0}$ & 35.66 & BB+PL & $<$200 & $<$34.3 \\
                  SGR 1900+14   &  12.5$\pm 1.7$  & D09 & 550$\pm$30 & 2.7$^{+1.5}_{-1.2}$ & 35.15 & 2BB & $<$150 &  $<$33.8 \\    
                  4U 0142+61   & 3.6$\pm 0.5$ & D06 & 560$\pm$20 & 2.5$^{+1.1}_{-1.0}$ & 35.42 & 2BB+PL &  $<$150 &  $<$33.8\\
                  1RXS J170849.0-400910 & 3.8$\pm 0.5$ & D06 & 730$\pm$10 & 0.7$^{+0.8}_{-0.6}$ & 34.86 &  2BB+PL &  $<$130 &  $<$33.6 \\
                  CXOU J010043.1-721134 & 60.6$\pm 3.8$ & K06 & 380$\pm$30 & 9.0$^{+5.0}_{-4.0}$ & 35.29 &  BB &     -  &      -\\
                  CXOU J171405.7-381031 & 13.2$\pm 0.2$ & T12 & 600$^{+70}_{-40}$ & 1.3$\pm$1.0 & 34.44 &  BB+PL &  $<$180 &  $<$34.1 \\
                  SGR 0526-66 & 49.7 & M06 & 490$^{+70}_{-60}$ & 3.9$^{+3.7}_{-2.7}$ & 35.05 & BB+PL &  $<$200 &  $<$34.3\\
                  PSR J1119-6127      & 8.4$\pm 0.4$ & C04  &  300$^{+40}_{-30}$ & 1.1$\pm$0.02          & 33.3 & BB     & $<$100 & $<$33.2  \T\B \\
                  PSR J1846-0258  & 6.0 & K08 & 970$\pm$30 & 0.5$\pm$0.2 & 34.51 & BB+PL & - & - \\ 
                  PSR J1622-4950$^\dagger$  & 9.0 & & - & - & $<$32.64 & - & - & - \\ 
                  Swift J1818.0-1607$^\dagger$  & 4.8 & & - & - & $<$33.81 & - & - & - \\ 
                  SGR J1830-0645$^{\dagger}$  & 10$^a$ & - & - & - & $<$34.30 & - & - & - \\ 
                  3XMM J185246.6+003317$^\dagger$  & 7.1$\pm$0.7 & Z14 & - & - & $<$30.78 & - & - & - \\ 
                  Swift J1555.2-5402$^\dagger$ & 5 & & - & - & $<$35.17 & - & - & - \\         
        \hline
       \multicolumn{9}{c}{Central Compact Objects (CCOs)} \\       
       \hline
        CXOU J185238.6+004020 &                   7.1 & F89 & 450$\pm$20 & 0.8$\pm$0.3 &   33.48 &      BB &  - &  - \\
        1E 1207.4-5209 &  2.1$^{+ 1.8}_{- 0.8}$ & G00 & 260$\pm$5 &  1.5$\pm$0.5 &   33.43 &    2BB &  - &  - \\ 
         RX J0822-4300 &  2.2$\pm 0.3$ & R95 & 520$^{+60}_{-40}$ &  1.7$^{+0.9}_{-0.7}$ &   33.60 &      2BB &  - &  - \\
         1E 161348-5055 & 3.3 & C75 & 530$\pm$20 & 0.6$\pm$0.3 & 33.86 & 2BB & - & - \\
         CXOU J085201.4-461753 & 0.7$\pm$0.3 & A15 & 400$\pm$9 & 0.24$\pm$0.08 & 32.26 & BB & - & -\\
  CXOU J232327.9+584842 &  3.4$^{+ 0.3}_{- 0.1}$ & RH95 & 400$^{+9}_{-8}$ &  0.8$\pm$0.2 &   33.32 &  BB+PL &  - &  - \\
        2XMM J104608.7-594306 & 2.4 & S06 & 138$\pm$4 & 2.8$^{+1.5}_{-1.3}$ & 32.57 & BB & - & -\\ 
     CXOU J160103.1-513353 & 5.0-11.0 & M01 & 450$\pm$30 & 0.6$^{+0.5}_{-0.3}$ & 32.28 & BB & - & - \\ 
        1WGA J1713.4-3949 & 1.3 & F03 & 540$^{+50}_{-30}$ & 0.2$\pm$0.1 & 33.06 & 2BB & - & -\\ 
      XMMU J172054.5-372652 & 4.5-10.7 & G08 & 520$\pm$10 & 0.9$^{+0.4}_{-0.3}$ & 33.89 & BB & - & - \\ 
      XMMU J173203.3-344518 & 3.2 & T08b & 500$\pm$4 & 0.9$\pm$0.2 & 33.86 & BB & - & -\\ 
      CXOU J181852.0-150213 & 10$^a$ & - & 440$^{+30}_{-20}$ & 0.6$^{+0.4}_{-0.3}$ & 33.32 & BB & - & - \\ 
         \hline
       \multicolumn{9}{c}{X-ray Dim Isolated Neutron Stars (XDINSs)} \\       
       \hline
        RX J0420.0-5022 &                   0.34 & P07 & 49$_{-4}^{+2}$ & 4.5$^{+2.5}_{-1.0}$  &     31.15 &          BB &     -  &      -\\
       RX J1856.5-3754 &          0.12$\pm0.01$ & W10 & 159$^{+29}_{-21}$  &  0.04$\pm$0.02 &    31.68 &          2BB &     -  &      -\\
       RX J2143.0+0654 &                   0.43 & P07 & 108$\pm$1 &  2.42$\pm$0.05 &     32.00 &          BB &     -  &      -\\
       RX J0720.4-3125 &  0.29$^{+0.03}_{-0.02}$ & P07 & 81.9$\pm$0.02 &   6.27$\pm$0.04 &      32.34 &          BB &     -  &      -\\
       RX J0806.4-4123 &                   0.25 & P07 & 91$\pm$1 &  1.95$\pm$0.03 &    31.49 &    BB &     -  &      -\\
       RX J1308.6+2127 &                  0.50 & TN10 & 87$\pm$1 &  2.6$\pm$0.2 &     31.34 &         BB &     -  &      -\\
       RX J1605.3+3249 &                  0.10 & T11  & 146$^{+36}_{-15}$ &  0.8$^{+0.5}_{-0.3}$ &    33.30 &      2BB &     -  &      -\\
       \hline
       \multicolumn{9}{c}{Rotation-Powered Pulsars (RPPs)} \\
       \hline
PSR J0538+2817      & 1.5$^{+0.4}_{-0.3}$           & N07 &  190$\pm$8         & 1.5$^{+0.3}_{-0.2}$   & 32.6 & BB     & $<$100  & $<$33.3  \T\B \\
PSR B1055-52        & 0.73$\pm0.15$          & P15 & 65$^{+4}_{-5}$    & 4.0$^{+1.4}_{-0.9}$   & 31.9 & 2BB+PL & $<$40 & $<$31.8  \T\B \\
PSR J0633+1746      & 0.25$^{+0.22}_{-0.08}$ & F07 & 40$\pm$2          & 7.4$^{+2.4}_{-0.1}$   & 31.6 & BB+PL  & $<$30 & $<$31.3  \T\B \\
PSR B1706-44        & 2.6$^{+ 0.5}_{- 0.6}$  & K95 & 170$\pm$20        & 1.7$^{+1.3}_{-0.5}$   & 32.5 & BB+PL  & $<$90 & $<$33.1  \T\B \\
PSR B0833-45        & 0.28$\pm0.02$          & D03 & 70$\pm$2          & 10.9$^{+1.7}_{-1.2}$  & 32.7 & 2BB+PL & $<$40 & $<$31.7  \T\B \\
PSR B0656+14        & 0.28$\pm0.03$          & B03 & 100$\pm$4         & 1.4$^{0.3}_{-0.2}$    & 31.4 & BB+PL  & $<$50 & $<$32.0  \T\B \\
PSR B2334+61        & 3.1$^{+ 0.2}_{- 2.4}$  & MG06 & 160$^{+50}_{-30}$ & 0.9$^{+3.3}_{-0.5}$   & 31.9 & BB     & $<$60 & $<$32.3  \T\B \\
PSR J1740+1000      & 1.4                    & R22 & 160$^{+19}_{-14}$ & 0.5$^{+0.3}_{-0.2}$   & 32.3 & 2BB    & $<$40 & $<$31.8  \T\B \\
PSR J0726-2612      & $<$1.0                 & S11 & 120$\pm$3         & 2.9$^{+0.4}_{-0.3}$   & 32.4 & BB     & $<$60 & $<$32.3  \T\B \\
PSR J1819-1458      & 3.6$\pm$0.9            & M13  & 130$\pm$6         & 12.7$^{+4.0}_{-2.3}$  & 33.7 & BB     & - & -  \T\B \\
PSR J1718-3718      & 4.5$^{+ 5.5}_{- 0.0}$  & K05 & 190$^{+30}_{-20}$ & 1.7$^{+1.7}_{-0.9}$   & 32.7 & BB     & - & -  \T\B \\
1RXS J1412+7922     & 0.3                    & H13 & 230$^{+20}_{-14}$ & 0.14$\pm$0.03         & 31.1 & 2BB    & - & - \T\B \\
PSR J1357-6429      & 2.5                    & C11 & 360$\pm$20        & 0.14$\pm$0.04         & 32.2 & BB     & - & - \T\B \\
PSR J1741-2054     & 0.8                    & M14 & 60$^{+5}_{-4}$    & 15.3$^{+11.0}_{-5.0}$ & 32.5 & BB+PL  & $<$60 &  $<$32.4  \T\B \\
PSR B1822-09        & 0.9                    & Z05 & 170$^{+20}_{-30}$ & 1.2$^{+1.9}_{-0.2}$   & 30.4 & BB     & - & -  \T\B \\
PSR J1957+5033      & 0.1-1.0                 & Z21 & 50$\pm$10         & 7.9$^{+30.3}_{-3}$    & 30.3 & BB+PL  & - & -  \T\B \\
PSR J0205+6449   & 3.2                    & K13 & 160$\pm$7         & 2.1$^{0.3}_{-0.2}$    & 32.5 & BB+PL  & - & - \T\B \\
PSR J0554+3107  & 4.5 & T22 & 140$\pm$90 & 1.9$^{+4.9}_{-1.8}$ & 33.3 & BB+PL & - & - \\
\hline
\hline
\end{longtable}

\endgroup

\clearemptydoublepage
\let\textcircled=\pgftextcircled
\chapter{Numerical Performance of \emph{MATINS}}
\label{appendix: numerical performance}

\initial{I}n this appendix, we present a comprehensive evaluation of the numerical performance of the \emph{MATINS} code. Our analysis centers around two key aspects. In both aspects, we consider a multipolar magnetic field topology featuring an initial poloidal dipolar field of $B_{pol}(t=0) = 10^{14}$\,G at the star's surface, and a maximum toroidal magnetic field of $B_{tor}(t=0) \sim 10^{15}$\,G, with a total average field in the system of $B_{avg}(t=0) = 10^{15}$\,G. The temperature remains fixed at $T=3\times 10^{8}$\,K throughout the entire evolution. The maximum magnetic Reynolds number at $t=0$ is approximately $R_m^{max}\sim 250$, while the average magnetic Reynolds number is around $R_m^{avg}\sim 80$.

Firstly, we delve into assessing the scalability of the \emph{MATINS} code using a specified number of processors and varying resolutions of the magnetic grid. This examination is illustrated in Table~\ref{tab: scalability of MATINS}, with the simulation extending over a duration of $100$\,kyr. Notably, the scalability of \emph{MATINS} aligns with a number of processors that are multiples of 6, mirroring the inherent structure of the cubed-sphere, which encompasses 6 distinct patches.

Additionally, we investigate the evaluation of the \emph{MATINS} code's numerical efficacy in relation to different Runge-Kutta time advancement methods. Referring to Table~\ref{tab: runge kutta scalability}, we explore various Runge-Kutta methods, specifically RK2, RK3, RK4, and RK6, while modifying the Courant number for each. Our exploration spans a Courant number range from $k_c=1$ to $k_c=20$, contingent upon the simulation's stability. In this assessment, we evolve the simulation until the code crashes because of numerical instabilities. This outcome is anticipated due to the extreme magnetic field intensity and the magnetic timestep utilized in this test. Our findings underscore that the fourth-order Runge-Kutta method emerges as optimal in terms of both simulation speed and numerical stability. This advantage is particularly pronounced at a Courant number of $k_c=5$, which delivers faster performance when compared to lower Courant number.

 \begin{table}
 \centering
  \caption{Numerical performance of the \emph{MATINS} 3D magnetic code, investigating various grid resolutions and processor counts over a $100$\,kyr of evolution. The initial Courant number is set at $k_c=0.01$ for this test. ``Number of Processors'' corresponds to the count of processors employed in a specific simulation. ``Number of Iterations'' indicates the count of magnetic iterations performed to achieve a $100$\,kyr evolution time. ``Simulation Time'' represents the duration required for execution on a desktop featuring an Intel Core i9-10900 processor ($2.80$\,GHz). ``Time per Iteration'' reflects the time needed for an individual magnetic iteration.
  }
  \label{tab: scalability of MATINS}
 \begin{tabular}{|l|c|c|c|c|} 
  \hline
Magnetic  & Number of    & Number of  & Simulation &  Time per\\
grid &Processors  & Iterations (for $100$\,kyr)  & Time&   Iteration  \\
 &  &  & [day(s)] &   [s]\\
 \hline
 \hline
$40\times 43\times 43\times 6$ &1  &2538000 & 15.8 & 0.538\\
$40\times 43\times 43\times 6$& 2 & 2538000& 10.2 &0.347\\
$40\times 43\times 43\times 6$& 3 &2538000 &8.28 & 0.282\\
$40\times 43\times 43\times 6$& 4 & 2538000&8.15 & 0.277\\
$40\times 43\times 43\times 6$& 6 &2538000 &6.93 & 0.236 \\
$40\times 43\times 43\times 6$& 8 & 2538000&6.91 &0.2352\\
$40\times 43\times 43\times 6$& 10 &2538000 &6.89 & 0.2348\\
  \hline
$30\times 31\times 31\times 6$&1 &1430000 &3.97 & 0.240 \\
$30\times 31\times 31\times 6$& 2 &  1430000 & 2.60 & 0.157\\
$30\times 31\times 31\times 6$& 3 &  1430000 & 2.22 & 0.134\\
$30\times 31\times 31\times 6$& 4&  1430000 & 2.05  & 0.124\\
$30\times 31\times 31\times 6$& 6 &  1430000 & 1.82 & 0.110\\
$30\times 31\times 31\times 6$& 8 &  1430000 & 1.84 & 0.111\\
$30\times 31\times 31\times 6$& 10&  1430000 & 1.90  & 0.115\\
  \hline
$20\times 23\times 23\times 6$ &1 & 636000 & 0.87 &0.118\\
$20\times 23\times 23\times 6$& 2 &  636000 & 0.57 &0.077\\
$20\times 23\times 23\times 6$& 3 &  636000 &0.49 & 0.066\\
$20\times 23\times 23\times 6$& 4 &  636000 &0.42 & 0.063\\
$20\times 23\times 23\times 6$& 6 &  636000 &0.42 & 0.057\\
$20\times 23\times 23\times 6$& 8 & 636000 &0.43 & 0.058 \\
$20\times 23\times 23\times 6$& 10 & 636000 &0.44  &  0.060\\
\hline
 \end{tabular}
 \end{table}

\begin{table}
\centering
 \caption{Numerical assessment of the 3D magnetic code employing various Runge-Kutta time advancement methods, e.g., RK2, RK3, RK4, and RK6. The magnetic grid resolution is set to $30\times31\times31\times6$, and the simulation is conducted using a single processor. Here, the heading "Evolution Time" corresponds to the neutron star's age when the simulation ends. For additional information regarding the remaining headings, please consult Table~\ref{tab: scalability of MATINS}.
 }
 \label{tab: runge kutta scalability}
 \begin{tabular}{|l|c|c|c|c|c|}
  \hline
Runge-Kutta  & Courant   & Evolution  & Number of  & Simulation & Time per\\
Method & Number $k_c$  & Time & Iterations &Time & Iteration\\
 &    & [yrs]& &[s] & [s] \\
\hline
\hline
RK2 & 1.0 & 1755 &  12298   &  5360 & 0.44 \\
RK2 & 2.0   & 1763 &  6372  & 2788 & 0.44\\ 
RK2 & 3.0  & 1843 &  4396 & 1906  & 0.43 \\ 
RK2 & 4.0  &  1056 & 1770 & 829  &  0.47\\ 
RK2 & 6.0  & 787 & 936 & 463 & 0.49\\
\hline
RK3 & 1.0 &  1752 & 12298 &  7414 &  0.60 \\ 
RK3 & 2.0   & 1762 &     6372   & 4044 & 0.63  \\ 
RK3 & 3.0  & 1767 &  4396   &  2784 & 0.63\\ 
RK3 & 4.0  & 1786 &  3186  &2098 & 0.66\\
RK3 & 6.0 & 1004 &   1170  &811 & 0.69\\
RK3 & 10.0  & 740 & 517 & 404 & 0.78\\
\hline
RK4 & 1.0  & 1751 &  12298  & 9917 & 0.81 \\ 
RK4 & 2.0   & 1754 & 6372  &5188 & 0.81\\ 
RK4 & 3.0   &1760 &   4396 &  3796 & 0.86\\ 
RK4 & 4.0  & 1765 &  3304 & 2849 & 0.86\\
RK4 & 6.0  & 1769 &   2184  & 1859 &0.85 \\
RK4 & 8.0  &  1783 &   1652 & 1434 &0.87 \\
RK4 & 10.0  & 1915 &  1457 &1325 & 0.91 \\
RK4 & 12.0   &7072 & 6123 & 4796 & 0.78\\
RK4 & 15.0   &3000 & 1674 &1590 &0.95\\
RK4 & 18.0  & 1870  &  832& 826 & 0.99\\
RK4 & 20.0  &  1129 & 391&369 & 0.94\\
\hline
RK6 & 1.0  &  1752 & 12298 & 17472 & 1.42 \\ 
RK6 & 2.0   & 1753 &  6372 & 8813 & 1.38\\
RK6 & 3.0  & 1755  &   4396 & 6751 & 1.54\\ 
RK6 & 4.0  &1755 & 3304 & 4845 & 1.47\\
RK6 & 6.0  &  1755 & 2184& 3196 & 1.46\\
RK6 & 8.0  & 1757 & 1652  & 2470 & 1.50\\
RK6 & 10.0  & 1762 &1316&1985 & 1.51\\
RK6 & 12.0  &1285 & 780   & 1366 & 1.75\\
RK6 & 15.0  &988 & 465 & 750 & 1.61\\
\hline
 \end{tabular}
\end{table}

\clearemptydoublepage
\bibliographystyle{apalike} 
\bibliography{thesisbiblio}
\clearemptydoublepage
%
%
\end{document}